\documentclass[a4paper,11pt]{book}
\usepackage[utf8]{inputenc}
\usepackage{{amsmath,amsfonts,latexsym, amstext, amssymb, amsthm}}
\usepackage{xspace} 
\usepackage{slashed}
\usepackage[bf]{caption}
\usepackage{subcaption}
\usepackage{comment}
\usepackage{cite}
\usepackage[pdfa=true]{hyperref}
\usepackage{dsfont}
\usepackage[english]{babel}
\usepackage{placeins}
\usepackage[dvips]{graphicx}
\usepackage[dvips]{color}
\usepackage{hyphenat}
\usepackage{cancel}
\usepackage{pstricks}
\usepackage{multirow,graphics}
\usepackage{titlesec}
\usepackage{import}
\usepackage{transparent}
\usepackage{tensor}
\hypersetup{
	    pdfstartpage=1, 
	    hidelinks,
	    pdfpagemode=UseOutlines,
	    bookmarks=true,
	    bookmarksnumbered=true,
	    bookmarksopen=false,
	    pdflang=English,
            pdfauthor={Laura Koster},
            pdftitle={Form factors and correlation functions in N=4 super Yang-Mills theory from twistor space},
            pdfkeywords={Super Yang-Mills theory, Form factors, On-shell methods, Twistors, Anomalous dimensions, Integrability}
}

\allowdisplaybreaks

\makeatletter

\divide\@tempdima\baselineskip
\@tempcnta=\@tempdima
\makeatother

\def\ps@plain{\let\@mkboth\@gobbletwo%
     \let\@oddfoot\@empty\def\@oddhead{\reset@font\hfil\thepage}%
     \let\@evenfoot\@empty\def\@evenhead{\reset@font\thepage\hfil}}

\usepackage{fancyhdr} 
\usepackage{amsfonts,amssymb,amsmath,amsthm,mathtools,hyperref,colonequals}
\usepackage{caption}
\usepackage[utf8]{inputenc}
\usepackage{tabularx}
\usepackage{wasysym}
\usepackage{bbm}
\usepackage{mathrsfs}
\usepackage{color}
\usepackage{framed}
\definecolor{shadecolor}{rgb}{1,0.9,0.7}
\usepackage{comment}
 \excludecomment{paper}
 \includecomment{forus}
 \excludecomment{additionalappendix}

\begin{forus}
\usepackage[colorinlistoftodos,textsize=tiny]{todonotes}
\end{forus}
\begin{paper}
\usepackage[colorinlistoftodos,textsize=tiny,disable]{todonotes}
\end{paper}
\usepackage{graphicx}
\usepackage{shuffle}
\usepackage{url}
\usepackage{verbatim}

\numberwithin{equation}{section}
\usepackage{ytableau}
\begin{forus}
\usepackage{rotating}
\end{forus}

\theoremstyle{definition}
\newcommand{\beqa}{\begin{eqnarray}}
\newcommand{\eeqa}{\end{eqnarray}}
\newcommand{\beq}{\begin{equation}}
\newcommand{\eeq}{\end{equation}}

\newcommand{\fP}{\mathsf{P}}
\newcommand{\fp}{\mathsf{p}}
\newcommand{\ft}{\mathsf{t}}
\newcommand{\ftau}{\mathsf{\tau}}
\newcommand{\fTau}{\mathsf{T}}
\newcommand{\fq}{\mathsf{q}}

\newcommand{\fs}{\mathsf{s}}

\newcommand{\calF}{\mathscr{F}}

\newcommand{\calS}{\mathcal{S}}

\newcommand{\calA}{\mathcal{A}}

\newcommand{\calO}{\mathcal{O}}
\newcommand{\calI}{\mathcal{I}}\newcommand{\calZ}{\mathcal{Z}}
\newcommand{\calN}{\mathcal{N}}
\newcommand{\calW}{\mathcal{W}}

\newcommand{\calZt}{\widetilde{\mathcal{Z}}}
\DeclareMathOperator{\Tr}{Tr}

\newcommand{\ver}[1]{\ensuremath{\mathbf{V}(#1)}}
\newcommand{\verop}[3]{\ensuremath{\mathbf{W}_{#1}(#2;#3)}}

\newcommand{\melement}[3]{\ensuremath{\left< #1\left| \, #2\,\right| #3\right>}}

\newcommand{\vac}[1]{\ensuremath{\left< \, #1\, \right>}}

\newcommand{\abra}[2]{\ensuremath{\langle #1 #2\rangle}}
\newcommand{\sbra}[2]{\ensuremath{[ #1 #2]}}
\newcommand{\bradot}[3]{\ensuremath{\langle #1|\, #2|#3]}}

\newcommand{\dd}{\mathrm{d}}
\newcommand{\DD}{\mathrm{D}}
\newcommand{\AAA}{\mathcal{A}}
\newcommand{\WWW}{\mathcal{W}}

\newcommand{\PPP}{\mathbf{F}}
\newcommand{\la}{\lambda}

\newcommand{\m}{\textbf{m}}\newcommand{\bm}{\bar{\textbf{m}}}
\newcommand{\n}{\textbf{n}}

\newcommand{\nalpha}{i (4\pi)^2}
\newcommand{\xdot}{\boldsymbol{\cdot}}
\newcommand{\notA}{B}

\newcommand{\e}{\operatorname{e}}
\newcommand{\ri}{i}
\newcommand{\rj}{j}

\newcommand{\eqndot}{\, .}
\newcommand{\eqncom}{\, ,}
\newcommand{\eqnsem}{\, ;}
\newcommand{\YM}{{\mathrm{\scriptscriptstyle YM}}}
\DeclareMathOperator{\phaneq}{\phantom{{}=}}

\newcommand{\bbA}{\textbf{A}}
\newcommand{\bbB}{\textbf{B}}
\newcommand{\bbC}{\textbf{C}}
\newcommand{\bbD}{\textbf{D}}

\numberwithin{equation}{section}

\usepackage{feynmp}
\makeatletter

\define@boolkey{FDiagram}{schannel}[true]{}
\define@boolkey{FDiagram}{tchannel}[true]{}
\define@boolkey{FDiagram}{xchannel}[true]{}
\define@boolkey{FDiagram}{leftSE}[true]{}
\define@boolkey{FDiagram}{rightSE}[true]{}
\define@boolkey{FDiagram}{long}[true]{}
\define@boolkey{FDiagram}{longup}[true]{}
\define@boolkey{FDiagram}{oldlabels}[true]{}
\define@key{FDiagram}{styleleftbottom}%
 {\def\FDiagramstyleleftbottom{#1}}
\define@key{FDiagram}{stylerightbottom}%
 {\def\FDiagramstylerightbottom{#1}}
\define@key{FDiagram}{stylelefttop}%
 {\def\FDiagramstylelefttop{#1}}
\define@key{FDiagram}{stylerighttop}%
 {\def\FDiagramstylerighttop{#1}}
\define@key{FDiagram}{stylemid}%
 {\def\FDiagramstylemid{#1}}
\define@key{FDiagram}{labelleftbottom}%
 {\def\FDiagramlabelleftbottom{#1}}
\define@key{FDiagram}{labelrightbottom}%
 {\def\FDiagramlabelrightbottom{#1}}
\define@key{FDiagram}{labellefttop}%
 {\def\FDiagramlabellefttop{#1}}
\define@key{FDiagram}{labelrighttop}%
 {\def\FDiagramlabelrighttop{#1}}
\define@key{FDiagram}{labelmid}%
 {\def\FDiagramlabelmid{#1}}
\presetkeys{FDiagram}{%
schannel=false,%
tchannel=false,%
xchannel=false,%
leftSE=false,%
rightSE=false,%
long=false,%
longup=false,%
oldlabels=false,%
labelleftbottom=$\phantom{0}$,%
labelrightbottom=$\phantom{0}$,%
labellefttop=$\phantom{0}$,%
labelrighttop=$\phantom{0}$,%
labelmid=$\phantom{0}$,%
styleleftbottom=plain,%
styleleftbottom=plain,%
stylerightbottom=plain,%
stylelefttop=plain,%
stylerighttop=plain,%
stylemid=plain}{}
\newcommand*\FDiagram[6][]{%
 \setkeys{FDiagram}{#1}%
\settoheight{\eqoff}{$\times$}%
\setlength{\eqoff}{0.5\eqoff}%
\addtolength{\eqoff}{-12.0\unitlength}%
\raisebox{\eqoff}{%
\fmfframe(2,2)(2,2){%
\begin{fmfchar*}(12,20)
\fmfbottom{vb1,vbb1,v1,vb2,vbb2,vb3,vb4,v2,vbb5,vb5}
\fmftop{vt1,vtt1,v3,vt2,vtt2,vt3,vt4,v4,vtt5,vt5}
\ifKV@FDiagram@schannel
\fmf{#2,left=0.3,tension=1}{v1,vc1}
\fmf{#3,right=0.3,tension=1}{v2,vc1}
\fmf{#4,tension=2}{vc1,vc2}
\fmf{#5,left=0.3,foreground=(0.65,,0.65,,0.65)}{vc2,v3}
\fmf{#6,right=0.3,foreground=(0.65,,0.65,,0.65)}{vc2,v4}
\else\fi
\ifKV@FDiagram@tchannel
\fmf{#2,left=0,tension=1}{v1,vc1}
\fmf{#3,right=0,tension=1}{v2,vc2}
\fmf{#4,tension=0}{vc1,vc2}
\fmf{#5,foreground=(0.65,,0.65,,0.65)}{vc1,v3}
\fmf{#6,foreground=(0.65,,0.65,,0.65)}{vc2,v4}
\else\fi
\ifKV@FDiagram@xchannel
\fmf{#2,left=0.25,tension=1}{v1,vc1}
\fmf{#3,right=0.25,tension=1}{v2,vc1}
\fmf{#5,left=0.25,foreground=(0.65,,0.65,,0.65)}{vc1,v3}
\fmf{#6,right=0.25,foreground=(0.65,,0.65,,0.65)}{vc1,v4}
\else\fi
\ifKV@FDiagram@leftSE
\fmf{#2,left=0,tension=1}{v1,vc1}
\fmf{#2,right=0,tension=1,foreground=(0.65,,0.65,,0.65)}{vc1,v3}
\fmf{#3,foreground=(0.65,,0.65,,0.65)}{v2,v4}
\fmfv{decor.shape=circle,decor.filled=full,decor.size=15}{vc1}
\else\fi
\ifKV@FDiagram@rightSE
\fmf{#3,left=0,tension=1}{v2,vc1}
\fmf{#3,right=0,tension=1,foreground=(0.65,,0.65,,0.65)}{vc1,v4}
\fmf{#2,foreground=(0.65,,0.65,,0.65)}{v1,v3}
\fmfv{decor.shape=circle,decor.filled=full,decor.size=15}{vc1}
\else\fi
\fmffreeze
\fmfposition
\fmfipath{p[]}
\fmfipair{vm[]}
\fmfcmd{pair verta, vertb, vertc, vertd, vertca, vertcb; verta = vloc(__v1); vertb = vloc(__v2); vertc = vloc(__v3); vertd = vloc(__v4); vertca = vloc(__vc1); vertcb = vloc(__vc2);}
\ifKV@FDiagram@oldlabels
\fmfiv{label=\FDiagramlabelleftbottom,l.a=+120,l.dist=0.07w}{verta}
\fmfiv{label=\FDiagramlabelrightbottom,l.a=+60,l.dist=0.07w}{vertb}
\ifKV@FDiagram@tchannel
\fmfiv{label=\FDiagramlabelmid,l.a=30,l.dist=0.18w}{vertca}
\else
\fmfiv{label=\FDiagramlabelmid,l.a=60,l.dist=0.20w}{vertca}
\fi
\fmfiv{label=\FDiagramlabellefttop,l.a=-120,l.dist=0.07w}{vertc}
\fmfiv{label=\FDiagramlabelrighttop,l.a=-60,l.dist=0.07w}{vertd}
\else
\fmfiv{label=\FDiagramlabelleftbottom,l.a=+120,l.dist=0.07w}{verta}
\fmfiv{label=\FDiagramlabelrightbottom,l.a=+60,l.dist=0.07w}{vertb}
\ifKV@FDiagram@tchannel
\fmfiv{label=\FDiagramlabellefttop,l.a=+120,l.dist=0.07w}{vertca}
\fmfiv{label=\FDiagramlabelrighttop,l.a=+60,l.dist=0.07w}{vertcb}
\else
\fi
\ifKV@FDiagram@schannel
\fmfiv{label=\FDiagramlabellefttop,l.a=+165,l.dist=0.22w}{vertcb}
\fmfiv{label=\FDiagramlabelrighttop,l.a=+15,l.dist=0.22w}{vertcb}
\else
\fi
\ifKV@FDiagram@xchannel
\fmfiv{label=\FDiagramlabellefttop,l.a=+158,l.dist=0.20w}{vertca+(0,0.025h)}
\fmfiv{label=\FDiagramlabelrighttop,l.a=+22,l.dist=0.20w}{vertca+(0,0.025h)}
\else
\fi
\ifKV@FDiagram@leftSE
\fmfiv{label=\FDiagramlabellefttop,l.a=+120,l.dist=0.07w}{verta+(0,0.6125h)}
\else
\fi
\ifKV@FDiagram@rightSE
\fmfiv{label=\FDiagramlabelrighttop,l.a=+60,l.dist=0.07w}{vertb+(0,0.6125h)}
\else
\fi
\fi
\fmfdraw
\ifKV@FDiagram@long
\fmf{plain,width=1mm}{vb1,vb5}
\else 
\fmf{plain,width=1mm}{v1,v2}
\fi
\ifKV@FDiagram@longup
\fmf{plain,width=1mm,foreground=(0.65,,0.65,,0.65)}{vt1,vt5}
\fi
\end{fmfchar*}%
}}%
}
\makeatother

\makeatletter
\DeclareRobustCommand*{\bfseries}{%
  \not@math@alphabet\bfseries\mathbf
  \fontseries\bfdefault\selectfont
  \boldmath
}

\makeatother

\title{Off-shell quantities from twistor space}
\author{Laura Koster}

\begin{document}
\nocite{*}
\begin{fmffile}{diagrams}
\fmfcmd{%
thin := 1pt; 
thick := 2thin;
arrow_len := 3mm;
arrow_ang := 15;
curly_len := 3mm;
dash_len :=0.3; 
dot_len := 0.75mm; 
wiggly_len := 2mm; 
wiggly_slope := 60;
zigzag_len := 2mm;
zigzag_width := 2thick;
decor_size := 5mm;
dot_size := 2thick;
}
\fmfcmd{%
marksize=7mm;
def draw_cut(expr p,a) =
  begingroup
    save t,tip,dma,dmb; pair tip,dma,dmb;
    t=arctime a of p;
    tip =marksize*unitvector direction t of p;
    dma =marksize*unitvector direction t of p rotated -90;
    dmb =marksize*unitvector direction t of p rotated 90;
    linejoin:=beveled;
    drawoptions(dashed dashpattern(on 3bp off 3bp on 3bp));
    draw ((-.5dma.. -.5dmb) shifted point t of p);
    drawoptions();
  endgroup
enddef;
style_def phantom_cut expr p =
    save amid;
    amid=.5*arclength p;
    draw_cut(p, amid);
    draw p;
enddef;
}
\fmfcmd{%
smallmarksize=4mm;
def draw_smallcut(expr p,a) =
  begingroup
    save t,tip,dma,dmb; pair tip,dma,dmb;
    t=arctime a of p;
    tip =smallmarksize*unitvector direction t of p;
    dma =smallmarksize*unitvector direction t of p rotated -90;
    dmb =smallmarksize*unitvector direction t of p rotated 90;
    linejoin:=beveled;
    drawoptions(dashed dashpattern(on 2bp off 2bp on 2bp) withcolor red);
    draw ((-.5dma.. -.5dmb) shifted point t of p);
    drawoptions();
  endgroup
enddef;
style_def phantom_smallcut expr p =
    save amid;
    amid=.5*arclength p;
    draw_smallcut(p, amid);
    draw p;
enddef;
}

\fmfcmd{%
style_def plain_ar expr p =
  cdraw p;
  shrink (0.6);
  cfill (arrow p);
  endshrink;
enddef;
style_def plain_rar expr p =
  cdraw p; 
  shrink (0.6);
  cfill (arrow reverse(p));
  endshrink;
enddef;
style_def dashes_ar expr p =
  draw_dashes p;
  shrink (0.6);
  cfill (arrow p);
  endshrink;
enddef;
style_def dashes_rar expr p =
  draw_dashes p;
  shrink (0.6);
  cfill (arrow reverse(p));
  endshrink;
enddef;
style_def dots_ar expr p =
  draw_dots p;
  shrink (0.6);
  cfill (arrow p);
  endshrink;
enddef;
style_def dots_rar expr p =
  draw_dots p;
  shrink (0.6);
  cfill (arrow reverse(p));
  endshrink;
enddef;
}

\fmfcmd{%
style_def phantom_cross expr p =
    save amid,ang;
    amid=.5*length p;
    ang= angle direction amid of p;
    draw ((polycross 4) scaled 8 rotated ang) shifted point amid of p;
enddef;
}

\fmfcmd{%
style_def plain_sarrow expr p =
  cdraw p;
  shrink (0.55); 
  cfill (arrow p);
  endshrink;
enddef;
style_def dashes_sarrow expr p =
  draw_dashes p;
  shrink (0.55);
  cfill (arrow p);
  endshrink;
enddef;
style_def dots_sarrow expr p =
  draw_dots p;
  shrink (0.55);
  cfill (arrow p);
  endshrink;
enddef;
style_def plain_srarrow expr p =
  cdraw p;
  shrink (0.55);
  cfill (arrow (reverse p));
  endshrink;
enddef;
style_def dashes_srarrow expr p =
  draw_dashes p;
  shrink (0.55);
  cfill (arrow (reverse p));
  endshrink;
enddef;
style_def dots_srarrow expr p =
  draw_dots p;
  shrink (0.55);
  cfill (arrow (reverse p));
  endshrink;
enddef;
}

\begingroup\parindent0pt
\vspace*{4em}
\centering
\begingroup\LARGE
\bf Form factors and correlation functions in \\ $\mathcal{N}=4$ super Yang-Mills theory from twistor space
\par\endgroup
\vspace{2.5em}
D i s s e r t a t i o n 
 \vspace{\baselineskip}

zur Erlangung des akademischen Grades 
 \vspace{\baselineskip}

d o c t o r \phantom{n} r e r u m \phantom{n} n a t u r a l i u m 

(Dr. rer. nat.) 
 \vspace{\baselineskip}

im Fach: Physik
 \vspace{\baselineskip}

Spezialisierung: Theoretische Physik
 \vspace{\baselineskip}
 
eingereicht an der 
 \vspace{\baselineskip}

Mathematisch-Naturwissenschaftlichen Fakultät

der Humboldt-Universität zu Berlin 
 \vspace{\baselineskip}

von 
 \vspace{\baselineskip}

Laura Rijkje Anne Koster M.Sc.
 \vspace{\baselineskip}

 \vspace{\baselineskip}

Präsidentin der Humboldt-Universität zu Berlin 

Prof. Dr.-Ing. Dr. Sabine Kunst
 \vspace{\baselineskip}

Dekan der Mathematisch-Naturwissenschaftlichen Fakultät

Prof. Dr. 
 Elmar Kulke
 \vspace{\baselineskip}

\begin{flushleft}
Gutachter/innen:    \begin{enumerate}
                     \item Prof. Dr. Matthias Staudacher
                     \item Prof. Dr. Lionel Mason
                     \item Dr. Valentina Forini
                    \end{enumerate}

Tag der mündlichen Prüfung: 12.07.2017
\end{flushleft}
\endgroup

\thispagestyle{empty}

\cleardoublepage

I declare that I have completed the thesis independently using only the aids and tools specified. I have not applied for a doctor's degree in the doctoral subject elsewhere and do not hold a corresponding doctor's degree. I have taken due note of the Faculty of Mathematics and Natural Sciences PhD Regulations, published in the Official Gazette of Humboldt-Universit\"{a}t zu Berlin no. 126/2014 on 18/11/2014.

\cleardoublepage
\phantomsection
\addcontentsline{toc}{chapter}{Zusammenfassung}
\chapter*{Zusammenfassung}
\vspace{-1.4cm}
\markboth{Zusammenfassung}{Zusammenfassung}
Das Standardmodell der Teilchenphysik hat sich bis heute, mit Ausnahme der allgemeinen Relativit\"{a}tstheorie, als erfolgreichste Theorie zur Beschreibung der Natur erwiesen. St\"{o}\-rungstheoretische Rechnungen f\"{u}r bestimmte Mengen in Quantenchromodynamik (QCD) haben bisher unerreicht pr\"{a}zise Vorraussagen erm\"{o}glicht, die experimentell nachgewiesen wurden. Trotz dieser Erfolge gibt es Teile des Standardmodells und Energieskalen bei denen die St\"{o}rungstheorie versagt und man nach Alternativen suchen muss. Vieles k\"{o}nnen wir hierbei verstehen, indem wir eine \"{a}hnliche Theorie untersuchen, die sogenannte planare $\calN=4$ Super Yang-Millstheorie in vier Dimensionen ($\calN=4$ SYM). Es existieren viele Indizien daf\"{u}r, dass die Theorie exakte L\"{o}sungen zul\"{a}sst. Dies l\"{a}sst sich zur\"{u}ckf\"{u}hren auf die Integrabilit\"{a}t der Theorie, eine unendlich dimensionale Symmetriealgebra, die die Theorie stark einschr\"{a}nkt. Neben besagter Integrabilit\"{a}t besitzt diese Theorie auch andere spezielle Eigenschaften. So ist sie des am besten verstandenen Beispiels der Eich-/Gravitations Dualit\"{a}t durch die AdS/CFT Korrespondenz. Au\ss erdem sind die Streuamplituden von Gluonen auf Baumgraphenniveau in $\calN=4$ SYM die selben wie in Quantenchromodynamik. Diese Streuamplituden besitzen eine elegante Struktur und stellen sich als deutlich simpler heraus, als die dazugeh\"{o}rigen Feynmangraphen vermuten lassen. Tats\"{a}chlich umgehen viele der zur Berechnung von Streuamplituden entwickelten Masseschalenmethoden die Feynmangraphen, indem sie vor\-r\"{u}bergehend manifeste Unitarit\"{a}t und Lokalit\"{a}t aufgeben und dadurch die Rechnungen stark vereinfachen. Alle diese Entwicklungen suggerieren, dass der konventionelle Formalismus der Theorie mit Hilfe der Wirkung im Minkowskiraum nicht der aufschlussreichste oder effizienteste Weg ist, die Theorie zu untersuchen. Diese Arbeit untersucht der Hypothese, ob dass stattdessen Twistorvariablen besser geeignet sind, die Theorie zu beschreiben. Der Twistorformalismus wurde zuerst von Roger Penrose eingef\"{u}hrt. Auf dem klassischen Level ist die holomorphe Chern-Simonstheorie im Twistorraum \"{a}quivalent zur klassischen selbst-dualen Yang-Mills L\"{o}sung in der Raumzeit. Die volle Twistorwirkung, welche eine St\"{o}rung um diesen klassisch integrablen Sektor ist und durch eine Eichbedingung auf die $\mathcal{N}=4$ SYM Wirkung reduziert werden kann, produziert unter einer anderen Eichbedingung alle sogenannten maximalhelizit\"{a}tsverletzenden 
(MHV) Amplituden auf Baumgraphenniveau. Durch die Einf\"{u}hrung eines Twistorpropagators konnten auch N$^k$MHV Amplituden effizient beschrieben werden. In dieser Arbeit erweitern wir den Twistorformalismus um auch Gr\"{o}\ss en, die sich nicht auf den Masseschalen befinden, beschreiben zu k\"{o}nnen. Wir untersuchen alle lokalen eichinvarianten zusammengesetzten Operatoren im Twistorraum und zeigen, dass sie alle Baumgraphenniveau-Formfaktoren des sogenannten MHV-Typs erzeugen. Wir erweitern diese Methode zu NMHV und h\"{o}her N$^k$MHW Level in Anlehnung an die Amplituden. Schlie\ss lich kn\"{u}pfen wir an die Integrabilit\"{a}t an, indem wir den ein-Schleifen Dilatationsoperator in dem skalaren Sektor der Theorie im Twistorraum berechnen.

\cleardoublepage
\phantomsection
\addcontentsline{toc}{chapter}{Abstract}
\chapter*{Abstract}
\markboth{Abstract}{Abstract}
\vspace{-\baselineskip}
The Standard Model of particle physics has proven to be, with the exception of general relativity, the most accurate description of nature to this day. Perturbative calculations for certain quantities in Quantum Chromo Dynamics (QCD) have led to the highest precision predictions that have been experimentally verified. However, for certain sectors and energy regimes, perturbation theory breaks down and one must look for alternative methods. Much can be learned from studying a close cousin of the standard model, called planar $\mathcal{N}=4$ super Yang-Mills theory in four dimensions ($\calN=4$ SYM), for which a lot of evidence exists that it admits exact solutions. This exact solvability is due to its quantum integrability, a hidden infinite symmetry algebra that greatly constrains the theory, which has led to a lot of progress in solving the spectral problem. Integrability aside, this non-Abelian quantum field theory is special in yet other ways. For example, it is the most well understood example of a gauge/gravity duality via the AdS/CFT correspondence. Furthermore, at tree level the scattering amplitudes in its gluon sector coincide with those of Quantum Chromo Dynamics. These scattering amplitudes exhibit a very elegant structure and are much simpler than the corresponding Feynman diagram calculation would suggest. Indeed, many on-shell methods that have been developed for computing these scattering amplitudes circumvent the tedious Feynman calculation, by giving up manifest unitarity and locality at intermediate stages of the calculation, greatly simplifying the work. All these developments suggest that the conventional way in which the theory is presented, i.e.\ in terms of the well-known action on Minkowski space, might not be the most revealing or in any case not the most efficient way. This thesis investigates whether instead twistor variables provide a more suitable description. The twistor formalism was first introduced by Roger Penrose. At the classical level, a holomorphic Chern-Simons theory on twistor space is equivalent to classically integrable self-dual Yang-Mills solutions in space-time. A quantum perturbation around this classically integrable sector reduces to the conventional $\calN=4$ SYM action by imposing a partial gauge condition. This action generates all so-called maximally helicity violating (MHV) amplitudes at tree level directly, when a different gauge was chosen. By including a twistor propagator into the formalism, also higher degree N$^k$MHV amplitudes can be described efficiently. In this thesis we extend this twistor formalism to encompass (partially) off-shell quantities. We describe all gauge-invariant local composite operators in twistor space and show that they immediately generate all tree-level form factors of the MHV type. We use the formalism to compute form factors at NMHV and higher N$^k$MHV level in parallel to how this was done for amplitudes. Finally, we move on to integrability by computing the one-loop dilatation operator in the scalar sector of the theory in twistor space.
\setcounter{tocdepth}{1}

\tableofcontents

\cleardoublepage
\phantomsection
\addcontentsline{toc}{chapter}{Publications}
\chapter*{Publications}
\markboth{Publications}{Publications}

This thesis is based on the following publications by the author:
\begin{itemize}
 \item[\cite{Koster:2014fva}]
L.~Koster, V.~Mitev and M.~Staudacher, 
``{A Twistorial Approach to Integrability in $\mathcal{N}=4$ SYM},''
\href{http://dx.doi.org/10.1002/prop.201400085}{{\em Fortsch. Phys.} {\bfseries 63} (2015) 142-147}, 
\href{http://arxiv.org/abs/1410.6310}{{\ttfamily arXiv:1410.6310 [hep-th]}}.
 \item[\cite{Koster:2016ebi}]
L.~Koster, V.~Mitev, M.~Staudacher and M.~Wilhelm, 
``{Composite Operators in the Twistor Formulation of $\mathcal{N}=4$ Supersymmetric Yang-Mills Theory},''
\href{http://dx.doi.org/10.1103/PhysRevLett.117.011601}{{\em Phys. Rev. Lett.} {\bfseries 117} (2016) 011601}, 
\href{http://arxiv.org/abs/1603.04471}{{\ttfamily arXiv:1603.04471 [hep-th]}}.
 \item[\cite{Koster:2016loo}]
L.~Koster, V.~Mitev, M.~Staudacher and M.~Wilhelm, 
``{All Tree-Level MHV Form Factors in $\mathcal{N}=4$ SYM from Twistor Space},''
\href{http://dx.doi.org/10.1007/JHEP06(2016)162}{{\em JHEP} {\bfseries 06} (2016) 162}, 
\href{http://arxiv.org/abs/1604.00012}{{\ttfamily arXiv:1604.00012 [hep-th]}}.
 \item[\cite{Koster:2016fna}]
L.~Koster, V.~Mitev, M.~Staudacher and M.~Wilhelm, 
``{On Form Factors and Correlation Functions in Twistor Space},''
\href{http://dx.doi.org/10.1007/JHEP03(2017)131}{{\em JHEP} {\bfseries 03} (2017) 131},
\href{http://arxiv.org/abs/1611.08599}{{\ttfamily arXiv:1611.08599 [hep-th]}}.
\end{itemize}

\cleardoublepage
\phantomsection
\addcontentsline{toc}{chapter}{Introduction}
\chapter*{Introduction}
\markboth{Introduction}{Introduction}
\label{chap: general introduction}
Over the past century our understanding of the universe has undergone a major transformation. Two areas of physics are responsible for this. On the one hand, the theory of general relativity that governs the universe at large scale. On the other hand, the theory of small scale phenomena, described by quantum mechanics. Reconciling these two seemingly incompatible theories of nature has been and remains to this day the biggest open problem of modern physics. This is particularly puzzling given the grand successes of both theories which have been tested and verified to a tremendous extent.
Quantum field theory in the context of the Standard Model of particle physics has made predictions that have been verified to astounding precision. The Standard Model combines three of the four fundamental forces, the weak and strong nuclear forces and the electromagnetic force as well as encompasses all the subatomic particles that we know. One of its biggest achievements is the successful prediction of the Higgs boson \cite{PhysRevLett.13.321,PhysRevLett.13.508}, which was observed at the Large Hadron Collider at CERN by two enormous collaborations almost half a century after its prediction \cite{Aad:2012tfa,Chatrchyan:2012xdj}.\newline\\
The textbook method for computing observables in a quantum field theory is via perturbation theory. This relies on the possibility of expanding solutions in powers of a certain small parameter, called the coupling constant. The higher order terms, although both more complicated and more numerous, yield smaller contributions. For the standard model, the coupling constants are not constant, but vary with the energy scale. Still, the electromagnetic interactions as well as the weak interactions can be accessed using perturbation theory. However, for the theory of strong interactions, which is called quantum chromodynamics (QCD) at low energies the coupling is so large that perturbation theory breaks down and one needs other non-perturbative methods to access the theory. It is for this reason that phenomena such as quark confinement cannot be studied using conventional perturbative methods. This provides one of the main motivations for searching non-perturbative methods in quantum field theories. It is sensible to first investigate these in a more accessible theory than the physical Standard Model, in which exact, i.e.\ non-perturbative results are deemed possible. One such theory that has many similarities with the standard model is the non-Abelian maximally supersymmetric Yang-Mills theory in four space-time dimensions ($\calN=4$ SYM) with gauge group SU$(N)$ in the limit where the rank of the gauge group goes to infinity. This limit is called the planar limit, or the `t Hooft limit\footnote{Even when not stated explicitly, in this thesis we will only consider $\calN=4$ SYM in this planar limit.}. Despite the fact that it is very different from the standard model that describes our world -- for instance, it is a supersymmetric theory while no evidence has been found to support existence of supersymmetry in nature so far -- still its gluon sector coincides at tree-level with QCD. Therefore,  also from a purely computational perspective it is justified to study this theory.
This maximally supersymmetric Yang-Mills theory was initially obtained from $\calN=1$ supersymmetric Yang-Mills theory in ten dimensions by dimensionally reducing to four space-time dimensions \cite{Brink:1976bc}. Since its discovery, many absolutely astounding properties of the theory have been discovered and are still being discovered. Indeed, $\calN=4$ SYM connects many very different branches of physics, among which are string theory, collider physics, and spin chains. Its first remarkable feature is the large symmetry group it admits, PSU$(2,2|4)$, which makes it a superconformal field  theory. This superconformal structure remains intact even at the quantum level, which means that there is no mass-scale and the interactions are scale independent. \newline\\
One way of obtaining results beyond the perturbative regime of $\calN=4$ SYM is via the AdS/CFT correspondence.
This correspondence was proposed by Maldacena in \cite{Maldacena:1997re} is a (conjectured) duality between a string theory in an anti-de Sitter space (AdS) background and a certain conformal field theory in one dimension less on the boundary of AdS. In fact, $\calN=4$ SYM with gauge group SU$(N)$ and `t Hooft coupling $\la$ is one side of the prime example of this duality. The dual theory is a type II B superstring theory in an AdS$_5\times$S$^5$ background with string coupling $g_s=\la/4\pi N$ and string tension $g=\sqrt{\la}/2\pi$ in the limit where $N$ tends to infinity. On the Yang-Mills side, the effective coupling is the `t Hooft coupling, which is related to the Yang-Mills coupling $g_{\text{YM}}$ via $\la\equiv g_{\text{YM}}^2N$ in the limit where $g_{\text{YM}}\rightarrow 0$ and $N\rightarrow \infty$, while $\la$ is kept fixed \cite{tHooft:1973alw}. Observables can be computed in either one of the two theories that are related via this duality. In the perturbative regime of $\calN=4$ SYM, on the AdS side, the string length is large compared to the curvature of the background and therefore perturbation theory on that side breaks down. Alternatively, in the strong coupling regime on the field theory side, where both $\la$ and $N$ are large, we can describe the string theory by a super gravity theory. This way, we can obtain non-perturbative results in $\calN=4$ SYM, but we cannot cross check the results on both sides because the two theories can only be accessed in different regimes. \newline\\
Another way of obtaining non-perturbative results in $\calN=4$ SYM is via its believed integrability in the planar theory, for which a lot of evidence has been presented over the past decade. Quantum integrability means that it should in principle be possible to solve the theory exactly, i.e.\ without resorting to perturbation theory. 
The fact that a four-dimensional quantum field theory is integrable at the quantum level is remarkable as integrability is very much a two dimensional feature. Integrability is due to the existence of a hidden infinite symmetry algebra, with infinitely many conserved charges. These conserved charges restrain the scattering matrix of an $n\rightarrow m$ scattering process in such a way that a) there is no on-shell particle production, hence $n=m$, and b) the only solutions to the symmetry constraints are the trivial ones, in which the set of initial momenta is conserved, but the individual momenta may be permuted. Confined to a two-dimensional plane, this implies that the scattering matrix factorizes into $2\rightarrow 2$ scattering matrices. All possible factorizations are required to be equivalent for consistency, which yields the so-called Yang-Baxter equation for the $2\rightarrow 2$ scattering matrix (S-matrix). Solving the $2\rightarrow 2$ S-matrix thus solves the theory and this can be achieved using so-called Bethe ansatz techniques. The two-dimensional nature of this integrability is crucial. It is therefore curious that an interacting quantum field theory in four dimensions such as $\calN=4$ SYM, could be quantum integrable as well. This is due to the nature of the planar limit that we mentioned briefly earlier. The Feynman diagrams that are of leading order in 1/N can be drawn on a plane, whereas the diagrams that are suppressed by factors of $1/N$ cannot\footnote{That is, they cannot be drawn on a genus zero surface.}. This is the two dimensionality that ultimately allows for the integrable structure of the theory. It is not at all obvious that $\calN=4$ SYM  is quantum integrable, certainly not from the explicit form of the action, which governs both the planar and the non-planar theory. It is therefore not surprising that it took about twenty-five years for people to first observe its integrable structure. Minahan and Zarembo noticed that the so-called one-loop dilatation operator, which is the matrix of the one-loop corrections to the scaling dimensions of the theory, in the scalar sector of the theory could be mapped to the Hamiltonian of an integrable spin chain \cite{Minahan:2002ve}. In the spin chain picture, the two-dimensional scattering problem is the scattering of spin waves propagating through the one-dimensional spin chain and in time, while only exchanging momenta. The physics is described by the Hamiltonian of an SO$(6)$ spin chain, which can be diagonalized using a Bethe ansatz. The Bethe ansatz is named after Hans Bethe who used this ansatz to solve the spectrum of a Heisenberg spin chain \cite{Bethe1931}. In the quantum field theory picture, the problem of finding and diagonalizing the dilatation operator to any order in perturbation theory is important as its eigenvalues, the anomalous dimensions, determine two-point correlation functions completely. Finding its eigenvectors as well is needed to also solve the three-point correlation functions. Together, these two families of observables in principle completely determine the whole theory as any higher point function can a priori be obtained via the so-called operator product expansion. The observation that the problem of diagonalizing the dilatation operator, first only in the SO$(6)$ sector, could be accessed by going to its spin chain Hamiltonian interpretation sparked an interest in using and further developing these Bethe ansatz techniques that had first appeared some seventy years earlier. A new field of integrability related research in the context of quantum field theories was born. Not long after the first paper about the SO$(6)$ spin chain, the full one-loop dilatation operator \cite{Beisert:2004ry} was identified with the Hamiltonian of a PSU$(2,2|4)$-integrable spin chain \cite{Beisert:2003yb}. The statement that integrability was present at all loop order \cite{Beisert:2005fw} paved the way to using an asymptotic all-loop Bethe ansatz for the full theory.
These Bethe-Ansatz related techniques have been used to compute observables to a precision that lies far beyond the reach of Feynman graph computations. Since the emergence of quantum integrability in $\calN=4$ SYM, the problem of solving the infinitely many Bethe equations has been reformulated several times before finally being cast into the form of the so-called Quantum Spectral Curve (QSC). This construction provides the state of the art results in the spectral problem of $\calN=4$ SYM. It has resulted in a complete numerical solution of the spectral problem of $\calN=4$ for finite coupling \cite{Gromov:2015wca} and analytic results for the anomalous dimensions of ten operators up to ten loops\footnote{This loop number is arbitrary in the sense that the precision reached by the QSC in principle solely depends on the power and the running time of the computer.}  \cite{Marboe:2014gma} including the Konishi anomalous dimension which has been compared against five-loop results obtained from field theory. 
Although progress in the development of computational techniques has been tremendous, the origin of the quantum integrable nature of the theory has remained somewhat obscure so far.\newline
\\
In parallel to the development of integrability of $\calN=4$, a seemingly unrelated approach to solving the theory was made by people studying its scattering amplitudes. Scattering amplitudes comprise the fundamental building blocks of cross sections of scattering processes. These scattering amplitudes can be obtained by computing Feynman diagrams after which one sets the external scattering particles on their mass-shell, $p^2=m^2$. Interestingly, a massless momentum can be written as a two-by-two Hermitian matrix that decomposes into a product of two conjugate two-spinors, $p^{\alpha}\bar p^{\dot\alpha}$. This decomposition is clearly invariant under the phase transformation $p^{\alpha}\rightarrow e^{i\phi} p^{\alpha}$. Any particle is now classified by an integer- or half-integer number, called the helicity, that labels its behavior under this transformation. Amplitudes are completely classified by the helicities of the external particles. This induces a classification of amplitudes into the degree to which they violate helicity conservation. This classification and these spinor-helicity variables reveal the hidden elegant structure of the amplitudes that via old-school Feynman computations would only emerge at the stage of the final physical answer. For example, to compute a scattering amplitude consisting of five gluons of which two are of negative helicity at tree level\footnote{This amplitude is of maximally helicity violating class (MHV).}, one ordinarily needs to sum over approximately two thousand Feynman diagrams. After setting the external legs on shell and expressing the answer in so-called spinor-helicity variables, the final and physical answer can be expressed on just one line as
\begin{equation}
\label{parketayloerintro}
\mathscr{A}^{\text{MHV}}_n(1^+, \dots , i^-, \dots, j^- , \dots n^+) =\frac{\delta^{4}(\sum_{i=1}^n \fp_i)\abra{i}{j}^4}{\prod_{k=1}^n\abra{k}{(k+1)}} \eqncom
\end{equation}
where the angular brackets are the contractions of the positive-helicity spinors of the external particles and $n$ denotes the number of external legs. This formula holds for any number of external particles. The intermediate stages of the computation, which by themselves are non-physical, clearly obscure the simplicity of the theory. This is yet another hint that the space-time action is neither the most efficient nor the most revealing formulation of the theory. Over the past decades, many techniques for computing amplitudes that rely on their on-shell nature have been developed. For example, using so-called Britto-Cachazo-Feng-Witten (BCFW) relations one breaks down amplitudes into two kinds of three-point amplitudes \cite{Britto:2004ap,PhysRevLett.94.181602}, of maximally helicity violating type and its conjugate. The Cachazo-Svrcek-Witten (CSW) formalism also decomposes amplitudes into smaller amplitudes, but instead, building blocks contain any number of external legs and are all of maximally helicity violating type. These on-shell techniques have led to, among others, the computation of all tree-level amplitudes \cite{Drummond:2008cr}.
\newline\\
Interestingly, $\calN=4$ SYM admits a certain duality under which scattering amplitudes get mapped to polygonal Wilson loops and vice versa by changing space-time coordinates to so-called regional coordinates and back. This was first described in \cite{Alday:2007hr,Berkovits:2008ic} for MHV amplitudes and later extended to amplitudes of arbitrary helicity configuration in \cite{CaronHuot:2010ek,Mason:2010yk,Bullimore:2011ni}. Under this duality, the edges of the polygonal Wilson loop are mapped to the momenta of the scattering amplitude and the Wilson loop is closed due to momentum conservation. The dual Wilson loop admits a superconformal symmetry which gives rise to a dual superconformal symmetry on the side of the amplitude. This dual superconformal symmetry supplemented by the ordinary superconformal symmetry generates a Yangian algebra. A Yangian algebra is an infinite symmetry algebra, that typically indicates existence of integrability. Tree-level color-ordered scattering amplitudes have been shown to be invariant under this Yangian \cite{Drummond:2009fd}. More recently, it was shown that the action is Yangian invariant \cite{Beisert:2017pnr}, in the sense that the equations of motion are invariant. This might be a step in relating the quantum integrability of the theory emerging in the spin chain picture with the Yangian invariance of amplitudes. 
\newline\\
Another approach of bringing together the on-shell scattering amplitudes and the off-shell correlation functions is by first studying a class of hybrids of these two quantities. For adapting and applying on-shell methods to off-shell quantities, so-called form factors, being in between scattering amplitudes and correlation functions, provide an ideal testing ground. A form factor of a local composite operator $\calO$ of momentum $\fq$, which is off shell, is the overlap of the state created by the operator from the vacuum and a state of $n$ on-shell particles $\Phi_i$:
\begin{equation}
\label{eq:formfactordefinition}
\calF_{\calO}(1^{\Phi_1},\ldots, n^{\Phi_n};\fq)= \int \frac{\dd^4x}{(2\pi)^4}\, \e^{-i\fq x}\melement{\Phi_1(\fp_1)\cdots \Phi_n(\fp_n)}{\calO(x)}{0}\eqncom
\end{equation}
where $\fp_i$ are the null momenta of the on-shell particles. Indeed, choosing $\calO=\text{Id}$ we recover an amplitude and alternatively, setting $n=0$ gives rise to a correlation function. Form factors appear in many physical scattering processes in which a correction to a vertex appears that is often too complicated to compute explicitly. The form of these vertex corrections however is greatly restricted by symmetry requirements and Ward identities. In the standard model for instance, a form factor describes the effective Higgs to gluon decay, where the Higgs couples to the gluons via a quark loop. In the large top-mass limit, the top quark loop can be integrated out yielding an effective vertex of the Higgs with two on-shell external gluons. This process can be part of a larger scattering process that produces the Higgs field, in which case the Higgs momentum is off-shell and the effective vertex is described by a form factor.
In $\calN=4$ SYM form factors were first introduced by van Neerven in \cite{vanNeerven:1985ja} who described the form factor of an operator consisting of two scalars with the minimal number of external legs and computed it up to two loops. Since then, a lot of progress has been made at weak coupling \cite{Brandhuber:2010ad, Bork:2010wf, Brandhuber:2011tv, Bork:2011cj, Henn:2011by, Gehrmann:2011xn, Brandhuber:2012vm, Bork:2012tt, Engelund:2012re, Johansson:2012zv, Boels:2012ew, Penante:2014sza, Brandhuber:2014ica, Bork:2014eqa, Wilhelm:2014qua, Nandan:2014oga, Loebbert:2015ova, Bork:2015fla, Frassek:2015rka, Boels:2015yna, Huang:2016bmv, Chicherin:2016qsf, Brandhuber:2016fni,  Broedel:2012rc, Bork:2016hst, Bork:2016xfn, He:2016dol, Caron-Huot:2016cwu, Brandhuber:2016xue, He:2016jdg,Yang:2016ear,Ahmed:2016vgl,Loebbert:2016xkw,Bork:2016egt} and at strong coupling \cite{Alday:2007he,Maldacena:2010kp,Gao:2013dza}. However, compared to amplitudes, progress in form factors has lagged behind. In parallel to amplitudes, form factors can be most easily described in spinor-helicity variables as they can also be classified according to their MHV degree. In fact, they look very similar to their amplitude cousins. For instance, MHV form factors contain the same Parke-Taylor denominator as amplitudes. Despite their partially off-shell nature, on-shell techniques have been successfully applied to form factors in certain cases. For example, BCFW was first applied to compute form factors in \cite{Brandhuber:2010ad}. In \cite{Brandhuber:2011tv} it was shown that the CSW recursion relations can be applied to form factors of the stress-tensor multiplet, and subsequently this was extended to half-BPS operators in \cite{Penante:2016ycx}. The latter work contains a pedagogical introduction to both CSW and BCFW in the context of amplitudes as well as form factors. Recently, a very interesting duality between form factors and Wilson loops was found in the Lorentz Harmonic Chiral (LHC) space \cite{Chicherin:2016ybl}. There, a form factor of an $m$-sided polygonal Wilson loop with $n$ external states is mapped to a form factor of an $n$-gonal Wilson loop with $m$ external legs.
\newline\\
We have argued that planar maximally supersymmetric Yang-Mills theory in four dimensions admits a lot of structure that is hidden by its space-time formulation. Therefore, we expect that there exists an alternative formulation of the theory that makes all this structure more manifest. This suspicion that space-time variables might not be the best variables for describing nature is not new. The idea of instead describing physics in terms of light-rays dates back to 1967.
In that year, Roger Penrose introduced bosonic twistor space and proposed that light rays rather than space-time points should be the fundamental variables \cite{Penrose:1967wn}. Roughly speaking, a point in space-time corresponds to a Riemann sphere in twistor space and a point (twistor) in twistor space corresponds to a light ray in Minkowski space. This turned out to be useful for describing massless fields. Twistor space admits a holomorphic Chern-Simons action which is in one-to-one correspondence with a self-dual Yang-Mills action. This self-dual theory is an integrable system of equations that can be reformulated into the zero-curvature condition of a Lax connection, see e.g.\ \cite{Mason:1991rf}. Thus at least classically, integrability of the Yang-Mills equations appears very naturally in twistor space. However, the bosonic twistor theory remained in a niche for many years after its conception, and it was not until more than three decades later that a wider circle became interested in the theory. \newline\\
In 2004 Witten proposed a twistor string theory as a holomorphic Chern-Simons theory on supertwistor space. This sparked a broader interest in twistor theory \cite{Witten:2003nn}. In the years that followed it was shown that the full $\calN=4$ SYM action could be obtained from an action functional for fields on twistor space in terms of a single super-field $\AAA$. Interestingly, the supersymmetry generators of PSU$(2,2|4)$ act linearly on twistor space, which is a first hint that the symmetry properties might be more manifest in this description. This twistor action is a perturbation around the previously mentioned self-dual sector \cite{Boels:2006ir}. It contains an infinite sum of interaction vertices of increasing valency. At first glance, this may seem like a disadvantage, but it turns out that this is in fact very efficient for describing the on-shell scattering amplitudes. Choosing a specific axial gauge and inserting on-shell external states directly into these interaction vertices precisely yields all tree-level (super)amplitudes of the previously mentioned maximally helicity violating (MHV) type. This works for any number of external particles. The next-to-maximally-helicity violating amplitude, or NMHV amplitude for short, is constructed from two such MHV amplitudes connected by a propagator. More generic so-called N$^{k}$MHV amplitudes are constructed from $k+1$ interaction vertices and $k$ twistor-space propagators \cite{Boels:2007qn}. In fact, these Feynman diagrams on twistor space have been shown to be equivalent to the CSW recursion relations in space-time.
\newline\\
Since the twistor formalism reproduces the CSW relations for amplitudes, one may wonder how these relations for form factors can be found from twistor space. Namely, for certain operators CSW recursion relations can be used to compute form factors using an off-shell continued form factor, called operator vertex, as a fundamental ingredient. Therefore, these operator vertices need to have a twistor space equivalent in order to make this picture and the mapping between twistor space and CSW complete\footnote{These operator vertices were independently constructed in the Lorentz Harmonic Chiral formalism, LHC for short, in  \cite{Chicherin:2016fac,Chicherin:2016fbj,Chicherin:2016qsf,Chicherin:2016ybl}. This formalism was argued to be closely related to the one for twistor space in \cite{Chicherin:2016soh,Chicherin:2016qsf}. }. The first goal of this thesis is to describe these operator vertices for all local composite operators in twistor space. 
From this, all tree-level MHV form factors should be straightforwardly derived. Subsequently, we extend the formalism to NMHV and higher N$^k$MHV level form factors. 
Furthermore, since the twistor action is a perturbation around a classically integrable sector, we investigate whether the integrable structure of $\calN=4$ SYM is more manifest in this formulation. To this end, we go completely off shell and compute the one-loop correlation functions in the SO$(6)$ sector. From these we extract the one-loop dilatation operator in this sector\footnote{This last part is based on the paper  \cite{Koster:2014fva} and appeared on the same day as the paper \cite{Wilhelm:2014qua}, which contains the derivation of the complete one-loop dilatation operator from unitarity. After these two papers appeared, the equivalent SO$(6)$-computation using MHV diagrams was presented in \cite{Brandhuber:2014pta}. Subsequently, the results of \cite{Wilhelm:2014qua} for the SO$(6)$- and SU$(2|3)$ sectors were reproduced from generalized unitarity in \cite{Brandhuber:2015boa}.}.

\cleardoublepage
\phantomsection
\addcontentsline{toc}{chapter}{Overview}
\chapter*{Overview}
\markboth{Overview}{Overview}
This thesis is divided into two parts. Part I is a two-chapter review, and part II presents in five chapters the main research results obtained by the author and collaborators. In Chapter~\ref{review1} we summarize some relevant basics of maximally supersymmetric Yang-Mills theory. We recall its field content, the local composite operators, the dilatation operator, scattering amplitudes and form factors. Chapter~\ref{chaptwistor} then reviews the construction of twistor space. We start by revisiting classical non-supersymmetric twistor space as it was first developed by Penrose in the sixties. Section~\ref{penrosetransform} reviews the so-called Penrose transform. Section~\ref{classint} concerns the correspondence between solutions to the self-dual Yang-Mills equations and holomorphic Chern-Simons theories on twistor space. Section~\ref{susytwistor} deals with the extension to supersymmetric twistor space and a twistor action that was introduced by Witten and Boels, Mason and Skinner. We review how the twistor action in a certain partial gauge reduces to the conventional space-time action of $\calN=4$ in Section~\ref{twistoraction}. This section contains an erratum to the paper \cite{Boels:2006ir}. In Section~\ref{amptwistor} we impose a different, axial gauge, from which tree-level MHV amplitudes are straightforwardly derived. This framework is then extended to NMHV level and this concludes the section. The last Section gives a brief summary of the review and presents the main motivation for Part II. \newline\\
In Chapter~\ref{chap:} we extend the formalism developed for amplitudes also to form factors by considering the operators that consists of two identical scalars. In the first section we find an expression for the vertex of the scalar field in twistor space and compute some of its MHV form factors. We combine these into the super MHV form factor for this operator in Section~\ref{superff}. In Section~\ref{NMHVscalar} we extend the construction to NMHV level. In the final section of this chapter we show how the operator vertex can be found from a generating Wilson loop. This chapter is based on \cite{Koster:2016ebi}. In Chapter~\ref{MHV} we extend the formalism to include all the rest of the field content of $\calN=4$ SYM as well. We find general expressions for all local composite operators in Section~\ref{sec:construction} and derive a general expression for all minimal (Section~\ref{sec:minff}) and non-minimal (Section~\ref{subsec: all MHV form factors}) tree-level MHV form factors of the theory. Section~\ref{app: derivation} contains a proof of this result on MHV form factors. This chapter is based on and contains overlap with \cite{Koster:2016loo}. In Chapter~\ref{NMHV} we build further on this work by considering form factors of higher NMHV degree in twistor space. In Section~\ref{subsec: inversesoftlimit} we prove an inverse soft limit for form factors. Section~\ref{NMHVformfactortwistorspace} deals with NMHV form factors and Section~\ref{sec:NkMHVtreelevelformfactors} with higher degree N$^k$MHV form factors. In Chapter~\ref{momentumspace} we translate some NMHV results to momentum space, in Section~\ref{eq:NMHV amplitudes in momentum twistor space} first for amplitudes, in Section~\ref{subsec:NMHV form factors in momentum twistor space} for NMHV form factors without $\dot\alpha$ indices. In Section~\ref{subsec: non-chiral operators} we treat an example of a general NMHV form factor. Chapters ~\ref{NMHV} and \ref{momentumspace} are based on and contain significant overlap with \cite{Koster:2016fna}.
Finally, in Chapter~\ref{corrfunc} we use our formalism to compute correlation functions. Correlation functions are completely off shell and therefore it is interesting to see that our formalism can also applied there. We compute the $1$-loop dilatation operator in the scalar sector of the theory by computing two-point correlation functions at one loop. The chapter is based on the work of \cite{Koster:2014fva} and \cite{Koster:2016fna} and contains some overlap with the last paper.

\makeatletter
\def\toclevel@part{0}
\makeatother

\part{Review}
\chapter{Planar $\calN=4$ super Yang-Mills theory}
\label{review1}
In this chapter we review some basics of planar $\calN=4$ super Yang-Mills theory and its observables in its conventional formulation in Minkowski space. The physical states of the theory are given by gauge-invariant local composite operators. In the first section of this chapter we discuss the field content, the action with its supersymmetry algebra, and finally the local composite operators. Furthermore, we comment on the planar limit which will be considered throughout this thesis. In the next section we review the computation of the one-loop dilatation operator in the so-called SO$(6)$ sector that was first performed by Minahan and Zarembo in \cite{Minahan:2002ve}. It was here that the integrability of $\calN=4$ SYM made its first appearance. We conclude the section by briefly sketching how this result is extended to the full one-loop dilatation operator. The section that follows contains a brief and basic introduction to scattering amplitudes. We explain the spinor-helicity formalism and discuss one particular recursive method for computing amplitudes, called CSW after Cachazo-Svrcek-Witten \cite{Cachazo:2004kj}, which we illustrate by computing an example. In the last section we introduce form factors, which constitute the main topic of this thesis. The section is concluded by discussing how the CSW recursion can be extended to form factors. 

\section{Field content, action and local composite operators}
\label{fieldsoperatorsaction}
In this section we review the field content, local composite operators and the action of $\calN=4$ SYM theory in the usual Minkowski space formulation. The theory contains one gauge field $A_{\alpha\dot\alpha}$, with $\alpha, \dot\alpha = 1,2$ both spinor indices, $6$ scalar fields $\phi_{ab}$, where $a,b = 1,\dots ,4$ are antisymmetric SU$(4)$ indices, four Weyl fermions $\bar\psi_{a}^{\dot\alpha}$ and four anti-fermions $\psi^a_{\alpha}$. 
These are referred to as elementary or fundamental fields. 
We choose to present the $\calN=4$ SYM action in spinorial representation and use antisymmetric SU$(4)$ indices instead of the (more) common fundamental SO$(6)$ indices for the scalar fields to make the connection with the next chapter easier. The action is given by \cite{Fokken:2017qpu}
\begin{align}
S=\int\dd^4 x \Tr\Big( &\frac{1}{2} (F_{\alpha\beta}F^{\alpha\beta}+\bar F_{\dot\alpha\dot\beta}\bar F^{\dot\alpha\dot\beta}) -\frac{1}{4}D_{\alpha\dot\alpha}\phi_{ab}D^{\alpha\dot\alpha}\phi^{ab}+\frac{1}{4}g_{\text{YM}}^2 [\phi_{ab},\phi_{cd}] [\phi^{ab},\phi^{cd}] \notag \\
&-\bar\psi_{a}^{\dot\alpha} D_{\alpha\dot\alpha}\psi^{ a\alpha}+g_{\text{YM}} \bar\psi_{a}^{\dot\alpha}[\phi^{ab},\bar\psi_{b\dot\alpha}]-g_{\text{YM}} \psi^{a \alpha} [\phi_{ab},\psi^{b}_{\alpha}]\Big)\eqncom
\end{align}
where $D_{\alpha\dot\alpha}$ is the covariant derivative, $\bar F^{\dot\alpha\dot\beta}$ and $F_{\alpha\beta}$ are the self-dual and anti-self-dual part of the field strength respectively and satisfy
\begin{equation} \label{fieldstrengthdef}
[D_{\alpha\dot\alpha},D_{\beta\dot\beta}] \propto\epsilon_{\alpha\beta}\bar F_{\dot\alpha\dot\beta}+\epsilon_{\dot\alpha\dot\beta}F_{\alpha\beta} \eqndot
\end{equation}
Throughout this thesis the gauge group is always SU$(N)$ and the elementary fields transform in the adjoint representation.  They can be expanded as $\Phi= \sum_{i=1}^N \Phi_i T^i,$ where $T^i$ are the SU$(N)$ generators in the adjoint representation, satisfying 
\begin{equation}
\Tr(T^a T^b)=\delta^{ab}\eqncom
\end{equation}
and the completeness relation
\begin{equation}
\sum_{a=1}^{N^2-1} (T^{a})^{ i}_j (T^{a})^{ k}_l= \delta^i_l\delta^k_j -\frac{1}{N}\delta^{i}_j\delta^k_l\eqncom
\end{equation}
where the sum is over all the $N^2-1$ generators of the gauge group SU$(N)$.

\paragraph{The superconformal algebra}The action is invariant under the global supersymmetry algebra $\mathfrak{psu}(2,2|4)$, whose generators are the Lorentz generators $\mathfrak{J}_{\alpha\beta}$ and $\mathfrak{J}_{\dot \alpha \dot \beta}$, the translations $\mathfrak{P}_{\alpha\dot\alpha}$, the dilatation $\mathfrak{D}$, the special conformal transformations $\mathfrak{K}_{\alpha\dot\alpha}$,  the internal R-symmetry generators $\mathfrak{R}^a_{\phantom{a}b}$, the super translations $\mathfrak{Q}_{\alpha a}$ and $\bar {\mathfrak{Q}}_{\phantom{a}\dot\alpha}^{a}$ and the special superconformal transformations $\mathfrak{S}^{\alpha a}$ and $\bar {\mathfrak{S}}^{\dot\alpha}_{\phantom{\alpha} a}$. Among the many commutation and anti-commutation relations they satisfy, we recall the commutation relations for the dilatation generator with the other generators 
\begin{align}
\label{commrel}
&[\mathfrak{D},\mathfrak{P}_{\alpha\dot\alpha}]=i\mathfrak{P}_{\alpha\dot\alpha}\eqncom &&[\mathfrak{D},\mathfrak{Q}_{\alpha a}]=\frac{i}{2}\mathfrak{Q}_{\alpha a}\eqncom\notag\\
&[\mathfrak{D},\mathfrak{K}_{\alpha\dot\alpha}]=-i\mathfrak{K}_{\alpha\dot\alpha}\eqncom &&[\mathfrak{D},\bar {\mathfrak{Q}}_{\phantom{a}\dot\alpha}^{a}]=\frac{i}{2}\bar {\mathfrak{Q}}_{\phantom{a}\dot\alpha}^{a}\eqncom\notag\\
&[\mathfrak{D},J_{\alpha\beta}]=0\eqncom &&[\mathfrak{D},\mathfrak{S}^{\alpha a}]=-\frac{i}{2}\mathfrak{S}^{\alpha a}\eqncom\notag\\
&[\mathfrak{D},J_{\dot\alpha\dot\beta}]=0\eqncom &&[\mathfrak{D},\bar{\mathfrak{S}}^{\dot\alpha}_{\phantom{\alpha} a}]=-\frac{i}{2}\bar{ \mathfrak{S}}^{\dot\alpha}_{\phantom{\alpha} a}\eqndot
\end{align}
These commutation relations show that $\mathfrak{P}$, $\mathfrak{Q}$ and $\bar{\mathfrak{Q}}$ act as raising operators with respect to the eigenvalues of $\mathfrak{D}$ whereas $\mathfrak{K}$, $\mathfrak{S}$ and $\bar {\mathfrak{S}}$ act as lowering operators.
Physical states are organized into unitary irreducible representations of the superconformal algebra $\mathfrak{psu}(2,2|4)$, also called superconformal multiplets. Because the rank of this symmetry algebra is six, a representation is labeled by a $6$-tuple of numbers, $[j_1,j_2,\Delta, R_1,R_2,R_3]$ which consists of the Lorentz spins $j_1$ and $j_2$, the conformal or scaling dimension $\Delta$ and the three $R$-symmetry labels $R_i$. According to the operator-state correspondence, the states of in a (Euclidean) conformal theory are obtained by acting on the vacuum with gauge-invariant local composite operators in a small neighborhood of the origin\footnote{To compute correlation functions we always Wick rotate to Euclidean space.}. These gauge invariant local composite operators of the theory are given by traces of products of (covariant derivatives of) fundamental fields 
$\Phi_\ri$ for $i=1,\dots, L$, as
\begin{equation}
\label{eq:composite operator1}
\mathcal{O}(x)=\Tr\left(D^{k_1}\Phi_1(x)D^{k_2}\Phi_2(x)\cdots D^{k_L}\Phi_L(x)\right)\eqndot
\end{equation}
Each field $\Phi_i$ is one of the six scalars $\phi_{ab}$, or one of the four fermions $\bar{\psi}_{a\dot{\alpha}}$, the four anti-fermions $\psi_{a\alpha}$, or the  self-dual or anti-self-dual part of the strength\footnote{The gauge field $A_{\alpha\dot\alpha}$ itself does not transform covariantly under gauge transformations, and therefore the gauge invariant operators can only contain the corresponding field strength \eqref{fieldstrengthdef}, which does transform covariantly.} $\bar F_{\dot\alpha\dot\beta}$ and $F_{\alpha\beta}$. The length of the operator $\mathcal{O}$ is denoted by $L$.
Note that the operator in \eqref{eq:composite operator1} is a single trace operator. Of course, one can in principle also consider multi-trace operators. The action of the dilatation generator $\mathfrak{D}$ on an operator $\calO$ of definite scaling dimension $\Delta$ is as follows. Upon rescaling space-time by $x\rightarrow \lambda x$ an operator scales as $\calO(x)\rightarrow \lambda^{\Delta}\calO(\lambda x)$, 
or
\begin{equation}
[\mathfrak{D},\calO(x)]=i(\Delta +x \frac{\partial}{\partial x})\mathcal{O}(x)\eqndot
\end{equation}
Recall that the operators $\mathfrak{P}$ and $\mathfrak{Q}$ act as raising operators with respect to the eigenvalues $i\Delta$ of $\mathfrak{D}$ and $\mathfrak{K}$ and $\mathfrak{S}$ as lowering operators. For example, if $\calO(0)$ is a non-primary state of scaling dimension $\Delta$, then $\mathfrak{K}$ creates a state of scaling dimension $\Delta -1$ via
\begin{equation}
[\mathfrak{D},[\mathfrak{K}_{\alpha\dot\alpha},\calO(0)]]=i(\Delta -1)\calO(0)\eqndot
\end{equation}
Because of unitarity the scaling dimension $\Delta$ is required to be greater than or equal to zero. Physically this is because the corresponding eigenstate is mapped to a string state of energy $\Delta$. Since the generators $\mathfrak{K}$ and $\mathfrak{S}$ decrease the conformal dimension by $1$ and $1/2$ respectively, and the other supersymmetry generators either raise the value of $\Delta$ by $1$ or $1/2$ or leave it invariant, we can start from any state in a certain multiplet, and by acting with lowering operators eventually obtain a state with the lowest conformal dimension. This state is called a conformal primary and corresponds to a lowest weight state\footnote{Mathematicians often prefer to work instead with highest weight states.}.
\section{Correlation functions and the dilatation operator}
\label{oneloopdilspacetime}
\begin{figure}[h!]\centering
\def\svgwidth{\linewidth}
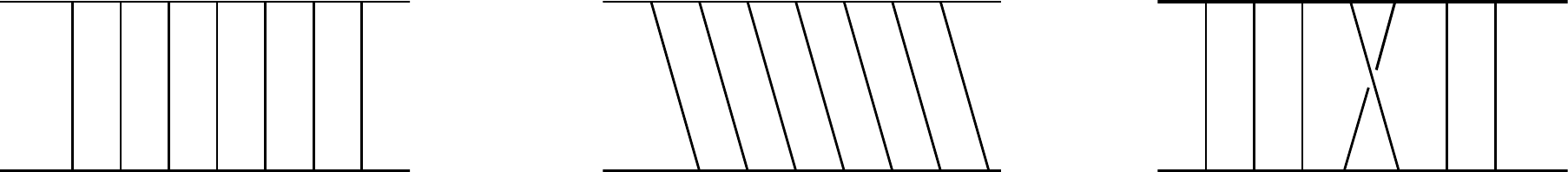
\caption{Tree-level Feynman diagrams of two single trace operators of equal length. Two planar graphs on the left and middle and a non-planar graph on the right. The horizontal lines represent the operator, with a field sitting on each intersection with a vertical line. Each of the vertical lines represents a propagator connecting two fundamental fields.}
\label{planarcorrelation}
\end{figure}

\begin{figure}[h!]\centering
\def\svgwidth{\linewidth}
 \scalebox{.3}{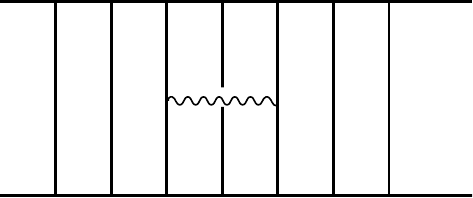}
\caption{Gluon exchange between two non-nearest-neighboring propagators corresponds to a non-planar one-loop diagram, which is suppressed in the planar limit.}
\label{nonplanar1loop}
\end{figure}

\begin{figure}[h!]\centering
\def\svgwidth{\linewidth}
 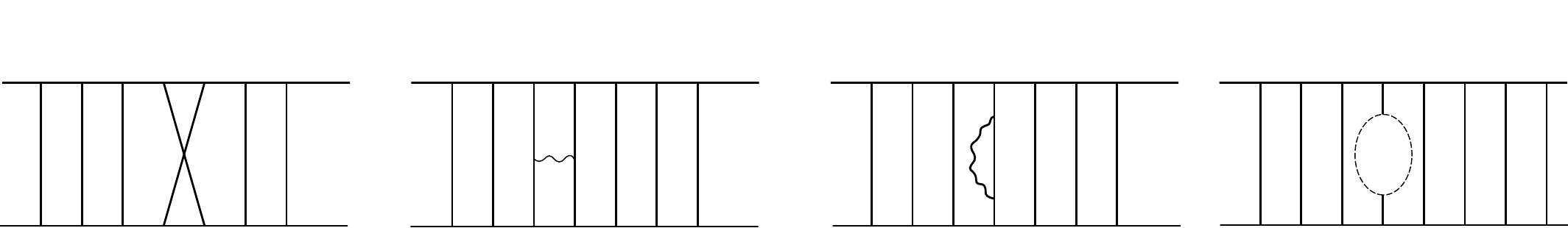
\caption{All planar one-loop diagrams. From left to right: a quartic scalar field interaction, a gluon exchange, a gluon self-interaction and a fermion loop.}
\label{planar1loop}
\end{figure}

The two- and three-point correlation functions of $\calN=4$ SYM constitute the building blocks for all other higher point functions of the theory via the so-called operator product expansion. Conformal symmetry greatly restricts the two-point correlation function of any two scalar primary operators of equal conformal dimension $\Delta$. For scalar operators it is of the form (up to a normalization constant)
\begin{equation}
\vac{\calO(x)\overline\calO'(y)}=\frac{1}{|x-y|^{2\Delta}}\eqndot
\end{equation}
Similar, slightly more complicated expressions hold for other non-scalar operators. Therefore, we can reformulate the problem of solving all the two-point functions for all operators of the theory as finding all their scaling dimensions $\Delta$. At tree-level, the scaling dimension is just the half-integer- or integer-valued classical scaling dimension, also called the bare dimension, and is denoted by $\Delta_0$. At higher loop orders, the scaling dimension generically receives quantum corrections. Solving the two-point functions to higher order in perturbation theory yields higher order corrections to the bare dimension, 
\begin{equation}
\Delta = \Delta_0+ g_{\text{YM}}^2\Delta_{1\text{-loop}}+\calO(g_{\text{YM}}^4).
\end{equation}
The quantum correction to the bare dimension $\gamma\equiv\Delta-\Delta_0$ is called the anomalous dimension. The anomalous dimensions for all states form  an infinite matrix. Together with the bare dimensions this matrix is called the dilatation operator. This dilatation operator is however not diagonal, in the sense that it mixes distinct operators at loop level. However, it is closed when restricted to certain sectors, meaning that operators within a specific sector mix with each other under the action of the dilatation operator but not with operators outside of that sector.
\paragraph{The one-loop dilatation operator in the SO$(6)$ sector}
In this paragraph we briefly summarize the basics of the computation of the one-loop dilatation operator in the so-called SO$(6)$ sector\footnote{The SO$(6)$ sector is not closed beyond one loop.}. We will not explicitly do the computation, but rather sketch how it can be done following the original paper \cite{Minahan:2002ve} and the review \cite{Minahan:2010js}. Many more details can be found in these two sources. The SO$(6)$ sector consists of local composite operators built exclusively out of scalar fields without covariant derivatives,
\begin{equation}
\label{eq:defoperatorO}
\mathcal{O}(x)\colonequals\, \Tr \,(\phi_{a_1b_1}\cdots \phi_{a_Lb_L})(x)\eqncom
\end{equation} where the indices $a_i,b_i$ are antisymmetric fundamental SU$(4)$ indices. Keeping the indices unspecified, the one-loop dilatation operator in this sector can be found by computing the one-loop correlation function of $\mathcal{O}$ with itself, because the scalar sector is closed at this loop order and mixing between operators of different number of traces is suppressed by a power of $1/N$.
Let us stress once more that we are considering the planar limit in which all the Feynman diagrams that contribute can be drawn on a plane, which we exemplify\footnote{In fact, due to the trace, one should imagine the pairs of horizontal lines to be parallel circles, and imagine the diagram to be like a barrel.} in Figure~\ref{planarcorrelation}.
At tree-level and in the planar limit, the two-point function is just
\begin{equation}
\vac{\calO(x)\overline\calO(y)}_{\text{tree}}=\sum_{\sigma}\prod_{i=1}^L\vac{\phi_{a_ib_i}(x)\overline\phi_{a_{\sigma(i)}b_{\sigma(i)}}(y)}_{\text{tree}}
\end{equation}
where the sum is over all cyclic permutations of the $L$ elements and the bar indicates Hermitian conjugation. Furthermore, the tree-level propagator between two scalar fields reads
\begin{equation}
\label{eq:treelevelpropagatorposspace}
\langle \phi_{ab}(x)\overline\phi_{a_{\sigma(i)}b_{\sigma(i)}}(y)\rangle_{\text{tree}}=\frac{1}{(2\pi)^2}\frac{\epsilon_{aba_{\sigma(i)}b_{\sigma(i)}}}{|x-y|^2} \eqncom
\end{equation} where we have suppressed the color indices.
Moving on to one-loop order, and as always in the planar limit, only nearest-neighbor and self-energy terms contribute, any other diagram is non planar, see  Figure~\ref{nonplanar1loop}.
Because only at most nearest-neighbor interaction can occur, it suffices to compute the first order correction to the subcorrelator 
\begin{equation}
\label{eq:oneloopsub}
\langle \phi_{ab}\phi_{cd}(x)\overline\phi_{a'b'}\overline\phi_{c'd'}(y)\rangle_{\text{1-loop}}\eqncom
\end{equation} 
where for brevity we denoted $a,b,c,d=a_i,b_i, a_{i+1},b_{i+1}$ and the primed indices
$a',b'c'd'=a_{\sigma(i)},b_{\sigma(i)},a_{\sigma(i+1)},b_{\sigma(i+1)}$, for some cyclic permutation $\sigma$.
For this subcorrelator the corresponding Feynman diagrams that appear are diagrams containing a four-vertex, a gluon exchange between two adjacent propagators and self-energy corrections, which are depicted in   Figure~\ref{planar1loop}. 
Computing these Feynman graphs and summing them gives \eqref{eq:oneloopsub}, which due to SO$(6)$ invariance
must be proportional to
\begin{equation}
\label{SU4structure}
(\textsf{A}\epsilon_{abcd}\epsilon_{a'b'c'd'}+\textsf{B}\epsilon_{aba'b'}\epsilon_{cdc'd'}+\textsf{C} \epsilon_{abc'd'}\epsilon_{a'b'cd})\times (\text{UV-divergent integral})\eqncom
\end{equation}
for some coefficients $\textsf{A},\textsf{B},\textsf{C}$ that are to be determined. Clearly, the right three diagrams of  Figure~\ref{planar1loop} only contribute to terms that preserve the SU$(4)$ index structure, which corresponds only to coefficient $\textsf{C}$. The left diagram contributes also to coefficient $\textsf{C}$, but must also completely give the values of the coefficients $\textsf{A}$ and $\textsf{B}$. Computing this left diagram results in $\textsf{A}=-1/2$ and $\textsf{B}=-1$. Now, we only need to determine the coefficient $\textsf{C}$. We can circumvent computing the remaining three Feynman diagrams by using a little trick. For some special operators, satisfying a certain unitarity bound, the scaling dimension is protected from quantum corrections. Therefore, their full scaling dimension is given by just the bare dimension and the anomalous dimension is zero to all loop orders. One such state is $\Tr(\phi_{12}^2)$. For this state in particular, the one-loop anomalous dimension must be zero. Setting the coefficients $a,b,c,d=1,2,1,2$ and $a',b',c',d'=3,4,3,4$ we straightforwardly read off that $\textsf{B}+\textsf{C}=0$ and hence $\textsf{C}=1$. 
Finally, we need to evaluate the logarithmically UV-divergent integral, which is (up to some normalization constant)
\begin{equation}
\int \dd^4 z \frac{1}{|x-z|^4}\frac{1}{|y-z|^4}\eqndot
\end{equation}
This divergent integral must be regularized using some convenient regularization scheme\footnote{Since the theory is conformal, the coupling is non-renormalized and the anomalous dimensions do not depend on the regularization scheme.} after which the one-loop dilatation operator can be found by equating \begin{align}
\label{eq:onelooppropagatorposspace}
 \langle \phi_{a_ib_i}\phi_{a_{i+1}b_{i+1}}(x)&\overline\phi_{a_{\sigma(i)}b_{\sigma(i)}}\overline\phi_{a_{\sigma(i+1)}b_{\sigma(i+1)}}(y)\rangle_{\text{UV, 1-loop}}\notag\\
 &= \mathcal{D}_{\text{1-loop}} \langle \phi_{a_ib_i}\phi_{a_{i+1}b_{i+1}}(x)\overline\phi_{a_{\sigma(i)}b_{\sigma(i)}}\overline\phi_{a_{\sigma(i+1)}b_{\sigma(i+1)}}(y)\rangle_{\text{tree}} \eqncom
\end{align}
and solving for $\mathcal{D}_{\text{1-loop}},$ where the lefthand side is found as indicated above\footnote{In fact, $\mathcal{D}_{\text{1-loop}}$ is the one-loop dilatation operator \textit{density}. To find the full dilatation operator one sums over all pairs of nearest neighbors.} and the correlator on the righthand side of the equation is given by \eqref{eq:treelevelpropagatorposspace}. Here, the lefthand side denotes the coefficient of the UV-divergence of the one-loop correlation function \eqref{eq:oneloopsub} and $\mathcal{D}_{\text{1-loop}}$ is
\begin{equation}
\mathcal{D}_{\text{1-loop}}=\frac{g_{YM}^2N}{8\pi^2}\left(\frac{1}{2}\epsilon_{abcd}\epsilon_{a'b'c'd'}-\epsilon_{aba'b'}\epsilon_{cdc'd'}+\epsilon_{abc'd'}\epsilon_{a'b'cd}\right)\eqndot
\end{equation}
The Levi-Civita tensor $\epsilon_{abcd}\epsilon_{a'b'c'd'}$ is called the Trace operator, as it traces over the SU$(4)$ indices and is denoted by $K$, the tensor $\epsilon_{aba'b'}\epsilon_{cdc'd'}$ is the Permutation operator $P$, because it permutes two pairs of indices and finally, $\epsilon_{abc'd'}\epsilon_{a'b'cd}$ is the identity operator, $\text{Id}$. The one-loop dilatation operator in the SO$(6)$ sector can thus be written as
\beq
\label{dilintermsoftr}
\Gamma_{\text{SO}(6)}=\frac{g_{YM}^2N}{8\pi^2}\sum_{\ell=1}^L\left(\text{Id}-P_{\ell,\ell+1}+\frac{1}{2}K_{\ell,\ell+1}\right)\eqncom
\eeq
where the sum is over all the elementary fields of the local composite operator. This operator can be interpreted as the Hamiltonian of a spin chain, and it was discovered by Minahan and Zarembo that this spin chain was integrable. The Hamiltonian density, which is written between parentheses acts on each site of the chain. Interestingly, the Trace, Permutation and Identity operators can be recombined as projectors onto irreducible representations. Namely, the tensor product of two antisymmetric SU$(4)$ representations is reducible and decomposes into the direct sum of the antisymmetric, the symmetric traceless and the singlet representation according to
\begin{equation} \begin{ytableau}
$ $\cr
$ $ \cr
\end{ytableau} \otimes  \begin{ytableau}
$ $\cr
$ $ \cr
\end{ytableau} =  
\begin{ytableau}
$ $& $ $\cr
$ $ \cr
$ $\cr
\end{ytableau}
\oplus
\begin{ytableau}
$ $ & $ $\cr
$ $ & $ $\cr
\end{ytableau}\oplus  \begin{ytableau}
$ $\cr
$ $ \cr
$ $\cr
$ $ \cr
\end{ytableau}\eqndot
\end{equation} 
 The Trace, Permutation and Identity operator are now combined into operators that project onto these three irreducible representations as
 \begin{align}
& \Pi^{\text{antisymmetric}}=\frac{1}{2}(\text{Id}-P)\eqncom\notag\\
&\Pi^{\text{symmetric, traceless}}=\frac{1}{2}(\text{Id}+P)-\frac{1}{6}K\eqncom\notag\\
&\Pi^{\text{singlet}}=\frac{1}{6}K\eqndot
 \end{align}
Then $\mathcal{D}_{\text{1-loop}} \propto 0\; \Pi^{\text{symmetric, traceless}}+ 2\;  \Pi^{\text{antisymmetric}} + 3 \;\Pi^{\text{singlet}}$. The coefficients $0,2,3$ are twice the first three so-called harmonic numbers, defined by $h_i=\sum_{j=1}^i 1/j$, and $h_0=0$.
\paragraph{The full one-loop dilatation operator}
The decomposition in terms of projectors is useful when one wants to lift the result to the full one-loop dilatation operator. Namely, extending beyond the SO$(6)$ sector, the elementary fields are in the singleton representation of the full superconformal algebra $\mathfrak{psu}(2,2|4)$. The tensor product of two such representations decomposes into an infinite sum of a numbered series of representations that starts with the symmetric traceless ($i=0$), followed by the antisymmetric ($i=1$), the singlet ($i=2$) and continuing ($i=3,\dots$) with representations of increasing scaling dimension and spin numbers. The full one-loop dilatation operator can then be written as the sum of the corresponding set of projectors $\Pi^i$ with respect to this expansion with coefficients given by twice the harmonic numbers \cite{Beisert:2003yb,Beisert:2003jj},
\begin{equation}
\label{fulloneloop}
\Gamma=\frac{g_{YM}^2N}{8\pi^2}\sum_{\ell=1}^L\sum_{i=0}^{\infty}2h_i \Pi_{\ell,\ell+1}^i \eqncom
\eeq
where the projector $\Pi_{\ell,\ell+1}^i$ acts on two adjacent spin sites and $L+1=1$.
This result was initially found by symmetry arguments and it was remarkably non trivial to find a field theoretic derivation. First it was observed that the one-loop dilatation operator is essentially given by the four point amplitude\footnote{Up to a regulating piece.} and that it admits a remarkably simple expression in so-called spinor-helicity variables or harmonic oscillators \cite{Zwiebel:2011bx}. This form of the dilatation operator was eventually proven using field theory rather than by symmetry arguments in \cite{Wilhelm:2014qua}. At higher loops and restricted to certain sectors, the dilatation operator has been computed at two \cite{Eden:2005bt,Belitsky:2005bu,Georgiou:2011xj}, three \cite{Beisert:2003ys,Eden:2005ta,Sieg:2010tz} and four loops \cite{Beisert:2007hz}.
\section{Scattering amplitudes in $\calN=4$ SYM}
\label{sec:scatteringamps}
Together with correlation functions scattering amplitudes are perhaps the most fundamental quantities that one can compute in a quantum field theory. An amplitude gives the expectation value of a state of $n$ on-shell particles with the vacuum and is depicted in   Figure~\ref{amplitudepic}. 
 \begin{figure}[h!]
 \fontsize{1cm}{1em}\centering
\def\svgwidth{\linewidth}
\scalebox{.5}{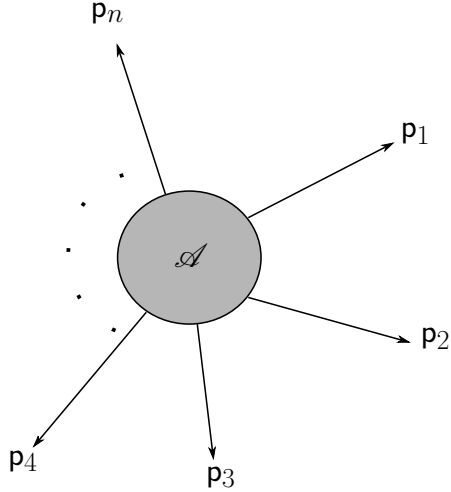}
\caption{An amplitude with $n$ on-shell external particles.}
\label{amplitudepic}
\end{figure}
They originate from correlation functions by setting the external legs on shell via the Lehmann-Symanzik-Zimmermann (LSZ) reduction \cite{Lehmann1955}.
The condition for the external particles to be on their mass shell is given by $p_i^2=m_i^2$. In the case of $\calN=4$ SYM all particles are massless so that the mass-shell condition reads $p_i^2=0$. Although amplitudes can be computed by summing Feynman diagrams and setting the external legs on shell, it turns out that this procedure is blind to a lot of the structure that amplitudes exhibit which renders this approach pretty inefficient. Feynman diagrams are manifestly local and unitary but this comes at the price of virtual particles and gauge redundancies. Giving up manifest locality and unitarity at intermediate stages of the calculation allows one to fully exploit the symmetries of the amplitude and uncover its simplicity. In this section we will review some of the basic concepts, definitions and techniques that can be used to compute amplitudes. Amplitudes admit several decompositions. First of all, at weak coupling they have a loop expansion. This allows us to compute the amplitude order by order in the powers of the coupling constant. In this thesis we will only be concerned with tree-level amplitudes and henceforth we will always imply tree-level unless explicitly stated otherwise. Second, they admit a color decomposition, which we discuss in the first paragraph. Then, we introduce the spinor-helicity variables, which are related to twistor variables and are an essential tool for exploiting the on-shell structure of both form factors and amplitudes. We will exemplify the simplicity of the amplitude by discussing amplitudes of a certain class, called maximally helicity violating, which have an extremely simple expression despite being the sum of a tremendous number of Feynman diagrams. Finally, we will briefly explain one of the techniques that have been created for on-shell objects over the past years. Excellent and more extended reviews on amplitudes are \cite{Alday:2008yw,Dixon:2013uaa}. 
\paragraph{Color ordering}
Scattering amplitudes can be stripped of their color structure. The result is called the color-ordered amplitude. In $\calN=4$ SYM, the color-ordered amplitude can be found by decomposing the full amplitude of $n$ external particles as
\begin{equation}
\begin{split}
 \mathscr{A}_n^{\text{tree}}(\{p_i,\epsilon_i, a_i\}) = g^{n-2}_{\text{YM}}\hspace{-0.3cm}\sum_{\sigma\in S_n/\mathbb{Z}_n} \hspace{-0.3cm}\mathscr{A}^{\text{tree}} _{\sigma}(\sigma(1),\dots, \sigma(n)) \Tr(T^{a_{\sigma(1)}}\cdots T^{a_{\sigma(n)}}) \\
 + \text{multi trace terms}\eqncom\end{split}
\end{equation}
where $p_i$ are the on-shell momenta, $\epsilon_i$ the polarizations, $a_i$ the color indices, $T^{a_{\sigma(i)}}$ are the generators of the gauge group SU$(N)$, and the sum is over all non-cyclic permutations $\sigma$. In the planar limit where $N\to \infty$, the multi-trace terms are suppressed. The color-ordered amplitude is given by the coefficient of the trace, $\mathscr{A}_{\sigma}$. In the rest of this work we will only concern ourselves with this reduced amplitude and with a slight abuse of notation just refer to it as amplitude. 
\paragraph{Spinor-helicity formalism}The color-orded amplitude is a function only of the kinematical data. This kinematical data can be conveniently expressed using the spinor-helicity formalism that we will explain in this paragraph. It is based on the group theoretical fact that SO$(3,1;\mathbb{C})$ is locally isomorphic to SL$(2,\mathbb{C})\times$SL$(2,\mathbb{C})$. The four dimensional Lorentz vector is in the $(\tfrac{1}{2},\tfrac{1}{2})$ of SL$(2,\mathbb{C})\times$SL$(2,\mathbb{C})$. Therefore, a Lorentz four-vector $p^{\mu}$ can be expressed as a bi-spinor $\fp_{\alpha\dot\alpha}$, carrying an undotted and a dotted SL$(2,\mathbb{C})$ index. Explicitly, this can be realized by contracting a four-vector $p_i^{\mu}$ as
\begin{equation}
\fp_{i\alpha\dot\alpha}=p_i^{\mu}\sigma_{\mu\alpha\dot\alpha}\eqncom
\end{equation} where $\sigma_{\mu}$ for $\mu=1,2,3$ are the Pauli matrices and $\sigma_0=\mathbb{Id}$. We will always denote a Lorentz four vector by $p^{\mu}$ and its corresponding $2\times 2$-matrix by $\fp_{\alpha\dot\alpha}$. 
For $p_i^{\mu}$ a real four-vector, its corresponding $2\times2$ matrix $\fp_i$ is Hermitian. 
Furthermore, for massless particles, the condition for a four-momentum to be on the mass-shell becomes $0=p_i^{\mu}p_{i\mu} =  \det (\fp_{i\alpha\dot\alpha})$. 
This means that the $2\times 2$ matrix $\fp_{i\alpha\dot\alpha}$ is not of maximal rank and factorizes into the product of two $2$-spinors $p_{i\alpha}$ and $\tilde p_{i\dot\alpha}$ of opposite chirality as\footnote{It is more common to denote spinor helicity variables by $\la_{\alpha}$ and $\tilde\la_{\dot\alpha}$. However, we choose to always denote these by $p$ $\tilde p$ to make the connection with the later part of this work more explicit. Furthermore, the spinor $\la$ is reserved for the closely related spinor that forms the top component of the twistor as will be introduced in Section~\ref{review2}.}
\begin{equation}
\label{decompvec}
\fp_{i\alpha\dot\alpha}=p_{i\alpha}\tilde p_{i\dot\alpha}\eqndot
\end{equation}
The repeated index $i$ is not summed over in this equation. Here, the spinor $p$ transforms in the $(1/2,0)$ and the spinor $\tilde p$ in the $(0,1/2)$. For real momenta the two spinors are related to each other depending on the signature of space-time. In Lorentzian signature, $\tilde p= \bar p$. 
Out of two spinors $p_1$ and $p_2$ we can form the invariant scalar 
\begin{equation}
\abra{1}{2}\equiv\abra{p_1}{p_2}=\epsilon^{\alpha\beta}p_{1\alpha}p_{2\beta}\eqncom
\end{equation}
where $\epsilon^{\alpha\beta}$ is the two-dimensional Levi-Civita tensor and the repeated indices $\alpha$ and $\beta$ are summed over.
Similarly, the two spinors $\tilde p_1$ and $\tilde p_2$ of the opposite chirality can be contracted using the two-dimensional Levi-Civita tensor $\epsilon^{\dot\alpha\dot\beta}$, 
\begin{equation}
[12]\equiv[\tilde p_1\tilde p_2]=\epsilon^{\dot\alpha\dot\beta}\tilde p_{1\dot\alpha}\tilde p_{2\dot\beta}\eqndot
\end{equation} 
These two contractions are antisymmetric under exchange of the two spinors and therefore vanish whenever the two spinors are proportional and hence
\begin{align}
&\abra{1}{1}=p_{1\alpha}p_1^{\alpha}=0\eqncom\notag\\
&[11]=\tilde p_{1\dot\alpha}\tilde p_1^{\dot\alpha}=0\eqndot
\end{align}
Furthermore, for two on-shell momenta $\fp_1$ and $\fp_2$, we can write
\begin{equation}
(p_1+p_2)^2=2 p_1\cdot p_2 = \abra{1}{2}[21]\eqndot
\end{equation}
Aditionally, the two contractions satisfy the so-called Schouten identity,
\begin{align}
&\abra{1}{2}\abra{3}{4}+\abra{1}{3}\abra{4}{2}+\abra{1}{4}\abra{2}{3}=0\eqncom\notag\\
&[12][34]+[13][24]+[14][23]=0\eqndot
\end{align}
The Schouten identity turns out to be very useful for algebraic manipulations when computing amplitudes and form factors. From the perspective of the field content, the fermions of the theory have polarization $p_{\alpha}$ and $\tilde p_{\dot\alpha}$ of helicity $-1/2$ and $+1/2$ respectively. The polarizations of the positive and negative helicity gluons can only be defined with the use of auxiliary spinors $\zeta$ and $\tilde\zeta$ via
\begin{equation}
\label{polvectors}
\epsilon^+_{\alpha\dot\alpha}=\frac{\zeta_{\alpha}\tilde p_{\dot\alpha}}{\abra{\zeta}{p}},\quad \quad \epsilon^-_{\alpha\dot\alpha}=\frac{p_{\alpha}\tilde\zeta_{\dot\alpha}}{[\tilde\zeta\tilde p]} \eqncom
\end{equation}
which are of helicity $+1$ and $-1$ respectively. It follows directly from this definition that $\fp^{\alpha\dot\alpha}\epsilon^+_{\alpha\dot\alpha} = \tilde p^{\dot\alpha}\tilde p_{\dot\alpha}=0$ and $\fp^{\alpha\dot\alpha}\epsilon^-_{\alpha\dot\alpha} =  p^{\alpha}p_{\alpha}=0$.
The decomposition of the momentum $\fp_{i\alpha\dot\alpha}$ into two two-spinors $p_{\alpha}$ and $\tilde p_{\dot\alpha}$ is not unique, as the rescaling
\begin{equation}\label{littlegroup}
p_{\alpha}\mapsto t p_{\alpha}, \quad\quad \tilde p_{\dot\alpha}\mapsto t^{-1}\tilde p_{\dot\alpha}\eqncom
\end{equation}
where $t$ is a nonzero complex number, leaves the four-momentum $\fp$ invariant. For $p^{\mu}$ a real four vector, i.e.\ $\tilde p =\bar p$, the complex number $t$ is a phase. The behavior of this rescaling for any particle is labeled by its helicity: for a particle of helicity $h$, its polarization scales under \eqref{littlegroup} as $t^{-2h}$. The color-ordered amplitude is exclusively determined by the helicities and momenta of its outgoing particles. The helicity configuration gives rise to a very useful classification of the amplitude that is called its MHV degree, which we explain in the next paragraph.
\paragraph{Maximally helicity violating (MHV) amplitude}Let us review the MHV degree as it plays an essential role also for the form factors that we will discuss later on. The condition that the total helicity of an $n$-particle scattering process is conserved is written as 
\begin{equation}
\sum_{i=1}^n h_i=0\eqncom
\end{equation}
where $h_i$ is the helicity of the $i$th particle considered as an outgoing particle. An incoming particle of momentum $p$ and helicity $h$ is considered an outgoing particle of momentum $-p$ and helicity $-h$. For example, in a process with three incoming particles of helicity $+1$ (equivalently, three outgoing particles of helicity $-1$) and three outgoing particles of helicity $+1$ the total helicity is conserved. An maximally helicity violating (MHV) amplitude of $n$ external particles is an amplitude with a helicity configuration that maximizes the sum $\sum_{i=1}^n h_i$, while the amplitude is nonzero. Clearly, the sum is maximal for a scattering process of $n$ gluons all of helicity $+1$. 
However, this amplitude as well as the next (i.e.\ the one with $n-1$ positive and one negative helicity gluon) vanish as we shall now demonstrate.
At tree level, each Feynman graph is a function of at most $n-2$ momenta and all $n$ polarization vectors\footnote{This is because these amplitudes are built out of $3$- and $4$-valent gluon vertices exclusively, of which only the $3$-vertices carry (one power of) momentum. Any $n$-legged Feynman graph contains at most $n-2$  3-vertices at tree level. All indices must be contracted and there are thus at most $n-2$ contractions between a momentum and a polarization vector.}. Therefore, the expression for the amplitude contains at least one contraction between a pair of polarization vectors, $\epsilon_i\cdot \epsilon_j$. For the all-plus amplitude this contraction is of the form $\epsilon_i^+\cdot \epsilon_j^+$ which vanishes trivially upon choosing $\zeta_i=\zeta_j$ for the auxiliary spinor $\zeta_i$ in the polarization vectors \eqref{polvectors}. For the next amplitude, i.e.\ the one with precisely one negative helicity gluon, say at position $1$, we can choose for the auxiliary spinors $\zeta_i=p_1$ for all $i\neq1$ and in addition $\zeta_1=p_2$. Now all contractions between polarization vectors vanish and hence the amplitude vanishes. For $\calN=4$ SYM these all-plus and all-but-one-plus amplitudes are in fact zero to all loop orders. The first non-vanishing amplitude has $n-2$ positive helicity gluons and $2$ negative helicity ones and is called the maximal helicity violating amplitude, or MHV amplitude. In fact, the same analysis can be done for the all-minus and all-but-one-minus amplitude, to see that these also vanish. The first non-vanishing amplitude is the $(n-2)$-minus amplitude, which is called $\overline{\text{MHV}}$.  \newline\\
Already a few decades ago, people realized that classifying amplitudes by the degree to which they violate conservation of helicity is the ``right'' thing to do, in the sense that amplitudes of vastly different number of external legs but equal MHV degree exhibit the same structure. Notably the amplitude of lowest MHV degree, the MHV amplitude, has a strikingly simple structure at tree level for any number of external particles. To illustrate this, let us consider an example of a tree-level scattering process of two gluons into three gluons, but bear in mind that similar considerations hold for all other types of scattering particles. Although the scattering process we are considering is of a very small number of particles, the Feynman diagram calculation involves around two thousand Feynman diagrams. However, all these thousands of terms taken together, the final result can be cast into a very simple form that fits on just one line! This expression is called the Parke-Taylor formula after its discoverers \cite{Parke:1986gb}, and reads
\begin{equation}
\label{eq:MHV amplitude5}
\mathscr{A}^{\text{MHV}}_5(1^+ , 2^-, 3^+, 4^- ,5^+) =\frac{\delta^{4}(\sum_{i=1}^5 \fp_i)\abra{2}{4}^4}{\abra{1}{2}\abra{2}{3}\abra{3}{4}\abra{4}{5}\abra{5}{1}} \eqncom
\end{equation}
where $\fp_i=p_{i\alpha}\tilde p_{i\dot\alpha}$ is the momentum of the $i$th particle and we recall that $\abra{i}{j}=\epsilon^{\alpha\beta}p_{i\alpha}p_{j\beta}$.
 In fact, one can straightforwardly generalize this formula to MHV amplitudes of any number $n$ of scattering gluons, two of which of negative helicity and the remaining ones of positive helicity, 
\begin{equation}
\label{eq:MHV amplitudePT}
\mathscr{A}^{\text{MHV}}_n(1^+, \dots , i^-, \dots, j^- , \dots n^+) =\frac{\delta^{4}(\sum_{i=1}^n \fp_i)\abra{i}{j}^4}{\prod_{k=1}^n\abra{k}{(k+1)}} \eqndot
\end{equation}
Let us emphasize once more that this formula holds for \textit{any} number of external legs $n$. In contrast, the number of corresponding Feynman diagrams that one would need to compute increases factorially with $n$. Even for the most powerful computers this number would very rapidly become too large. Interestingly, the corresponding $\overline{\text{MHV}}$ amplitude can be obtained by replacing all $p_i$'s by $\tilde p_i$, or equivalently, replacing all angular brackets $\langle a b \rangle $ by square brackets $[ab]$  in \eqref{eq:MHV amplitudePT}. Another striking feature is that this elegant Parke-Taylor formula also holds for tree-level gluon amplitudes in quantum chromodynamics (QCD). Indeed, at tree level the gluon scattering amplitudes of QCD and $\calN=4$ SYM  coincide. Computing these many-gluon amplitudes in QCD is extremely important for experiments that are done at colliders. For example at the LHC, in order to discover new physics, it is imperative to have extremely high precision control over the background physics, which is dominated by QCD. These scattering amplitudes, though identical in $\calN=4$ and QCD, are more easily computed in the framework of $\calN=4$ SYM due to the much bigger symmetry group there. Of course, at loop level the amplitudes of the two theories diverge from one another, but still many techniques that were developed in the context of loop amplitudes of $\calN=4$ are now standard for computing loop level amplitudes at the LHC. These so-called on-shell techniques that have been developed over the last few decades for scattering amplitudes in $\calN=4$ SYM include the BCFW recursion \cite{Britto:2004ap,PhysRevLett.94.181602}, the CSW recursion \cite{Cachazo:2004kj} and (generalized) unitarity \cite{Bern:1994zx,Bern:1994cg,Bern:1997sc,Britto:2004nc,Bern:1995db,Bern:1996fj,Bern:1996je,Bern:2004cz,Cachazo:2008vp} . All of these techniques share the fact that they circumvent the inefficient Feynman diagram computations and exploit the underlying simplicity of the amplitude.
\paragraph{CSW recursion for amplitudes}Let us finish this section by discussing an on-shell method that is a crucial motivation for the rest of this thesis.
It is called CSW recursion\footnote{This method also goes by the name ``MHV diagrams''.} after Cachazo, Svr\v{c}ek and Witten, who described it in \cite{Cachazo:2004kj}. According to this method a tree-level N$^k$MHV amplitude is decomposed into so-called $k+1$ MHV vertices, which are off-shell continuations of MHV amplitudes.
More precisely, an MHV vertex is obtained from an MHV amplitude, where one on-shell momentum is replaced by an off-shell momentum, and its corresponding spinors replaced by off-shell spinors. Let us explain how one defines an off-shell spinor from an off-shell momentum. Let $\boldsymbol{\ell}$ be an off-shell momentum, then we define a corresponding off-shell spinor by 
\begin{equation}
\label{offhsellcontin}
\ell_{\alpha} =\boldsymbol{\ell}_{\alpha\dot\alpha}\xi^{\dot\alpha}\eqncom
\end{equation} where $\xi^{\dot\alpha}$ is an arbitrary non-zero reference spinor.
The N$^k$MHV amplitude is constructed from $k+1$ off-shell MHV vertices that are connected via $k$ propagators, $1/\boldsymbol{\ell}_i^2$ for $i=1,\dots, k$. In this sense, the MHV amplitude and propagators are the fundamental building blocks from which all higher level amplitudes can be obtained. \newline\\
Let us illustrate this method by giving a simple example. We consider an amplitude with four external gluons. The $4$-legged MHV amplitude has $2$ positive and $2$ negative helicity gluons. Therefore, the next-to-MHV amplitude must have $3$ negative and $1$ positive helicity gluon. Let us assume the positive helicity gluon is the $4$th particle.
According to CSW we can split this diagram into two off-shell MHV vertices which are connected by a propagator. 
This can be done in precisely two ways, depicted in Figure~\ref{CSWamp}, which need to be summed.
\begin{figure}[h!]\centering
\def\svgwidth{\linewidth}
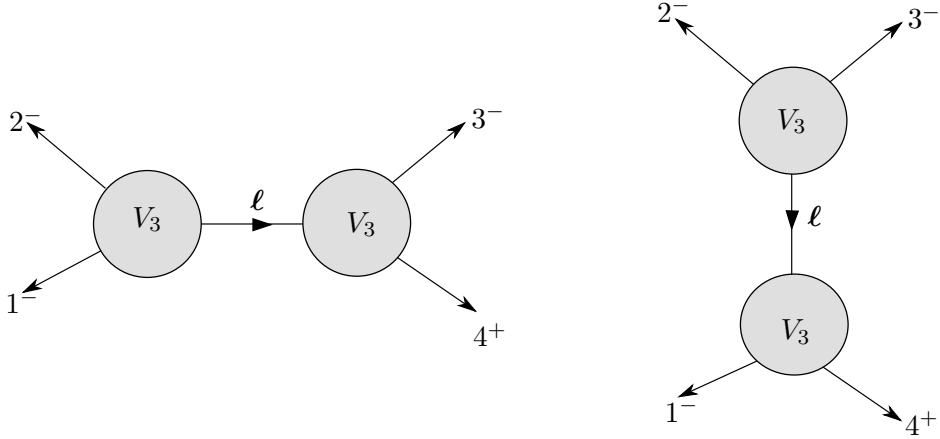
\caption{The decomposition of $A_4^{\text{NMHV}}(1^-,2^-,3^-,4^+)$ into MHV vertices.}
\label{CSWamp}
\end{figure}
We compute the diagram on the lefthand side of  Figure~\ref{CSWamp}, where a $3$-valent MHV vertex with two external negative helicity states of momentum $\fp_1$ and $\fp_2$ is connected by the propagator $1/{\boldsymbol{\ell}}^2$ to another $3$-valent MHV vertex with one negative and one positive helicity external states of momentum $\fp_3$ and $\fp_4$ respectively. At each vertex there is momentum conservation of the three momenta that enter the vertex. 
The expression for the left diagram is thus
\begin{equation}
V_3(1^-,2^-,\boldsymbol{\ell}^+)\frac{1}{\boldsymbol{\ell}^2}V_3(\boldsymbol{\ell}^-,3^-,4^+)= \frac{\abra{1}{2}^4}{\abra{1}{2}\abra{2}{{\ell}}\abra{{\ell}}{1}}\frac{1}{\boldsymbol{\ell}^2} \frac{\abra{{\ell}}{3}^4}{\abra{{\ell}}{3}\abra{3}{4}\abra{4}{{\ell}}}\delta^4(\fp_1+\fp_2+\boldsymbol{\ell})\delta^4(\fp_3+\fp_4-\boldsymbol{\ell})\eqndot
\end{equation}
Using momentum conservation, we can replace $\boldsymbol{\ell} = -\fp_1-\fp_2=\fp_3+\fp_4$. This gives,
\begin{equation}
 \frac{\abra{1}{2}^4}{\abra{1}{2}\abra{2}{{\ell}}\abra{{\ell}}{1}}\frac{1}{\boldsymbol{\ell}^2} \frac{\abra{{\ell}}{3}^4}{\abra{{\ell}}{3}\abra{3}{4}\abra{4}{{\ell}}}=\frac{[4\xi]^3\abra{3}{4}}{[1\xi][2\xi][3\xi][12]}\eqndot
\end{equation}
Furthermore, the two momentum conserving delta functions give rise to the overall momentum conserving delta function $\delta^4(\sum_{k=1}^4\fp_k)$.
The diagram on the righthand side can be obtained completely analogously and reads
\begin{equation}
\frac{[4\xi]^3\abra{2}{3}}{[1\xi][2\xi][3\xi][41]} \delta^4(\sum_{k=1}^4\fp_k)\eqndot
\end{equation}
After summing these, the total expression for the amplitude is
\begin{equation}
\mathscr{A}_4^{\text{NMHV}}(1^-,2^-,3^-,4^+)=\frac{[4\xi]^3}{[1\xi][2\xi][3\xi][12][41]} \left(\abra{3}{4}[41]+\abra{3}{2}[21]\right)\delta^4(\sum_{k=1}^4\fp_k),
\end{equation}
which vanishes 
\begin{equation}
\left(\abra{3}{4}[41]+\abra{3}{2}[21]\right)\delta^4(\sum_{k=1}^4\fp_k) = -\left(\abra{3}{3}[31]+\abra{3}{1}[11]\right)\delta^4(\sum_{k=1}^4\fp_k)=0\eqndot
\end{equation}
This result was expected since the $\overline{\text{MHV}}$ amplitude has (at least) two positive helicity gluons for the same reason that the MHV amplitude has at least two negative helicity gluons. Therefore, the only nonzero four-point amplitude is the amplitude with two positive and two negative helicity gluons. This amplitude is both MHV and $\overline{\text{MHV}}$. This concludes our discussion of CSW recursion for amplitudes. In the next section we introduce form factors and recapitulate how the CSW recursion can be applied to certain form factors as well. In Section~\ref{amptwistor} we review how CSW correspond to the Feynman rules in twistor space.
\section{Form factors in $\calN=4$ SYM}
\label{secformfactor}
The different techniques that were developed for amplitudes all use the on-shell character of the scattering amplitude at least to a certain extent. In the past few years a lot of progress has been made in adapting many of these techniques for computing quantities that are partially or completely off shell as well. Especially, so-called form factors of local composite operators, sharing some of the on-shell structure with amplitudes, can be computed using certain on-shell techniques. In this section we introduce form factors and describe how the CSW recursion can be extended to compute them. \newline\\
Let us start by introducing the form factor. The definition of a form factor\footnote{It is common to Fourier transform to momentum space.} of a local composite operator $\calO$ of off-shell momentum $\fq$ is the expectation value of the state created by the operator from the vacuum and a state of $n$ on-shell particles $\Phi_i$:
\begin{equation}
\label{eq:formfactordefinition}
\calF_{\calO}(1^{\Phi_1},\ldots, n^{\Phi_n};\fq)= \int \frac{\dd^4x}{(2\pi)^4}\, \e^{-i\fq x}\melement{\Phi_1(\fp_1)\cdots \Phi_n(\fp_n)}{\calO(x)}{0}\eqncom
\end{equation}
where $\fp_i$ with $\fp_i^2=0$ are the momenta of the on-shell particles and $n\ge L$ at tree level, and $L$ is the length of the operator. The form factor is schematically depicted in \ref{formfac1}.
\begin{figure}[h!]\centering
\def\svgwidth{\linewidth}
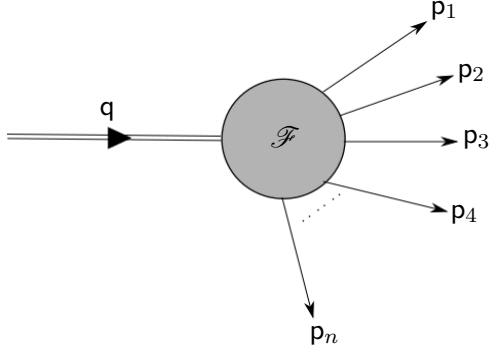
\caption{The form factor of an operator of off-shell momentum $\textbf{q}$ and $n$ on-shell external particles with momenta $\fp_i$.}
\label{formfac1}
\end{figure}
The local composite operators that we consider are trace operators which are color singlets and do not play a role in the color ordering of the external states. 
Therefore, in analogy to the amplitude, we can strip off the color structure from the external particles and consider the color-ordered form factor.\paragraph{Minimal form factors}A form factor with the lowest possible number of external legs is appropriately called a minimal form factor. The minimal form factor has the same number of outgoing external particles as fundamental fields that constitute the local composite operator. In $\calN=4$ SYM it was first defined and computed for the operator $\tfrac{1}{2}\Tr(\phi^2)$ by van Neerven \cite{vanNeerven:1985ja} up to two loops and is also called the Sudakov form factor\footnote{In fact, it is not uncommon in the literature to consider $\Tr(\phi^2)$ instead of $\tfrac{1}{2}\Tr(\phi^2)$ and discard the $2$ cyclic ways of Wick contracting the operator to the scattering states. Here, however we choose to keep track of all factors.}. 
The minimal form factor of $\tfrac{1}{2}\Tr(\phi^2)$ can be obtained from just the vertex of the theory with two outgoing scalars on shell. At tree-level, this minimal form factor is simply
\begin{equation}
\label{minimalformfactorscalar}
\mathscr{F}_{\tfrac{1}{2}\Tr(\phi^2)}^{\text{MHV}}(1^{\phi},2^{\phi};\fq)= 1\times \delta^4(\fq-\fp_1-\fp_2)\eqndot
\end{equation}
\paragraph{Non-minimal MHV form factors}
Next, one can compute non-minimal form factors for which the number of external on-shell particles is larger than the length of the operator. Analogously to amplitudes, form factors can be classified by the degree to which they violate helicity conservation. The MHV form factor is such that the sum of the helicity of the operator and the helicity of the external particles is maximal. Interestingly, MHV form factors exhibit a very similar structure to MHV amplitudes. For example, the MHV form factor of the operator $\tfrac{1}{2} \Tr(\phi^2)$, consisting of two identical scalars, with $(n-2)$ external positive helicity gluons is given by
\begin{equation}
\label{eq:formfactor2phi2}
\calF_{\tfrac{1}{2}\Tr(\phi^2)}(1^{+},\ldots, i^{\phi_{ab}},\ldots, j^{\phi_{ab}},\ldots, n^{+};\fq)
=-\frac{\abra{i}{j}^2\delta^4(\fq-\sum_{k=1}^n\fp_k)}{\abra{1}{2}\cdots \abra{n}{1}}\eqndot
\end{equation} 
One recognizes the Parke-Taylor denominator that also appeared in the expression for the MHV amplitude \eqref{eq:MHV amplitudePT}.
\begin{figure}[h!]\centering
\def\svgwidth{\linewidth}
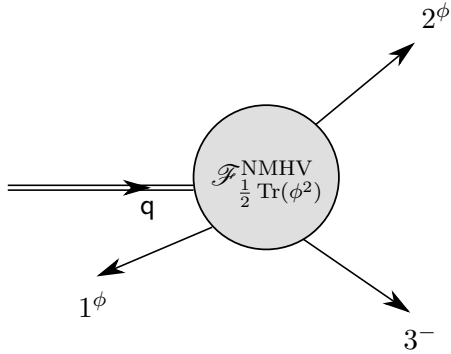
\caption{The NMHV form factor of the operator $\frac{1}{2}\Tr(\phi^2)$ with two external scalars and one negative helicity gluon.}
\label{formfactornmhv}
\end{figure}
\paragraph{CSW for NMHV form factors}
Although form factors are partially off shell, some techniques that were developed in the context of amplitudes can be extended to form factors. For example, in \cite{Brandhuber:2011tv} the MHV diagrams method was extended to form factors of the stress-tensor multiplet, by supplementing the MHV vertices and propagators that were the fundamental building blocks for amplitudes, by so-called \textit{operator vertices} (which are obtained from setting at least one of the external states of an MHV form factor off shell). We denote the operator vertices in momentum space by $W_n$, in close analogy to their twistor analogues that will be introduced later on in this thesis. Via the CSW recursion an NMHV form factor is constructed by gluing together an MHV form factor and an MHV vertex via an (off-shell) propagator.\newline\\
As an example, let us compute the NMHV form factor of the operator $\tfrac{1}{2}\Tr(\phi^2)$ with two external scalars and a negative helicity gluon, see   Figure~\ref{formfactornmhv}.
This NMHV form factor admits two decompositions as shown in   Figure~\ref{CSWFF}.
We compute the diagram on the lefthand side. 
\begin{figure}[h!]\centering
\def\svgwidth{\linewidth}
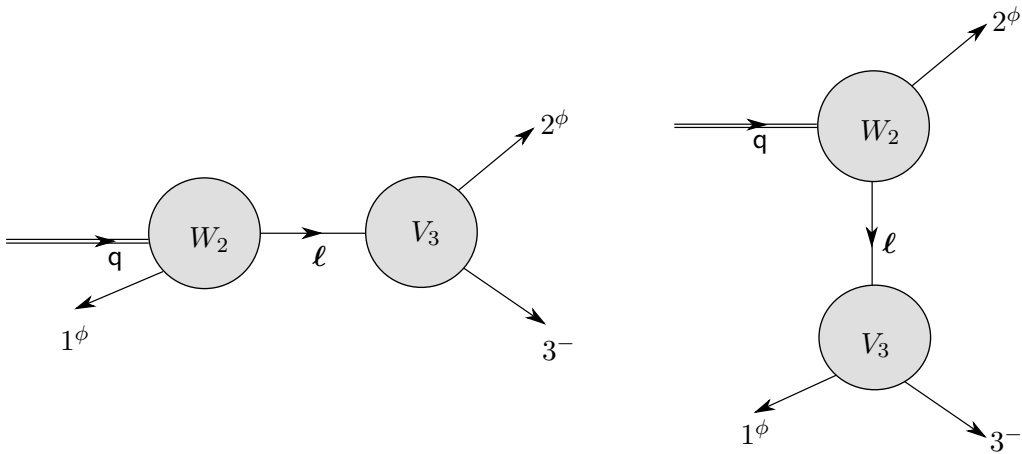
\caption{The two decompositions of the NMHV form factor $\mathscr{F}_{\frac{1}{2}\Tr(\phi^2)}(1^{\phi},2^{\phi},3^-;\fq)$ with two external scalars and one negative helicity gluon into an operator vertex and an MHV vertex. The internal momentum $\boldsymbol{\ell}$ and the momentum of the operator $\fq$ are off shell.}
\label{CSWFF}
\end{figure}
The operator vertex $W_2$ is connected via a propagator $1/\boldsymbol{\ell}^2$ to a 3-valent MHV vertex, $V_3$,
\begin{equation}
\label{NMHVforphi}
W_2(1^{\phi},\boldsymbol{\ell};q) \frac{1}{\boldsymbol{\ell}^2}V_3(\boldsymbol{\ell},2^{\phi},3^-)\eqncom
\end{equation}
where the operator vertex $W_2$ is the off-shell continued minimal form factor of $\tfrac{1}{2}\Tr(\phi^2)$ which equals $1$ after stripping off the momentum conserving delta function. Momentum conservation at both vertices implies that the off-shell momentum $\boldsymbol{\ell}$ satisfies 
\begin{equation}
\label{momentumconservation1}
\boldsymbol{\ell}=-\fp_1+\fq=\fp_2+\fp_3\eqndot
\end{equation}
Furthermore, we can define the spinor $\ell_{\alpha}\equiv \boldsymbol{\ell}_{\alpha\dot\alpha}\xi^{\dot\alpha}$, where $\xi^{\dot\alpha}$ is a reference spinor. In this way, \eqref{NMHVforphi} stripped of overall momentum conservation is written as
\begin{equation}
1\times \frac{1}{\abra{2}{3}[32]} \frac{\abra{2}{3}^2\abra{\ell}{3}^2}{\abra{\ell}{2}\abra{2}{3}\abra{3}{\ell}}\eqndot
\end{equation} Now, \eqref{NMHVforphi} reduces to
\begin{equation}
\frac{[2\xi]}{[23][3\xi]}
\end{equation} after using \eqref{momentumconservation1}.
A completely analogous computation shows that the righthand side of   Figure~\ref{CSWFF} stripped of momentum conservation equals
\begin{equation}
\label{righthandsideNMHV}
\frac{[1\xi]}{[13][3\xi]}\eqndot
\end{equation}
Taking the two expressions together and using the Schouten identity we find
\begin{equation}
\mathscr{F}^{\text{NMHV}}_{\frac{1}{2}\Tr(\phi^2)}(1^{\phi},2^{\phi},3^-;\fq)=\frac{[12]}{[13][32]}\delta^4(\fq-\fp_1-\fp_2-\fp_3)\eqndot
\end{equation}
Note that the dependence of the reference spinor has dropped out, as is required for a gauge-invariant physical quantity. Interestingly, this is precisely the $\overline{\text{MHV}}$ form factor, which can also be obtained by conjugating
\begin{equation}
\mathscr{F}^{\text{MHV}}_{\frac{1}{2}\Tr(\phi^2)}(1^{\phi},2^{\phi},3^+;\fq)=\frac{\abra{1}{2}}{\abra{1}{3}\abra{3}{2}}\delta^4(\fq-\fp_1-\fp_2-\fp_3)\eqndot
\end{equation}
In this section we reviewed the extension of CSW to form factors of certain operators. In addition to the fundamental MHV vertices that are obtained from MHV amplitudes one needs fundamental operator vertices that can be obtained from MHV form factors by setting an on-shell leg off shell. In the previous section we mentioned that the CSW recursion rules for amplitudes can be identified with the Feynman rules in twistor space. If this mapping between MHV rules and Feynman rules in twistor space is indeed correct, then the operator vertices needed for CSW for form factors should have an analogue in twistor space as fundamental and indecomposable building blocks. In Chapter~\ref{chap:} we illustrate this by performing the analogous computations that were done in the current section using the twistor space formalism. First, we compute MHV form factors and then NMHV form factors. Before we are ready to set up this formalism, we need some twistor basics, which will be given in the following chapter.

\chapter{Twistor space}
\label{chaptwistor}
In the previous chapter we reviewed some basics of $\calN=4$ SYM. We discussed the field content, the states, their correlation functions, the dilatation operator, and finally introduced amplitudes and form factors. At various instances, for example in the discussion of the complete dilatation operator \eqref{fulloneloop} or in the explicit form of MHV amplitudes \eqref{eq:MHV amplitudePT} it appeared that the final physical result exhibited a much simpler and more symmetric structure than the intermediate calculation would suggest. This simplicity of the physical observables is somehow concealed by the formulation of the theory in space-time variables. Indeed, one may wonder whether the theory allows for an alternative formulation that reveals all of this structure and thereby allows one to skip all the tedious intermediate steps and arrive at the simple final physical answer right away. Over the past decade people have discovered that at least for amplitudes such a formulation indeed exists in so-called twistor space.
Twistor space was introduced by Sir Roger Penrose in 1967 as an alternative to Minkowski space-time in which one uses light rays rather than space-time points as coordinates, initially in the hope that it could be used to unify quantum mechanics and general relativity. We review this original bosonic twistor space in the first section of this chapter. In Section~\ref{classint} we describe the concept of classical integrability of the self-dual Yang-Mills equations and how they are mapped bijectively to holomorphic Chern-Simons theories in non-supersymmetric twistor space. The fact that classical integrability is very naturally described in twistor space is one of the main motivations for studying quantum integrability of planar $\calN=4$ SYM via the one-loop dilatation operator in the final chapter of this thesis. This ends our discussion of bosonic classical twistor space, for which for the first thirty years after its initial introduction interest was relatively modest. The paper \cite{Witten:2003nn} by Witten caused a revival of interest in twistor techniques in the realm of quantum field theory, and more specifically scattering amplitudes. We review this extension to supersymmetric twistor space in Section~\ref{susytwistor}. In \cite{Boels:2006ir} the full twistor action was introduced and shown to reduce to the $\calN=4$ SYM action in usual space-time when a certain partial gauge is imposed. We review this derivation in Section~\ref{twistoraction}. In the last section we discuss how the twistor action generates all tree-level MHV amplitudes quite trivially and review the extension to NMHV level. The construction given in this section will be very similar to how we obtain form factors in twistor space later on in this thesis. \section{Classical bosonic twistor space}
\label{review2}
In this section we introduce twistor space and its relation to Minkowski space. We follow the line of the original paper on twistor space by Penrose of 1967 \cite{Penrose:1967wn}. Twistors are closely related to the spinor-helicity formalism that we reviewed in Section~\ref{sec:scatteringamps}. Let $n^{\mu}$ be a four-vector in Minkowski space. As in the spinor-helicity formalism, instead of the four Lorentz indices $\mu=0,1,2,3$ we can conveniently express it using spinor indices $\alpha=1,2$ and $\dot\alpha=1,2$ by contracting the four-vector $n_{\mu}$ with $\sigma_{\mu}$. As before, $\sigma^{\mu}$ are the Pauli matrices supplemented by the $2\times2$ identity matrix. The resulting $2\times 2$ matrix $\mathsf{n}^{\alpha\dot\alpha}$ takes the form
\begin{equation}
\mathsf{n}^{\alpha\dot{\alpha}}\equiv  n_{\mu}\sigma^{\mu\alpha\dot\alpha} =
\left( \begin{array}{cc}
n_0+n_3&n_1-in_2\\
n_1+in_2&n_0-n_3\end{array}\right)^{\alpha\dot{\alpha}}\eqndot\end{equation}
The spinor indices $\alpha$ and $\dot\alpha$ are raised and lowered using the two-dimensional Levi-Civita tensors $\epsilon^{\alpha\beta}$ and $\epsilon^{\dot\alpha\dot\beta}$, which satisfy $\epsilon^{12}=-\epsilon_{12}=1$. Recall that for any null vector $n^{\mu}$, we have that $0=n_{\mu}n^{\mu}=\det (\mathsf{n}_{\alpha\dot\alpha})$. This means that we can express the $2\times 2 $ matrix as the product of two $2$-vectors, $\mathsf{n}_{\alpha\dot\alpha}=\la_{\alpha}\tilde\la_{\dot\alpha}$ according to \eqref{decompvec}. When $n^{\mu}$ is real and future pointing $\mathsf{n}_{\alpha\dot\alpha}$ can be expressed as $\la_{\alpha}\bar\la_{\dot\alpha}$, where $\bar\la$ is the conjugate of $\la$. In Lorentzian signature, the conjugate of $\la=(\la_0,\la_1)$ is defined as
\begin{equation}
\bar\la =(\bar\la_0,\bar\la_1)\eqndot
\end{equation}
If we are interested only in the null \textit{direction}, then $\la$ can be identified up to scaling, i.e.\ as an element of $\mathbb{CP}^1$. To $\la$ we add a point in space-time, which in spinor notation is written as $x^{\alpha\dot\alpha}$. Then, the following element of $\mathbb{CP}^3$, denoted by $Z$ and called a twistor, determines the light ray $L$ in Minkowski space that passes through $x$ and is in the direction $\mathsf{n}_{\alpha\dot\alpha}=\la_{\alpha}\bar\la_{\dot\alpha}$: the element of $\mathbb{CP}^3$ given by the four homogeneous coordinates
\begin{equation}\label{twistor}
Z^I\equiv(Z_0,Z_1,Z_2,Z_3)=(\la_{\alpha},\mu^{\dot\alpha})\eqncom
\end{equation} where $\mu$ satisfies the equation
\begin{equation}\label{inc0}
\mu^{\dot\alpha}=ix^{\alpha\dot\alpha}\la_{\alpha}\eqndot
\end{equation}
The latter equation is called the incidence relation for $x$. For a fixed space-time point $x^{\alpha\dot\alpha}$ we can consider the set of all twistors satisfying its incidence relation. Clearly, for each $\la$ in $\mathbb{CP}^1$ there exists a corresponding $\mu$ by contracting $\la$ with $i x$ and hence associated to each $x$ is a Riemann sphere $\mathbb{CP}^1$ in $\mathbb{CP}^3$ parametrized by $\la_{\alpha}$. There are thus two interpretations of $x$, either as a point in Minkowski space, or as a Riemann sphere in twistor space. \newline\\
One might be tempted to think that $\mathbb{CP}^3$ is a fiber bundle over $\mathbb{M}$, however this is incorrect. Although at each separate $x\in \mathbb{M}$, the corresponding set of twistors in twistor space is just $x\times \mathbb{CP}^1$, twistor space is not locally a product space. Any twistor $Z$ incident with $x$ and corresponding to the light ray $L$ is also incident with any other space-time point on the light ray. The correct picture is a so-called double fibration, which is depicted in Figure \ref{doublefibration}.
\begin{figure}[h!]\centering
\def\svgwidth{\linewidth}
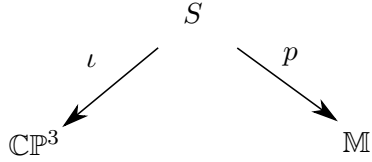
\caption{The twistor space as a double fibration of space-time. The space $S$ is called the correspondence space with coordinates $(\la_{\alpha},x^{\alpha\dot\alpha})$ and maps to twistor space via the incidence relation $\iota$ and to Minkowski space by the projection $p$.}
\label{doublefibration}
\end{figure}
In this diagram, the space $S$ is called the correspondence space, satisfying $S\cong \mathbb{CP}^1\times \mathbb{M}$ which has coordinates $(\la_{\alpha},x^{\alpha\dot\alpha})$. The map $p$ is just the projection onto the space-time variables and $\iota$ is given by the incidence relation $\iota (\la_{\alpha},x^{\alpha\dot\alpha})= (\la_{\alpha},ix^{\alpha\dot\alpha}\la_{\alpha})$.\newline\\
Note that adding any multiple of $\mathsf{n}$ to $x$ does not change the twistor \eqref{twistor} or \eqref{inc0}, so indeed we could have chosen any other point on the light ray to define the same twistor. Thus we also have two interpretations of $L$: either as a light ray in Minkowski space, or as an element of the projective space $\mathbb{CP}^3$. For $x^{\mu}$ a real four-vector, the corresponding matrix $x^{\alpha\dot\alpha}$ is Hermitian and we have 
\begin{equation}\label{reality}\text{Re}(\la_{\alpha}\bar\mu^{\alpha})=0\eqncom\end{equation} 
where $\bar\mu=(\bar\mu^0,\bar\mu^1)$.
This condition is also sufficient for the existence of a real point $x$ relating $\la$ and $\mu$. Twistors that satisfy this reality condition are called null twistors. Note that even if the space-time point $x$ is real, the twistors that satisfy the incidence relation still form a (complex) Riemann sphere $\mathbb{CP}^1$ and thus the corresponding twistors are complex objects.
We can also consider the condition under which two light-rays $L$ and $\tilde L$ intersect. In space-time this obviously corresponds to the existence of an intersection point $ x$. Since we can use any point along the light-ray to play the role of the space-time point that is used in the incidence relation \eqref{inc0} to define the corresponding twistor, we can in particular choose the intersection point $ x$ for both light rays. Hence, $L$ is given by the twistor $Z_I=(\la, i x^{\alpha\dot\alpha}\la_{\alpha})$ and $\tilde L$ by $\tilde Z_{I}= (\tilde \la, i x^{\alpha\dot\alpha}\tilde \la_{\alpha})$. We find that $\tilde \la_{\alpha}\bar\mu^{\alpha}=\tilde\la_{\alpha}(-i) x^{\alpha\dot\alpha}\bar\la_{\dot\alpha}=-\tilde\mu^{\dot\alpha}\bar\la_{\dot\alpha}$. This means that a necessary condition for the two light rays to intersect is 
\begin{equation}
\label{neccond}
\tilde Z_{I}\bar{ Z}^{I}=0\eqncom
\end{equation}
where in Lorentzian signature,
\begin{equation} 
\bar Z^I= (\bar\mu^{\alpha},\bar\la_{\dot\alpha})\eqndot
\end{equation} For $\la$ and $\tilde\la$ not proportional to each other this is also a sufficient condition as one can recover $ x$ via
\begin{equation}\label{incidence1}
 x_{\alpha\dot\alpha}=\frac{-i}{\abra{\tilde\la}{\la}}(\la_{\alpha}\tilde\mu_{\dot\alpha}-\tilde\la_{\alpha}\mu_{\dot\alpha})\eqncom
\end{equation} where we remind the reader of the spinor bracket $\abra{\tilde\la}{\la}\equiv\tilde\la^{\alpha}\la_{\alpha}$. \newline\\
If instead $L$ and $\tilde L$ are parallel light rays, then we can think of this as the pair intersecting at a point at infinity. In fact, any other line that is also parallel to $L$ and $\tilde L$ intersects at the same point at infinity. Thus, we can supplement $\mathbb{M}$ by all these points at infinity, one for each null hyperplane of parallel light rays. These additional points form a light cone at infinity. Furthermore, we can add one more point at infinity to $\mathbb{M}$, denoted by $I$, which is not on any light ray that goes through a finite point.  This additional point at infinity is the vertex of the light cone at infinity. Any twistor on the light cone at infinity can be written in the form $Z=(0,\mu^{\dot\alpha})$. The compactified space that is obtained by adding this closed null cone to Minkowski space is denoted by $\mathbb{M}^{\#}$. \newline\\
Another way of establishing the relationship between complexified compactified Minkowski space and twistor space is via the so-called Klein correspondence\footnote{This was named after the German mathematician Felix Klein, not the Swedish physicist Oskar Klein.}.
Complexified compactified Minkowski space $\mathbb{CM}^{\#}$ has a description as the Klein quadric in $\mathbb{CP}^5$. This can be seen by choosing six homogeneous coordinates on $\mathbb{CP}^5$,
\begin{equation}
X_{IJ}=-X_{JI}\eqncom
\end{equation} where $I$, $J$ are fundamental SL$(4)$ indices and the coordinates are defined up to complex rescaling. Subsequently, $\mathbb{CM}^{\#}$ is identified with the quadric 
\begin{equation}
\epsilon^{IJKL}X_{IJ}X_{KL}=0\eqncom
\end{equation} 
where $\epsilon^{IJKL}$ is the $4$-dimensional Levi-Civita tensor. In this construction the action of the conformal group becomes manifest. The complexified conformal group is isomorphic to PGL$(4,\mathbb{C})$ and acts on $X$ by $X\mapsto A X A^{T}$ for $A$ in the group GL$(4,\mathbb{C})$, where the action is defined up to scaling. 
As a $4\times4$ matrix, $X_{IJ}$ can be naturally viewed as a map on twistor space. The incidence relation now becomes the statement that twistors incident with $X$ belong to its kernel,
\begin{equation}
\label{incidence2}
Z \text{ incident with } X \iff X^{IJ}Z_J=0\eqndot
\end{equation} Considering this equation non projectively, the condition that $X\cdot X=0$ implies that $X$ is not of maximal rank. Recalling furthermore that the rank of a skew symmetric matrix is always even, the kernel is necessarily of dimension $2$ or $4$. Since we are in fact considering $X$ projectively, it is not the null matrix. Therefore, projectively, $X$ as a map acting on $\mathbb{CP}^3$ has a $1$-complex dimensional kernel, in agreement with the incidence relation \eqref{inc0}.
Furthermore, for any two twistors $Z$ and $\tilde Z$ that are both incident with $X$, we can reconstruct $X$ by taking their antisymmetric product
\begin{equation}
\label{cross product}
X^{IJ}=Z^{[I}\tilde Z^{J]}\eqndot
\end{equation}
This immediately implies the necessary condition \eqref{neccond} for the two light rays to intersect.
Two elements $X$ and $X'$ on the Klein quadric are null-separated as points in Minkowski space, $x$ and $x'$ respectively, if there exists a light ray that passes through both. In other words, if there exists a twistor $\tilde Z$ that satisfies both their incidence relations. This twistor satisfies
\begin{equation}
X^{IJ}=Z^{[I}\tilde Z^{J]} \quad \text{and}\quad X^{'IJ}\tilde Z_{J}=0\eqncom
\end{equation}
which imply that $X\cdot X'=0$. The correspondence between space-time points and Riemann spheres, and light rays and twistors as well as the incidence relation are depicted in Figure~\ref{twistorcorrespondence}.~Now let us make the connection with the previous construction \eqref{incidence1} explicit.
\begin{figure}[h!]\centering
\def\svgwidth{\linewidth}
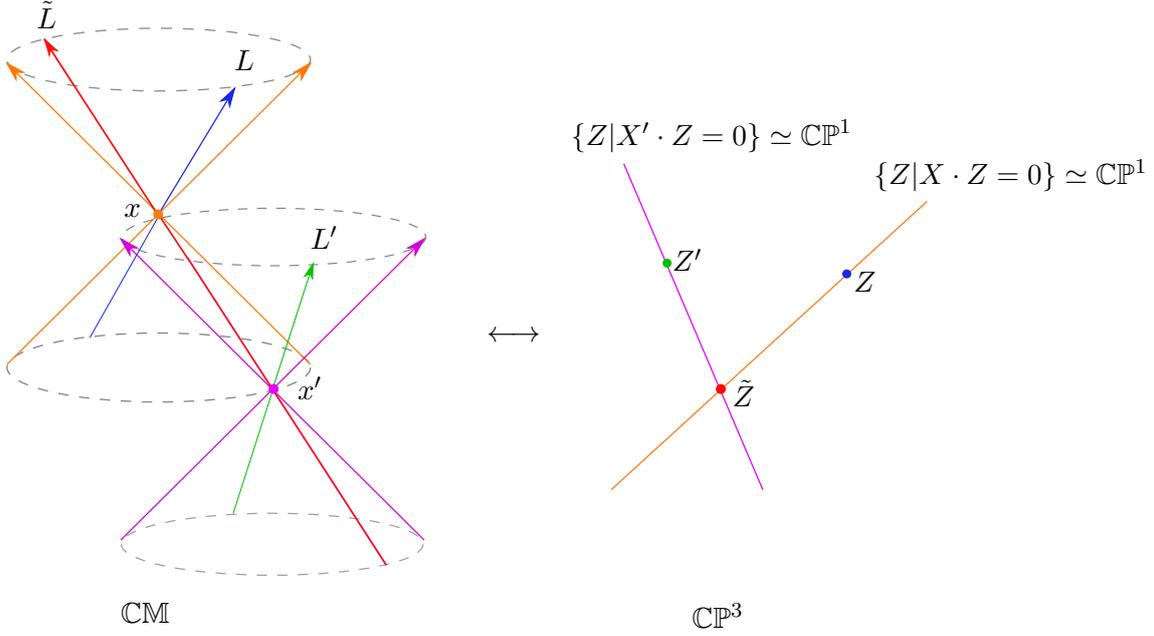
\caption{The correspondence between $\mathbb{CM}$ and $\mathbb{CP}^3$. The light cone of $x$ in Minkowski space depicted in orange on the left  corresponds to a $\mathbb{CP}^1$ depicted as a line on the right. The blue light ray $L$ on the left corresponds to the twistor $Z$ on the line given by $X$ on the right. Similarly, for $x'$, the fuchsia colored light cone corresponds to the line given by $X'$. On the left hand side, the light cone of $x'$ contains the light ray $L'$ that corresponds to the twistor $Z$ on the line on the right, both colored green. The space-time points $x$ and $x'$ are light-like separated as there exists a light ray $\tilde L$ that connects them on the left hand side. On the right hand side this translates to the intersection of the two $\mathbb{CP}^1$'s in $\tilde Z$.}
\label{twistorcorrespondence}
\end{figure}In order to do so we need to choose inhomogeneous coordinates for compactified Minkowski space, thus breaking conformal symmetry and obtaining non-compactified Minkowski space. Recall that compactified Minkowski space was obtained by supplementing Minkowski space by a light cone at infinity. Conversely, we mod out the Klein quadric $\mathbb{CM}^{\#} $ by the light cone of a point at infinity $I^{IJ}$, 
\begin{equation}
\mathbb{CM}=\mathbb{CM}^{\#}/\{ X\in\mathbb{CM}^{\#} \:|\: X\cdot I =0\} \,.
\end{equation}
The point at infinity then can be used to construct a metric on $\mathbb{CM}$ via
\begin{equation}
\label{eq:defmatric}
g(X,Y)=\frac{X\cdot Y}{(I\cdot X)(I\cdot Y)}\,.
\end{equation}
The fundamental SL$(4,\mathbb{C})$ indices are decomposed into spinor SL$(2,\mathbb{C})\times \text{SL}(2,\mathbb{C})$ indices $\alpha=1,2$ and $\dot\alpha=1,2$, 
\begin{equation}
Z^I=(\lambda_{\alpha},\mu^{\dot{\alpha}})\,.
\end{equation} 
This point at infinity  may be chosen as
\begin{equation}
I_{IJ}=
\left( \begin{array}{cc}
\epsilon^{\alpha\beta}&0\\
0&0\end{array}\right),
\qquad I^{IJ}=
\left( \begin{array}{cc}
0&0\\
0&\epsilon^{\dot{\alpha}\dot{\beta}}\end{array}\right),
\eeq
allowing us to introduce coordinates $x^{\alpha\dot{\alpha}}$ on $\mathbb{CM}$ as
\begin{equation} X^{IJ}=
\left( \begin{array}{cc}
\epsilon_{\alpha\beta}&-ix_{\alpha}^{\phantom{\alpha}\dot{\beta}}\\
ix^{\phantom{\beta}\dot{\alpha}}_{\beta}&-\frac{1}{2}x^2\epsilon^{\dot{\alpha}\dot{\beta}}\end{array}\right)\quad  \text{ or } \quad X_{IJ}=
\left( \begin{array}{cc}
-\frac{1}{2}x^2\epsilon^{\alpha\beta}&ix^{\alpha}_{\phantom{\alpha}\dot{\beta}}\\
-ix_{\phantom{\beta}\dot{\alpha}}^{\beta}& \epsilon_{\dot{\alpha}\dot{\beta}}\end{array}\right),
\end{equation}
where 
\begin{equation}
\label{explicitformofx}x^{\alpha\dot{\alpha}}=
\left( \begin{array}{cc}
x_0+x_3&x_1-ix_2\\
x_1+ix_2&x_0-x_3\end{array}\right)^{\alpha\dot{\alpha}}\,.\end{equation}
 In these coordinates, the incidence relation \eqref{incidence2} reduces to \eqref{inc0} and from \eqref{cross product} we find \eqref{incidence1}.
 For real Minkowski space we impose the reality condition
 \begin{equation}
 Z\cdot \bar Z=0\eqncom
 \end{equation}
 which is equivalent to \eqref{reality}. 
\section{The Penrose transform}
\label{penrosetransform}
Using the incidence relation \eqref{inc0} one can construct solutions to the massless field equations in twistor space very naturally. Suppose we want to find a solution to the Klein-Gordon equation $\square \phi =0$ on space-time from a field on twistor space. This can be done according to a theorem by Penrose \cite{Penrose:1969ae}, which states that for an open set $U'\subset \mathbb{CP}^3$, and $U\subset\mathbb{M}$ the corresponding open subset in space-time, there is an isomorphism between elements in the Dolbeault cohomology class $H^{0,1}(U', \calO(2h-2))$, which are the equivalence classes of homogeneous $(0,1)$-forms of degree $2h-2$ on $U'$, and zero-rest-mass fields of helicity $h$ on $U$.
In our example, to construct a helicity-$0$ field, we use a twistor field $\phi$ that is homogeneous in $\la$ of degree $-2$. We construct the corresponding massless scalar field on space-time by integrating out the degrees of freedom associated with the Riemann sphere. The twistor field $\phi$ is a $(0,1)$-form which can be integrated against the $(1,0)$-form $\DD\la\equiv\abra{\la}{\dd \la}$ which is of homogeneous degree $2$,
\begin{equation}\label{Penrosetransformscalar}
\phi_{\text{space-time}}(x)=\int_{\mathbb{CP}^1}\DD\la\phi(\la^{\alpha},ix^{\alpha\dot\alpha} \la_{\alpha})\eqndot
\end{equation}
This transform is called the Penrose transform. Clearly, 
\begin{equation}
\phi \rightarrow \phi + \bar\partial f
\end{equation} leaves the transform invariant such that indeed $\phi$ is a representative of a cohomology class. To see that the space-time field satisfies the Klein-Gordon equation, we act with the d'Alembert operator $\square$ under the integral,
\begin{equation}
\frac{\partial^2}{\partial x^{\alpha \dot\alpha}\partial x_{\alpha \dot\alpha}} \phi(\la^{\alpha},ix^{\alpha\dot\alpha} \la_{\alpha})=-\abra{\la}{\la}\frac{\partial^2 \phi}{\partial \mu^{\dot\alpha\alpha}\partial\mu_{\dot\alpha}}=0\eqndot
\end{equation}
The Penrose transform for fields on twistor space of different degree of homogeneity or different helicity on space-time is defined for $n> 0$ as 
\begin{align}
&\phi^{\text{space-time}}_{h=n/2,\; \dot\alpha_1\dots\dot\alpha_n}(x)=\int_{\mathbb{CP}^1}  \DD\la\; \frac{\partial}{\partial\mu^{\dot\alpha_1}}\cdots \frac{\partial}{\partial\mu^{\dot\alpha_n}}\phi(\la^{\alpha},ix^{\alpha\dot\alpha} \la_{\alpha})\notag\\
&\phi^{\text{space-time}}_{h=-n/2,\;\alpha_1\dots\alpha_n}(x)=\int_{\mathbb{CP}^1} \DD\la\; \la_{\alpha_1}\cdots \la_{\alpha_n}\phi(\la^{\alpha},ix^{\alpha\dot\alpha} \la_{\alpha})
\eqncom
\end{align}
for positive and negative helicity fields respectively. These fields are completely symmetric in their spinor indices. Clearly, for positive helicity fields, to obtain a degree $0$ integrand, which is necessary for performing the integration, $\phi$ must be of homogeneity $n-2$ and for negative helicity fields, $\phi$ must be of homogeneity $-n-2$ . These fields satisfy the field equations
\begin{align}
&\partial^{\alpha_1\dot\alpha_1}\phi^{\text{space-time}}_{h=n/2,\;\alpha_1\dots\alpha_n}(x) =0\eqncom\notag\\
&\partial^{\alpha_1\dot\alpha_1}\phi^{\text{space-time}}_{h=-n/2,\; \dot\alpha_1\dots\dot\alpha_n}(x)(x) =0\eqndot
\end{align}
\section{Self-dual Yang-Mills and twistor space}
\label{classint}
In this section we discuss a particular class of classically integrable models, namely self-dual Yang-Mills theories with gauge group GL$(N)$ or SL$(N)$. We will see that these theories can be uniquely identified with holomorphic Chern-Simons theories on twistor space. Let us first introduce self-dual Yang-Mills.
Let $E\rightarrow \mathbb{M}$ be a principal fiber bundle with structure group SU$(N)$ over Minkowski space. We can define a connection $D=d+A$ with $A$ some $(0,1)$-form. The curvature $F=\dd A+ A\wedge A$  of the connection in four dimensions can be decomposed as $F= F^++F^-$, where $F^+$ is the self-dual part, which satisfies $\star F^+ =F^+$, where $\star$ is the Hodge dual. The connection $F$ is said to be a Yang-Mills connection if it satisfies the Yang-Mills equation,
\begin{equation}
D\star F=0\eqndot
\end{equation}
For $F$ self dual, i.e.\ $F=F^+$ and $F^-=0$, the Yang-Mills equations are automatically satisfied due to the Bianchi identity,
\begin{equation}
D F=0\eqndot
\end{equation}
In spinor notation, the full space-time curvature $F_{\alpha\beta\dot\alpha\dot\beta}$,
\begin{equation}
F_{\alpha\beta\dot\alpha\dot\beta}\propto [D_{\alpha\dot\alpha},D_{\beta\dot\beta}]\eqncom
\end{equation} where $D_{\alpha\dot\alpha}$ is the space-time covariant derivative, can be decomposed into a self-dual and anti-self-dual part as
\begin{equation}
F_{\alpha\beta\dot\alpha\dot\beta}=\epsilon_{\alpha\beta}\bar F_{\dot\alpha\dot\beta}+\epsilon_{\dot\alpha\dot\beta}F_{\alpha\beta}\eqncom
\end{equation} 
where  $\epsilon_{\alpha\beta}\bar F_{\dot\alpha\dot\beta}=\tfrac{1}{2}(F+(\star F))$ and $\epsilon_{\dot\alpha\dot\beta}F_{\alpha\beta}=\tfrac{1}{2}(F-(\star F))$. The symmetric tensors $\bar F_{\dot\alpha\dot\beta}$ and $F_{\alpha\beta}$ are called the self-dual and anti-self-dual field strength respectively. For the self-dual theory, $F=\star F$ and $F_{\alpha\beta}$ vanishes. \newline \\
So far this discussion was restricted to self-dual Yang-Mills theories in $4$-dimensional space-time. The connection to twistor space is given by the Ward correspondence \cite{WARD197781}. This is a one-to-one correspondence between solutions of the self-dual equations on an open region of space-time $U\subset \mathbb{M}$ with gauge group GL$(N,\mathbb{C})$, and holomorphic vector bundles $E'\rightarrow \mathbb{CP}^3$ whose restriction to points $x$ in $U$, $E'|_x$ is trivial. We shall not prove this rigorously, but rather sketch both sides of the correspondence briefly.\newline\\
\begin{figure}[h!]\centering
\def\svgwidth{\linewidth}
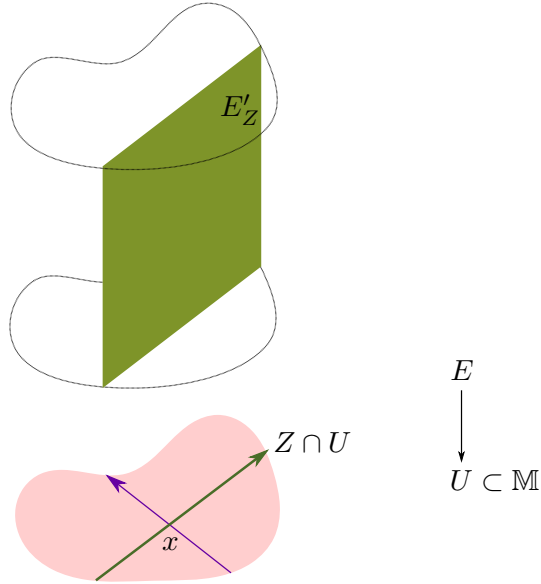
\caption{The space-time bundle $E\rightarrow U$ with $U\subset \mathbb{M}$. Through every space-time point $x\in U$ there pass light rays. To the light ray $Z$ which passes through $x$ the associated fiber is the span of the covariantly constant sections of the bundle $E$ restricted to $Z\cap U$, viewed as  the subset of $U$ containing the space-time points on the light ray $Z$.  }
\label{fiber bundle}
\end{figure}
Let us start with a rank $N$ bundle $E\rightarrow U$ on an open subset $U$ of space-time admitting a self-dual connection $D$. Let $\Gamma(U,E)$ denote the space of sections of $E$ over $U$. Then, for each $Z\in\mathbb{CP}^3$ we can define a space of sections $E'_Z$ by
\begin{equation}\label{sections}
E'_Z\equiv \{s\in \Gamma(Z\cap U, E) | D s_{|Z\cap U}=0\}\eqncom
\end{equation} which is just the restriction of sections in $\Gamma(U,E)$ to sections in $\Gamma(Z\cap U, E)$, where one views $Z\cap U$ as the subset of space-time points in $U$ that are on the light ray $Z$. 
The self duality of the connection $D$ means that its restriction to every light ray has zero curvature. Therefore, the covariantly constant sections in \eqref{sections} are constant and thus fixed by their value at one point $x\in U$ on the light ray $Z$. Therefore, we can define $E'\rightarrow \mathbb{CP}^3$ as the holomorphic vector bundle with fibers $E'_Z$ vector spaces of complex dimension $N$. \newline\\
Going in the other direction, we start with a holomorphic vector bundle $E'\rightarrow U'\subset \mathbb{CP}^3$, which is trivial on restriction to every $x\cong \mathbb{CP}^1\subset \mathbb{CP}^3$, hence $E'|_x\cong\mathbb{C}^N\times x$. This means that there exists a holomorphic frame $h_x(\la)$ of $E'|_x$ which is unique up to global (on $x\cong \mathbb{CP}^1$) gauge transformations, $h_x(\la)\rightarrow h_x(\la)g_x$. This holomorphic frame will play an important role in the construction of local composite operators later on in this thesis. The fibers of the bundle $E\rightarrow U$ can then be defined by $E_x\equiv\Gamma(x,E')$. The sections are holomorphic maps from $x$ to $\mathbb{C}^N$, which are constant and therefore $E_x$ is an $N$-dimensional vector space. Furthermore, one can show that the vector bundle $E\rightarrow U$ admits a unique connection $D_{\alpha\dot\alpha}$ that is self dual, see \cite{Bullimore:2013jma,Wolf:2010av,Mason:1991rf} for more details.
Briefly, the holomorphic vector bundle $E'\rightarrow \mathbb{CP}^3$ admits an almost complex structure $\bar D=\bar\partial +a$ with components $\bar D_{\alpha}=\bar\partial_{\alpha}+a_{\alpha}$ and $\bar D_{\dot\alpha}=\la^{\alpha}\partial_{\alpha\dot\alpha}+a_{\dot\alpha}$. Here, $a$ is the gauge field and the almost complex structure satisfies $[\bar D, \bar D]=0$. On the twistor side this implies the holomorphicity of the gauge field $a$, 
\begin{equation}
\bar \partial a+ [a, a] =0\eqncom
\end{equation}
which is the equation of motion for the holomorphic Chern-Simons action
\begin{equation}
\int_{\mathbb{CP}^3} \DD^{(3,0)} Z\; a\wedge(\bar\partial +\frac{2}{3}a )\wedge a\eqncom
\end{equation}
where $a$ and $ \bar\partial$ are $(0,1)$-forms. It will be extended to a fully supersymmetric twistor action in Section~\ref{susytwistor}.
On the space-time bundle $E$, the almost complex structure $\bar D$  induces a space-time covariant derivative $\partial_{\alpha\dot\alpha}+A_{\alpha\dot\alpha}$, with $A_{\alpha\dot\alpha}$ the space-time gauge field. Pulling back to the correspondence space, the $\dot\alpha$ component of the almost complex structure $\bar D$ satisfies
\begin{equation}
[D_{\dot\alpha},D_{\dot\beta}]=\la^{\alpha}\la^{\beta}\big(\epsilon_{\alpha\beta}\bar F_{\dot\alpha\dot\beta}+\epsilon_{\dot\alpha\dot\beta}F_{\alpha\beta}\big)= \epsilon_{\dot\alpha\dot\beta}\la^{\alpha}\la^{\beta}F_{\alpha\beta}\eqndot
\end{equation}
Therefore $F_{\alpha\beta}=0$ and the connection satisfies the self-duality equations. Now, $D_{\dot\alpha}=\la^{\alpha}\partial_{\alpha\dot\alpha}+a_{\dot\alpha}$ is a so-called Lax connection and its vanishing curvature means that the self-dual theory is classically integrable. This is interesting as, as we shall see later on, the full twistor action is a quantum perturbation around the self-dual sector. In this sense, the quantum integrability might appear more natural in twistor space. 
\section{Supersymmetric twistor action}
\label{susytwistor}
In Section~\ref{classint} we discussed the relationship between (bosonic) self-dual Yang-Mills theories and holomorphic Chern-Simons theories on twistor space, which is given by the Ward correspondence. Interestingly, this correspondence extends to a correspondence between the self-dual sector of maximally supersymmetric $\mathcal{N}=4$ Yang-Mills theory and a holomorphic Chern-Simons theory on supertwistor space. As we shall review in this section, the full self-dual and anti-self-dual $\calN=4$ SYM action can even be obtained from a twistor action by supplementing an additional piece to the holomorphic Chern-Simons term.\newline\\
Let us start by extending twistor space to supertwistor space and reviewing its relation to super Minkowski space.
Super Minkowski space can be described in coordinates by appending the bosonic coordinate $x^{\alpha\dot\alpha}$ with eight additional Gra\ss mann coordinates $\theta^{\alpha a}$. Similarly, twistor space can be extended to supertwistor space $\mathbb{CP}^{3|4}$ by adding four Gra\ss mann coordinates to the bosonic twistor $Z$ to obtain the supertwistor $\calZ^I=(Z,\chi)$. As in the case of bosonic twistor space, we can relate supertwistors that are incident with an element in super Minkowski space $(x,\theta)$ via the incidence relation 
\beq
\label{eq:definitioncalZonaline}
\mathcal{Z}^I=(\lambda_{\alpha},\mu^{\dot{\alpha}},\chi^a)=(\lambda_{\alpha},ix^{\alpha\dot{\alpha}}\lambda_{\alpha},i\theta^{\alpha a}\lambda_{\alpha})\,,
\eeq where the index $a$ runs from $1$ to $4$.
From now on we will continue to use calligraphic $\mathcal{Z}$ to denote the supertwistor and ordinary $Z$ to denote its bosonic part. The supersymmetric analogue of the bosonic twistor field $a(Z)$ of the previous section is the supersymmetric $(0,1)$-form $\AAA(\mathcal{Z})$ in
$\Omega_{\mathbb{CP}^{3|4}}^{(0,1)}(\text{End}(\mathcal{E}))$, where $\mathcal{E}\rightarrow \mathbb{CP}^{3|4}$ is a vector bundle of rank $N$, whose structure group is the complexification of the gauge group SU$(N)$. Here, $\Omega_{\mathbb{CP}^{3|4}}^{(0,1)}(\text{End}(\mathcal{E}))$ denotes the space of smooth $(0,1)$-forms with coefficients $\text{End}(\mathcal{E})$-valued functions of $\mathbb{CP}^{3|4}$. In fact, we require that $\AAA$ is a $(0,1)$-form with components only in the bosonic directions, $\AAA=\AAA_I\dd \bar{Z}^I$.
As for the purely bosonic theory, the twistor field admits a holomorphic Chern-Simons action\footnote{See also \cite{Sokatchev:1995nj} for some earlier work, in which a Chern-Simons action for self-dual $\calN=4$ SYM theory is presented using Lorentz harmonics.} \cite{Witten:2003nn}, which we will call $\calS_1$,
\begin{equation}\label{holchern}
\calS_1=\int_{\mathbb{CP}^{3|4}}\DD\calZ^{3|4} \AAA\wedge (\bar\partial +\frac{2}{3}\AAA)\wedge \AAA\eqncom
\end{equation}
with measure $ \DD^{3|4}\mathcal{Z}\equiv  \frac{1}{4!}\epsilon_{IJKL}Z^I\dd Z^J\dd Z^K\dd Z^L\dd^4\chi$.
The superfield $\AAA$ can be expanded in the Gra{\ss}mann variables $\chi$ as \cite{Nair:1988bq}
\begin{equation}\label{expans}
\AAA(\calZ)=g^+(Z)+\chi^{a}\bar{\psi}_a(Z)+\frac{1}{2}\chi^{a}\chi^b\phi_{ab}(Z)+\frac{1}{3!}\epsilon_{abcd}\chi^{a}\chi^b\chi^c\psi^d(Z)+\chi^1\chi^2\chi^3\chi^4 g^-(Z),
\end{equation} where the fields $g^+$, $\bar{\psi}$, $\phi$, $\psi$ and $g^-$ depend only on the bosonic part of the supertwistor $\calZ=(Z,\chi)$. The field $\AAA$ is homogeneous of degree $0$ in $\la$, whereas $\chi=\theta^{\alpha a}\la_{\alpha}$ is homogeneous of degree $1$. Thus the homogeneities of the components $g^+$, $\bar\psi$, $\phi$, $\psi$ and $g^-$ can be easily read off. Upon expanding \eqref{holchern} and integrating out the degrees of freedom associated to $\mathbb{CP}^1$, one can recover the self-dual part of the $\calN=4$ SYM action. We will pursue this in the next section. To recover the full action, the action \eqref{holchern} is completed by an additional non-local part\footnote{A similar non-local action of $\log\det$ form appeared before in \cite{ZUPNIK1987175} for the anti-selfdual interactions of an $\calN=2$ SYM theory in harmonic superspace.}, introduced in \cite{Boels:2006ir}, 
\begin{equation}
\calS_2=-g_{YM}^2\int \dd^4 z\:\dd^8\theta \:\log\det \,(\bar{\partial}+\AAA)_{(z,\theta)}\eqndot
\end{equation}
The complete twistor action, $\calS_1+\calS_2$ reduces to the $\calN=4$ SYM action when a particular partial gauge is chosen. This will be reviewed in the next section. 
Having seen that the action and field content can be formulated in twistor space, we move on to the supersymmetry generators.
Indeed, the $\mathfrak{psu}(2,2|4)$ generators have a very natural description in twistor variables, namely as
\begin{equation}
\calZ_I\frac{\partial}{\partial \calZ_J}\eqncom
\end{equation}
which in spinor components reads
\begin{align}
&\mathfrak{P}_{\alpha \dot\alpha}=  \la_{\alpha}\frac{\partial}{\partial \mu^{\dot\alpha}}\eqncom\quad && \mathfrak{K}^{\alpha\dot\alpha}=\mu^{\dot\alpha}\frac{\partial}{\partial \la_{\alpha}\eqncom}\notag\\
&\mathfrak{J}_{\alpha\beta}=\frac{i}{2}\left( \la_{\alpha}\frac{\partial}{\partial \la^{\beta}}+\la_{\beta}\frac{\partial}{\partial \la^{\alpha}}\right)\eqncom\quad &&\mathfrak{J}_{\dot\alpha\dot\beta}=\frac{i}{2} \left(\mu_{\dot\alpha}\frac{\partial}{\partial \mu^{\dot\beta}}+\mu_{\dot\beta}\frac{\partial}{\partial \mu^{\dot\alpha}}\right)\eqncom\notag\\
&\mathfrak{D}=\frac{i}{2}\left(\la_{\alpha} \frac{\partial}{\partial \la_{\alpha}} -\mu^{\dot\alpha}\frac{\partial}{\partial \mu^{\dot\alpha}}\right)\eqncom \quad && \mathfrak{R}^a_{\phantom{a}b}=\chi^a\frac{\partial}{\partial \chi_b}\eqncom\notag\\
&\mathfrak{Q}_{\alpha a}=i \la_{\alpha }\frac{\partial}{\partial \chi^a}\eqncom \quad && \bar {\mathfrak{Q}}_{\phantom{a}\dot\alpha}^{a}= i\chi^a \frac{\partial}{\partial \mu^{\dot\alpha}}\eqncom \notag\\
&\mathfrak{S}^{\alpha a}= i\chi^a \frac{\partial}{\partial \la_{\alpha}}\eqncom \quad && \bar {\mathfrak{S}}^{\dot\alpha}_{\phantom{\dot\alpha}a}= i\mu^{\dot\alpha} \frac{\partial}{\partial \chi^{a}}\eqndot 
\end{align}
From here, the $\mathfrak{psu}(2,2|4)$ commutation relations can be straightforwardly verified.
\section{Reduction to the $\calN=4$ SYM action}
\label{twistoraction}
In this section, we review how one can obtain the $\calN=4$ SYM action from the full supersymmetric twistor action $\calS_1+\calS_2$ after fixing extra gauge degrees of freedom and integrating over $\mathbb{CP}^1$ for each space-time point.
Up until now we have been working in Lorentzian signature. The connection between twistor space and Euclidean space $\mathbb{E}$ is however more easily made, because in Euclidean signature twistor space is a fibration over $\mathbb{E}$ rather than the double fibration that corresponds to the Lorentzian signature.
The Euclidean conjugation is
\begin{equation}
\la_{\alpha}=(\la_0,\la_1)\rightarrow \hat\la_{\alpha}=(\bar\la_1,-\bar\la_0)\eqncom \quad \mu^{\dot\alpha}=(\mu^0,\mu^1)\rightarrow \hat\mu^{\dot\alpha}=(-\bar\mu^1,\bar\mu^0)\eqndot
\end{equation}
The conjugation $\sigma$ of the twistor is $\sigma(Z_I)= \hat Z_I=(\hat\la,\hat\mu)$, which squares to minus the identity $\sigma^2=-\text{Id}$. This means that there are no twistors that obey the reality condition. However, there are lines that are preserved by the reality structure. These fixed lines are given by $X^{IJ}=[Z^I,\hat  Z^J]$. This relation associates a unique real point in Euclidean space to each element in twistor space, or rather to each pair of a twistor and its conjugate. Hence, twistor space is a fiber bundle $\mathbb{CP}^3\rightarrow \mathbb{E}$ over Euclidean space and the double fibration \ref{doublefibration} is redundant. This simplifies the mapping from twistor space quantities to space-time quantities and it is therefore in this signature that the reduction of the twistor action to the $4$-dimensional action is most easily established, which we review in this section. In Euclidean signature one can define a basis of $(0,1)$-forms
\begin{equation}\label{basis}
\bar{e}^0=\frac{\hat{\lambda}^{\alpha}\dd\hat{\lambda}_{\alpha}}{\abra{\lambda}{\hat{\lambda}}^2},\quad \bar{e}^{\dot{\alpha}}=\frac{\dd x^{\dot{\alpha}\alpha}\hat{\lambda}_{\alpha}}{\abra{\lambda}{ \hat{\lambda}}}
\end{equation}
and a dual basis of differential operators
\begin{equation}
\bar\partial_0 =\abra{\la}{\hat\la} \la_{\alpha} \frac{\partial}{\partial \hat\la_{\alpha}}, \quad \bar\partial_{\dot\alpha} =\la^{\alpha}\frac{\partial}{\partial x^{\alpha\dot\alpha}}\eqndot
\end{equation}
In addition to the expansion \eqref{expans}, we can write the $(0,1)$-form $\AAA$, or each of its components, in the basis \eqref{basis},
\begin{equation}
\AAA=\AAA_0\bar{e}^0+\AAA_{\dot\alpha}\bar{e}^{\dot{\alpha}}.
\end{equation} 
We can think of $\AAA_0\bar{e}^0$ as the restriction of $\AAA$ along the fiber $\mathbb{CP}^1$ and $\AAA_{\dot\alpha}\bar{e}^{\dot\alpha}$ the component perpendicular to that.
Recall that on supertwistor space Witten introduced a supersymmetric action, which we denoted by $\calS_1$ that describes the kinetic terms and anti self-dual interactions of $\calN=4$ SYM
\begin{equation}
\calS_1=\frac{i}{2\pi}\int D^{3|4}\mathcal{Z} \:\:\Tr\,(\AAA\wedge\bar{\partial}\AAA+\frac{2}{3}\AAA\wedge\AAA\wedge\AAA)\,, 
\end{equation} 
with measure $ D^{3|4}\mathcal{Z}\equiv  \frac{1}{4!}\epsilon_{IJKL}Z^I\dd Z^J\dd Z^K\dd Z^L\dd^4\chi$. Note that both the superfield $\AAA(\calZ)$ and $\bar\partial$ are $(0,1)$-forms on twistor space. Therefore, the Lagrangian is a $(0,3)$-form which complements the $(3,0)$-form that is explicitly written out in the measure. The resulting $(3,3)$-volume form can be integrated over.
The holomorphic Chern-Simons action $\calS_1$ was supplemented by an additional term, denoted by $\calS_2$,
\begin{equation}
\calS_2=-g_{YM}^2\int \dd^4 z\:\dd^8\theta \:\log\det \,(\bar{\partial}+\AAA)_{(z,\theta)}\,,
\end{equation}
which by expanding the $ \log\det$ generates an infinite sum of interaction vertices
\begin{equation}
\label{eq:interactionexpanded}
\calS_2=-g_\YM^2\int\displaylimits_{\mathbb{M}^{4|8}} \frac{\dd^4 z\:\dd^8\theta }{(2\pi)^4} \Tr(\ln \bar\partial_{(z,\theta)})+ 
\sum_{n=1}^{\infty}\frac{1}{n} \int\displaylimits_{(\mathbb{CP}^1)^n}\frac{\DD\la_1\DD\la_2\cdots \DD\la_n \Tr \big(\AAA(\la_1)\cdots\AAA(\la_n)\big)}{ \abra{\lambda_1}{\lambda_2} \cdots \abra{\lambda_{n-1}}{ \lambda_{n}} \abra{\lambda_{n}}{\lambda_{1}}}
\eqncom
\end{equation}
where $\DD \la_i := \abra{\la_i}{\dd\la_i}$. The $\Tr(\ln\bar\partial_{(z,\theta)})$ is a constant that does not depend on the fields and will therefore be dropped henceforth.\newline\\
Let us first remark that the two respective pieces $\calS_1$ and $\calS_2$ look very different from each other. Indeed, the first part consists only of two terms, whereas the expansion of the $\log \det$ piece is an infinite sum of interaction terms of ever increasing length. Moreover, $\calS_1$ is a local action, i.e.\ each interaction term is localized onto a single twistor $\calZ$, whereas in $\calS_2$ each interaction term consists of fields positioned at a different twistor at the same line in twistor space $\mathbb{CP}^1$, corresponding to the space-time point $(z,\theta)$. 
In this section we will review how the twistor action $\calS_1+\calS_2$ reduces to the usual space-time action for $\calN=4$ SYM in a partial gauge. In the next section we review how it gives the MHV rules that we discussed in Section~\ref{sec:scatteringamps} when we choose an axial gauge instead. 
\paragraph{The holomorphic Chern-Simons term}
Since  the $3$-complex dimensional twistor space is larger than our usual $4$-dimensional space it also exhibits more gauge freedom. Compared to the gauge freedom in $\calN=4$ SYM some of this gauge freedom is therefore redundant and in order to recover the full un-gauge-fixed space-time action these extra gauge degrees of freedom need to be fixed. This was done in \cite{Boels:2006ir} and we review this procedure here briefly, including some corrections to what was done there.\\
\newline
The gauge condition that removes the residual gauge degrees of freedom is the condition
\begin{equation}\label{gauge}
\bar{\partial}_{(z,\theta)}^* \AAA_0=0,
\end{equation}
where $\bar{\partial}_{(z,\theta)}$ is the $\bar{\partial}$ operator restricted to the fiber $(z,\theta)$ and the $^*$ means the Hermitian conjugate.\footnote{The corresponding ghost sector decouples from the theory, see \cite{Boels:2006ir}.} We will refer to this gauge choice as harmonic gauge.
In addition to the gauge \eqref{gauge}, $\AAA$ is assumed to have only holomorphic dependence on $\chi$. This implies that the harmonic gauge must hold for each of the components of $\AAA_0$ in the expansion \eqref{expans} individually. Said in a different way, all the components are co-closed along the fibers $(z,\theta)$. Forms of degree $(0,1)$ on a $1$-complex dimensional manifold are automatically $\bar\partial$-closed, since
\begin{equation}
\bar{\partial} \alpha =\bar{\partial} (f(z,\bar{z}) d\bar{z})=\frac{\partial f}{\partial \bar{z}}\dd\bar{z}\wedge\dd\bar{z}=0.
\end{equation}
Therefore, all the components, $g^+, \bar\psi,\phi,\psi, g^-$ are harmonic (i.e.\ closed and co-closed) when restricted to the fibers, and hence one can apply the Hodge theorem. This implies that when restricted to the fibers, the component fields are in the first \v{C}ech cohomology group $H^1(\mathbb{CP}^1, \mathcal{O}(n))$. Here, $n$ is the homogeneity of the component (or, the degree of the line bundle) and takes the values $0$, $-1$, $-2$, $-3$ and $-4$ for $g^+_0,\dots g^{-}_0$. This is proven in Appendix~\ref{appA}. In this appendix, we also prove that $H^1(\mathbb{CP}^1, \mathcal{O}(-1))=H^1(\mathbb{CP}^1, \mathcal{O}(0))=0$. Therefore, in this harmonic gauge, $g^+_0=0$ and $\bar\psi_0=0$ and $\AAA_0$ is proportional to $\chi^2$. Furthermore, for $f\in H^1(\mathbb{CP}^1, \mathcal{O}(n))$ a $(0,1)$-form with values in the line bundle $\mathcal{O}(n)$ with $n\le -2$,
\begin{equation}
f=\frac{1}{\abra{\la}{\hat\la}^{-n}}{{-n-1}\choose{-n-2}}f^{\alpha_1\dot \alpha_{-n-2}}\hat{\la}_{\alpha_1}\cdots \hat{\la}_{\alpha_{-n-2}} \abra{\hat\la}{\dd\hat\la}\eqncom
\end{equation} with $f^{\alpha_1\dot \alpha_{-n-2}}$ a $(-n-2)$-symmetric tensor, see \cite{Wells:1979, Woodhouse:1985id}. It turns out that every harmonic form on $\mathbb{CP}^1$ with values in $\mathcal{O}(n)$ can be expressed in this way.
This means that for the remaining components of $\AAA_0$ we can write $\phi_{0 ab}=\phi_{ab}(x)$, $\psi_{a 0}=2 \psi(x)_{a}^{\alpha}\hat\la_{\alpha}/\abra{\la}{\hat\la}$ and $g_{0}^-=3 G_{\alpha\beta}(x)\hat\la_{\alpha}\hat\la_{\beta}/\abra{\la}{\hat\la}^2$.
Writing the component fields in the basis of Eq.\eqref{basis} gives schematically
\begin{align}\label{eucexpansion}
g^+&=g^+_{\dot{\alpha}}\bar{e}^{\dot{\alpha}},\notag\\
\bar\psi_a&=\bar\psi_{a\dot{\alpha}}\bar{e}^{\dot{\alpha}},\notag\\
\phi_{ab}&=\phi_{ab}(x)\bar{e}^0+\phi_{ab\dot{\alpha}}\bar{e}^{\dot{\alpha}},\notag\\
\psi^{a}&=2 \frac{\psi(x)_{a}^{\alpha}\hat\la_{\alpha}}{\abra{\la}{\hat\la}}\bar{e}^0+\psi^{a}_{\dot{\alpha}}\bar{e}^{\dot{\alpha}},\notag\\
g^-&=3\frac{G_{\alpha\beta}(x)\hat\la_{\alpha}\hat\la_{\beta}}{\abra{\la}{\hat\la}^2}\bar{e}^0+g_{\dot{\alpha}}\bar{e}^{\dot{\alpha}}.
\end{align}\newline
So starting from the action $\mathcal{S}_1$, we integrate over the four Gra\ss mann variables $\chi$, so that only terms with precisely four different $\chi$'s survive. These are schematically $g^+ \bar\partial g^-$, $\bar \psi \bar\partial \psi$, $\phi \bar\partial \phi$, $\psi\bar\partial \bar\psi$ and $g \bar\partial g^+$. Similarly, for the interaction term in $\calS_1$ we find schematically $g^- [g^+,g^+]$, $\bar\psi [\psi,g^+]$, $\phi [g^+,\phi]$ and $\bar\psi [\phi, \bar\psi]$. Then, we expand in the basis \eqref{basis} and use the substitutions \eqref{eucexpansion}. Subsequently, we integrate out the $\mathbb{CP}^1$ degrees of freedom, leaving only the Euclidean space integration $\int\dd^4z$. We will not go into full detail here, but rather refer the reader to \cite{Boels:2006ir}. 
After some manipulations $\calS_1$ equals
\begin{equation}
\calS_1 =\int \dd^4 z \Tr(G_{\alpha\beta}F^{\alpha\beta}+\bar\psi^a_{\dot\alpha} D^{\alpha\dot\alpha}\psi_{a\alpha}-\frac{1}{4}D_{\alpha\dot\alpha}\phi_{ab}D^{\alpha\dot\alpha}\phi^{ab}+\bar\psi^{a\dot\alpha}[\phi_{ab},\bar\psi^b_{\dot\alpha}])\eqncom
\end{equation}where $ F_{\alpha\beta}$ is the self-dual part of the field strength and $D_{\alpha\dot\alpha}=\frac{d}{d z^{\alpha\dot\alpha}}+A_{\alpha\dot\alpha}$ is the covariant derivative. Note that the self-duality condition of the field strength $F_{\alpha\beta}=0$ is now the equation of motion for the auxiliary field $G_{\alpha\beta}$.
\paragraph{The log det term} Now we study the non-local part of the action, $\calS_2$, a bit closer. Recall that 
\begin{equation}
\label{eq:interactionexpanded2}
\calS_2=-g_\YM^2\int\displaylimits_{\mathbb{M}^{4|8}} \frac{\dd^4 z\:\dd^8\theta }{(2\pi)^4} 
\sum_{n=1}^{\infty}\frac{1}{n} \int\displaylimits_{(\mathbb{CP}^1)^n}\frac{\DD\la_1\DD\la_2\cdots \DD\la_n \Tr \big(\AAA(\la_1)\cdots\AAA(\la_n)\big)}{ \abra{\lambda_1}{\lambda_2} \cdots \abra{\lambda_{n-1}}{ \lambda_{n}} \abra{\lambda_{n}}{\lambda_{1}}}
\eqndot
\end{equation}
Because of the measure $\dd^4z$ all the $\AAA_{\dot\alpha}$ components in \eqref{eucexpansion} do not appear in this part of the action, so we only have $\AAA_0$ components here. Now, since $g^+_0=\bar\psi_0=0$, the component $\calA_0$ is proportional to $\chi\chi$ and the expansion \eqref{eq:interactionexpanded2} terminates after the fourth term. This is reassuring since the conventional $\calN=4$ SYM action has at most quartic interactions. Furthermore, since we are working with gauge group SU$(N)$, the first term in the sum is trivially zero, since the matrices in the corresponding algebra are traceless. The previous two arguments together mean that the only terms that survive in this axial gauge are the terms of length $n=2,3,4$. Let us now consider these three terms in some more detail, particularly to make a correction to the derivation in \cite{Boels:2006ir}. The $n=2$ term can only be formed of two negative helicity gluons. Other terms simply would not reach the number of $8$ $\chi$'s required for the integration over $\dd^8\theta$. Performing the fermionic integration gives
\begin{equation}
-\frac{1}{2}g_{\text{YM}}^2 \int \dd^4 z \DD \la_1\DD\la_2 \frac{\abra{1}{2}^2}{\abra{1}{\hat 1}^2\abra{2}{\hat 2}^2}3 G_{\alpha\beta}\hat\la_1^{\alpha}\hat\la_1^{\beta}3 G_{\rho\sigma}\hat\la_2^{\rho}\hat\la_2^{\sigma}\eqncom
\end{equation}
We can write out the angular brackets in the numerator in components and use equation (A.7) of \cite{Boels:2006ir},
\begin{equation}
\int\dd^4z \frac{ \DD\la \DD\hat\la }{\abra{\la}{\hat\la}^4} S_{\alpha_1\dots\alpha_m} T_{\beta_1\dots\beta_m} \frac{\la^{\alpha_1}\dots \la^{\alpha_m}\hat\la^{\beta_1}\dots \hat\la^{\beta_m}}{\abra{\la}{\hat\la}^m}=-\frac{2\pi i}{m+1}\int\dd^4 z \:S_{\alpha_1\dots\alpha_m} T^{\alpha_1\dots\alpha_m} \eqndot
\end{equation}
to obtain
\begin{equation}
-\int\dd^4 z\frac{g_{\text{YM}^2}}{2}G_{\alpha\beta}G^{\alpha\beta}\eqndot
\end{equation}
Now, the $n=3$ term goes analogously. Expanding the $\calA^3$ we obtain six terms that all contain a scalar and two fermions. By cyclicity of the trace we can rewrite this as six times the same term. Integrating over the fermionic components gives us
\begin{equation}
-\frac{1}{3}g_{\text{YM}}^2 \int \dd^4 z \DD \la_1\DD\la_2 \DD\la_3 6\Tr\left( \frac{2\psi_{\alpha}^a\hat\la_{1\alpha}}{\abra{1}{\hat 1}^2} \frac{\phi_{ab}}{\abra{ 2}{\hat 2}^2}\frac{2 \psi^{b\beta}\hat\la_{3\beta}}{\abra{ 3}{\hat3}^2}\right) \abra{1}{3}\eqncom
\end{equation}which is
\begin{equation}
-2 g_{\text{YM}}\Tr( \psi^{a \alpha} \phi_{ab},\psi^{b}_{\alpha})=- g_{\text{YM}}\Tr( \psi^{a \alpha} [\phi_{ab},\psi^{b}_{\alpha}])\eqndot
\end{equation}
Computing the four-vertex term is a bit more involved. One first expands the superfields into components
\begin{equation}\label{phiint}
-\kappa \int \dd^4 z\:\dd^8\theta\: \frac{1}{(2\pi \mathrm{i})^4}\int \prod_{r=1}^4 \frac{K_r}{\abra{r}{r+1}}\chi^a_1\chi^b_1\chi^c_2\chi^d_2\chi^{a'}_3\chi^{b'}_3\chi^{c'}_4\chi^{d'}_4\frac{1}{2^4}\Tr(\phi_{ab}\phi_{cd}\phi_{a'b'}\phi_{c'd'})\eqncom
\end{equation} 
where $K_i$ denotes the K\"{a}hler metric on $\mathbb{CP}^1$,
\begin{equation}
K_i=\frac{\abra{\la_i}{\dd\la_i}\abra{\hat\la_i}{\dd\hat\la_i}}{\abra{\la_i}{\hat\la_i}^2}\eqndot
\end{equation}
Because of the antisymmetry of the flavor indices of $\phi_{ab}$, we can easily see that 
\begin{equation}
\int \prod_{r=1}^4 \frac{K_r}{\abra{r}{r+1}}\chi^a_1\chi^b_1\chi^c_2\chi^d_2\chi^{a'}_3\chi^{b'}_3\chi^{c'}_4\chi^{d'}_4
\end{equation}
 must have the following flavor structure
\begin{equation}
A \epsilon_{abcd}\epsilon_{a'b'c'd'}+B\epsilon_{aba'b'}\epsilon_{cdc'd'}+C\epsilon_{abc'd'}\epsilon_{cda'b'}.
\end{equation}
Therefore all we need to determine are the coefficients $A$, $B$, and $C$. To do so, we choose explicit values for $a,b,c,d,a',b',c'$ and $d'$. For example, for $(a,b,c,d)=(1,2,3,4)$ and $(a'b'c'd')=(1,2,3,4)$ we find
\begin{equation}
A+C=\int \prod_{r=1}^4 \frac{K_r}{\abra{r}{r+1}}\abra{1}{3}^2\abra{2}{4}^2.
\end{equation}
In order to perform the integration over the fibers we use the following formula\footnote{Note the difference with Eq.(3.26) of \cite{Boels:2006ir}. }
 which is proven in Appendix~\ref{kahlerintegral}\begin{equation}\label{joehoe}
\int K_1 \frac{\lambda_{1{\alpha}}\lambda_{1\beta}}{\abra{1}{2} \abra{4}{1}}= -\frac{\pi i}{\abra{2}{4}}\left(\frac{\lambda_{2{\alpha}}\hat{\lambda}_{2\beta}+\hat{\lambda}_{2{\alpha}}\lambda_{2\beta}}{\abra{2}{\hat 2}} - \frac{\lambda_{4{\alpha}}\hat{\lambda}_{4\beta}+\hat{\lambda}_{4{\alpha}}\lambda_{4\beta}}{\abra{4}{\hat 4}}  \right).
\end{equation}
We expand 
\begin{equation}
\abra{1}{3}^2\abra{2}{4}^2=\lambda_{1{\alpha}}\lambda_{3\beta}\lambda_{1\gamma}\lambda_{3\delta}\epsilon^{{\alpha}\beta}\epsilon^{\gamma\delta} \abra{2}{4}^2.
\end{equation}
Now we use formula \eqref{joehoe} to integrate over $\la_1$ and $\la_3$ and obtain,
\begin{align}\label{joe}
&-\int K_2 K_4 \pi^2 2\left(\frac{\lambda_{2{(\alpha}}\hat{\lambda}_{2\gamma)}}{\abra{2}{\hat 2}} - \frac{\lambda_{(4{\alpha}}\hat{\lambda}_{4\gamma)}}{\abra{4}{\hat 4}}  \Big)\Big(\frac{\lambda_{(4\beta}\hat{\lambda}_{4\delta)}}{\abra{4}{\hat 4}} - \frac{\lambda_{(2\beta}\hat{\lambda}_{2\delta)}}{\abra{2}{\hat 2}}  \right)\epsilon^{{\alpha}\beta}\epsilon^{\gamma\delta}\notag\\
&=\int K_2 K_4 \pi^2 2 \left(\frac{\abra{2}{\hat 2}^2}{\abra{2}{\hat 2}^2}+\frac{\abra{4}{\hat 4}^2}{\abra{4}{\hat 4}^2}\right)\eqncom
\end{align}
where in the first line the parentheses around the indices indicate that they are symmetrized. In order to obtain the second line we have used equation (A.7) of \cite{Boels:2006ir},
\begin{equation}
\int\dd^4z \frac{ \DD\la \DD\hat\la }{\abra{\la}{\hat\la}^4} S_{\alpha_1\dots\alpha_m} T_{\beta_1\dots\beta_m} \frac{\la^{\alpha_1}\dots \la^{\alpha_m}\hat\la^{\beta_1}\dots \hat\la^{\beta_m}}{\abra{\la}{\hat\la}^m}=-\frac{2\pi i}{m+1}\int\dd^4 z \:S_{\alpha_1\dots\alpha_m} T^{\alpha_1\dots\alpha_m} \eqndot
\end{equation}
This straightforwardly gives that 
\begin{equation}
A+C =4\pi^2\int K_2 K_4 =-16\pi^4\eqncom
\end{equation}
Similarly, the coefficient $B=16\pi^4$. Using the cyclicity of the trace 
we now find that \eqref{phiint} equals
\begin{equation}
g_{\text{YM}}^2 \int \dd^4 z \frac{1}{4}\Tr\big([\phi_{ab},\phi_{cd}][\phi^{ab},\phi^{cd}]\big)\eqndot
\end{equation}
Taking $\calS_1$ and $\calS_2$ together, we find that in the harmonic gauge, the twistor action reduces to
\begin{align}
\calS=&\int \dd^4 z  \Tr\Big(G_{\alpha\beta}F^{\alpha\beta}+\bar\psi^a_{\dot\alpha} D^{\alpha\dot\alpha}\psi_{a\alpha}-\frac{1}{4}D_{\alpha\dot\alpha}\phi_{ab}D^{\alpha\dot\alpha}\phi^{ab}+\bar\psi^{a\dot\alpha}[\phi_{ab},\bar\psi^b_{\dot\alpha}] \notag\\
&-\frac{g_{\text{YM}^2}}{2}G_{\alpha\beta}G^{\alpha\beta} - g_{\text{YM}}^2 \psi^{a \alpha} [\phi_{ab},\psi^{b}_{\alpha}]+ \frac{g_{\text{YM}}^2}{4}[\phi_{ab},\phi_{cd}][\phi^{ab},\phi^{cd}]\Big)\eqndot
\end{align}
After rescaling $\bar\psi \rightarrow g_{\text{YM}}\bar\psi$ and $\psi\rightarrow  \psi/g_{\text{YM}}$, integrating out the auxiliary field $G$, and adding a topological term $\sim F^2-\bar F^2$, this yields the well known action for $\calN=4$ super Yang-Mills theory.
\section{Amplitudes from twistor space}
\label{amptwistor}

\subsection{From interaction vertices to tree-level MHV amplitudes}
Recall from the previous chapter that the twistor action is the sum of two parts, $\calS_1+\calS_2$, where $\calS_1$, introduced in  \cite{Witten:2003nn}, describes the self-dual part and $\calS_2$ is referred to as the interaction part \cite{Boels:2006ir}.
In the previous section we imposed a harmonic gauge in which the infinite sum of $\calS_2$ was truncated after the fourth term. After some manipulations the usual space-time action of $\calN=4$ SYM was recovered. Instead, one can impose a different gauge, the so-called CSW gauge, which will turn out to be extremely suitable for computing tree-level amplitudes, see \cite{Adamo:2011cb}, as well as form factors in later chapters. 
The gauge condition reads
 \begin{equation}\label{gaugecondition}
\overline{Z_{\star} \cdot \frac{\partial}{\partial Z}}\:\lrcorner\:\AAA=0\, ,
\end{equation} 
where $\lrcorner$ denotes the interior product and $Z_{\star}$ is a reference twistor which is on the light cone at infinity.
We will give a brief review of  \cite{Adamo:2011cb} in this section as the tools that are introduced there will play an important role in the derivation and calculation of NMHV form factors in Chapter~\ref{NMHV}. Due to the gauge condition \eqref{gaugecondition}, the $(0,1)$-form $\AAA$ has only two independent components and therefore the cubic term in $\calS_1$ vanishes and all interactions come from the $\calS_2$ part of the action.
Recall that expanding $\calS_2$ gives
\begin{equation}
\label{eq:interactionexpanded}
\calS_2=-g_\YM^2\int_{\mathbb{M}^{4|8}} \frac{\dd^4 z\:\dd^8\theta }{(2\pi)^4} \sum_{n=1}^{\infty}\frac{1}{n} \int_{(\mathbb{CP}^1)^n}\DD\la_1\DD\la_2\cdots \DD\la_n\frac{\Tr\big(\AAA(\la_1)\cdots\AAA(\la_n)\big)}{ \abra{\lambda_1}{\lambda_2} \cdots \abra{\lambda_{n-1}}{ \lambda_{n}} \abra{\lambda_{n}}{\lambda_{1}}}
\eqncom
\end{equation}
where $\abra{\la_i}{\la_j} =\epsilon^{\alpha\beta}\la_{i\alpha} \la_{j\beta}=\la_{i}^\alpha \la_{j\alpha} $, $\DD\la_i\equiv\frac{\abra{\la_i}{\dd\la_i}}{2\pi i}$ and $\AAA(\la_i)\equiv \AAA(\calZ_{z}(\la_i))$. In our conventions, $\epsilon^{\alpha\beta}=\epsilon^{\dot\alpha\dot\beta}$ and $\epsilon^{12}=\epsilon_{21}=1$.
From the interaction part $\calS_2$, one can derive the (color-stripped) vertices \cite{Adamo:2011cb},
\begin{equation}
\label{eq:definitionoftheamplitudevertex 2.1}
\ver{\calA_1,\ldots, \calA_{n}}=
\int_{\mathbb{M}^{4|8}} \frac{\dd^4 z\:\dd^8\vartheta}{(2\pi)^4} 
\int_{(\mathbb{CP}^1)^n}\DD\la_1\DD\la_2\cdots \DD\la_n\frac{\AAA_1(\la_1)\AAA_2(\la_2)\cdots\AAA_n(\la_n)}{ \abra{\lambda_1}{\lambda_2} \cdots \abra{\lambda_{n-1}}{ \lambda_{n}} \abra{\lambda_{n}}{\lambda_{1}}}
 \,, 
\end{equation}
where $\DD\lambda\equiv\frac{\abra{\lambda}{\dd\lambda}}{2\pi i }$ and $n\geq2$.
As every propagator increases the MHV degree by one, the tree-level $n$-point MHV amplitudes are directly obtained from the corresponding $n$-legged vertex in \eqref{eq:definitionoftheamplitudevertex 2.1}, see \cite{Boels:2007qn}.
In order to obtain amplitudes (or form factors), we need external states. An external state in position twistor space is \cite{Adamo:2011cb}
\begin{equation}
\label{eq: external state in position twistor space}
 \calA_{\calZt}(\calZ)=2\pi i \,\bar{\delta}^{3|4}(\calZ,\calZt)\eqncom
\end{equation}
where we denote the external supertwistors as $\calZt$. We are of course interested in obtaining amplitudes and form factors in momentum space, so that we instead need to insert the on-shell momentum representation%
\footnote{Note that the $\lambda_\alpha$ are components of the supertwistor $\mathcal{Z}$ in \eqref{eq:definitioncalZonaline}, which are integrated over at each vertex \eqref{eq:definitionoftheamplitudevertex 2.1}. In particular, they are not spinor-helicity variables parametrizing momenta.
When inserting the supermomentum eigenstate \eqref{eq:definitiononshellmomentumeigenstates}, however, the $s$ and $\lambda$ integration in combination with the delta function effectively replace $\lambda_\alpha$ with the spinor-helicity variable $p_\alpha$. 
}
\begin{equation}
 \label{eq:definitiononshellmomentumeigenstates}
\calA_{\fP}(\calZ)=2\pi i \int_{\mathbb{C}}\frac{\dd s}{s}\e^{s(\sbra{\bar{p}}{\mu}+\{\chi\eta\})}
\bar{\delta}^2(s\lambda_{\alpha}-p_{\alpha})\, , \qquad \calZ=(\lambda_{\alpha},\mu^{\dot{\alpha}}, \chi^a)\eqnsem
\end{equation}
see for instance \cite{Cachazo:2004kj,Adamo:2011cb}.
We denote the on-shell supermomenta in terms of super-spinor-helicity variables as $\fP=(\fp_{\alpha\dot{\alpha}},\eta_a)\equiv (p_{\alpha},\bar{p}_{\dot{\alpha}},\eta_a)$, i.e.\ we write an on-shell momentum as $\fp_{\alpha\dot{\alpha}}=p_{\alpha}\bar{p}_{\dot{\alpha}}$.
 In \eqref{eq:definitiononshellmomentumeigenstates}, we used the bracket $\sbra{\mu}{\mu'}=\epsilon_{\dot \alpha\dot\beta}\mu^{\dot \alpha}{\mu'}^{\dot \beta}=\mu_{\dot \alpha}{\mu'}^{\dot\alpha}$ and the short-hand notation$\{\chi\eta\}=\chi^a\eta_a$.
The delta function $\bar{\delta}^2(\lambda)= \bar{\delta}^1(\lambda_1)\bar{\delta}^1(\lambda_2)$ is obtained from
 \begin{equation} \bar{\delta}^1(z)=\frac{1}{2\pi i}\dd z\,\bar\partial \big(\frac{1}{z}\big)\eqncom
 \end{equation}
 which is the $\delta$ function on the complex plane. 
Inserting the on-shell states \eqref{eq:definitiononshellmomentumeigenstates} into the $n$\textsuperscript{th} summand of \eqref{eq:interactionexpanded} and taking into account the $n$ ways of cyclically attaching them yields the tree-level $n$-point MHV amplitude:
\begin{equation}
\label{eq:MHV amplitude}
\begin{split}
\mathscr{A}^{\text{MHV}}(\fP_1, \dots , \fP_n) &= n\int\frac{\dd^4z\dd^8\theta}{(2\pi)^4}\frac{1}{n}\int \DD\lambda_1\cdots \DD\lambda_n\frac{\calA_{\fP_1}(\calZ_z(\lambda_1))\cdots \calA_{\fP_1}(\calZ_z(\lambda_n))}{\abra{\lambda_1}{\lambda_n}\cdots \abra{\lambda_n}{\lambda_1}}\\
&=\int\frac{\dd^4z\dd^8\theta}{(2\pi)^4}\frac{\e^{iz\sum_{j=1}^np_j\tilde{p}_j+i\theta\sum_{j=1}^np_j\eta_j}}{\abra{1}{2}\cdots \abra{(n-1)}{n}\abra{n}{1}}=\frac{\delta^{4|8}(\sum_{i=1}^n \fP_i)}{\prod_{k=1}^n\abra{k}{(k+1)}} \eqncom
\end{split}
\end{equation}
with $\abra{i}{j}\equiv \abra{p_i}{p_j}$. We remark that the integrations over $s_k$ from the definition of $\calA_{\fP_k}(\calZ_z(\la_k))$ as well as over its corresponding $\la_k$ effectively cancels $s_k$ and replaces $\la_k\rightarrow p_k$, $\mu_k^{\dot{\alpha}}\rightarrow iz^{\alpha \dot{\alpha}}p_{k,\alpha}$ and $\chi_k^a\rightarrow i\theta^{\alpha a }p_{k,\alpha}$ due to the $\bar{\delta}^2$ function and the parametrization \eqref{eq:definitioncalZonaline}. 
Here and in the rest of this thesis, we have moreover dropped the explicit integration range.  

\subsection{NMHV Amplitudes in twistor space}
\label{NMHVamp}
In the twistor-space formalism, the amplitude of MHV degree $k$ directly translates to twistor-space diagrams involving $k$ propagators at tree level. 
In \cite{Boels:2007qn} it was shown that the MHV rules for decomposing an N$^k$MHV amplitude as $k+1$ MHV amplitudes via $k$ propagators are in fact the Feynman rules obtained from the twistor action in the axial gauge we chose in the previous section.
At the NMHV level, we therefore connect two MHV vertices by one propagator. Let us first review this construction for amplitudes as was presented in \cite{Adamo:2011cb}. The NMHV amplitude is constructed by connecting two vertices  \eqref{eq:definitionoftheamplitudevertex} via a twistor propagator,
\begin{equation}
\label{propagator}
\Delta(\calZ_1,\star,\calZ_2)=i(4\pi)^2\bar\delta^{2|4}(\calZ_1,\star,\calZ_2)\eqncom
\end{equation}
where $\star$ denotes the reference twistor from the gauge condition \eqref{gaugecondition}, which has no Gra\ss mann components.
The projective delta function $\bar\delta^{2|4}(\calZ_1,\calZ_2,\calZ_3)$ is defined as
\begin{equation}
\bar\delta^{2|4}(\calZ_1,\calZ_2,\calZ_3)=\int_{\mathbb{C}}\frac{\dd s}{s}\int_{\mathbb{C}}\frac{\dd t}{t} \bar\delta^{4|4}(\calZ_1+s\calZ_2+t\calZ_3)\eqncom
\end{equation}
and forces the three twistors to lie on one projective line $\mathbb{CP}^1$.
The most general such diagram is shown\footnote{From now on, in figures we shall refrain from writing $\calZt_{j}$ for the supertwistors of the external on-shell states and just label them by their index $j$.} in  Figure~\ref{fig: general NMHV amplitude diagram}.%
\begin{figure}[tbp]
 \centering
  \includegraphics[height=3.5cm]{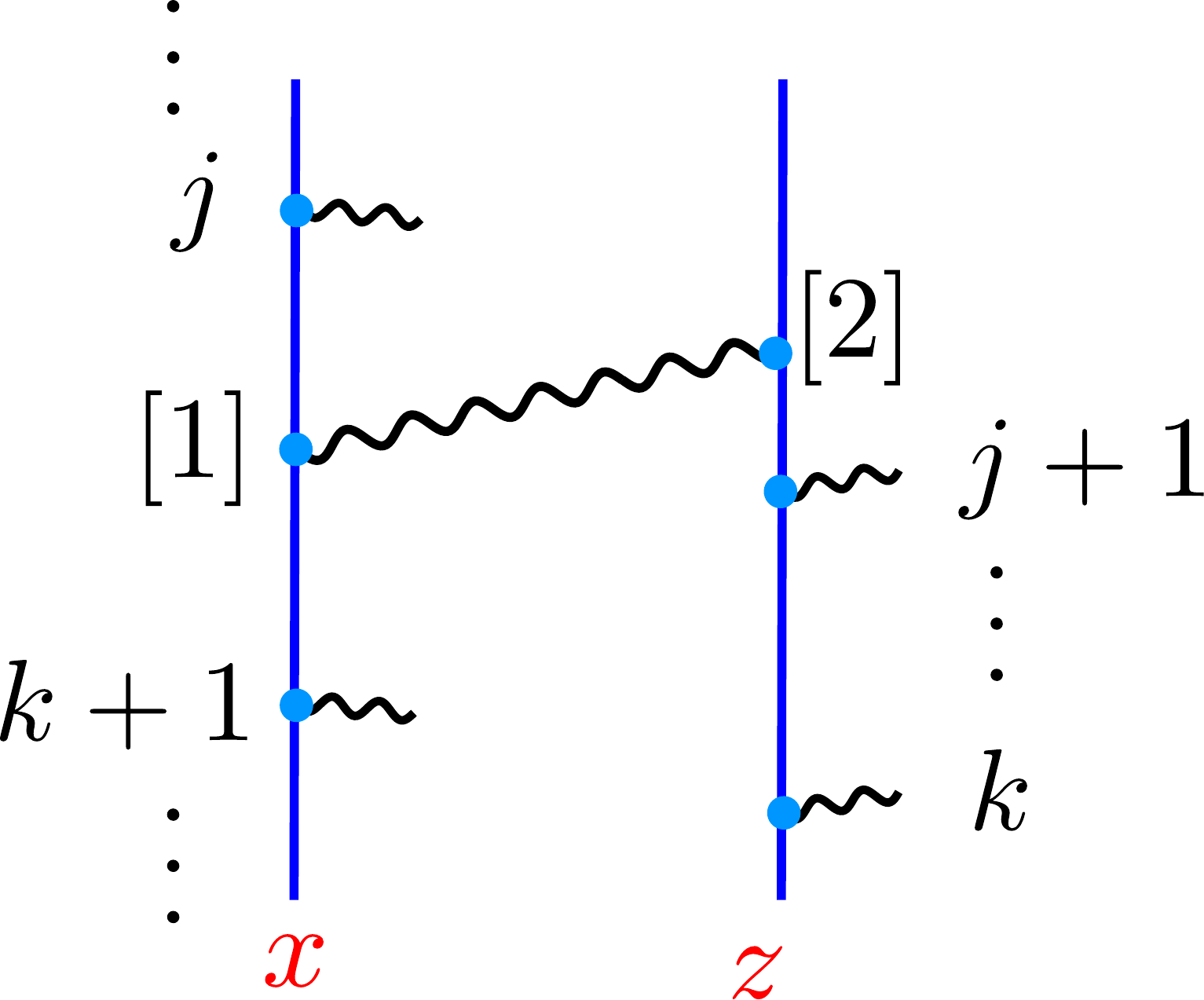}
  \caption{
  \it 
  The most general diagram for amplitudes at NMHV level in position twistor space. The points $z_1$ and $z_2$ are integrated over as required by the definition of the vertex $\textbf{V}$.}
\label{fig: general NMHV amplitude diagram}
\end{figure}
The interaction vertex $\mathbf{V}$ can involve two or more fields.  
However, the two-point vertex vanishes after substituting an external on-shell field, see \cite{Adamo:2011cb}. 
Hence, we can restrict ourselves to the case of at least three fields in $\mathbf{V}$.
Using \eqref{eq:amplitudeverticesinversesoftlimit}, we find that the general expression for the NMHV amplitude of $n$ external particles in twistor space reads
\begin{align}
\label{eq: NMHV amplitude in position twistor space}
&\mathbb{A}_{n}^{\text{NMHV}}=\\
& \sum_{1\leq j<k\leq n}\notag\int \DD^{3|4}\calZ_{[1]}\DD^{3|4}\calZ_{[2]}\ver{\ldots, \calZt_j,\calZ_{[1]},\calZt_{k+1},\ldots}\Delta(\calZ_{[1]},\calZ_{[2]})\ver{\ldots, \calZt_{k},\calZ_{[2]},\calZt_{j+1},\ldots}\eqndot
\end{align}
This expression can be simplified by expressing the $n$-point vertices as $(n-1)$-point vertices and a delta function using the so-called inverse soft limit which we will repeat here for the reader's convenience.
\paragraph{Inverse soft limit for amplitudes}
Via the so-called inverse soft limit, the $n$-point twistor-space vertices of the elementary interactions can be expressed in terms of $(n-1)$-point vertices. This procedure  plays a crucial role in the calculation of amplitudes beyond MHV level in position twistor space \cite{Adamo:2011cb}, which we shall see in the remainder of this section.
In section~\ref{subsec: inversesoftlimit}, we find a similar recursion for the Wilson loop vertices, which will play an equally important role in the calculation of form factors beyond MHV level.
\begin{figure}[h!]\centering
\def\svgwidth{\linewidth}
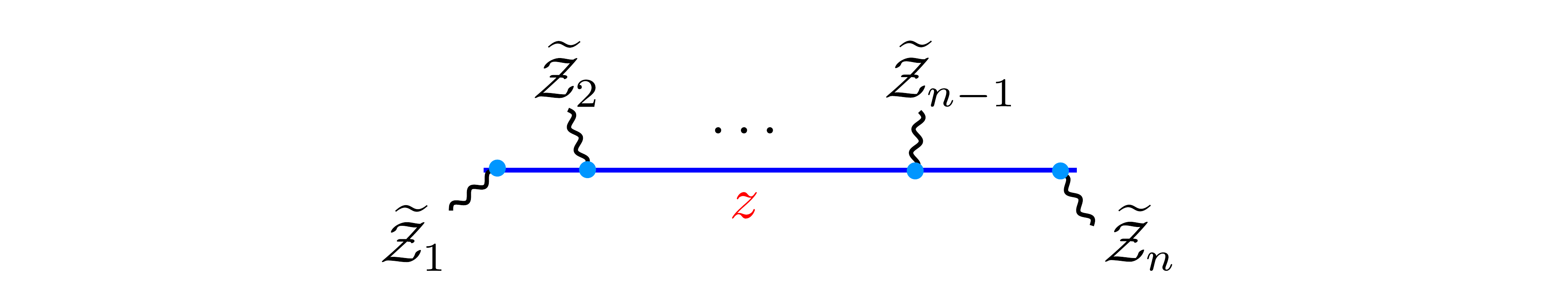
\caption{The $n$-point vertex \eqref{eq:definitionoftheamplitudevertex 2.1} from the action $\calS_2$. Throughout this thesis, the vertices are labeled clockwise.}
\label{fig: picture for the vertices}
\end{figure}
 Here, we review the case of the interaction vertices. The $n$-point vertices
\begin{equation}
\label{eq:definitionoftheamplitudevertex}
\ver{\calZt_1,\ldots, \calZt_{n}}\equiv\ver{\calA_{\calZt_1},\ldots, \calA_{\calZt_{n}}}
\end{equation}
of \eqref{eq:definitionoftheamplitudevertex 2.1} (see Figure~\ref{fig: picture for the vertices})
can be reduced to $(n-1)$-point vertices via recursion relations \cite{Adamo:2011cb}. Repeated use of this recursion allows us to break them all the way down to the product of two-point vertices\footnote{The two-point vertex only vanishes after an external field has been inserted, it does not vanish in the inverse soft limit.}. Specifically, the inverse soft limit is derived by first parametrizing the $\lambda_{j\alpha}$ as $(1,\sigma_j)_\alpha$ so that $\DD\lambda_j=\tfrac{\dd\sigma_j}{2\pi i }$ and $\abra{\lambda_j}{\lambda_{j+1}}=\sigma_{j+1}-\sigma_{j}$. We now choose a $k\in \{1,\ldots, n\}$ and replace $\sigma_k$ by 

\begin{equation}
s=\tfrac{\sigma_{k}-\sigma_{k+1}}{\sigma_{k-1}-\sigma_{k}}\eqncom\quad  \text{such that }\quad\tfrac{\dd s}{s}=-\tfrac{(\sigma_{k+1}-\sigma_{k-1})\dd\sigma_k}{(\sigma_{k+1}-\sigma_k)(\sigma_k-\sigma_{k-1})}\eqncom
\end{equation} and 
\begin{equation}\calZ_z(\sigma_k)=\tfrac{1}{1+s}\left[s\calZ_z(\sigma_{k-1})+\mathcal{Z}_z(\sigma_{k+1})\right]\eqndot\end{equation}
 Finally, we use 
 \begin{equation}
 \int \tfrac{\dd s}{s}\bar{\delta}^{3|4}(s\calZ_z(\sigma_{k-1})+\calZ_z(\sigma_{k+1}),\calZt_k)=-\bar \delta^{2|4}(\calZt_{k-1},\calZt_k,\calZt_{k+1})\eqncom
 \end{equation}
  where we have used the $\bar \delta^{3|4}$ to replace the $\mathcal{Z}_z(\sigma_{k\pm 1})$ by $\tilde{\calZ}_{k\pm 1}$, to obtain 
\begin{equation}
\label{eq:amplitudeverticesinversesoftlimit}
\ver{\calZt_1,\ldots, \calZt_{n}}=\ver{\calZt_1,\ldots,\calZt_{k-1},\calZt_{k+1},\ldots \calZt_{n}}\bar{\delta}^{2|4}(\calZt_{k-1},\calZt_{k},\calZt_{k+1})\eqndot
\end{equation}
Equation \eqref{eq:amplitudeverticesinversesoftlimit} implies that the vertex can be written as the product
\begin{equation}
\label{eq:inversesoftlimitamplitudes}
\ver{\calZt_1,\ldots, \calZt_{n}}=\ver{\calZt_1,\calZt_2}\prod_{i=3}^n\bar{\delta}^{2|4}(\calZt_1,\calZt_{i-1},\calZt_i)\eqndot
\end{equation}
There are different ways of writing the $n$-point amplitude that are related by identities of the kind 
$\ver{\calZt_1,\calZt_2,\calZt_3}\bar{\delta}^{2|4}(\calZt_1,\calZt_{3},\calZt_4)=\ver{\calZt_2,\calZt_3,\calZt_4}\bar{\delta}^{2|4}(\calZt_2,\calZt_{4},\calZt_1)$, which come from the total antisymmetry of the $\bar\delta^{2|4}$ and from the different choices of points $k$ to remove using \eqref{eq:amplitudeverticesinversesoftlimit}.\newline\\
\paragraph{The general NMHV amplitude in twistor space}
Let us use the inverse soft limit to simplify each summand in the expression for the general amplitude \eqref{eq: NMHV amplitude in position twistor space},
\begin{equation}
\begin{aligned}
&\int \DD^{3|4}\calZ_{[1]}\DD^{3|4}\calZ_{[2]}\ver{\ldots, \calZt_j,\calZ_{[1]},\calZt_{k+1},\ldots}\Delta(\calZ_{[1]},\calZ_{[2]})\ver{\ldots, \calZt_{k},\calZ_{[2]},\calZt_{j+1},\ldots}\\
&=\nalpha\ver{\ldots, \calZt_{j},\calZt_{k+1},\ldots}\ver{\ldots, \calZt_k,\calZt_{j+1},\ldots}\\
&\phaneq\times \int \DD^{3|4}\calZ_{[1]}\DD^{3|4}\calZ_{[2]}
\bar{\delta}^{2|4}(\calZt_{j},\calZ_{[1]},\calZt_{k+1})
\bar{\delta}^{2|4}(\calZ_{[1]},\star,\calZ_{[2]})
\bar{\delta}^{2|4}(\calZt_{k},\calZ_{[2]},\calZt_{j+1})\\
&=\nalpha \ver{\ldots, \calZt_{j},\calZt_{k+1},\ldots}\ver{\ldots, \calZt_k,\calZt_{j+1},\ldots}[\calZt_{k+1},\calZt_j,\star,\calZt_{j+1},\calZt_{k}]\eqndot
\end{aligned}
\end{equation}
The five-bracket $[\cdot,\cdot,\cdot,\cdot,\cdot]$, also called the R-invariant, was first introduced in \cite{Drummond:2008vq} and plays an important role in the rest of this work. It is defined as
\begin{equation}
\begin{aligned}
\label{eq:fivebracket}
[\calZ_{1},\calZ_{2},\calZ_{3},\calZ_{4},\calZ_{5}]&\equiv
\bar{\delta}^{0|4}(\calZ_{1},\calZ_{2},\calZ_{3},\calZ_{4},\calZ_{5})=
\frac{\prod_{a=1}^4(\chi_{1}(2 3 4 5)+\mathrm{cyclic})^a}{(1234)(2 3 4 5)(3451)(4512)(5123)}
\end{aligned}
\end{equation} 
with the four-bracket $(ijkl)\equiv \det(Z_i,Z_j,Z_k,Z_l)$ given by the determinant of four bosonic twistors, interpreted as non-homogeneous four-vectors. Note that the five-bracket, like $\bar\delta^{i|4}$, is totally antisymmetric in its five arguments.
Summing over all diagrams of this type yields the NMHV amplitude 
\begin{equation}
\label{eq: NMHV amplitude in position twistor space final}
\begin{aligned}
\mathbb{A}^{\text{NMHV}}= -\nalpha \sum_{1\leq j<k\leq n}&\ver{\ldots, \calZt_{j},\calZt_{k+1},\ldots}\\& \times [\calZt_{j},\calZt_{j+1},\star,\calZt_{k},\calZt_{k+1}]\ver{\ldots, \calZt_k,\calZt_{j+1},\ldots}\eqncom
\end{aligned}
\end{equation}
where we have used said antisymmetry.
In Section~\ref{eq:NMHV amplitudes in momentum twistor space}, we shall derive the usual expression for the NMHV amplitudes in momentum twistor space from \eqref{eq: NMHV amplitude in position twistor space final} by inserting the momentum eigenstates and computing the integrals that are left implicit in $\mathbf{V}$.

\section{Review summary and motivation}
In Chapter~\ref{review1} we reviewed three important classes of observables in $\calN=4$ SYM: correlation functions of local composite operators, form factors and amplitudes. In this chapter we discussed the formulation of the theory in twistor space.
We saw that the twistor action reduces to the standard $\calN=4$ SYM action when one fixes all the additional gauge degrees of freedom that appear in twistor space. Alternatively, one can impose a different, axial gauge with the help of a reference twistor in which the Feynman rules in twistor space translate to the CSW rules for amplitudes. At MHV level, the MHV vertices straightforwardly generate all MHV amplitudes in momentum space. At NMHV level, connecting two MHV vertices with a twistor propagator gives the general expression for the NMHV amplitude in twistor space as was shown in Subsection~\ref{NMHVamp}. Having discussed how amplitudes can be more naturally computed in twistor space, in the rest of this work we investigate how we can calculate also form factors and finally correlation functions in this language.
\makeatletter
\def\toclevel@part{0}
\makeatother
\part{Form factors and correlation functions from twistor space}
\chapter{The scalar sector}
\label{chap:}
In the previous chapter we reviewed twistor space and the twistor action. We saw that in a partial gauge the twistor action reduces to the well-known action for $\calN=4$ SYM. Imposing a different, axial gauge, one obtains all MHV amplitudes straightforwardly by inserting external on-shell states. The CSW rules, that were discussed in Section~\ref{sec:scatteringamps}, appeared as the Feynman rules in twistor space by connecting MHV amplitudes via twistor propagators in Section~\ref{NMHVamp}. In this gauge, the MHV vertices and the propagator are the fundamental building blocks for amplitudes. The discussion of the extension of CSW to form factors of scalar operators in Section~\ref{secformfactor} showed that additional vertices, in the form of off-shell-extended MHV form factors, must be added to the MHV vertices as fundamental building blocks. This means that also in twistor space, there must be a fundamental building block analogue of the operator vertex. In order to find this, we first describe local composite operators in twistor space. Subsequently, one can compute its MHV form factors by inserting on-shell states and move on to higher levels in MHV degree, which parallels the construction for NMHV amplitudes. 
In this chapter, to motivate our construction as well as familiarize the reader with the concepts and notation, we start by considering an operator consisting only of scalars, $\tfrac{1}{2}\Tr(\phi^2)$. In the first section we see that the expression for the scalar field as was known from the literature is incomplete. We modify this expression and show that it yields all tree-level MHV form factors of $\tfrac{1}{2}\Tr(\phi^2)$ in much the same way as the action vertices yielded all MHV amplitudes. Having gained some confidence after this initial success at MHV level, we compute an NMHV form factor of the same operator in the subsequent section. This calculation provides motivation for the section that follows, and serves as warm up for Chapter~\ref{NMHV}. In the last section we show how our operator vertex can be obtained from acting with a suitable derivative operator on a Wilson loop. This section touches on the main concepts and ideas that will be fully detailed in the next chapter, where we will extend the Wilson loop construction to all local composite operators. This chapter is based on and contains overlap with \cite{Koster:2016ebi}.
\section{The operator vertex for $\tfrac{1}{2}\Tr(\phi^2)(x)$ }
\label{examplescalar}
In this section we find the expression for the scalar field in twistor space. More precisely phrased, we define the \textit{vertex} for the scalar field, in analogy to the interaction vertex in the twistor action. As we shall see, it is not the local composite operators that are the fundamental building blocks in our formalism, but rather vertices containing these operators. The vertex of an operator $\calO(x)$ in twistor space will be denoted by $\textbf{W}_{\calO(x)}$. Recall the expansion of the super twistor field $\AAA$, which packages the on-shell degrees of freedom of $\calN=4$ SYM -- the two helicity $\pm 1$ gluons $g^{\pm}$, the four helicity $\frac{1}{2}$ fermions $\bar{\psi}_a$ and their antiparticles $\psi^a$ and the six scalars $\phi_{ab}$ -- as
\begin{equation}
\label{eq:expansionAAA}
\AAA(\mathcal{Z})=g^+(Z) +\chi^a\bar{\psi}_a(Z)+\frac{1}{2}\chi^a\chi^b\phi_{ab}(Z)+\frac{1}{3!}\chi^a\chi^b\chi^c\psi^d(Z)\epsilon_{abcd}+\chi^1\chi^2\chi^3\chi^4 g^-(Z)\eqncom
\end{equation}
where the component fields no longer depend on the Gra\ss mann variables $\chi$, but do on $Z$. Recall furthermore that the Penrose transform relates cohomology classes $H^{0,1}(U, \mathcal{O}(2h-2))$ on an open region of twistor space $U\subset \mathbb{CP}^3$ and mass-less fields of helicity $h$. This means that the component of $\AAA$ with degree of homogeneity $-2$, which is $\phi$, can be related via the Penrose transform to a massless field of helicity $0$, i.e.\ a scalar field. This might suggest that to extract a space-time scalar field from \eqref{eq:expansionAAA}, one merely needs to act with the derivatives $\partial/\partial \chi^a$ and $\partial/\partial \chi^b$ on $\AAA$, set $\theta=0$, and apply the Penrose transform \eqref{Penrosetransformscalar}. Let us investigate this ansatz
\begin{equation}
\label{eq:naturalfirstattempt}
\textbf{W}_{\phi_{ab}(x)}\stackrel{?}=\int \DD\la \, h_{(x,\theta)}^{-1}(\la)\frac{\partial^2\calA(\lambda)}{\partial\chi^a\partial\chi^b}h_{(x,\theta)}(\la)\rvert_{\theta=0}\eqndot
\end{equation} 
We have written a question mark above the equation to indicate that this is an ansatz that we will soon see is wrong.
Here $h_{(x,\theta)}$ is a holomorphic frame that trivializes the restriction of the bundle $\mathcal{E}\rightarrow \mathbb{CP}^{3|4}$ to each point $(x^{\alpha\dot\alpha},\theta^{\alpha a})$ in super Minkowski space in analogy to the holomorphic frame of Section~\ref{classint} associated to the bundle over bosonic twistor space. For brevity, we will simply denote the point in super Minkowski space $(x,\theta)$, as well as the associated line (Riemann sphere) in twistor space by $x$ whenever there is no risk of confusion. The frame $h$ trivialises the gauge connection $\AAA$ along the line $x$ and is included to guarantee gauge covariance of the scalar field. In the remainder of this chapter we will not write out the flavor indices for the scalar field explicitly, but simply denote it by $\phi$. Using the ansatz \eqref{eq:naturalfirstattempt}, we build the vertex for the simplest local composite operator $\tfrac{1}{2}\Tr(\phi^2)$,
\begin{equation}
\label{eq:operator}
\textbf{W}_{\frac{1}{2}\Tr(\phi^2)(x)}\stackrel{?}= \frac{1}{2}\int \DD\la \DD{\la'}\,\Tr\Big[\frac{\partial^2\calA(\la)}{\partial\chi^a\partial\chi^b}U_x(\la,\la')\frac{\partial^2\calA(\la')}{\partial\chi'^a\partial\chi'^b}U_x(\la',\la)\Big]_{\big{|}\theta=0}\,,
\end{equation} where 
\begin{equation}\label{propagatorU} U_x(\la,\la')=h_{x}(\la)h^{-1}_x(\la')
\end{equation}is a parallel propagator from $\la$ to $\la'$ on the line $x$ and ensures gauge invariance of the operator. 
This parallel propagator $U_x(\la,\la')$ admits an expansion in the superfield $\AAA$ as follows
\begin{equation}
\label{eq:frameUdefinitionpart1}
 U_{x}(\la,\la')\equiv 
  U_{x}(\calZ_x(\la),\calZ_x(\la'))=1+\sum_{m=1}^\infty\int \frac{\abra{\lambda}{\lambda'}\DD\tilde{\la}_1\cdots \DD\tilde{\la}_m\;\AAA(\tilde{\la}_1) \cdots \AAA(\tilde{\la}_m)}{\abra{\lambda}{\tilde{\la}_1}\abra{\tilde{\la}_1}{\tilde{\la}_2}\cdots \abra{\tilde{\la}_m}{\lambda'}}\eqncom
\end{equation}
where all $\AAA(\tilde\la_j)\equiv\AAA(\calZ(\tilde\la_j,x,\theta))$. Inserting \eqref{eq:frameUdefinitionpart1} into the ansatz \eqref{eq:operator}, one obtains
\begin{align}
\label{eq:operatorexp}
&\textbf{W}_{\frac{1}{2}\Tr(\phi^2)(x)}\stackrel{?}= \frac{1}{2}\sum_{m,n=0}^\infty \int \frac{\DD\la \DD{\la'} \DD\tilde{\la}_1\cdots \DD\tilde{\la}_m\DD\hat{\la}_{1}\cdots \DD\hat{\la}_{n}\abra{\la}{\la'}\abra{\la'}{\la}}{\abra{\lambda}{\tilde{\la}_1}\abra{\tilde{\la}_1}{\tilde{\la}_2}\cdots \abra{\tilde{\la}_m}{\lambda'}\abra{\lambda'}{\hat\la_1}\abra{\hat{\la}_1}{\hat{\la}_2}\cdots \abra{\hat{\la}_{n}}{\lambda}}\notag\\
&\times\Tr\Big[\frac{\partial^2\calA(\la)}{\partial\chi^a\partial\chi^b} \AAA(\tilde{\la}_1) \cdots \AAA(\tilde{\la}_m)\frac{\partial^2\calA(\la')}{\partial\chi'^a\partial\chi'^b} \AAA(\hat{\la}_{1}) \cdots \AAA(\hat{\la}_{n})\Big]_{\big{|}\theta=0}\eqndot
\end{align} By relabeling the variables $\la$, $\la'$, $\tilde\la_i$, $\hat\la_k$ and using the cyclicity of the trace we can write this as
\begin{align}
\label{eq:operatorexp}
&\textbf{W}_{\frac{1}{2}\Tr(\phi^2)(x)}\stackrel{?}= \frac{1}{2}\sum_{\substack{n-j+i=1\\ j-i=1}}^\infty \int \frac{\DD\la_1\cdots \DD{\la}_{n}\abra{\la_i}{\la_j}\abra{\la_j}{\la_i}}{\abra{\lambda_1}{{\la}_2}\cdots \abra{{\la}_n}{\lambda_1}}\notag\\
&\times\Tr\Big[ \AAA({\la}_1) \cdots \frac{\partial^2\calA(\la_i)}{\partial\chi_i^a\partial\chi_i^b}\AAA({\la}_{i+1})\cdots \frac{\partial^2\calA(\la_j)}{\partial\chi_j^a\partial\chi_j^b} \AAA({\la}_{j})\AAA(\la_{j+1}) \cdots \AAA({\la}_{n})\Big]_{\big{|}\theta=0}\,\eqndot
\end{align}
Note that this expression shares similarities with the expression for the interaction vertices in the expansion of the twistor action. 
We can now probe the ansatz by computing tree-level MHV form factors of $\tfrac{1}{2}\Tr(\phi^2)$ using expression \eqref{eq:operatorexp} and comparing the results to the literature. The computation is completely analogous to the computation of MHV amplitudes from the interaction vertices of the twistor action. 
Recall that a form factor is the expectation value of a state created from the vacuum by the operator $\tfrac{1}{2}\Tr(\phi^2)$ and a state of $n$ on-shell particles of helicity $h_i$,
\begin{equation}
\calF_{\tfrac{1}{2}\Tr(\phi^2)}= \langle1^{h_1},\dots, n^{h_n}| \textbf{W}_{\tfrac{1}{2}\Tr(\phi^2)(x)} |0 \rangle\eqncom
\end{equation}
where we recall \eqref{eq:definitiononshellmomentumeigenstates}, the external on-shell momentum eigenstates $\calA_{\fP}$ of supermomentum $\fP=(\fp_{\alpha\dot\alpha},\eta_a)= (p_{\alpha},\bar{p}_{\dot \alpha},\eta_a)$ \cite{Adamo:2011cb} given by
\begin{equation}
\calA_{\fP}(\calZ)=2\pi i \int_{\mathbb{C}}\frac{\dd s}{s}\e^{s(\mu^{\dot{\alpha}}\bar{p}_{\dot{\alpha}}+\chi^a\eta_{a})}
\bar{\delta}^2(s\lambda-p)\eqndot
\end{equation}
The external states are cyclicly Wick contracted with the fields that constitute the operator vertex $\textbf{W}_{\frac{1}{2}\Tr(\phi^2)(x)}$. Therefore, only the terms in \eqref{eq:operatorexp} of length $n$ contribute.
To set the external eigenstates on shell, we perform an LSZ reduction which effectively amounts to just replacing every field in \eqref{eq:operatorexp} by an on-shell state \eqref{eq:definitiononshellmomentumeigenstates} in a cyclic manner.\newline\\
Up to now, we have not specified the helicities of the outgoing states. 
As a first example, let us compute the minimal form factor of $\tfrac{1}{2}\Tr(\phi)^2$, which has two outgoing on-shell scalar fields. This process is maximally helicity violating and has a very simple expression, namely just the momentum conserving delta function of the two on-shell momenta minus the off-shell momentum of the operator.
As was mentioned before, the process of Wick contracting the fields only yields a nonzero result when the length of the term in the expansion \eqref{eq:operatorexp} equals the number of outgoing particles. In other words, we only pick up the term of length two and replace the two twistor fields by the on-shell states \eqref{eq:definitiononshellmomentumeigenstates}, which gives us
\begin{align}
& \frac{1}{2} \int \DD\la_1 \DD{\la}_{2} \Tr\Big[ \frac{\partial^2\calA(\la_1)}{\partial\chi_1^a\partial\chi_1^b} \frac{\partial^2\calA(\la_2)}{\partial\chi_2^a\partial\chi_2^b} \AAA({\la}_{2})\Big]_{\big{|}\AAA(\la_i)=\AAA_{\fP_i}(\la_i),\,\theta=0}\notag\\
&= \int \DD\la_1 \DD{\la}_{2}2\pi i \int_{\mathbb{C}}\frac{\dd s_1}{s_1}2\pi i \int_{\mathbb{C}}\frac{\dd s_2}{s_2}\Tr\Big[ \frac{\partial^2\e^{s_1(\mu_1^{\dot{\alpha}}\bar{p}_{1\dot{\alpha}}+\chi_1^a\eta_{1a})}}{\partial\chi_1^a\partial\chi_1^b} \frac{\partial^2\e^{s_2(\mu_2^{\dot{\alpha}}\bar{p}_{2\dot{\alpha}}+\chi_2^a\eta_{2a})}}{\partial\chi_2^a\partial\chi_2^b}\Big]_{\big{|}\theta=0} \notag\\
&\times \bar{\delta}^2(s_1\lambda_1-p_1)\bar{\delta}^2(s_2\lambda_2-p_2)\notag\\
&= \eta_{1a}\eta_{1b}\eta_{2a}\eta_{2b}\exp(ix^{\alpha\dot\alpha}(\fp_1+\fp_2)_{\alpha\dot\alpha})\eqncom
\end{align}
where $\fp_{i\alpha\dot\alpha}=p_{i\alpha}\bar p_{\dot\alpha}$.
Note that integrating over $s_i$ and the corresponding $\la_i$ cancels $s_i$ and replaces $\la_{i\alpha}\rightarrow p_{i\alpha}$, $\mu^{\dot{\alpha}}\rightarrow ix^{\alpha \dot{\alpha}}p_{i\alpha}$ and $\chi_i^a\rightarrow i\theta^{\alpha a }p_{i\alpha}$ due to the $\bar{\delta}^2$ function.
Collecting the coefficient of $ \eta_{1a}\eta_{1b}\eta_{2a}\eta_{2b}$ and performing the Fourier transformation of the operator
$\int \dd^4 x\, e^{-i\fq x}$, we find indeed the desired result
\begin{equation}
\calF_{\tfrac{1}{2}\Tr(\phi^2)(x)}(1^{\phi},2^{\phi};\fq)= \delta^4(\fq-\fp_1-\fp_2)\eqndot
\end{equation}
This is exactly the expression for the minimal form factor \eqref{minimalformfactorscalar}.
Although this is reassuring, we have not tested the validity of the appearance of the infinite expansion of twistor fields in the ansatz. Let us proceed by computing the form factor of the same operator, with two outgoing scalars and in addition $n-2$ outgoing positive helicity gluons,
\begin{equation}
\label{eq:thefirstformfactor}\calF_{\tfrac{1}{2}\Tr(\phi^2)}(1^{+},\ldots, i^{\phi_{ab}},\ldots, j^{\phi_{ab}},\ldots, n^{+};\fq)\eqndot
\end{equation} To be precise, this is the (color-ordered) form factor of $\tfrac{1}{2}\Tr(\phi^2)$ with $n$ external particles: two scalars $\phi_{ab}$ at position $i$ and $j$ and $n-2$ positive-helicity gluons. Again, only the terms in \eqref{eq:operatorexp} with the appropriate number of $\calA$'s contribute, namely those with $j-i-1$ from one $U_x$ and $n+i-j-1$ from the other. Inserting the on-shell states \eqref{eq:definitiononshellmomentumeigenstates} into \eqref{eq:operatorexp} and subsequently integrating over the variables $s_i$ with the corresponding $\la_i$, leads to the dropping out of all $s_i$ and replacements $\la_{i\alpha}\rightarrow p_{i\alpha}$, $\mu_i^{\dot{\alpha}}\rightarrow ix^{\alpha \dot{\alpha}}p_{i\alpha}$ and $\chi_i^a\rightarrow i\theta^{\alpha a }p_{i\alpha}$ just as in the previous example. This also effectively gives $\partial\chi_i^a\rightarrow \eta_i^a$. To find the form factor with a scalar at positions $i$ and $j$, we select the coefficient of $\eta_{i a}\eta_{i b}\eta_{j a}\eta_{j b}$. Fourier transforming in $x$ as $\int \frac{\dd^4x}{(2\pi)^4} \e^{-i\fq x}$ yields the desired form factor 
\begin{equation}
\label{eq:formfactor2phi}
\calF_{\tfrac{1}{2}\Tr(\phi^2)}(1^{+},\ldots, i^{\phi_{ab}},\ldots, j^{\phi_{ab}},\ldots, n^{+};\fq)
=-\frac{\abra{i}{j}^2\delta^4(\fq-\sum_{k=1}^n\fp_k)}{\abra{1}{2}\cdots \abra{n}{1}}\,,
\end{equation} where $\abra{i}{j}\equiv \abra{\la_i}{\la_j}$.
This is precisely the expression found in \cite{Brandhuber:2010ad}. Let us continue to perform a final check for the ansatz by computing the MHV form factor
\beq
\label{eq:formactorpsisquared}
\calF_{\tfrac{1}{2}\Tr(\phi^2)}(1^{\bar{\psi}_a},2^{\bar{\psi}_b},3^{\phi_{ab}};\fq)
\eeq
which was first calculated in \cite{Brandhuber:2011tv}.  Since the form factor has three external states, we select the two terms in $\textbf{W}_{\frac{1}{2}\Tr(\phi^2)(x)}$ of length three, which (after conveniently relabeling the $\la$'s) read
\begin{align}
\label{eq:operatorexpl3}
&\textbf{W}_{\frac{1}{2}\Tr(\phi^2)(x)}\stackrel{?}= 2\times \frac{1}{2}\int \frac{\DD\la_1\DD\la_2 \DD{\la}_{3}\abra{\la_1}{\la_3}\abra{\la_3}{\la_1}}{\abra{\lambda_1}{{\la}_2} \abra{{\la}_2}{\lambda_3}\abra{{\la}_3}{\lambda_1}}\Tr\Big[ \frac{\partial^2\calA(\la_1)}{\partial\chi_1^a\partial\chi_1^b}\AAA({\la}_{2}) \frac{\partial^2\calA(\la_3)}{\partial\chi_3^a\partial\chi_3^b}\Big]_{\big{|}\theta=0}\,.
\end{align}
The form factor that we are interested in is the coefficient of $\eta_{1a}\eta_{2b}\eta_{3a}\eta_{3b}$ after inserting external states \eqref{eq:definitiononshellmomentumeigenstates} into \eqref{eq:operatorexpl3} and integrating over the delta functions. However, this trivially gives zero, because the four Gra\ss mann derivatives yield $\eta_{1a}\eta_{1b}\eta_{3a}\eta_{3b}$. However, this form factor is known to be nonzero from the literature. Therefore, despite the suggestive examples we did earlier, we must conclude that our ansatz \eqref{eq:naturalfirstattempt} is not correct, or at least not complete.\newline\\
From our failed proposal for the vertex \eqref{eq:operatorexpl3} we learn that the expression must be completed such that it allows for the decay of each of the elementary fields $\phi$ into different types of fields on the condition that this decay preserves the MHV degree. In other words, at MHV and tree level, a scalar field is allowed to decay into two fermions and arbitrary many positive helicity gluons, or a scalar field and an arbitrary number of positive helicity gluons. Any other process is of higher MHV degree. More specifically, we should include terms with two twistor fields with each one $\chi$-derivative connected by a parallel propagator $U$, as in
\begin{align}
\label{eq:secondattempt}
\textbf{W}_{\phi(x)}=&\int \DD\la\,  h_x^{-1}(\la)\frac{\partial^2\calA(\lambda)}{\partial\chi^a\partial\chi^b}h_x(\la)_{\big{|}\theta=0}\notag\\&+\int \frac{\DD\la\DD\la'}{\abra{\la}{\la'}} h_x^{-1}(\la)\frac{\partial\calA(\lambda)}{\partial\chi^a}U_x(\la,\la')\frac{\partial\calA(\lambda')}{\partial{\chi'}^b}h_x(\la')_{\big{|}\theta=0}\notag\\&-(a\leftrightarrow b)\eqncom
\end{align}
where the fraction $\tfrac{1}{\abra{\la}{\la'}}$ ensures that the integrand is homogeneous of degree $0$ in both $\la$ and $\la'$.
This expression is schematically shown in   Figure~\ref{fig:Splitting}.\begin{figure}[htbp]
 \centering
  \includegraphics[height=3.5cm]{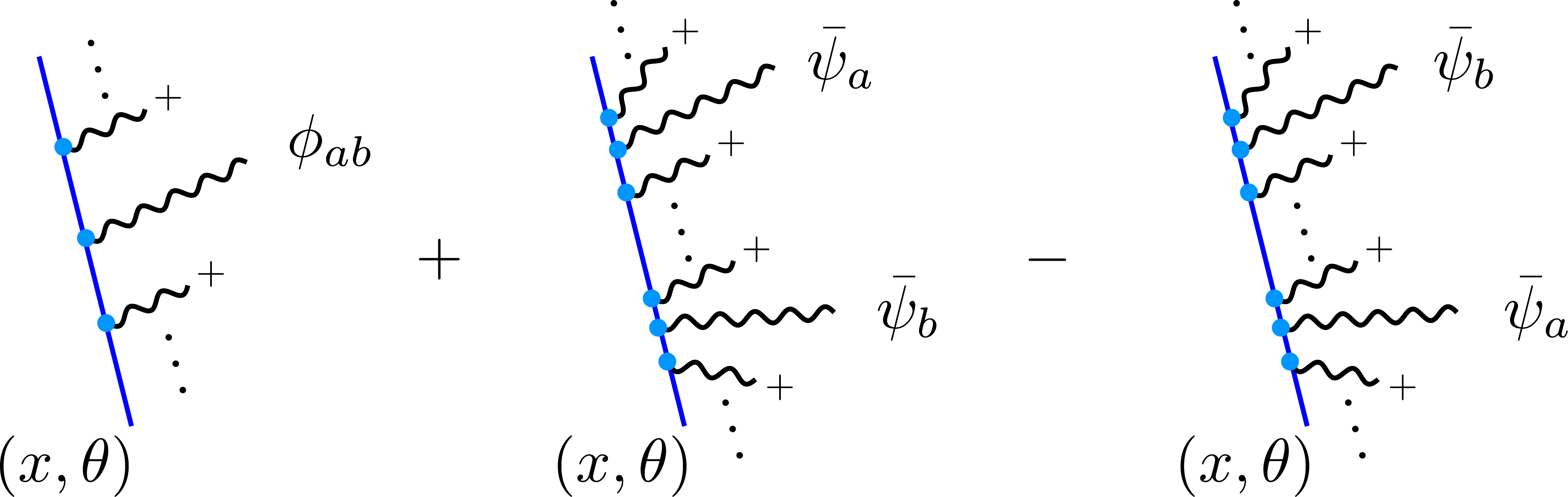}
  \caption{\it The vertex of an operator containing a scalar $\phi_{ab}$ includes all its MHV-preserving splitting terms.}
  \label{fig:Splitting}
\end{figure} From this, the corrected expression for the vertex of the operator $\tfrac{1}{2}\Tr(\phi^2)(x)$ could be easily obtained by writing out $\tfrac{1}{2}\Tr(\textbf{W}_{\phi(x)}\textbf{W}_{\phi(x)})$. Indeed, after inserting external momentum-eigenstates \eqref{eq:definitiononshellmomentumeigenstates} into the length 3 terms in $\textbf{W}_{\frac{1}{2}\Tr(\phi^2)(x)}$ and selecting the coefficient of $\eta_{1a}\eta_{2b}\eta_{3a}\eta_{3b}$ we find 
\begin{equation}
\label{formfactorscalarwithfermions}
\calF_{\tfrac{1}{2}\Tr(\phi^2)}(1^{\bar{\psi}_a},2^{\bar{\psi}_b},3^{\phi_{ab}};\fq)=\frac{\delta^4(\fq-\sum_{k=1}^3\fp_k)(-1)\abra{1}{3}\abra{2}{3}}{\abra{1}{2}\abra{2}{3}\abra{3}{1}}=\frac{\delta^4(\fq-\sum_{k=1}^3\fp_k)}{\abra{1}{2}}\,,
\end{equation}
which precisely matches the result of \cite{Brandhuber:2011tv}.

\section{The MHV super form factor of $\frac{1}{2}\Tr(\phi^2)$}
\label{superff}
Now looking at the result \eqref{formfactorscalarwithfermions}, together with the corresponding $\eta$'s,
\begin{equation}
\calF_{\tfrac{1}{2}\Tr(\phi^2)}(1^{\bar{\psi}_a},2^{\bar{\psi}_b},3^{\phi_{ab}};\fq)\eta_{1a}\eta_{3a}\eta_{2b}\eta_{3b}=\frac{\delta^4(\fq-\sum_{k=1}^3\fp_k)}{\abra{1}{2}\abra{2}{3}\abra{3}{1}}\abra{1}{3}\abra{2}{3}\eta_{1a}\eta_{3a}\eta_{2b}\eta_{3b}\eqncom
\end{equation}
we realize that this can be straightforwardly extended to
\begin{align}
&\calF_{\tfrac{1}{2}\Tr(\phi^2)}(1^{g^+},\dots,i^{\bar{\psi}_a},(i+1)^{g^+},\dots,j^{\bar{\psi}_b},\dots,l^{\phi_{ab}}, (l+1)^{g^+},\dots, n^{g^+};\fq)\eta_{ia}\eta_{la}\eta_{jb}\eta_{lb}\notag\\\
&=\frac{\delta^4(\fq-\sum_{k=1}^n\fp_k)}{\abra{1}{2}\cdots\abra{i}{i+1}\cdots\abra{j}{j+1}\cdots\abra{l}{l+1}\cdots \abra{n}{1}}\abra{i}{l}\abra{j}{l}\eta_{ia}\eta_{la}\eta_{jb}\eta_{lb}\eqncom
\end{align}
since any additional external positive helicity gluon will only change the Parke-Taylor denominator.
This can be combined with \eqref{eq:formfactor2phi} multiplied by its corresponding $\eta$'s,
\begin{equation}
\calF_{\tfrac{1}{2}\Tr(\phi^2)}(1^{+},\ldots, i^{\phi_{ab}},\ldots, j^{\phi_{ab}},\ldots, n^{+};\fq)
=\frac{\delta^4(\fq-\sum_{k=1}^n\fp_k)}{\abra{1}{2}\cdots \abra{n}{1}}\abra{i}{j}^2\eta_{ia}\eta_{ja}\eta_{ib}\eta_{jb}\,,
\end{equation}
into the super form factor of $\tfrac{1}{2}\Tr(\phi^2)$
\begin{equation}
\label{eq:calFsusy}
\mathscr{S}\calF_{\tfrac{1}{2}\Tr(\phi^2)}(1,\dots,n;\fq)\\=\frac{\delta^4(\fq-\sum_{k=1}^n\fp_k)\prod_{c=a,b}\left(\sum_{i<j}\abra{i}{j}\eta_{ic}\eta_{jc}\right)}{\abra{1}{2}\abra{2}{3}\cdots \abra{n}{1}}\eqndot
\end{equation}
Using this expression, one can obtain all tree-level MHV form factors by simply taking derivatives with respect to four suitable $\eta$'s.
\section{An NMHV form factor of $\frac{1}{2}\Tr(\phi^2)$}
\label{NMHVscalar}
After the successful computation of all tree-level MHV form factors of $\frac{1}{2}\Tr(\phi^2)$, the next step is to see whether we can also compute some NMHV form factors. This section will show a step by step computation of such a form factor in close analogy to the computation that was done in Section~\ref{secformfactor} using CSW. It serves also as a motivation for the next section and as a warm up for what is to come in Chapter~\ref{NMHV}.
We consider the same NMHV form factor that was calculated in Section~\ref{secformfactor} using the CSW recursion in momentum space, namely the NMHV form factor $\calF_{\Tr(\phi^2)}(1^{\phi},2^{\phi},3^-)$ of the operator $\frac{1}{2}\Tr(\phi^2)$ with three outgoing fields, one of which a negative helicity gluon and the other two scalar fields. It is the coefficient of $\eta_{1a}\eta_{1b}\eta_{2a}\eta_{2b}(\eta_{3})^4$ of the super NMHV form factor. Note that the NMHV form factor comes with $4$ extra $\eta$'s. This is general: the N$^k$MHV form factor comes with $4k$ additional $\eta$'s with respect to the MHV form factor. Here, the two diagrams that contribute to the process are depicted in Figure~\ref{NontrivNMHV}.
\begin{figure}[h!]\centering
\def\svgwidth{\linewidth}
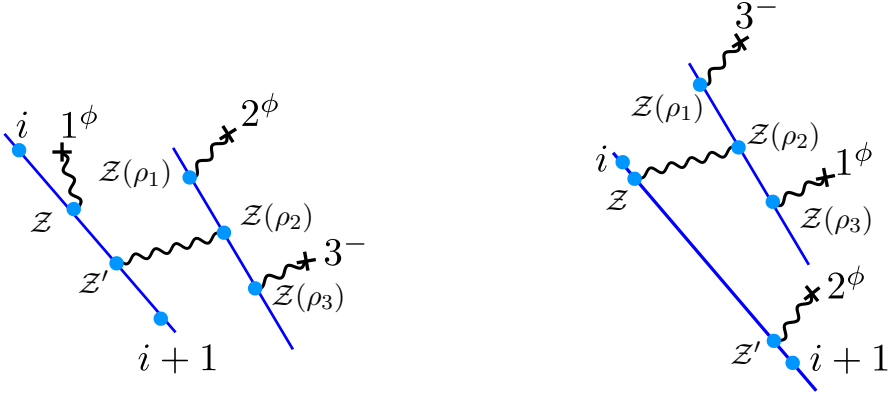
\caption{The two diagrams that contribute to the form factor $\calF_{\Tr(\phi^2)}(1^{\phi},2^{\phi},3^-)$.}
\label{NontrivNMHV}
\end{figure}
Let us start by computing the diagram on the lefthand side. The expression for the this diagram reads
\begin{align}
&\int \dd^4z\, \dd^8\vartheta\, \DD\la\,\DD\la'\,\DD\rho_1\,\DD\rho_2\,\DD\rho_3\,\frac{\partial^2\AAA_{\fP_1}(\calZ_x(\la))}{\partial\chi^a\partial\chi^b}\bigg\rvert_{\theta=0} \notag\\
&\times\frac{\partial^2}{\partial\chi'^{a}\partial\chi'^b} \bar\delta^{2|4}(\calZ_x(\la'),\star,\calZ_z(\rho_1))\bigg\rvert_{\theta=0} \frac{\AAA_{\fP_2}(\calZ_z(\rho_2)\AAA_{\fP_3}(\calZ_z(\rho_3))}{\abra{1}{2}\abra{2}{3}\abra{3}{1}}\eqncom
\end{align}
where the super twistors $\calZ_x(\la)$ and $\calZ_x(\la')$ are on the line $(x,\theta)$ and the super twistors $\calZ_z(\rho_i)$ are on the interaction line $(z,\vartheta)$. Now, we may use the following local coordinate chart for $\rho_1$: $
\rho_{1\alpha}=\rho_{2\alpha}+r\rho_{3\alpha}$, such that $\tfrac{\abra{\rho_1}{\dd\rho_1}}{\abra{\rho_1}{\rho_2}\abra{\rho_3}{\rho_1}}=-\tfrac{\dd r}{r \abra{\rho_2}{\rho_3}}$. This is in fact just the inverse soft limit for the interaction vertex that was reviewed in Section~\ref{NMHVamp}. We can do something similar for the spinor $\la'$. Choose two fixed twistors $\calZ_1$ and $\calZ_{i+1}$ on the line $x$, and write\footnote{This is an example of the inverse soft limit for operator vertices that will be generalized in Section~\ref{subsec: inversesoftlimit}.} $\la'= \la_{i\alpha} +t\la_{i+1\alpha}$. This replaces
\begin{equation}
\int_{\mathbb{CP}^1} \DD\la' \frac{\partial^2}{\partial\chi'^{a}\partial\chi'^{b}}\rightarrow \int_{\mathbb{C}} \frac{\dd t}{t}\abra{\la_{i}}{\la_{i+1}}\frac{\partial^2}{\partial\chi_i^a\partial\chi_{i+1}^b}\eqndot
\end{equation} Doing these substitutions yields
\begin{align}
&\int \DD\la\, \dd^4 z\, \dd^8 \vartheta\,  \abra{\la_{i}}{\la_{i+1}} \frac{\partial^2}{\partial\chi_i^a\partial\chi_{i+1}^b} [\calZ_x(\la_i),\calZ_x(\la_i+1),\star,\calZ_z(\rho_2),\calZ_z(\rho_3)]\bigg\rvert_{\theta=0}\notag\\
&\times \frac{-1}{\abra{\rho_2}{\rho_3}^2}\frac{\partial^2\AAA_{\fP_1}(\calZ_x(\la))}{\partial\chi^a\partial\chi^b}\bigg\rvert_{\theta=0}\AAA_{\fP_2}(\calZ_z(\rho_2))\AAA_{\fP_3}(\calZ_z(\rho_3))\\
&=\int \DD\la\, \dd^4 z\, \dd^8\vartheta\, \DD\rho_2\,\DD\rho_3\,\abra{\la_{i}}{\la_{i+1}} \frac{-1}{\abra{\rho_2}{\rho_3}^2}\frac{\partial^2\AAA_{\fP_1}(\calZ_x(\la))}{\partial\chi^a\partial\chi^b}\bigg\rvert_{\theta=0}\AAA_{\fP_2}(\calZ_z(\rho_2))\AAA_{\fP_3}(\calZ_z(\rho_3))\notag\\
& \frac{\partial^2}{\partial\chi_i^a\partial\chi_{i+1}^b} \frac{\bar\delta^{0|4}(\langle \la_{i+1}|x-z|\zeta]\chi_{i}- \langle \la_{i}|x-z|\zeta]\chi_{i+1} + \langle \rho_{3}|x-z|\zeta]\chi_{2} - \langle \rho_{2}|x-z|\zeta]\chi_{3})}{\langle \la_{i+1}|x-z|\zeta] \langle \la_{i}|x-z|\zeta] \langle \rho_{3}|x-z|\zeta]\langle \rho_{2}|x-z|\zeta]|x-z|^2\abra{\la_i}{\la_{i+1}}\abra{\rho_2}{\rho_3}}\bigg\rvert_{\theta=0}\eqncom
\end{align}
where the derivatives on the R-invariant are now straightforward. After performing these derivatives explicitly, we obtain
\begin{equation}\begin{split}
\int \DD\la\, \dd^4 z\, \dd^8\vartheta\, \DD\rho_2\,\DD\rho_3 \frac{-1}{\abra{\rho_2}{\rho_3}^2}\frac{\partial^2\AAA_{\fP_1}(\calZ_x(\la))}{\partial\chi^a\partial\chi^b}\bigg\rvert_{\theta=0}\AAA_{\fP_2}(\calZ_z(\rho_2))\AAA_{\fP_3}(\calZ_z(\rho_3))\notag\\
\frac{\bar\delta^{0|2}( \langle \rho_{3}|x-z|\zeta]\chi_{2} - \langle \rho_{2}|x-z|\zeta]\chi_{3})_{ab}}{\langle \rho_{3}|x-z|\zeta]\langle \rho_{2}|x-z|\zeta]|x-z|^2\abra{\rho_2}{\rho_3}}\bigg\rvert_{\theta=0}
\end{split}
\end{equation}
Now we can insert on-shell external states \eqref{eq:definitiononshellmomentumeigenstates} $\AAA_{\fP_1}(\la)$, $\AAA_{\fP_2}(\rho_2)$ and $\AAA_{\fP_3}(\rho_3)$ into this expression to obtain
\begin{equation}\begin{split}
\int\dd^4 z\, \dd^8\vartheta \exp(i x\fp_1)\eta_{1a}\eta_{1b} \frac{\bar\delta^{0|2}(\langle p_2|x-z|\zeta]i\vartheta p_3-\langle p_3|x-z|\zeta]i\vartheta p_2)}{\langle p_2|x-z|\zeta]\langle p_3|x-z|\zeta]|x-z|^2\abra{2}{3}^3}\notag\\
\exp(iz (\fp_2+\fp_3)+i\vartheta(p_2\eta_2+p_3\eta_3)))\eqncom\end{split}
\end{equation}
where $\abra{2}{3}=\abra{p_2}{p_3}$ and $\fp_i\equiv p_i \bar p_i$.
Subsequently, we integrate over $\dd^8\vartheta$, while keeping in mind that we need to pick up the term that has $\eta_{1a}\eta_{1b}\eta_{2a}\eta_{2b}(\eta_{3})^4$.
This results in
\begin{align}
&\int \dd^4 z \frac{\langle p_3|x-z|\zeta]\abra{2}{3}}{\langle p_2|x-z|\zeta]|x-z|^2}\exp(iz (\fp_2+\fp_3)+i\vartheta(p_2\eta_2+p_3\eta_3))) \exp(i x\fp_1)\eta_{1a}\eta_{1b}\eta_{2a}\eta_{2b}(\eta_{3})^4\eqndot
\end{align}
This is a tricky integral of Fourier type that is computed using \eqref{eq:importantFourieridentity} in Appendix~\ref{app:mainfourier}. It results in replacing $z$ by $\fp_2+\fp_3$, which, after stripping off the $\eta$'s, gives us
\begin{equation}
\frac{[2\zeta]}{[3\zeta][23]}  \exp(i x(\fp_1+\fp_2+\fp_3))\eqndot
\end{equation}
After Fourier transforming using $\int\dd x \exp(-i\fq)$, this is
\begin{equation}
\frac{[2\zeta]}{[3\zeta][23]}  \delta^4(\fq-\sum_{k=1}^3\fp_k)\eqndot
\end{equation}
Similarly, the diagram on the righthand side gives
\begin{equation}
\frac{[1\zeta]}{[3\zeta][13]}  \delta^4(\fq-\sum_{k=1}^3\fp_k)\eqndot
\end{equation}
After summing these two expressions and using the Schouten identity, we find that the minimal NMHV form factor of $\Tr(\phi^2)$ with two outgoing scalar fields and an outgoing negative helicity gluon equals
\begin{equation}
\label{eq:nontrivialnmhvformfactor}
\calF_{\Tr(\phi^2)}(1^{\phi},2^{\phi},3^-)=\frac{[12]}{[23][31]} \delta^4(\fq-\sum_{k=1}^3\fp_k)\eqncom
\end{equation}
which is precisely the conjugate of the MHV form factor \eqref{eq:formfactor2phi}, with $i=1$, $j=2$ and $n=3$. Note that the dependence on the reference twistor $\calZ_{\star}=(0,\zeta)$ has dropped out completely as is to be expected of a gauge invariant quantity. We have shown that our formalism can be successfully applied to compute an NMHV form factor. However, a couple of remarks are in order. First of all, the form factor we considered is a minimal NMHV form factor, and therefore relatively simple. The number of diagrams increases significantly when one considers more external legs. Secondly, we needed to make use of a non-trivial Fourier integral that we generalized in Appendix~\ref{app:mainfourier}. However, as we shall see in Chapter~\ref{momentumspace}, for more complicated operators the Fourier trick will no longer hold. Thirdly, it would be desirable if instead of doing these kind of calculations for a specific operator and external states, we could do it for generic operators and external states and only at the very end specify which operator we are dealing with and select the appropriate external state. For this, we need a more sophisticated construction of the operator vertex that can be generalized to the rest of the field content. This is the topic of the next section, where we show how we can construct the scalar operator vertex by acting with derivatives on a Wilson loop. This construction will then be extended to all types of fields in Chapter~\ref{MHV}. The study of more general NMHV form factors is momentarily left and will be picked up again in Chapter~\ref{NMHV}.
\section{Light-like Wilson loop as a generating object}
\label{secwilsonscalar}
Given expression \eqref{eq:secondattempt} for the vertex of the scalar field, it follows that an operator of length $L$ consisting only of scalars will involve $3^L$ terms. This number will rapidly become too large to be of any practical use. Therefore, we search for a more compact way of expressing \eqref{eq:secondattempt}, or the operator vertex of any scalar operator and eventually of any local composite operator. The crucial observation is that the operator vertex of the scalar field in twistor space \eqref{eq:secondattempt} resembles the structure of the product rule of taking two anti-commuting derivatives. 

To be more precise, acting on the parallel propagator $U$ with the derivative with respect to $\theta^{\alpha a}_{\ri}$ gives\footnote{This can be shown by explicitly expanding $U$.}
 \cite{Adamo:2011dq}
\begin{equation}
\label{derivativeonUU}
\frac{\partial}{\partial \theta^{\alpha a}}U_{x}(\la,\la')=i\int \frac{\DD\tilde{\la} \abra{\la}{\la'}}{\abra{\la}{\tilde{\la}}\abra{ \tilde{\la}}{\la'}} U_{x}(\la,\tilde{\la})\tilde{\la}_{\alpha}\frac{\partial \calA(\tilde\la)}{\partial {\tilde{\chi}}^a}U_{x}(\tilde{\la},\la')\eqncom
\end{equation}as can be seen from expanding the parallel propagators $U$.
Now acting with another fermionic derivative yields
\begin{align}
\label{eq:doublethetaderivativeofU}
&\frac{\partial^2U_{x}(\la,\la')}{\partial \theta^{\alpha a}\partial \theta^{\beta b}}
=i^2\!\int
\DD\tilde{\la}_1\frac{\abra{\la}{\la'}}{\abra{\la}{\tilde{\la}_1}\abra{\tilde{\la}_1}{\la'}}U_{x}(\la,\tilde{\la}_1)\tilde{\la}_{1\alpha}\tilde{\la}_{1\beta}\frac{\partial^2\calA(\tilde{\la}_1)}{\partial {\tilde{\chi}_1}^a\partial {\tilde{\chi}_1}^b}U_{x}(\tilde{\la}_1,\la)\\
&+i^2\!\int
\frac{\DD\tilde{\la}_1 \DD\tilde{\la}_2\abra{\la}{\la'}}{\abra{\lambda}{\tilde{\la}_1}\abra{\tilde{\la}_1}{\tilde{\la}_2}\abra{\tilde{\la}_2}{\la'}U_{x}}(\la,\tilde{\la}_1)\tilde{\la}_{1\alpha}\frac{\partial\calA(\tilde{\la}_1)}{\partial{\tilde{\chi}_1}^a}U_{x_{\ri}}(\tilde{\la}_1,\tilde{\la}_2)\tilde{\la}_{2\beta}\frac{\partial\calA(\tilde{\la}_2)}{\partial{\tilde{\chi}_2}^b}U_{x_{\ri}}(\tilde{\la}_2,\la')\notag\\
&-\binom{\alpha\leftrightarrow\beta}{a\leftrightarrow b}\,.\nonumber
\end{align}
To extract the scalar field $\phi_{ab}$ from this, we need to get rid of the indices $\alpha$ and $\beta$ and also keep the expression homogeneous in $\la_{\ri}$ and $\la'_{\ri}$ of degree $0$. This can be achieved by contracting \eqref{eq:doublethetaderivativeofU} with 
\begin{equation}
\label{eq: contraction prefactor}
\frac{\la^{\alpha}{\la'}^{\beta}}{\langle\la \la'\rangle} \eqndot
\end{equation}
To summarize, we find that vertex of the scalar field of \eqref{eq:secondattempt} equals
\begin{equation}
\label{onetoothscalar}
-\frac{\la^{\alpha}{\la'}^{\beta}}{\langle\la \la'\rangle} \frac{\partial^2U_{x}(\la,\la')}{\partial \theta^{\alpha a}\partial \theta^{\beta b}}|_{\theta=0}\eqndot
\end{equation}
The vertex for the scalar field is like a super curvature on a line in twistor space. That the expression for a local field needs two twistors $\calZ$ and $\calZ'$, can be understood from the fact that a point in Minkowski space corresponds to a $\mathbb{CP}^1$ in twistor space, which is defined by two distinct twistors. To define a local field in Minkowski space one therefore requires a non-local object $U$ in twistor space.
Now, the vertex of the composite operator $\Tr(\phi^L)(x)$ can be constructed from \eqref{onetoothscalar} via
\begin{equation}
\textbf{W}_{\Tr(\phi^L)(x)}= \Tr\left( \prod_{i=1}^L (-1)^L \frac{\la_{\ri}^{\alpha}{\la'}_{\ri}^{\beta}}{\langle\la_{\ri}\la'_{\ri}\rangle}\frac{\partial^2U_{x}(\la_i,\la_{i+1})}{\partial \theta^{\alpha a}\partial \theta^{\beta b}}|_{\theta=0}\right)\eqncom
\end{equation} where we write $\la_i'=\la_{i+1}$ and the trace is over the gauge group indices. However, for practical purposes it would be better to pull out the derivatives. In its current form however, the parallel propagators form a closed loop on a single space-time point $x$, which is just the identity. Hence, in order to pull out the derivatives, we need to put all the parallel propagators on a different space-time point $x_i$, which also ensures that each derivative only acts on a separate Wilson line $U$. Therefore, we point-split $(x,\theta)$, or equivalently, line-split the line corresponding to $x$, 
\begin{equation}
(x,\theta)\rightarrow \{ (x_1,\theta_1),\dots, (x_L,\theta_L)\}\eqncom
\end{equation}such that $x_{i-1}$ and $x_{i}$ are light-like separated for all $i$ and $x_{L+1}=x_{1}$. Due to this light-like separation, any pair of the associated successive lines $x_{i-1}$ and $x_{i}$ in twistor space intersects in a twistor, which we denote by $\calZ_{i}$. 
This yields the closed light-like Wilson loop
\begin{align}
\label{eq: polygonal Wilson loop}
& \WWW(x_1,\ldots, x_L)= \Tr\left( U_{x_1}(\calZ_1,\calZ_2) U_{x_2}(\calZ_2,\calZ_3) \cdots U_{x_L}(\calZ_L,\calZ_1)   \right)\eqncom
\end{align}
or pictorially,
\begin{figure}[htbp]
 \centering
  \includegraphics[height=3.5cm]{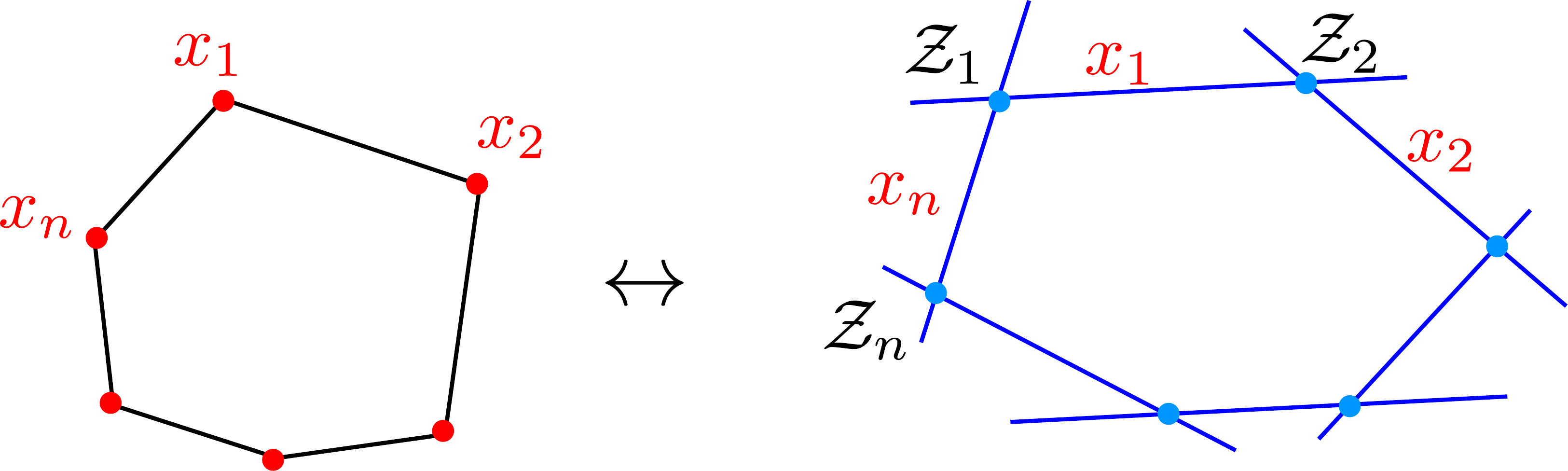}
  \caption{\it A supersymmetric $n$-gonal Wilson loop in position space and its twistor space analogue.
  }
  \label{fig:LgonalWilson}
\end{figure}

On this operator we can now act with our \textit{forming operator}
\begin{equation}
 \label{eq:formingopscalar}
\PPP_{\Tr(\phi^L)}= \prod_{i=1}^L\left(-\frac{\la_{\ri}^{\alpha}\la'_{\ri}{}^{\beta}}{\abra{ \la_{\ri}}{\la'_{\ri}}} 
(-i\xi_{\ri}^{a}\partial_{\ri,\alpha a})(-i\xi_{\ri}^b\partial_{\ri,\beta b})\right)|_{\theta_i=0}
\eqncom
\end{equation}
where both SU$(4)$ indices $a$, $b$ of each scalar $\phi_{ab}$ are removed by contracting the scalar field with fermionic polarization vectors $-\xi_{\ri}^a\xi_{\ri}^b$. Finally, as we want to use these fields to construct \textit{local} composite operators, we have to take the limit where all space-time points go to the same point $x$, see   Figure~\ref{fig:operatorlimit}. 
\begin{figure}[htbp]
 \centering
  \includegraphics[height=3.5cm]{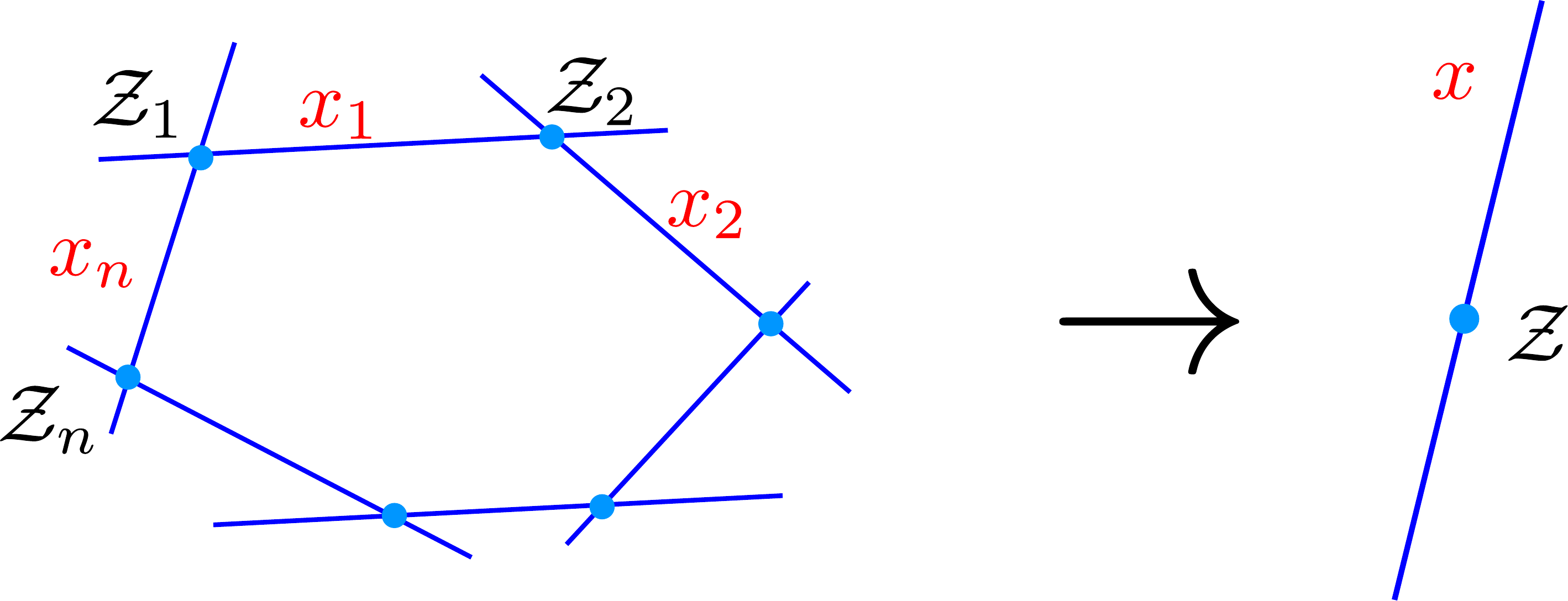}
  \caption{\it This figure sketches the limit procedure which sends the light-like Wilson loop to a point $x$, or in twistor space, to a line. The details are considerably more complicated, see Appendix~\ref{app:geometry}. 
  }
  \label{fig:operatorlimit}
\end{figure}
The resulting expression should only depend on the point $x$ in space-time and not on the geometry of the Wilson loop that we used to construct it. In particular, it should not depend on the coordinates $\la_{\ri}$ and $\la'_{\ri}$ of the corners of the loop. This independence can be achieved by taking the limit $\la_{\ri}\parallel\la'_{\ri}$, and furthermore choosing the normalization such that $\la'_{\ri}\rightarrow\la_{\ri}$. However, this raises a problem. Namely, we act with derivatives on each edge of the loop of  Figure~\ref{fig:LgonalWilson} and then take the limit such that $\la_\ri\rightarrow \la_{\ri+1}$ for each $\ri$. In this limit, the geometry would be somewhat ill-defined, since at least two different twistors are needed to define a line in twistor space. To circumvent this issue, we \textit{add extra edges} to the loop, which we describe in full detail in Section~\ref{sec:construction} and Appendix~\ref{app:geometry}. 
We call the limit of shrinking the Wilson loop to a point and $\la_{\ri}\rightarrow \la'_{\ri}$ \textit{the operator limit} and write it symbolically as $\hexagon\rightarrow \xdot$. 
Combing the forming operator, the Wilson loop and finally the operator limit gives us the vertex of the local composite operator $\Tr(\phi^L)(x)$
\begin{equation}
\textbf{W}_{\Tr(\phi^L)(x)}=\text{lim}_{\hexagon\rightarrow \xdot} \PPP_{\Tr(\phi^L)} \WWW(x_1,\ldots, x_L)\eqndot
\end{equation}
In the next chapter we will extend this construction to all local composite operators, obtaining all tree-level MHV form factors.

\chapter{All local composite operators and their tree-level MHV form factors}
\label{MHV}
In this chapter we extend the construction of scalar local composite operators and their form factors to arbitrary operators. In Section~\ref{sec:construction} we describe the forming operators for all fundamental fields with $n$ number of covariant derivatives. Acting with these forming operators on edges of a polygonal Wilson loop yields all operator vertices of the theory. The Wilson loop on which the forming operators act is briefly discussed in this section as well, although all details of its geometry as well as the so-called operator limit are relegated to Appendix~\ref{app:geometry}. In Section~\ref{sec:minff} we test our construction by computing all minimal MHV form factors of local composite operators at tree level. In Section~\ref{subsec: all MHV form factors} we then give an expression for all tree-level MHV form factors of any operator and any number of external particles. The proof of this expression in the last section concludes this chapter. This chapter is based on and contains overlap with \cite{Koster:2016loo}.
\section{Vertices for all local composite operators}\label{sec:construction}
In the previous chapter we proposed and tested the expression for the vertex of a scalar field in twistor space. Furthermore, we showed how a light-like Wilson loop generated this operator vertex by acting with derivative operators on it. In this section we extend the method to the rest of the field content of the theory. 
Recall how composite operators are constructed in Minkowski space-time, see Section~\ref{fieldsoperatorsaction}.
In the planar limit, multi-trace operators are suppressed and so we only look at single-trace operators $\mathcal{O}$. We denote the covariant derivative  by $D$ and construct these operators by tracing over products of covariantly transforming fields $D^{k_\ri}\Phi_\ri$ with $k_i=0,1,2,\ldots$ positioned at the same space-time point $x$:
\begin{equation}
\label{eq:composite operator}
\mathcal{O}(x)=\Tr\left(D^{k_1}\Phi_1(x)D^{k_2}\Phi_2(x)\cdots D^{k_L}\Phi_L(x)\right)\eqncom
\end{equation}
where the fields $\Phi_i$ are any of the six scalars $\phi_{ab}$, or any of the four anti-fermions $(\psi_{bcd})_{\alpha}=\epsilon_{abcd}\psi^d_{\alpha}$, the four fermions $\bar{\psi}_{a\dot{\alpha}}$, or the field strength $F_{\mu\nu}$, and all space-time spinorial indices have been suppressed.
By contracting with the Pauli matrices $\sigma^{\mu}_{\alpha\dot{\alpha}}$, we remove the Lorentz indices $\mu, \nu$ indices and will henceforth only use the spinor ones $\alpha$ and $\dot{\alpha}$.
The field strength is thus split into a self-dual and an anti-self-dual part as
\begin{equation}
F_{\mu\nu}(\sigma^\mu)_{\alpha\dot{\alpha}}(\sigma^\nu)_{\beta\dot{\beta}}\propto \epsilon_{\dot{\alpha}\dot{\beta}}F_{\alpha\beta}+\epsilon_{\alpha\beta}\bar{F}_{\dot{\alpha}\dot{\beta}} \eqndot
\end{equation}
When considering covariant derivatives
$D_{\alpha\dot{\alpha}}$
that act on the fields, we use the equations of motion of the fields, the definition of the field strength and the Bianchi identity for the field strength to replace any antisymmetric combination of the spinor indices $\alpha$, $\dot{\alpha}$ by a product of terms that are individually fully symmetric in the spinor indices. 
All composite operators can therefore be built from the following set of fields:
\begin{align}
\label{eq: alphabet of fields}
 D^k\Phi\in\{ 
 &D_{(\alpha_1\dot\alpha_1}\cdots D_{\alpha_k\dot\alpha_k}\bar{F}_{\dot\alpha_{k+1}\dot\alpha_{k+2})},\hspace{-.5cm}&
 &D_{(\alpha_1\dot\alpha_1}\cdots D_{\alpha_k\dot\alpha_k}\bar{\psi}_{\dot\alpha_{k+1})a}, \; D_{(\alpha_1\dot\alpha_1}\cdots D_{\alpha_k\dot\alpha_k)}\phi_{ab},&\notag\\
 &D_{(\alpha_1\dot\alpha_1}\cdots D_{\alpha_k\dot\alpha_k}\psi_{\alpha_{k+1})abc},&
 &D_{(\alpha_1\dot\alpha_1}\cdots D_{\alpha_k\dot\alpha_k}F_{\alpha_{k+1}\alpha_{k+2})} \}\eqncom 
\end{align}
where the parentheses denote symmetrization in the spinor indices $\alpha$, $\dot{\alpha}$ and the fields are antisymmetric in the flavor indices $a$.
\newline\\
Explicitly symmetrizing the spinor indices is quite cumbersome. This can be avoided by contracting each field with a 
light-like polarization vector $\ftau^{\alpha\dot\alpha}=\tau^{\alpha}\bar{\tau}^{\dot\alpha}$. 
Moreover, we include a Gra\ss mann variable $\xi^a$ which combines with  $\ftau$ to form the superpolarization vector  $\fTau=(\tau^{\alpha},\bar{\tau}^{\dot\alpha}, \xi^a)$.
We can then write the fields in \eqref{eq: alphabet of fields}, with slight abuse of notation, as
\begin{equation}
\label{eq: alphabet of fields with polarizations}
\begin{aligned}
 D^k\Phi\in\{ 
 &+\tau^{\alpha_1}\dots\tau^{\alpha_{k\phantom{+1}}}\bar{\tau}^{\dot\alpha_1}\dots\bar{\tau}^{\dot\alpha_{k+2}}\phantom{\xi^a}\phantom{\xi^a}\phantom{\xi^a}\phantom{\xi^a}
 D_{\alpha_1\dot\alpha_1}\cdots D_{\alpha_k\dot\alpha_k}\bar{F}_{\dot\alpha_{k+1}\dot\alpha_{k+2}},\\
 &+\tau^{\alpha_1}\dots\tau^{\alpha_{k\phantom{+1}}}\bar{\tau}^{\dot\alpha_1}\dots\bar{\tau}^{\dot\alpha_{k+1}}\xi^a\phantom{\xi^a}\phantom{\xi^a}\phantom{\xi^a}
 D_{\alpha_1\dot\alpha_1}\cdots D_{\alpha_k\dot\alpha_k}\bar{\psi}_{\dot\alpha_{k+1}a},\\
 &-\tau^{\alpha_1}\dots\tau^{\alpha_{k\phantom{+1}}}\bar{\tau}^{\dot\alpha_1}\dots\bar{\tau}^{\dot\alpha_{k\phantom{+1}}}\xi^a\xi^b\phantom{\xi^a}\phantom{\xi^a}
 D_{\alpha_1\dot\alpha_1}\cdots D_{\alpha_k\dot\alpha_k}\phi_{ab},\\
 &-\tau^{\alpha_1}\dots\tau^{\alpha_{k+1}}\bar{\tau}^{\dot\alpha_1}\dots\bar{\tau}^{\dot\alpha_{k\phantom{+1}}}\xi^a\xi^b\xi^c\phantom{\xi^a}
 D_{\alpha_1\dot\alpha_1}\cdots D_{\alpha_k\dot\alpha_k}\psi_{\alpha_{k+1}abc},\\
 &+\tau^{\alpha_1}\dots\tau^{\alpha_{k+2}}\bar{\tau}^{\dot\alpha_1}\dots\bar{\tau}^{\dot\alpha_{k\phantom{+1}}}\xi^a\xi^b\xi^c\xi^d
 D_{\alpha_1\dot\alpha_1}\cdots D_{\alpha_k\dot\alpha_k}F_{\alpha_{k+1}\alpha_{k+2}abcd}
 \}\eqncom
\end{aligned}
\end{equation}
where $F_{\alpha_{k+1}\alpha_{k+2}abcd}=\frac{1}{4!}\epsilon_{abcd}F_{\alpha_{k+1}\alpha_{k+2}}$.
Note that an independent superpolarization vector can be chosen for each field $D^{k_\ri}\Phi_\ri$ of \eqref{eq:composite operator}. The elements in \eqref{eq: alphabet of fields} can be recovered by taking suitable derivatives of \eqref{eq: alphabet of fields with polarizations}
with respect to the superpolarization vector.
For example,
\begin{equation}
\label{eq: specifying components by derivatives}
 \begin{aligned}
  D_{(\alpha_1\dot\alpha_1}\cdots D_{\alpha_k\dot\alpha_k)}\phi_{ab}
  &=
  \frac{1}{k! k! 2!}
  \frac{\partial}{\partial \tau^{\alpha_1}}\cdots\frac{\partial}{\partial \tau^{\alpha_k}} 
  \frac{\partial}{\partial \bar{\tau}^{\dot\alpha_1}}\cdots\frac{\partial}{\partial \bar{\tau}^{\dot\alpha_k}} 
  \frac{\partial}{\partial \xi^{a}}\frac{\partial}{\partial \xi^{b}} \\
  &\phaneq(-1)\tau^{\beta_1}\dots\tau^{\beta_{k}}\bar{\tau}^{\dot\beta_1}\dots\bar{\tau}^{\dot\beta_{k}}\xi^c\xi^d
 D_{\beta_1\dot\beta_1}\cdots D_{\beta_k\dot\beta_k}\phi_{cd}
  \eqndot
 \end{aligned}
\end{equation}
In order to obtain vertices for the composite operator $\mathcal{O}$ \eqref{eq:composite operator}, we act on a Wilson loop $\mathcal{W}$ with a differential operator, which we call the forming operator. The Wilson loop will be explained later on in this section. The forming operator of $\mathcal{O}$ is the product of respective forming operators for the irreducible fields $D^{k_{\ri}}\Phi_{\ri}$, out of which $\mathcal{O}$ is built:
\begin{equation}
\label{formingop}
 \PPP_{\mathcal{O}}=\prod_{\ri=1}^L\PPP_{D^{k_{\ri}}\Phi_{\ri}}\eqndot
\end{equation}
The forming operators of the irreducible fields \eqref{eq: alphabet of fields with polarizations} are
\begin{equation}
 \label{eq:definitionformingfactoronshellstates}
\PPP_{D^{k_{\ri}}\Phi_{\ri}}= -\frac{\la_{\ri}^{\alpha}\la'_{\ri}{}^{\beta}}{\abra{ \la_{\ri}}{\la'_{\ri}}} 
\big(-i\tau^\gamma_{\ri}\bar{\tau}_{\ri}^{\dot\gamma}\partial_{\ri,\gamma\dot\gamma}\big)^{k_{\ri}}
\left\{\begin{array}{ll} 
(-i\bar{\tau}_{\ri}^{\dot\alpha}\partial_{\ri,\alpha\dot\alpha})(-i\bar{\tau}_{\ri}^{\dot\beta}\partial_{\ri,\beta\dot{\beta}})& 
\text{ for } A_{\ri}=\bar{F}\\
(-i\bar{\tau}_{\ri}^{\dot\alpha}\partial_{\ri,\alpha\dot{\alpha}})(-i\xi_{\ri}^{a}\partial_{\ri,\beta a})& 
\text{ for } A_{\ri}=\bar{\psi}\\
(-i\xi_{\ri}^{a}\partial_{\ri,\alpha a})(-i\xi_{\ri}^b\partial_{\ri,\beta b})& 
\text{ for } A_{\ri}=\phi\\
(-i\xi_{\ri}^{a}\partial_{\ri,\alpha a})(-i\xi_{\ri}^{b}\partial_{\ri,\beta b})(-i\tau_{\ri}^{\gamma}\xi_{\ri}^{c}\partial_{\ri,\gamma c})&
\text{ for }A_{\ri}=\psi\\
(-i\xi_{\ri}^{a}\partial_{\ri,\alpha a})(-i\xi_{\ri}^{b}\partial_{\ri,\beta b})(-i\tau_{\ri}^{\gamma}\xi_{\ri}^{c}\partial_{\ri,\gamma c})^2& 
\text{ for } A_{\ri}=F
\end{array}\right. \eqncom
\end{equation}
where we used the abbreviations
\begin{equation}
 \partial_{\ri,\alpha\dot\alpha}\equiv \frac{\partial}{\partial x_{\ri}^{\alpha\dot\alpha}}\,,\qquad \partial_{\ri,\alpha a} \equiv \frac{\partial}{\partial \theta_{\ri}^{\alpha a}}\,\eqncom
\end{equation} and $D^{k_i}$ are $k_i$ covariant derivatives. Restricting to Gra\ss mann derivatives $\partial_{i\alpha a}$ we recover the forming operator for the scalar field \eqref{eq:formingopscalar}.
The next ingredient was the light-like Wilson loop on which the forming operator acts. In the previous chapter we argued that for the operator $\Tr(\phi^L)(x)$ one needs a Wilson loop with $2L$ edges. In general however, we should also include the possibility of taking covariant derivatives on any operator. This variation of the space-time point translates to a variation of the corresponding line in twistor space. This variation must not break the closed Wilson loop. For the precise details of the construction of this Wilson loop we refer to Appendix~\ref{app:geometry}. For now, we just state that we can construct a Wilson loop that works, which is shaped like a cogwheel and has $3L$ edges for any operator of length $L$, see   Figure~\ref{fig:CogwheelBig}.
\begin{figure}[tbp]
 \centering
  \includegraphics[height=3.5cm]{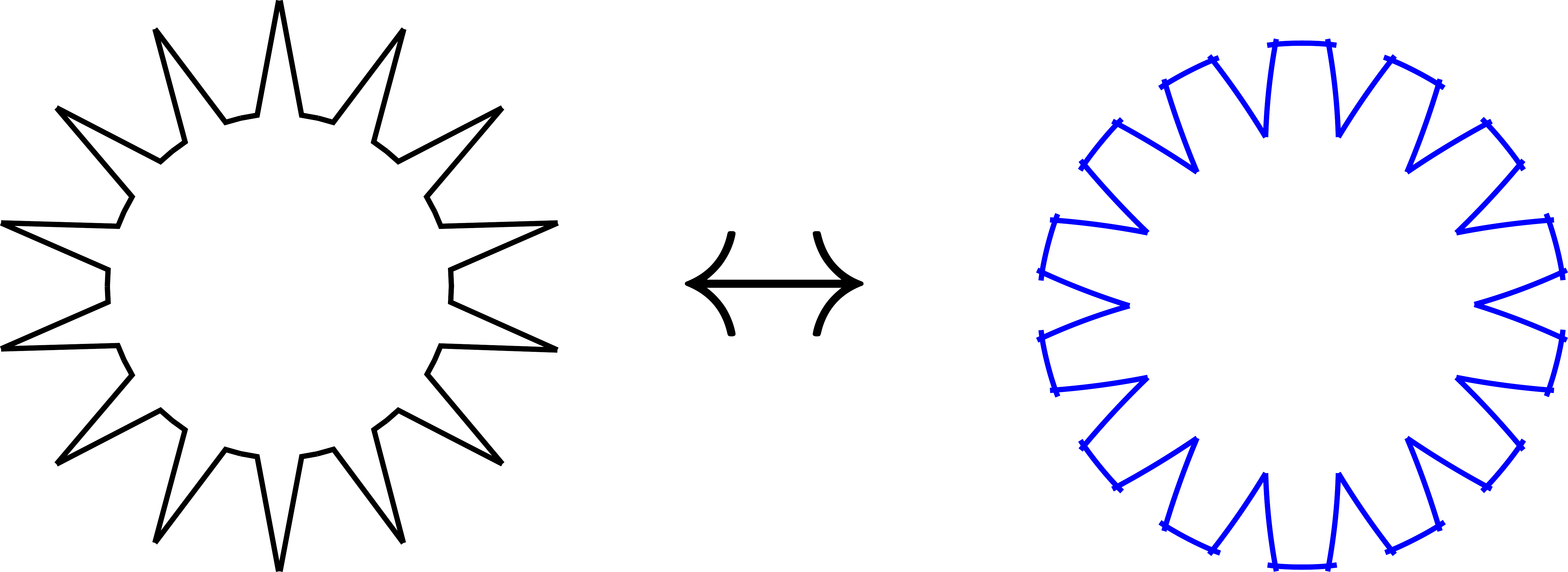}
  \caption{\it A composite operator of length  $L$  is constructed using a cogwheel Wilson loop of $3L$ vertices in space-time (left), or $3L$ edges in twistor space (right).
  }
  \label{fig:CogwheelBig}
\end{figure}\newline
Recall that we are interested in finding an expression for the local composite operator $\mathcal{O}(x)=\Tr(D^{k_1}\Phi_1(x)\cdots D^{k_L}\Phi_L(x))$ of length $L$. 
Acting with the forming operator  $\PPP$ \eqref{eq:definitionformingfactoronshellstates} corresponding to the operator $\mathcal{O}(x)$ on this Wilson loop $\mathcal{W}$ and  subsequently taking the operator limit, $\hexagon\rightarrow \xdot$, gives all local composite operators:
\beq\label{formingonwilson}
\textbf{W}_{\mathcal{O}(x)}=\lim_{\hexagon\rightarrow \xdot}\PPP_{\mathcal{O}}\, \mathcal{W}_{\big{|}\theta=0}\,.
\eeq
The vertex for each of the fundamental fields in the composite operator is of the general form 
\begin{align}
\label{eq: minimal vertices}
 \textbf{W}_{D^{k_{\ri}}\Phi_{\ri}(x)}&=
 \int \DD\la_{\ri} h^{-1}_x(\la_{\ri})
 \left\{\begin{array}{ll}
 \abra{\tau_{\ri}}{\la_{\ri}}^{k_{\ri}\phantom{+1}}(\bar{\tau}_{\ri}^{\dot\alpha}\partial_{i,\dot{\alpha}})^{k_{\ri}+2}\phantom{(\xi_{\ri}^a\partial_{i,a})^2}\AAA(\la_{\ri})& 
 \text{ for } A_{\ri}=\bar{F}\\
 \abra{\tau_{\ri}}{\la_{\ri}}^{k_{\ri}\phantom{+1}}(\bar{\tau}_{\ri}^{\dot\alpha}\partial_{i,\dot\alpha})^{k_{\ri}+1}(\xi_{\ri}^a\partial_{i,a})^{\phantom{1}}\AAA(\la_{\ri})& 
 \text{ for } A_{\ri}=\bar{\psi}\\
 \abra{\tau_{\ri}}{\la_{\ri}}^{k_{\ri}\phantom{+1}}(\bar{\tau}_{\ri}^{\dot\alpha}\partial_{i,\dot\alpha})^{k_{\ri}\phantom{+1}}(\xi_{\ri}^a\partial_{i,a})^2\AAA(\la_{\ri})& 
 \text{ for } A_{\ri}=\phi\\
 \abra{\tau_{\ri}}{\la_{\ri}}^{k_{\ri}+1}(\bar{\tau}_{\ri}^{\dot\alpha}\partial_{i,\dot\alpha})^{k_{\ri}\phantom{+1}}(\xi_{\ri}^a\partial_{i,a})^3\AAA(\la_{\ri})& 
 \text{ for } A_{\ri}=\psi\\
\abra{\tau_{\ri}}{\la_{\ri}}^{k_{\ri}+2}(\bar{\tau}_{\ri}^{\dot\alpha}\partial_{i,\dot\alpha})^{k_{\ri}\phantom{+1}} (\xi_{\ri}^a\partial_{i,a})^4\AAA(\la_{\ri})& 
 \text{ for } A_{\ri}=F
\end{array}
\right\}h_x(\la_{\ri})
\nonumber\\
&\phaneq+\text{terms at least quadratic in }\calA \eqncom 
\end{align}
where we used the abbreviations
\begin{equation}
\partial_{\ri,\dot\alpha}\equiv \frac{\partial}{\partial \mu_{\ri}^{\dot\alpha}}\,,\qquad
\partial_{\ri,a}\equiv \frac{\partial}{\partial \chi_{\ri}^{a}}\,.
\end{equation}
In addition to the explicitly shown terms, the vertices contain further terms which are at least quadratic in the field $\AAA$. They have partial derivatives with respect to $\mu$ and $\chi$, acting on up to $k_{\ri}+2$, $k_{\ri}+2$, $k_{\ri}+2$, $k_{\ri}+3$ and $k_{\ri}+4$ different $\AAA$, respectively.
The higher order terms are straightforwardly obtained using the product rule.
The previously treated example of the scalar field was already given in \eqref{eq:secondattempt}.
The vertex for an operator $\mathcal{O}(x)=\Tr(D^{k_1}\Phi_1(x)\cdots D^{k_L}\Phi_L(x))$ is then given by 
\begin{equation}
 \label{eq: operator vertex}
 \textbf{W}_{\mathcal{O}(x)}=\Tr(\textbf{W}_{D^{k_1}\Phi_1(x)}\cdots\textbf{W}_{D^{k_L}\Phi_L(x)})\eqncom
\end{equation}
where the frames $h_x$ and inverse frames $h_x^{-1}$ in \eqref{eq: minimal vertices} combine to parallel propagators $U_x$ according to \eqref{propagatorU}.
Note that we can also refrain from setting $\theta=0$ to obtain the vertex for the chiral part of a supermultiplet of which $\mathcal{O}$ is the lowest component. 
To test our expressions we compute the tree-level minimal and non-minimal MHV form factors corresponding to the general operator \eqref{eq:composite operator} in the following sections.

\section{All minimal tree-level MHV form factors}
\label{sec:minff}
As a warm up we compute the minimal tree-level form factor for our generic operator \eqref{eq:composite operator}.
Since we are only interested in the minimal form factor, we need only the lowest order terms in the expansion of \eqref{formingonwilson}, or in other words, the one that is precisely of length $L$. These are explicitly shown in \eqref{eq: minimal vertices}. This gives
\begin{multline}
\label{eq: vertex for minimal valency}
 \textbf{W}_{\Tr(D^{k_1}\Phi_1\cdots D^{k_L}\Phi_L)(x)}\Big|_{L\text{-valent}}\\
 =\prod_{\ri=1}^L \int \DD\la_{\ri}  \left\{\begin{array}{ll}
 \abra{\tau_{\ri}}{\la_{\ri}}^{k_{\ri}\phantom{+1}}(\bar{\tau}_{\ri}^{\dot\alpha}\partial_{\ri,\dot{\alpha}})^{k_{\ri}+2}& 
 \, \text{ for }\Phi_{\ri}=\bar{F}\\
 \abra{\tau_{\ri}}{\la_{\ri}}^{k_{\ri}\phantom{+1}}(\bar{\tau}_{\ri}^{\dot\alpha}\partial_{\ri,\dot\alpha})^{k_{\ri}+1}(\xi_{\ri}^a\partial_{\ri,a})& 
 \, \text{ for }\Phi_{\ri}=\bar{\psi}\\
 \abra{\tau_{\ri}}{\la_{\ri}}^{k_{\ri}\phantom{+1}}(\bar{\tau}_{\ri}^{\dot\alpha}\partial_{\ri,\dot\alpha})^{k_{\ri}\phantom{+1}}(\xi_{\ri}^a\partial_{\ri,a})^2& 
 \, \text{ for }\Phi_{\ri}=\phi\\
 \abra{\tau_{\ri}}{\la_{\ri}}^{k_{\ri}+1}(\bar{\tau}_{\ri}^{\dot\alpha}\partial_{\ri,\dot\alpha})^{k_{\ri}\phantom{+1}}(\xi_{\ri}^a\partial_{\ri,a})^3& 
 \, \text{ for }\Phi_{\ri}=\psi\\
\abra{\tau_{\ri}}{\la_{\ri}}^{k_{\ri}+2}(\bar{\tau}_{\ri}^{\dot\alpha}\partial_{\ri,\dot\alpha})^{k_{\ri}\phantom{+1}} (\xi_{\ri}^a\partial_{\ri,a})^4& 
 \, \text{ for }\Phi_{\ri}=F
\end{array}
\right\}\AAA(\la_{\ri})_{\big{|}\theta=0} \eqncom
\end{multline}
where the product is understood to be ordered and $\AAA(\la_{\ri})\equiv \AAA(\calZ_{x}(\la_{\ri}))$. \par
From the operator vertices \eqref{eq: vertex for minimal valency}, we can straightforwardly derive the minimal tree-level position-space form factors $\calF_{\calO}(1,\ldots, L;x)$ by connecting the $\AAA(\la_\ri)$ with the on-shell states $\AAA_{\fP_\rj}(\la_\ri)$ \eqref{eq:definitiononshellmomentumeigenstates}. There are $L$ distinct but cyclically related ways to planarly connect the $L$ $\AAA$'s in \eqref{eq: vertex for minimal valency} to external (super)momentum eigenstates \eqref{eq:definitiononshellmomentumeigenstates}. They are entirely determined by choosing to connect the field $1$ to the external state $\rj+1$, $2$ to state $\rj+2$ and so on. The integration over the spinors then completely factorizes. As an example, consider the vertex for $\bar F$ of helicity $+1$: the contributing factor is (dropping the superfluous indices)
\begin{multline}
\label{eq:deltaintegration}
 \int \DD\la \abra{\tau}{\la}^k(\bar{\tau}^{\dot\alpha}\partial_{\dot \alpha})^{k+2}2\pi i\int_{\mathbb{C}} \frac{\dd s}{s}\e^{s(\mu^{\dot\alpha}\bar{p}_{\dot\alpha}+\chi^a\eta_a)}\bar\delta^2(s\lambda-p)_{\big{|}\theta=0}\\
 =\int_{\mathbb{C}^2} \dd u_1\dd u_2 \abra{\tau}{u}^k\sbra{\bar p}{\bar\tau}^{k+2}\e^{ix^{\alpha\dot\alpha}u_\alpha \bar{p}_{\dot\alpha}}\bar\delta^2(u-p)=\abra{\tau}{p}^k\sbra{\bar p}{\bar \tau}^{k+2}\e^{ix\fp}\,,
\end{multline}
where we parametrized $\la=(1,u_2)$, renamed $s\rightarrow u_1$ and rescaled $u_2\rightarrow u_1u_2^{-1}$.
From this and similar calculations for the other cases we deduce that the insertion of \eqref{eq:definitiononshellmomentumeigenstates} and the subsequent integration over $\calS_\ri$ and $\la_\ri$ effectively replaces
\begin{equation}
 \lambda_{\ri,\alpha}\longrightarrow p_{\ri+\rj,\alpha} \eqncom \qquad 
 \partial_{\ri,\dot\alpha} \longrightarrow \bar{p}_{\ri+\rj,\dot\alpha}\eqncom \qquad 
\partial_{\ri,a}\longrightarrow \eta_{\ri+\rj,a}
\end{equation}
 in \eqref{eq: vertex for minimal valency}.
Finally, the minimal tree-level form factor in momentum space is obtained after Fourier transforming: 
\begin{equation}
 \label{eq: minimal form factor in momentum space}
\begin{multlined}
\calF_{\Tr(D^{k_1}\Phi_1\cdots D^{k_L}\Phi_L)}(1,\dots,L;\fq)=\int \frac{\dd^4 x}{(2\pi)^4}\e^{-ix\fq} \calF_{\Tr(D^{k_1}\Phi_1\dots D^{k_L}v_L)}(1,\dots,L;x)\\
=\sum_{j=0}^{L-1}\prod_{\ri=1}^L   \left\{\begin{array}{ll}
 {\abra{\tau_{\ri}}{p_{\ri+\rj}}}^{k_{\ri}\phantom{+1}}{\sbra{\bar{p}_{\ri+\rj}}{\bar{\tau}_{\ri}}}^{k_{\ri}+2} & \text{ for } \Phi_{\ri}=\bar{F}\\
 \abra{\tau_{\ri}}{p_{\ri+\rj}}^{k_{\ri}\phantom{+1}}\sbra{\bar{p}_{\ri+\rj}}{\bar{\tau}_{\ri}}^{k_{\ri}+1}\{\xi_{\ri}\eta_{\ri+\rj}\}& \text{ for }\Phi_{\ri}=\bar{\psi}\\
 \abra{\tau_{\ri}}{p_{\ri+\rj}}^{k_{\ri}\phantom{+1}}\sbra{\bar{p}_{\ri+\rj}}{\bar{\tau}_{\ri}}^{k_{\ri}\phantom{+1}}\{\xi_{\ri}\eta_{\ri+\rj}\}^2& \text{ for } \Phi_{\ri}=\phi\\
 \abra{\tau_{\ri}}{p_{\ri+\rj}}^{k_{\ri}+1}\sbra{\bar{p}_{\ri+\rj}}{\bar{\tau}_{\ri}}^{k_{\ri}\phantom{+1}}\{\xi_{\ri}\eta_{\ri+\rj}\}^3& \text{ for } \Phi_{\ri}=\psi\\
 \abra{\tau_{\ri}}{p_{\ri+\rj}}^{k_{\ri}+2}\sbra{\bar{p}_{\ri+\rj}}{\bar{\tau}_{\ri}}^{k_{\ri}\phantom{+1}}\{\xi_{\ri}\eta_{\ri+\rj}\}^4& \text{ for } \Phi_{\ri}=F
\end{array}
\right\} \delta^4\left(\fq-\sum_{\ri=1}^L \fp_{\ri}\right)\eqndot
\end{multlined}
\end{equation}
This perfectly agrees with the result originally obtained in \cite{Wilhelm:2014qua}.

For the minimal form factor of the operator $\frac{1}{2}\Tr(\phi_{ab}^2)$ in Section~\ref{examplescalar}, we find from \eqref{eq: minimal form factor in momentum space}:
\beq
\begin{split}
\calF_{\frac{1}{2}\Tr(\phi_{ab}^2)}(1,2;\fq)&=\frac{1}{2}\frac{\partial^2}{\partial \xi_1^a\partial \xi_1^b}\frac{1}{2}\frac{\partial^2}{\partial \xi_2^a\partial \xi_2^b}\frac{1}{2}\left(\{\xi_1\eta_1\}^2\{\xi_2\eta_2\}^2+\{\xi_1\eta_2\}^2\{\xi_2\eta_1\}^2\right)\\
&\times \delta^4\left(\fq-\fp_1-\fp_2\right)\\
&=\eta_{1a}\eta_{1b} \eta_{2a}\eta_{2b}\,\delta^4\left(\fq-\fp_1-\fp_2\right)
\,,
\end{split}
\eeq
where we used \eqref{eq: specifying components by derivatives}.
Hence, the only possible outgoing states in this case are two scalars $1^{\phi_{ab}}$ and $2^{\phi_{ab}}$ and we obtain the correct result for the minimal form factor of $\frac{1}{2}\Tr(\phi_{ab}^2)$.

\section{All tree-level MHV form factors}\label{subsec: all MHV form factors}
Using the vertices constructed in Section~\ref{sec:construction}, we can now derive the general tree-level $n$-point MHV form factors of the  composite operators.
We treated the example of the $n$-point MHV form factor of the operator $\frac{1}{2}\Tr(\phi_{ab}^2)$ in Section~\ref{examplescalar}.
Let us now calculate the $n$-point MHV form factor for a generic single-trace operator containing the fields in \eqref{eq: alphabet of fields with polarizations}. 
We first write down the result and give the proof in the next section.

The $n$-point tree-level MHV form factor of a generic single-trace operator  $\calO$ \eqref{eq:composite operator} in $\calN=4$ SYM reads
\beq
\begin{split}
\label{eq: general MHV form factor}
\calF_{\mathcal{O}}(1,\dots,n;\fq)
 &=\frac{\delta^4(\fq-\sum_{i=1}^n \fp_i)}{\prod_{i=1}^n\abra{i}{(i+1)}} \sum_{\{\notA_{a,b}\}}
 \prod_{j=1}^L 
 \abra{p_{\notA_{j,N_j}}}{p_{\notA_{j+1,1}}}
\left(\prod_{k=2}^{N_j-1}\abra{\tau_j}{p_{\notA_{j,k}}}\right)\\
 &\phaneq
 \sum_{\sigma\in S_{N_j}}
 \frac{\{\xi_j \eta_{\sigma(\notA_{j,1})}\}\cdots\{\xi_j \eta_{\sigma(\notA_{j,n_{\theta_j}})}\}\sbra{\bar{p}_{\sigma(\notA_{j,n_{\theta_j}+1})}}{\bar\tau_j}\cdots\sbra{\bar{p}_{\sigma(\notA_{j,N_j})}}{\bar\tau_j}}{M(\{\notA_{j,1},\dots,\notA_{j,N_j}\})!}
\\
 &\phaneq
 +\text{cyclic permutations}
 \eqndot
\end{split}
\eeq
We use the following notation:
\begin{enumerate}
\item  The total number of indices $\dot\alpha$ and $a$ of the field $D^{k_i}v_i$ in \eqref{eq: alphabet of fields with polarizations} is denoted by $N_i$, which is equal to the number of bosonic and fermionic derivatives required to create the field $D^{k_i}\Phi_i$ in \eqref{eq:definitionformingfactoronshellstates}. Specifically, for $\Phi_i=\bar F, \bar \psi $ or $\bar \phi$, we have $N_i=k_i+2$. Otherwise, $N_i=k_i+3$ for $\Phi_i=\psi$ and $N_i=k_i+4$ for $\Phi_i=F$. 
Moreover, the number of indices $a$ of the field $D^{k_i}\Phi_i$ in \eqref{eq: alphabet of fields with polarizations} is denoted by $n_{\theta_\ri}$, which is equal to the number of $\frac{\partial}{\partial\theta_\ri}$ derivatives required to generate the field $D^{k_i}\Phi_i$ in \eqref{eq:definitionformingfactoronshellstates}. Concretely, $n_{\theta_\ri}=0,1,2,3,4$ for $\Phi_i=\bar F, \bar \psi, \phi, \psi, F $, respectively.
\item The sum in \eqref{eq: general MHV form factor} is over all the sets $\{B_{i,j}|i=1,\ldots,L, j=1,\ldots, N_i\}$ with $1\leq \notA_{1,1}\leq \dots \leq \notA_{1,N_1}<\dots<\notA_{L,1}\leq \dots \leq \notA_{L,N_L}\leq n$. 
\item 
We denote by $M(\{\notA_{i,1},\notA_{i,2}, \dots , \notA_{i,N_i}\})$ the set of multiplicities of the entries in the original set. For example, $M(\{2,3,3,7,9,9,9\})=\{1,2,1,3\}$. 
The factorial of a set is defined as $\{a,b,c,\dots\}!= a!b!c!\cdots$, for example $ M(\{2,3,3,7,9,9,9\})!=1!2!1!3!=12$.
Note that for $n_{\theta_\ri}=0$ or $n_{\theta_\ri}=N_\ri$ the sum over all permutations reduces to $N_{\ri}!$ and the total numeric prefactor in the second line of \eqref{eq: general MHV form factor} becomes a multinomial coefficient.
\end{enumerate}

Before we discuss the proof of \eqref{eq: general MHV form factor}, let us make some remarks. To start, we emphasize that the result \eqref{eq: general MHV form factor}, which first appeared in \cite{Koster:2016loo}, is the first complete computation of the MHV tree-level form factors of \textit{all} $\calN=4$ SYM composite operators. 
In particular, \eqref{eq: general MHV form factor} is consistent with all available computations for specific operators -- it agrees with the results of \cite{Brandhuber:2011tv} for the operators in the stress-tensor supermultiplet, with those of \cite{Engelund:2012re} for operators in the SU$(2)$ sector and for twist-two operators in the SL$(2)$ sector. Furthermore, \eqref{eq: general MHV form factor} was verified using Lorentz Harmonic Chiral space in \cite{Chicherin:2016qsf}.
Let see how \eqref{eq: general MHV form factor} reduces to the result for scalars in some more detail.
For scalar fields, the second line in \eqref{eq: general MHV form factor} reduces to 
\begin{equation}
\label{eq: X factor scalar}
\begin{aligned}
 \{\xi_i \eta_{\notA_{i,1}}\} \{\xi_i \eta_{\notA_{i,2}}\} (\delta_{\notA_{i,1}=\notA_{i,2}} + 2\delta_{\notA_{i,1}\neq \notA_{i,2}})\eqndot 
 \end{aligned}
\end{equation}
Upon taking derivatives with respect to $\xi_i^a$ as specified in \eqref{eq: specifying components by derivatives}, one is essentially left with the expression found in \cite{Engelund:2012re}.
Moreover, we have checked a wide range of cases which are not available in the literature using Feynman diagrams.

We can derive \eqref{eq: general MHV form factor} in two different ways.
The first way closely follows our derivation of the MHV form factor of $\calO'=\frac{1}{2}\Tr(\phi_{ab}\phi_{ab})$ in Chapter~\ref{chap:}.
For every given field $\Phi_i$ in \eqref{eq:definitionformingfactoronshellstates}, we can apply the derivatives to the Wilson loop and perform the operator limit to obtain the corresponding vertices, as was done for several examples in Appendix~\ref{app:formfactordatamine}.
We can then insert external momentum eigenstates \eqref{eq:definitiononshellmomentumeigenstates} and perform the Fourier transformation to arrive\footnote{Heuristically, the MHV denominator in \eqref{eq: general MHV form factor} stems from the combined parallel propagators $U$. 
The second term stems from the numerators of the $U$'s between the different irreducible fields $D^{k_i}\Phi_i$.
The third term stems from the prefactors of the derivatives and the second line accounts for the combinatorics of acting with the derivatives.} at \eqref{eq: general MHV form factor}.%
The second way to derive \eqref{eq: general MHV form factor} is to insert momentum eigenstates into the Wilson loop vertex to compute the form factor of the Wilson loop, then act with the derivatives and perform the operator limit in the end. The second derivation is given in full detail in the next section. 

Finally, let us remark that one could also obtain form factors of the chiral parts of the supermultiplets that contain the operator $\mathcal{O}$ as lowest component by taking a suitable fermionic Fourier transformation with respect to $\theta$ instead of setting it to zero in analogy to what was done for the stress-tensor supermultiplet in \cite{Brandhuber:2011tv}. 

\section{Proof of the MHV form factor formula}
\label{app: derivation}

In this section, we prove our result \eqref{eq: general MHV form factor} for the tree-level $n$-point MHV form factors of all composite operators using the second strategy sketched at the end of the previous section. Here, we use the Wilson loop that is detailed in Appendix~\ref{app:geometry} and is shown in Figure~\ref{fig:CogwheelZoom2}. \begin{figure}[htbp]
 \centering
  \includegraphics[height=3.2cm]{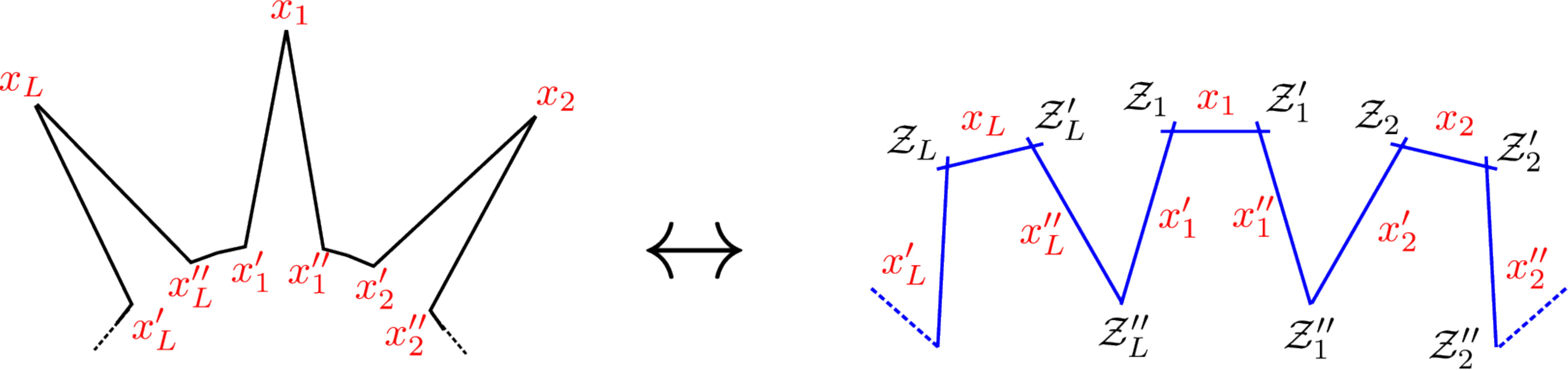}
  \caption{\it The geometry of the light-like Wilson loop. }
  \label{fig:CogwheelZoom2}
\end{figure}It has a cogwheel shape of $3L$ edges, where $L$ is the length of the operator, given by the expression
\begin{equation}
\label{eq:finaldefinitionWilsonloop1}
\begin{split}
 \WWW(x_1',x_1,x_1'',\ldots, x_L',x_L,x_L'')= \Tr\Big[&U_{x_1'}(\calZ_L'',\calZ_1) U_{x_1}(\calZ_1,\calZ_1')U_{x_1''}(\calZ_1',\calZ_1'')\\&  U_{x_2'}(\calZ_1',\calZ_2)U_{x_2}(\calZ_1,\calZ_1')U_{x_2''}(\calZ_1',\calZ_2'') \cdots \\&\cdots U_{x_L'}(\calZ_{L-1}'',\calZ_L) U_{x_L}(\calZ_{L},\calZ_L') U_{x_L''}(\calZ_{L}',\calZ_L'') \Big]\eqndot
\end{split}
\end{equation} 
For each term in \eqref{eq:finaldefinitionWilsonloop1}, we use the expression \eqref{eq:frameUdefinitionpart1} for the parallel propagators $U$.
Let $\calO=\Tr(D^{k_1}A_1\cdots D^{k_L}A_L)$ be our local operator with fields defined in \eqref{eq: alphabet of fields with polarizations} and let $n$ be the total number of external on-shell fields.
\begin{figure}[htbp]
 \centering
  \includegraphics[height=4.5cm]{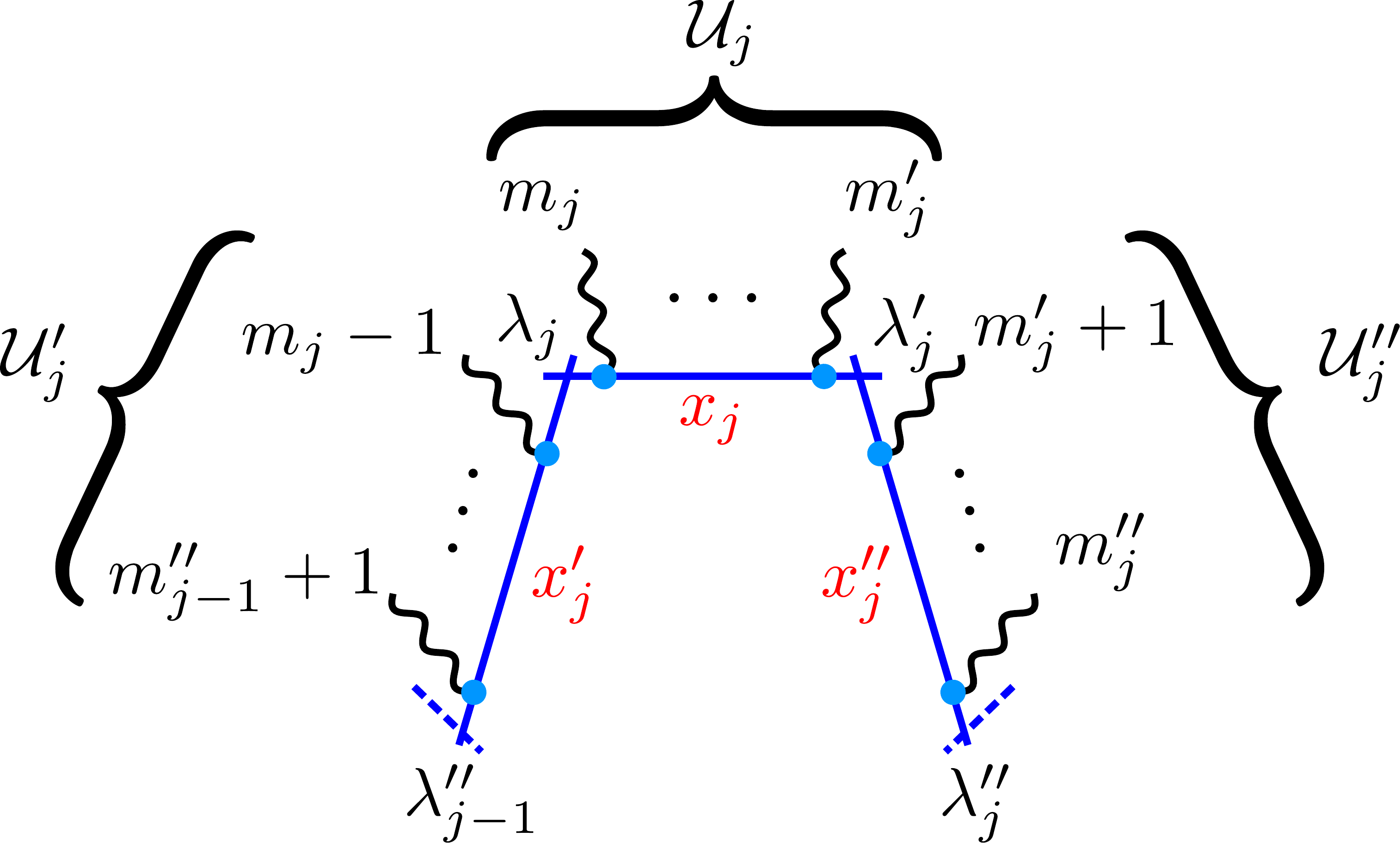}
  \caption{\it This figure shows the labels of the on-shell fields attached to each tooth of the cogwheel Wilson loop. Since the operator bearing edge $x_j$ must have at least one external field attached, we have the cyclic constraints $\cdots \leq m_{j-1}''< m_j\leq m_j'\leq m_j''< \cdots $. We denote by $\mathcal{U}_j,\cdots$ the contributions to the form factors from the different sides of the cog.
  }
  \label{fig:CogwheelPiece}
\end{figure}
We consider each cog (or tooth) of the cogwheel Wilson loop separately, see  Figure~\ref{fig:CogwheelPiece}.  We first look at the edges of the loop that carry the irreducible fields $D^{k_j}\Phi_j$ of $\calO$, i.e.\ at the parallel propagators $U_{x_j}(\la_j,\la_j')$.  
We take the term in \eqref{eq:frameUdefinitionpart1} with $(m_j'-m_j+1)$ $\calA$'s and insert external on-shell fields labeled $m_j,\dots,m_j'$ into it. We must have $m_j\leq m_j'$, since we want to emit at least one on-shell particle from the edges that carry the irreducible fields; otherwise, acting with the derivatives yields zero. 
The calculation is similar to those previously done in \eqref{eq:MHV amplitude} and \eqref{eq:deltaintegration}, and it gives
\begin{align}
\label{formula14.4}
 \mathcal{U}_j(m_j,m_j')&=\int_{}\frac{\abra{\lambda_j}{\lambda'_{j}}\DD\tilde{\la}_1 \cdots  \DD\tilde{\la}_{m_j'-m_j+1}}{\abra{\lambda_j}{\tilde{\la}_1}\abra{\tilde{\la}_1}{\tilde{\la}_2}\cdots \abra{\tilde{\la}_{m_j'-m_j+1}}{\lambda_{j}'}}\AAA_{\fP_{m_j}}(\calZ_{x_j}(\tilde{\la}_1)) \cdots \AAA_{\fP_{m_j'}}(\calZ_{x_j}(\tilde{\la}_{m_j'-m_j+1}))\notag\\
&= \frac{\abra{ \la_j}{\la_{j}'}}{\abra{\lambda_j}{p_{m_j}}\prod_{k=m_j}^{m_j'-1}\abra{p_k}{p_{k+1}}\abra{p_{m_j'}}{\lambda_j'}}\e^{ i\sum_{k=m_j}^{m_j'}(x_j\fp_k+\theta_j p_k\eta_k)}\,.
\end{align}
We then act on this with the forming operator \eqref{eq:definitionformingfactoronshellstates} for the irreducible field $D^{k_j}A_j$, take the operator limit and set $\theta=0$:
\begin{equation}
 \label{eq: formingfactoronshellstates on wilson line}
\begin{multlined}
\mathcal{I}_j(m_j,m_j')=\lim_{\hexagon\rightarrow \xdot}\PPP_{D^{k_j}\Phi_j} \mathcal{U}_j(m_j,m_j')_{\big{|}\theta=0}\,,
\end{multlined}
\end{equation}
which we write explicitly as
\begin{equation}
 \label{eq: formingfactoronshellstates on wilson line operator limit}
\begin{multlined}
\mathcal{I}_j(m_j,m_j')=\frac{-1}{\abra{\tau_j}{p_{m_j}}\prod_{k=m_j}^{m_j'-1}\abra{p_k}{p_{k+1}}\abra{p_{m_j'}}{\tau_j}}
\Bigg(\sum_{k=m_j}^{m_j'}\abra{\tau_j}{p_k}\sbra{\bar{p}_k}{\bar{\tau}_j}\Bigg)^{k_i}\\
\left\{\begin{array}{ll}
\big(\sum_{k=m_j}^{m_j'}\abra{\tau_j}{p_k}\sbra{\bar{p}_k}{\bar{\tau}_j}\big)^2 & \text{ for } \Phi_j=\bar F\\
\big(\sum_{k=m_j}^{m_j'}\abra{\tau_j}{p_k}\sbra{\bar{p}_k}{\bar{\tau}_j}\big)\big(\sum_{k=m_j}^{m_j'}\abra{\tau_j}{p_k}\{\xi_j\eta_k\}\big)& \text{ for }\Phi_j=\bar{\psi}\\
\big(\sum_{k=m_j}^{m_j'}\abra{\tau_j}{p_k}\{\xi_j\eta_k\}\big)^2& \text{ for } \Phi_j=\phi\\
\big(\sum_{k=m_j}^{m_j'}\abra{\tau_j}{p_k}\{\xi_j\eta_k\}\big)^3&\text{ for }\Phi_j=\psi\\
\big(\sum_{k=m_j}^{m_j'}\abra{\tau_j}{p_k}\{\xi_j\eta_k\}\big)^4& \text{ for } \Phi_j=F
\end{array}\right\} \e^{ i\sum_{k=m_j}^{m_j'}x\fp_k} \eqndot
\end{multlined}
\end{equation}
Recalling the notation introduced in Section~\ref{subsec: all MHV form factors}, we can write \eqref{eq: formingfactoronshellstates on wilson line operator limit} as
\beq
\label{eq: formingfactoronshellstates on wilson line operator limit 2}
\mathcal{I}_j(m_j,m_j')=-\frac{\left(\sum_{k=m_j}^{m_j'}\abra{\tau_j}{p_k}\sbra{\bar{p}_k}{\bar{\tau}_j}\right)^{N_j-n_{\theta_j}}\left(\sum_{k=m_j}^{m_j'}\abra{\tau_j}{p_k}\{\xi_j\eta_k\}\right)^{n_{\theta_j}}}{\abra{\tau_j}{p_{m_j}}\prod_{k=m_j}^{m_j'-1}\abra{p_k}{p_{k+1}}\abra{p_{m_j'}}{\tau_j}}\e^{ i\sum_{k=m_j}^{m_j'}x\fp_k}\,.
\eeq
Using the identity
\beq
\left(\sum_{k=1}^m\mathcal{C}_k\right)^N=\sum_{1\leq k_1\leq k_2\leq \cdots \leq k_N\leq m}\frac{N!}{M(\{k_1,\ldots, k_N\})!}\mathcal{C}_{k_1}\cdots \mathcal{C}_{k_N}\,,
\eeq
we can rewrite \eqref{eq: formingfactoronshellstates on wilson line operator limit 2} as 
\beq
\begin{aligned}
\label{eq: formingfactoronshellstates on wilson line operator limit 3}
\mathcal{I}_j(m_j,m_j')&=-\frac{\e^{ i\sum_{k=m_j}^{m_j'}x\fp_k}}{\abra{\tau_j}{p_{m_j}}\prod_{k=m_j}^{m_j'-1}\abra{p_k}{p_{k+1}}\abra{p_{m_j'}}{\tau_j}}\sum_{m_j\leq B_{j,1}\leq \cdots \leq B_{j,N_j}\leq m_j'}
\left(\prod_{k=1}^{N_j}\abra{\tau_j}{p_{\notA_{j,k}}}\right)\\
  &\phaneq\phantom{-}\times\sum_{\sigma\in S_{N_j}}
  \frac{\{\xi_j \eta_{\sigma(\notA_{j,1})}\}\cdots\{\xi_j \eta_{\sigma(\notA_{j,n_{\theta_j}})}\}\sbra{\bar{p}_{\sigma(\notA_{j,n_{\theta_j}+1})}}{\bar\tau_j}\cdots\sbra{\bar{p}_{\sigma(\notA_{j,N_j})}}{\bar\tau_j}}{M(\{\notA_{j,1},\dots,\notA_{j,N_j}\})!}
\,.
\end{aligned}
\eeq

In addition to the factor \eqref{eq: formingfactoronshellstates on wilson line operator limit 3}, a contribution from the two edges of the Wilson loop on the left and right of $x_j$ occurs, which are not acted on by derivative operators. 
Similarly to $\mathcal{U}_j$, we can compute the contributions from the two other sides of the cogwheel tooth, see  Figure~\ref{fig:CogwheelPiece}. Specifically, we find
\beq
\mathcal{U}_j'(m_{j-1}'',m_j)= \frac{\abra{ \la_{j-1}''}{\la_{j}}}{\abra{\lambda_{j-1}''}{p_{m_{j-1}''+1}}\prod_{k=m_{j-1}''+1}^{m_j-2}\abra{p_k}{p_{k+1}}\abra{p_{m_j-1}}{\lambda_j}}\e^{ i\sum_{k=m_{j-1}''+1}^{m_j-1}x_j'\fp_k}
\eeq
if $m_{j-1}''<m_j-1$ and $\mathcal{U}_j'=1$ if $m_{j-1}''=m_j-1$. Finally, we get
\beq
\mathcal{U}_j''(m_j',m_j'')= \frac{\abra{ \la_j'}{\la_{j}''}}{\abra{\lambda_j'}{p_{m_j'+1}}\prod_{k=m_j'+1}^{m_j''-1}\abra{p_k}{p_{k+1}}\abra{p_{m_j''}}{\lambda_{j}''}}\e^{ i\sum_{k=m_j'+1}^{m_j''}x_j''\fp_k}
\eeq
if $m_j'<m_j''$ and $\mathcal{U}_j''=1$ if $m_j'=m_j''$.
Now we must take the operator limit, which is detailed in Appendix~\ref{The operator limit} and summarized here, see Figure~\ref{fig:operatorlimitapp1}. First, we take the limit where
\begin{equation}
\label{eq:oplimit11}
\calZ_i\rightarrow (\lambda_i, ix\lambda_i, i\theta \lambda_i)\eqncom \qquad \calZ_i'\rightarrow (\lambda_i', ix\lambda_i',i \theta\lambda_i')\eqncom\qquad \calZ_i''\rightarrow (\lambda_i'', ix\lambda_i'', i\theta \lambda_i'')\eqncom
\end{equation}
i.e.\ all the intersection twistors lie on the same line.
\begin{figure}[htbp]
 \centering
  \includegraphics[height=4.5cm]{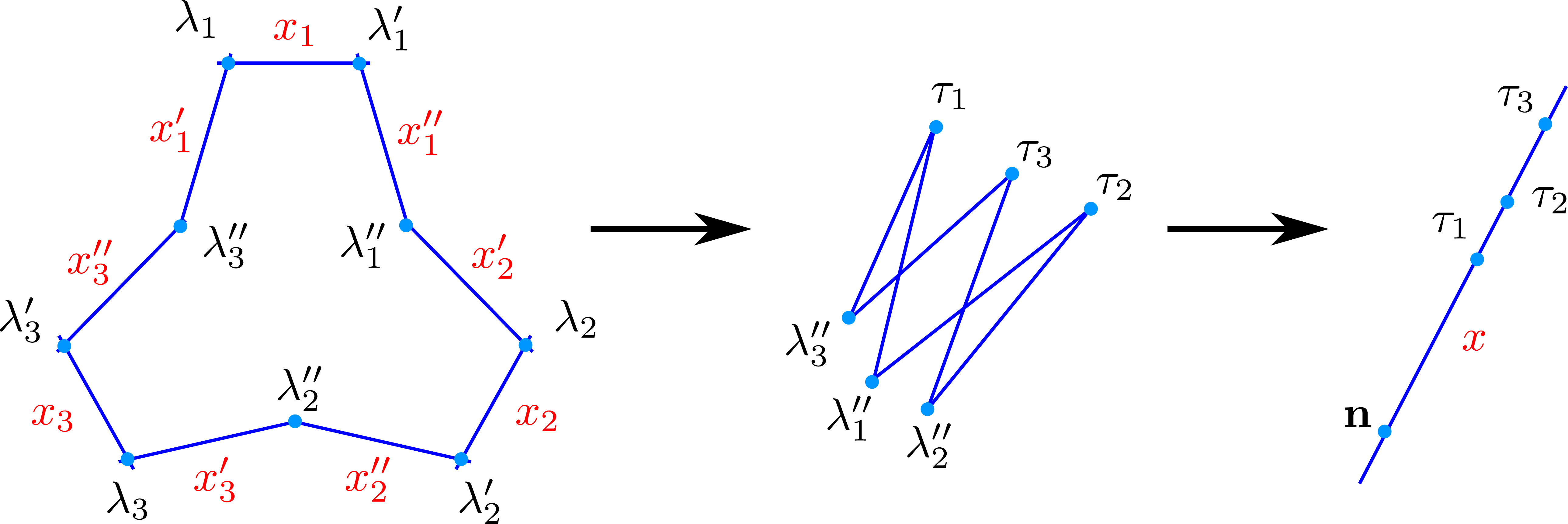}
  \caption{\it This figure illustrates the operator limit for $L=3$. To visualize the process a bit better, as a first step, we set $\lambda_i=\la_i'=\tau_i$ while bringing the $x_i$ closer to each other. The second step then just sends all $x_i$ to $x$ and $\lambda_{i}''\rightarrow \n$. 
  }
  \label{fig:operatorlimitapp1}
\end{figure}
Then, we take the limit $\lambda_i\parallel\lambda_i'\parallel\tau_i$ which is depicted in the middle diagram of Figure~\ref{fig:operatorlimitapp1}.
Projectively, this means $\lambda_i$, $\lambda_i'$ and  $\tau_i$ are equal. Summarizing, we have in the operator limit:
\begin{equation}
\label{eq:oplimit21}
\lambda_i\rightarrow \tau_i\,,\qquad  \lambda_i' \rightarrow \tau_i\,,\qquad \lambda_i''\rightarrow \n \eqndot
\end{equation}
After taking the operator limit \eqref{eq:oplimit11} and \eqref{eq:oplimit21}, we obtain
\beq
\label{eq: empty wilson line operator limit 1}
\mathcal{I}_j'(m_{j-1}'',m_j) =\frac{\abra{ \n}{\tau_{j}}}{\abra{\n}{p_{m_{j-1}''+1}}\prod_{k=m_{j-1}''+1}^{m_j-2}\abra{p_k}{p_{k+1}}\abra{p_{m_j-1}}{\tau_j}}\e^{ i\sum_{k=m_{j-1}''+1}^{m_j-1}x\fp_k}
\eeq
and
\beq
\label{eq: empty wilson line operator limit 2}
\mathcal{I}_j''(m_j',m_j'')= \frac{\abra{ \tau_j}{\n}}{\abra{\tau_j}{p_{m_j'+1}}\prod_{k=m_j'+1}^{m_j''-1}\abra{p_k}{p_{k+1}}\abra{p_{m_j''}}{\n}}\e^{ i\sum_{k=m_j'+1}^{m_j''}x\fp_k}
\eqndot
\eeq
As for $\mathcal{U}_j'$ and $\mathcal{U}_j''$, we have by definition that $\mathcal{I}_j'(m_{j-1}'',m_j)=1$ if $m_{j-1}''=m_j-1$ and that $\mathcal{I}_j''(m_j',m_j'')=1$ if $m_j'=m_j''$. 
The position-space form factor of the operator $\calO$ is now obtained by taking the product of all \eqref{eq: formingfactoronshellstates on wilson line operator limit 2}, \eqref{eq: empty wilson line operator limit 1} and \eqref{eq: empty wilson line operator limit 2} with $1\leq j\leq L$ and then summing over all possible indices $m_j$, $m_j'$ and $m_j''$. Since the irreducible fields are placed on the edges $x_j$, we see from  Figure~\ref{fig:CogwheelPiece} that we can have $\cdots \leq m_{j-1}''< m_j\leq m_j'\leq m_j''< \cdots $. Hence, the sum is cyclic over 
$1\leq m_1\leq m_1'\leq m_1''< m_2\leq m_2'\leq m_2''<m_3\leq \cdots <m_L\leq m_L'\leq m_L''=m_1+n-1$
 and we write for the form factor:
\begin{equation}
\label{eq: general MHV form factor in position space}
\calF_{\calO}(1,\dots,n;x)=\sum_{\{m_j,m_j',m_j''\}} \prod_{j=1}^L \mathcal{I}_j'(m_{j-1}'',m_j) \mathcal{I}_j(m_j,m_j')\mathcal{I}_j''(m_j',m_j'')\,.
\end{equation}
Let us now for simplicity denote by $\tilde{\mathcal{I}}_j$, $\tilde{\mathcal{I}}_j'$ and $\tilde{\mathcal{I}}_j''$ the contributions \eqref{eq: formingfactoronshellstates on wilson line operator limit 2}, \eqref{eq: empty wilson line operator limit 1} and \eqref{eq: empty wilson line operator limit 2} stripped off the exponential factors. We can almost immediately perform the sum over the $m_j$, $m_j'$ and $m_j''$, leaving in \eqref{eq: general MHV form factor in position space} only the sums over the $B_{i,j}$ that are contained implicitly in the $\mathcal{I}_j$. In order to do that, it turns out to be useful to rewrite the MHV prefactor of \eqref{eq: formingfactoronshellstates on wilson line operator limit 3} as
\begin{align}
&\frac{1}{\abra{\tau_j}{p_{m_j}}\prod_{k=m_j}^{m_j'-1}\abra{p_k}{p_{k+1}}\abra{p_{m_j'}}{\tau_j}}=\notag\\
&\frac{\abra{p_{B_{j,N_j}}}{\tau_j}}{\prod_{k=B_{j,N_j}}\abra{p_k}{p_{k+1}}\abra{p_{m_j'}}{\tau_j}}\frac{\abra{\tau_j}{p_{B_{j,1}}}}{\abra{\tau_j}{p_{m_j}}\prod_{k=m_j}^{B_{j,1}-1}\abra{p_k}{p_{k+1}}}\frac{1}{\abra{\tau_j}{p_{B_{j,1}}}\prod_{k=B_{j,1}}^{B_{j,N_j}-1}\abra{p_k}{p_{k+1}}\abra{p_{B_{j,N_j}}}{\tau_j}}\eqndot
\label{eq: denominator decomposition}
\end{align}
\begin{figure}[tbp]
 \centering
  \includegraphics[height=3.5cm]{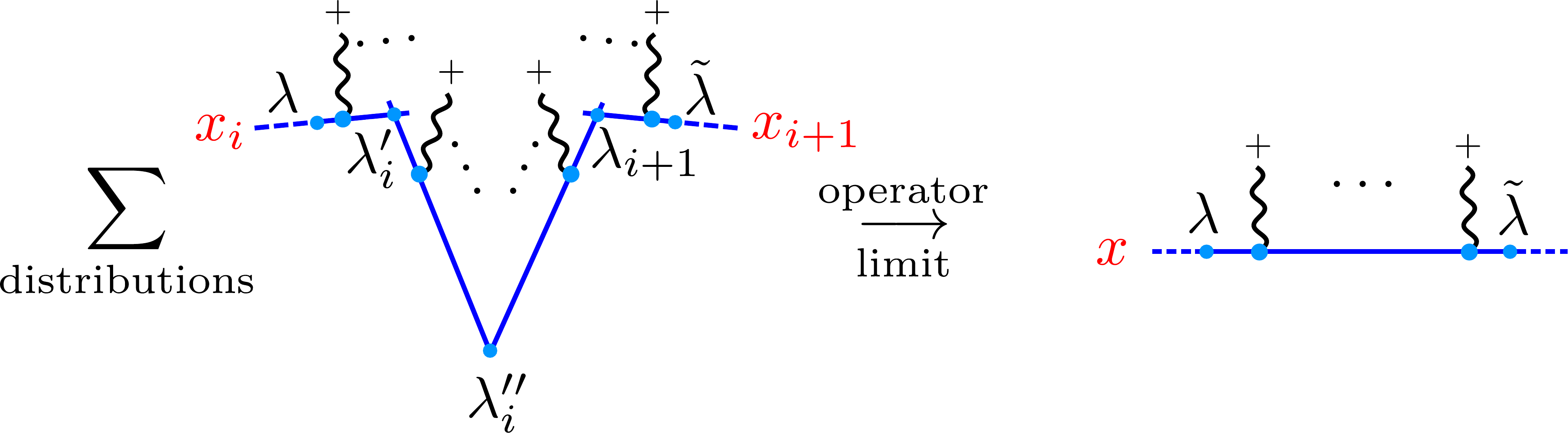}
  \caption{\it The contributions to the emissions of positive helicity gluons from the different edges combine in the operator limit.
  }
  \label{fig:CogwheelGaugeInvariance}
\end{figure}
Repeatedly using the following telescopic Schouten identity
\beq
\label{eq:telescopedSchouten}
\frac{\abra{a}{b}}{\abra{a}{1}\cdots \abra{m}{b}}+\sum_{k=1}^{m-1}\frac{\abra{a}{b}\abra{b}{c}}{\abra{a}{1}\cdots \abra{k}{b}\abra{b}{(k+1)}\cdots \abra{m}{c}}+\frac{\abra{b}{c}}{\abra{b}{1}\cdots \abra{m}{c}}=\frac{\abra{a}{c}}{\abra{a}{1}\cdots \abra{m}{c}}\,,
\eeq 
we can then show that
\begin{multline}
\label{eq:sumoversidesofcogs}
\sum_{B_{j,N_j}\leq m_j'\leq m_j''<m_{j+1}\leq B_{j+1,1}}\frac{\abra{p_{B_{j,N_j}}}{\tau_j}\tilde{\mathcal{I}}_j''(m_j',m_j'')\tilde{\mathcal{I}}_{j+1}'(m_{j}'',m_{j+1})\abra{\tau_{j+1}}{p_{B_{j+1,1}}}}{\prod_{k=B_{j,N_j}}\abra{p_k}{p_{k+1}}\abra{p_{m_j'}}{\tau_j}\abra{\tau_{j+1}}{p_{m_{j+1}}}\prod_{k=m_{j+1}}^{B_{j+1,1}-1}\abra{p_k}{p_{k+1}}}
\\=\frac{\abra{p_{B_{j,N_j}}}{p_{B_{j+1,1}}}}{\prod_{k=B_{j,N_j}}^{B_{j+1,1}-1}\abra{p_k}{p_{k+1}}}\,.
\end{multline}
Thus, the auxiliary spinor $\n$ in $\tilde{\mathcal{I}}_j'$ and $\tilde{\mathcal{I}}_{j}''$, which may not be part of our final result, drops out.
 In fact, the identity \eqref{eq:sumoversidesofcogs} is the direct consequence of the following identity for the parallel propagators $U_{x}(\lambda,\lambda_j')U_{x}(\lambda_j',\lambda_j'')U_{x}(\lambda_j'',\lambda_{j+1})U_{x}(\lambda_{j+1},\tilde\lambda)=U_{x}(\lambda,\tilde\lambda)$, see  Figure~\ref{fig:CogwheelGaugeInvariance}, which holds after taking the operator limit.

Hence, using \eqref{eq: general MHV form factor in position space}, the expression \eqref{eq: formingfactoronshellstates on wilson line operator limit 3} and the identity \eqref{eq:sumoversidesofcogs}, we obtain after Fourier transforming the claimed result \eqref{eq: general MHV form factor}.
In particular, the first and the last term in the product at the end of the first line of \eqref{eq: formingfactoronshellstates on wilson line operator limit 3} always cancel with corresponding terms in the denominator on the right hand side of \eqref{eq: denominator decomposition} and the global sign in \eqref{eq: formingfactoronshellstates on wilson line operator limit 3}. 
\chapter{N$^k$MHV form factors in twistor space}
\label{NMHV}
In the previous chapter we saw that all tree-level MHV form factors emerged from operator vertices in twistor space in complete analogy to the appearance of all tree-level MHV amplitudes from the twistor action vertices. More general N$^k$MHV amplitudes have been obtained from twistor space by gluing together MHV amplitudes via off-shell twistor propagators. These ``Feynman'' rules in twistor space are equivalent to CSW recursion in momentum space. To extend CSW recursion to form factors as well, as was reviewed in Section~\ref{secformfactor}, one needed to supplement the MHV action vertices by off-shell continued form factors, called operator vertices. Having constructed these operator vertices in twistor space in the last chapter, we can now construct higher N$^k$MHV form factors by connecting them to MHV vertices via a twistor propagator. The actual computation of such form factors heavily relies on a similar inverse soft limit that was shown to exist for amplitudes. In Section~\ref{subsec: inversesoftlimit} we therefore define and prove this inverse soft limit as a preliminary result. This inverse soft limit for the Wilson loop vertices already appeared in Section~\ref{NMHVscalar}. In Section~\ref{sec:NkMHVtreelevelformfactors} we give a general expression for NMHV form factors in twistor space, which is followed by a more general and less technical discussion on N$^k$MHV form factors in the last section.
This chapter is based on and contains significant overlap with \cite{Koster:2016fna}.
\section{The inverse soft limit for form factors}
\label{subsec: inversesoftlimit}

Via the so-called inverse soft limit, the $n$-point twistor-space vertices of the elementary interactions can be expressed in terms of $(n-1)$-point vertices. This procedure  played a crucial role in the calculation of amplitudes beyond MHV level in position twistor space \cite{Adamo:2011cb}. We reviewed this in Section~\ref{NMHVamp}. 
Here, we find a similar recursion for the Wilson loop vertices, which will play an equally important role in the calculation of form factors beyond MHV level.
In space-time, MHV form factors can be constructed via inverse soft limits as well \cite{Nandan:2012rk}. In our twistor-space formulation, this construction is based on the parallel propagator \eqref{eq:frameUdefinitionpart1}.
Using the external states in position twistor space \eqref{eq: external state in position twistor space},
the parallel propagator $U_x(\calZ_1,\calZ_2)$ in \eqref{eq:frameUdefinitionpart1} leads to the following Wilson line vertex: 
\begin{equation}
\label{eq: Wilson line vertex in position space}
 \begin{aligned}
\verop{x}{\calZ_1,\calZ_{2}}{\calZt_1,\dots,\calZt_m}
&\equiv\verop{x}{\calZ_1,\calZ_{2}}{\calA_{\calZt_1},\ldots, \calA_{\calZt_{m}}}\\
&=
\int  \frac{\abra{\lambda_1}{\lambda_{2}}
\prod_{j=1}^m \calA_{\calZt_j}(\calZ_x(\tilde\lambda_j))\DD\tilde\lambda_j}{\abra{\lambda_1}{\tilde\lambda_1} \abra{\tilde\lambda_1}{\tilde\lambda_2}\cdots \abra{\tilde\lambda_{m}}{\lambda_{2}} }
\eqncom
\end{aligned}
\end{equation}
see  Figure~\ref{fig: picture for the vertices W}.
In particular, for $m=0$ we set $\verop{x}{\calZ_1,\calZ_{2}}{}\equiv1$. Here, we used the Wilson loop that is detailed in Appendix~\ref{app:geometry} and is shown in Figure~\ref{fig:CogwheelZoom1}. One tooth of the cogwheel is detailed in Figure~\ref{fig: picture for the vertices W}, where the labels are shown as they are used in this section.
\begin{figure}[htbp]
 \centering
  \includegraphics[height=3.2cm]{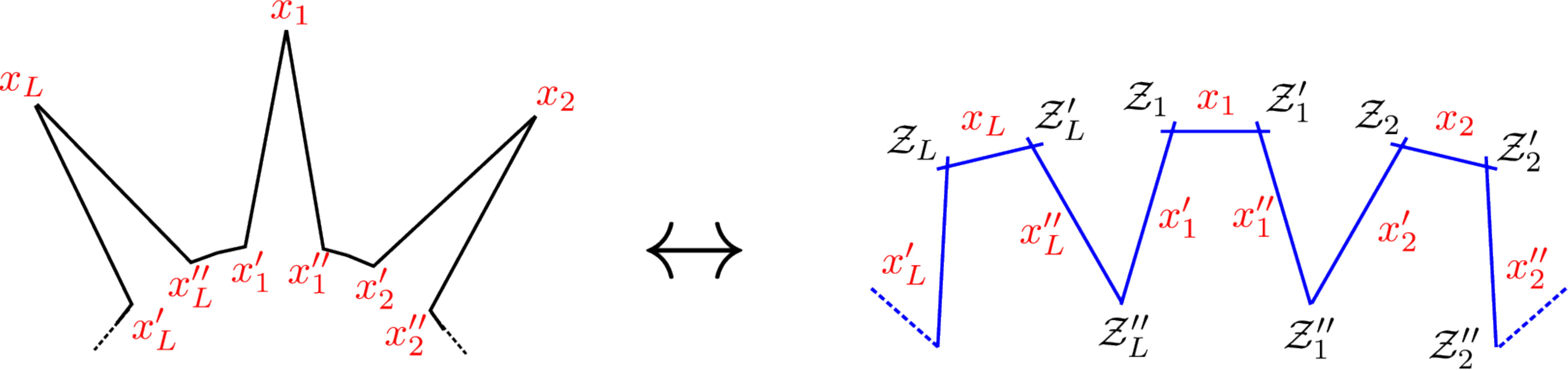}
  \caption{\it The geometry of the light-like Wilson loop. }
  \label{fig:CogwheelZoom1}
\end{figure}We denote by $\calZ_1$ and $\calZ_2$ the two twistors at which the line $x$ intersects with its neighboring edges in the Wilson loop, see  Figure~\ref{NMHVformfactortwistorspace}. Furthermore, we write $\calZt_i$ for the twistors on the line $x$ where on-shell states or propagators are to be attached.
In particular, the smallest non-trivial vertex is
\begin{equation}
\label{eq: minimal Wilson line vertex}
\begin{aligned}
\verop{x}{\calZ_1,\calZ_{2}}{\calZt}&=\int\DD\tilde{\lambda}\frac{\abra{\lambda_1}{\lambda_2}}{\abra{\lambda_1}{\tilde{\lambda}}\abra{\tilde{\lambda}}{\lambda_2}}2\pi i \,\bar{\delta}^{3|4}(\calZt,\calZ_x(\tilde{\la}))\\&=\int_{\mathbb{C}}\frac{\dd u}{u}\bar{\delta}^{3|4}(\calZ_1+u\calZ_2,\calZt)=\bar{\delta}^{2|4}(\calZ_1,\calZt,\calZ_{2})\eqncom
\end{aligned}
\end{equation}
where in the second line we used the parametrization of the line $x$ given by $\calZ_x(\tilde{\la})=\calZ_1+u\calZ_2$, which implies that 
$\abra{\lambda_1}{\tilde{\la}}=u \abra{\lambda_1}{\lambda_2}$, $\abra{\tilde{\la}}{\lambda_2}=\abra{\lambda_1}{\lambda_2}$ and $\DD\tilde{\la}=\abra{\lambda_1}{\lambda_2}\tfrac{\dd u}{2\pi i }$.

Using now the same method as for the vertices $\textbf{V}$,
we arrive at the identity
\begin{multline}
\label{eq: inverse soft limit frame}
 \verop{x}{\calZ_{1},\calZ_{2}}{\calZt_1,\dots,\calZt_m}
 \\= \verop{x}{\calZ_{1},\calZ_{2}}{\calZt_1,\dots,\calZt_{k-1},\calZt_{k+1},\ldots, \calZt_{m}}
 \bar{\delta}^{2|4}(\calZt_{k-1},\calZt_{k},\calZt_{k+1})\eqncom
\end{multline}
where $k\in \{1,\ldots, m\}$ and it is understood that $\calZt_{0}\equiv\calZ_1$ and $\calZt_{m+1}\equiv\calZ_2$.
Using the  relation \eqref{eq: inverse soft limit frame}, we can reduce every Wilson line vertex to the minimal Wilson line vertex \eqref{eq: minimal Wilson line vertex}:
\begin{equation}
 \label{eq: total inverse soft limit frame}
\begin{aligned}
  \verop{x}{\calZ_{1},\calZ_{2}}{\calZt_1,\dots,\calZt_m}&=\prod_{j=1}^m\bar{\delta}^{2|4}(\calZt_{j-1},\calZt_{j},\calZ_{2}) \eqndot
\end{aligned}
\end{equation}
The expression \eqref{eq: total inverse soft limit frame} has a geometric interpretation, which makes its origin obvious and goes as follows.
The vertices $\verop{x}{\calZ_1,\calZ_{2}}{\calZt_1,\dots,\calZt_m}$ in \eqref{eq: Wilson line vertex in position space} describe a Wilson line in position twistor space. Hence, all position twistors $\calZ_1,\calZ_{2},\calZt_1,\dots,\calZt_m$ have to lie on one line.
This is expressed in \eqref{eq: total inverse soft limit frame} by enforcing each of the first $m$ pairs of neighboring position twistors to be collinear with the last position twistor. However, there exist many equivalent ways to express the collinearity of all $m+2$ position twistors by forcing $m$ triplets of them to be collinear. 
\begin{figure}[tbp]
 \centering
  \includegraphics[height=1.6cm]{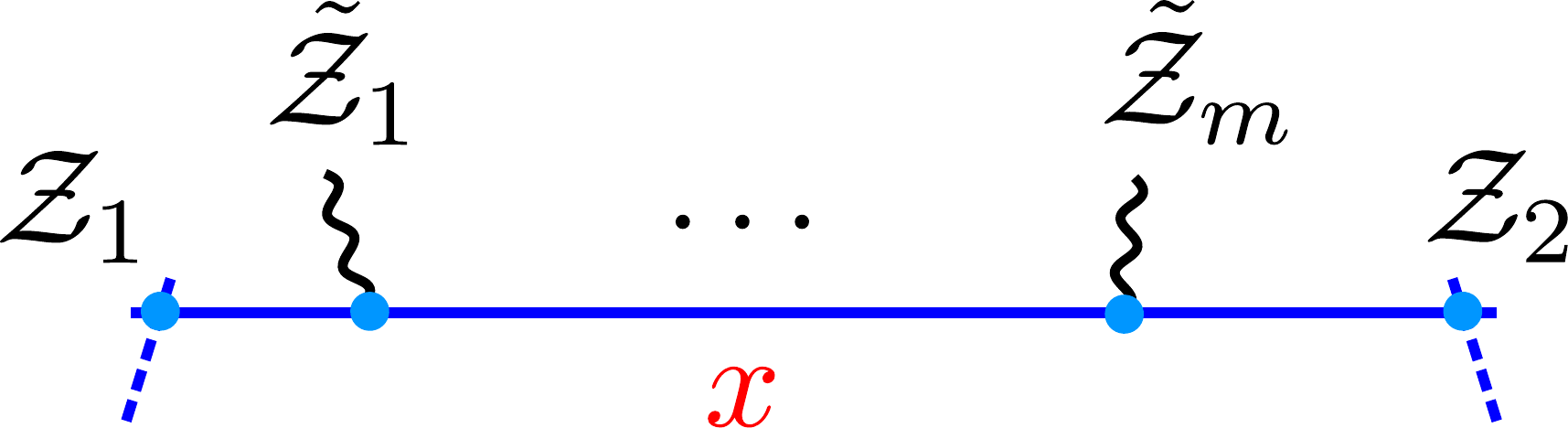}
  \caption{\it The $m$-point vertex \eqref{eq: Wilson line vertex in position space} from the parallel propagator $U_x(\calZ_1,\calZ_2)$. }
\label{fig: picture for the vertices W}
\end{figure}
\section{NMHV form factors}
\label{NMHVformfactortwistorspace}
Let us now return to form factors and calculate the NMHV form factors of general composite operators.
The propagator in this case has to connect a Wilson line vertex \eqref{eq: Wilson line vertex in position space} corresponding to an edge of the Wilson loop \eqref{eq:finaldefinitionWilsonloop1} with an interaction vertex \eqref{eq:definitionoftheamplitudevertex}.
In the end, we have to sum over all such edges and over all possible combinations to distribute the external fields on the interaction vertex $\mathbf{V}$ and the Wilson line vertices $\mathbf{W}$ of the different edges.

To simplify the notation and keep the presentation transparent, we will consider a particular edge $X$ with vertex $\verop{X}{\calZ_{X1},\calZ_{X2}}{\calZt_{a+1},\dots,\calZt_b}$, where $\calZ_{X1}$ ($\calZ_{X2}$) denotes the twistor corresponding to the intersection with the previous (next) edge.
Concretely, $(X,\calZ_{X1},\calZ_{X2})$ is an element of the union 
\begin{equation}
\{(x_i,\calZ_i,\calZ_i')|i=1,\dots,L\}\cup
\{(x_i'',\calZ_i',\calZ_i'')|i=1,\dots,L\}\cup
\{(x_i',\calZ_{i-1}'',\calZ_i)|i=1,\dots,L\}\eqndot\end{equation}
Moreover, we let $\calZt_{a+1}$ ($\calZt_b$) denote the first (last) external field on that edge.
\begin{figure}[tbp]
 \centering
  \includegraphics[height=2.3cm]{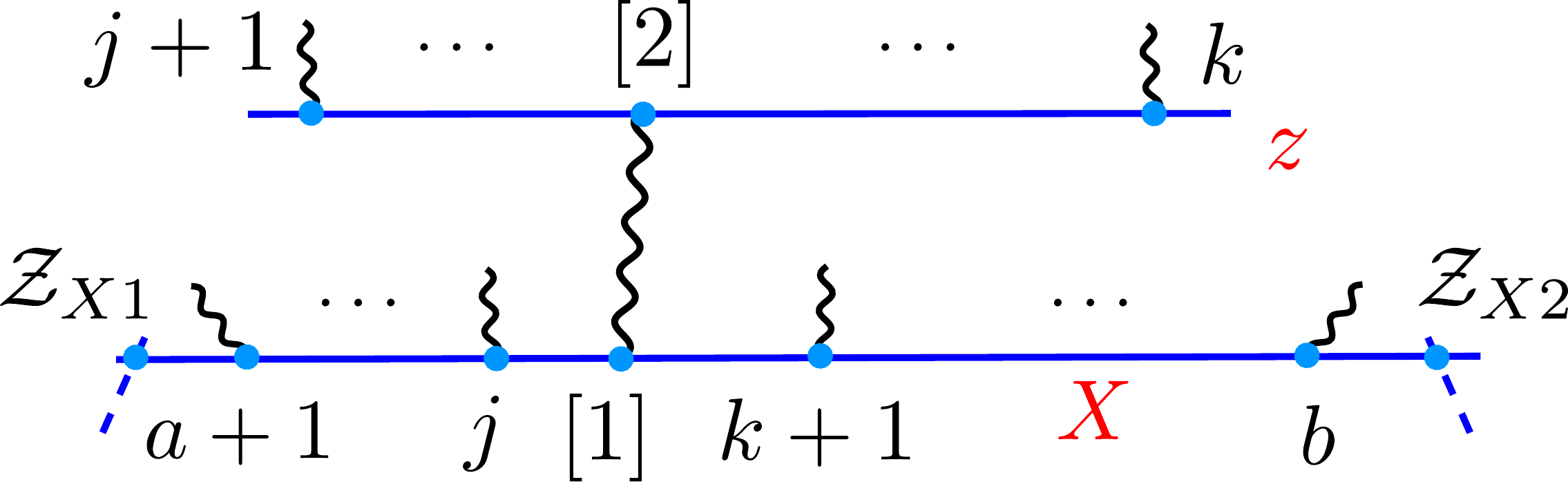}
   \caption{\it A generic NMHV diagram in which an interaction vertex $\textbf{V}$ is connected to a Wilson line vertex $\textbf{W}_X$. The line $X$ in twistor space contains the two twistors $\calZ_{X1}$ and $\calZ_{X2}$ at which it intersects the lines of the cogwheel Wilson loop adjacent to it.} 
  \label{fig:AllTermsNMHVGeneric}
\end{figure}
For a generic distribution of the external fields, shown in  Figure~\ref{fig:AllTermsNMHVGeneric}, we then find 
\begin{equation}
\label{eq: NMHV diagram for form factors}
 \begin{aligned}
&\int \DD^{3|4}\calZ_{[1]}\DD^{3|4}\calZ_{[2]}\verop{X}{\calZ_{X1},\calZ_{X2}}{\calZt_{a+1},\ldots, \calZt_j,\calZ_{[1]},\calZt_{k+1},\ldots,\calZt_b}\Delta(\calZ_{[1]},\calZ_{[2]})\\
&\phaneq\times\ver{\ldots, \calZt_{k},\calZ_{[2]},\calZt_{j+1},\ldots}\\
&=\nalpha\verop{X}{\calZ_{X1},\calZ_{X2}}{\calZt_{a+1},\ldots, \calZt_{j},\calZt_{k+1},\ldots,\calZt_b}\ver{\ldots, \calZt_k,\calZt_{j+1},\ldots}\\
&\phaneq\times \int \DD^{3|4}\calZ_{[1]}\DD^{3|4}\calZ_{[2]}\bar{\delta}^{2|4}(\calZt_{j},\calZ_{[1]},\calZt_{k+1})
\bar{\delta}^{2|4}(\calZ_{[1]},\star,\calZ_{[2]})
\bar{\delta}^{2|4}(\calZt_{k},\calZ_{[2]},\calZt_{j+1})\\
&=\nalpha\verop{X}{\calZ_{X1},\calZ_{X2}}{\calZt_{a+1},\ldots, \calZt_{j},\calZt_{k+1},\ldots,\calZt_b}\ver{\ldots, \calZt_k,\calZt_{j+1},\ldots}
\\&\phaneq\times[\calZt_{k+1},\calZt_j,\star,\calZt_{j+1},\calZt_k]\eqncom
\end{aligned}
\end{equation}
which is completely analogous to \eqref{eq: NMHV amplitude in position twistor space} except that we have also used the inverse soft limit \eqref{eq: inverse soft limit frame} for the Wilson line vertex.
We have to consider two special distributions.
If $a=j$, the external supertwistor $\calZt_j$ is not on the edge $X$ and we have to use $\calZ_{X1}$ for the inverse soft limit. We then need to replace $\calZt_j\rightarrow \calZ_{X1}$ in the five-bracket of the last line of \eqref{eq: NMHV diagram for form factors}.
Similarly, if $b=k$, the external supertwistor $\calZt_{k+1}$ is not on the edge $X$, which leads to a replacement $\calZt_{k+1}\rightarrow \calZ_{X2}$ in the five-bracket. In order to write the result in a condensed form, we denote by 
 \begin{equation} 
 \label{eq:definitionreplacementrule}
 \calZt_j\stackrel{j=a}{\longrightarrow} \calZ_{X1}
 \end{equation} 
  the supertwistor that is $\calZt_j$ if $j\neq a$ and becomes equal to $\calZ_{X1}$ if $j=a$.

It follows from the preceding discussion that the total position-twistor-space form factor is
\begin{equation}
\label{eq:NMHVWilsonloopformfactor}
\begin{aligned}
\mathbb{F}^{\text{NMHV}} &=-\nalpha
\sum_{X}\Big\{\sum_{a\leq j<k\leq b}
\verop{X}{\calZ_{X1},\calZ_{X2}}{\calZt_{a+1},\ldots, \calZt_{j},\calZt_{k+1},\ldots,\calZt_b}\\
&\phaneq\times \ver{\ldots, \calZt_k,\calZt_{j+1},\ldots}\big[\calZt_j\stackrel{j=a}{\longrightarrow} \calZ_{X1},\calZt_{j+1},\star,\calZt_k,\calZt_{k+1}\stackrel{k=b}{\longrightarrow} \calZ_{X2}\big]\Big\}\\
&\phaneq \times 
\prod_{X'\neq X}\verop{X'}{\calZ_{X1'},\calZ_{X2'}}{\ldots}\,,
\end{aligned}
\end{equation}
where we have kept the sum over all distributions of the remaining $n-b+a$ external fields on the $3L-1$ remaining edges implicit.

To compute the NMHV form factor $\mathbb{F}^{\text{NMHV}}_{\calO}$ of a specific operator $\calO$ \eqref{eq:composite operator}, we must now act with the forming factor, do the integral and take the operator limit as in \eqref{formingonwilson}. 
We will do the integral and rephrase the result in momentum twistor space in several cases in Chapter~\ref{momentumspace}.

\section{\texorpdfstring{N$^k$MHV}{NkMHV} form factors}
\label{sec:NkMHVtreelevelformfactors}

Having shown how to compute the NMHV form factors in position twistor space, let us now consider the case of arbitrary high MHV degree. The discussion parallels the one for amplitudes in \cite{Adamo:2011cb}, and hence we shall mostly concern ourselves with highlighting the differences. There are three different kinds of twistor-space diagrams for amplitudes, namely {\it generic, boundary} and {\it boundary-boundary}, and the same applies to form factors. The types of diagrams differ in the relative positions of the propagators in the (interaction) vertices.

\paragraph{Generic:} For these diagrams, no adjacent propagators occur at any vertex such that they can be calculated in complete analogy to the NMHV diagrams. The corresponding contribution to the amplitude/form factor is given by products of twistor-space vertices and of R-invariants of the kind $[\calZt_{a_i},\calZt_{a_j},\star,\calZt_{a_k},\calZt_{a_l}]$ with the $\calZt_{a_n}$ indicating some external supertwistors. For some diagrams, there are no external particles to the left (right) of a propagator on the Wilson line vertex $\textbf{W}_x$.
In this case, we need to replace the external twistors $\calZt_j$ ($\calZt_{k+1}$) by the appropriate twistors fixing the line $x$, cf.~\eqref{eq:definitionreplacementrule}.

\paragraph{Boundary:} In this case, some propagators are inserted next to each other, but each vertex for which that happens is either a Wilson line vertex $\mathbf{W}$ or it is an interaction vertex $\mathbf{V}$ but has at least two external particles, see  Figure~\ref{fig:BoundaryTerm} for one example of each. 
This allows us to use the inverse soft limit and the resulting delta functions to do all twistor integrals. 
The diagram in position twistor space thus evaluates to a product of R-invariants and twistor-space vertices $\textbf{V}$ and $\textbf{W}$. However, unlike in the generic case, the supertwistors entering the R-invariants are not the external ones but rather are obtained by simple geometric means, namely intersecting a line and a plane, as explained in the following.
Assuming that there is a propagator between the line given by the twistors 
$\calZt_{a_1}$ and $\calZt_{a_2}$ and another line given by $\calZt_{b_1}$ and $\calZt_{b_2}$, the contribution is given by the R-invariant
\begin{equation}
\label{eq:ruleforboundaryRinvariant}
[\calZt_{a_1},\skew{5}{\widehat}{\calZt}_{a_2},\star,\calZt_{b_1},\skew{5}{\widehat}{\calZt}_{b_2}]\,.
\end{equation}
Here, $\skew{5}{\widehat}{\calZt}_{a_2}=\calZt_{a_2}$ if there is no propagator to the right of the insertion twistor on the line $L_{\calZt_{a_1}\calZt_{a_2}}$ given by the twistors $\calZt_{a_1}$ and $\calZt_{a_2}$. Otherwise, $\skew{5}{\widehat}{\calZt}_{a_2}$ is the intersection of the line $L_{\calZt_{a_1}\calZt_{a_2}}$ and the plane $\langle \star, \calZt_{c_1},\calZt_{d_1}\rangle$ spanned by the twistors $\calZt_{c_1}$, $\calZt_{d_1}$ and the reference twistor $\star$. Here, $\calZt_{c_1}$ and $\calZt_{d_1}$ define the line that is connected with a propagator to the twistor on the right of the insertion twistor on the line $L_{\calZt_{a_1}\calZt_{a_2}}$:
\begin{equation}
\label{eq:intersectionlineplane}
\skew{5}{\widehat}{\calZt}_{a_2}=L_{\calZt_{a_1}\calZt_{a_2}}\cap\langle \star,\calZt_{c_1},\calZt_{d_1}\rangle=(\star,\calZt_{c_1},\calZt_{d_1},\calZt_{a_1})\calZt_{a_2}-(\star,\calZt_{c_1},\calZt_{d_1},\calZt_{a_2})\calZt_{a_1}\eqncom
\end{equation}
see  Figure~\ref{fig:BoundaryTermExample}. We further remark that in any computation in which some of the external twistors have to be replaced by the twistors on the corners of the Wilson loop as in \eqref{eq:definitionreplacementrule}, the replacement rule just carries through the computation, i.e.\ we simply have to replace the respective twistors in \eqref{eq:ruleforboundaryRinvariant} and \eqref{eq:intersectionlineplane} via \eqref{eq:definitionreplacementrule}.

\begin{figure}[tbp]
 \centering
  \includegraphics[height=3.9cm]{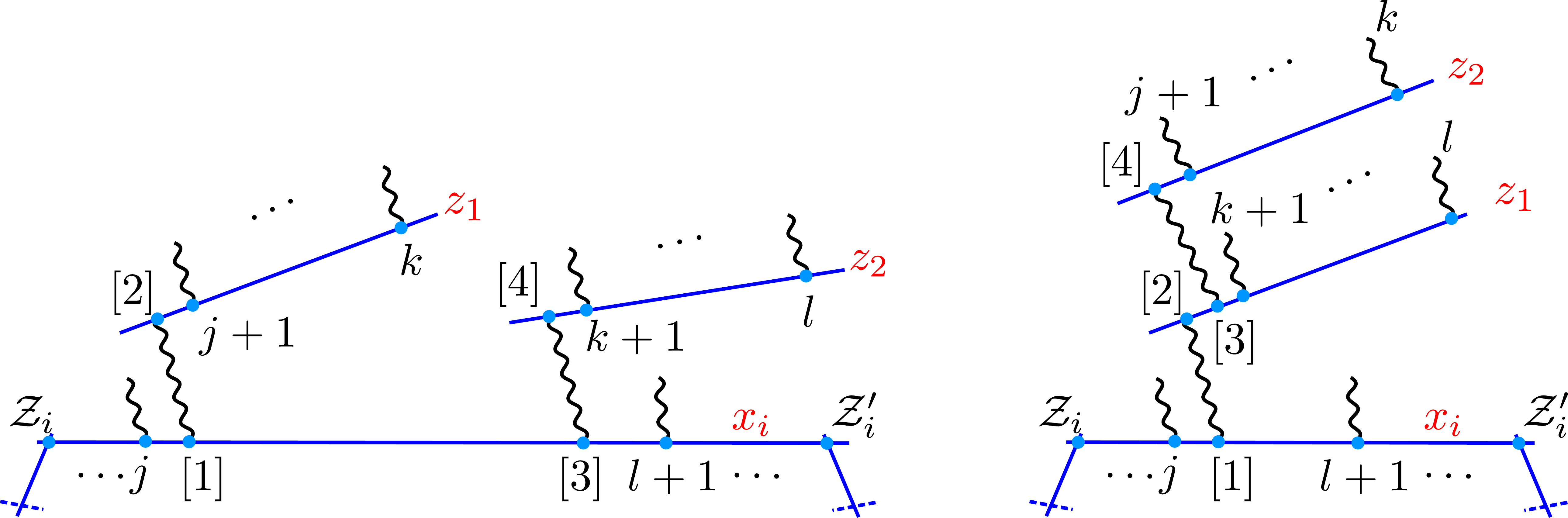}
  \caption{\it This figure depicts two boundary cases. On the left, the propagator insertions at the twistors $[1]$ and $[3]$ are adjacent, but they are on the line $x_i$ corresponding to an edge of the Wilson loop. On the right, the  twistors $[2]$ and $[3]$ are adjacent but there are at least two external particles on the line $z_1$ if $l>k+1$. 
  We denote the two interaction lines by $z_1$ and $z_2$, respectively. 
  }
  \label{fig:BoundaryTerm}
\end{figure}
\begin{figure}[tbp]
 \centering
  \includegraphics[height=3.5cm]{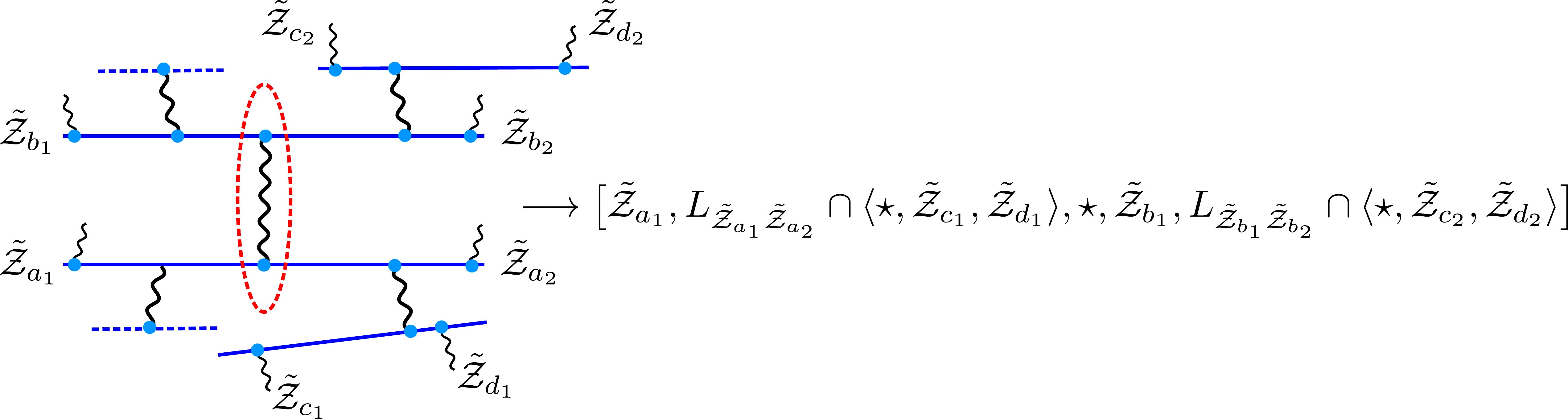}
  \caption{\it 
  The propagator encircled in red is replaced by the R-invariant \eqref{eq:ruleforboundaryRinvariant} with two twistors replaced by intersections \eqref{eq:intersectionlineplane} as indicated.
  }
\label{fig:BoundaryTermExample}
\end{figure}

\paragraph{Boundary-boundary:} This case is similar to the boundary one, except that there are now less than two external particles at a vertex $\mathbf{V}$. Thus, we cannot use the inverse soft limit to do all twistor integrals, as the inverse soft limit cannot be applied to the two-point vertex.
It seems difficult to obtain an expression built only out of R-invariants and MHV vertices for this case, but, as mentioned in \cite{Adamo:2011cb}, the boundary-boundary case is still fully described by the twistor formalism.
As no boundary-boundary case can occur at the Wilson line vertices $\mathbf{W}$,%
\footnote{Recall that we can always use the inverse soft limit for Wilson line vertices.}
the boundary-boundary diagrams for form factors are however no more difficult than for amplitudes.

\chapter{NMHV amplitudes and form factors in momentum twistor space}
\label{momentumspace}
In this chapter we translate our results of the previous chapter on NMHV form factors to momentum space. This allows us to compare our results to the literature. In Section~\ref{eq:NMHV amplitudes in momentum twistor space} we first prove that the NMHV amplitudes in twistor space can be translated to known results in momentum twistor space. In Section~\ref{subsec:NMHV form factors in momentum twistor space} we follow the same procedure but this time for NMHV form factors without $\dot\alpha$ indices. The idea relies on first computing the form factor using the Wilson loop construction, without specifying the operator or the type of external particles. After employing the inverse soft limit and Fourier transforming we act with the forming operator at the final step of the calculation only. The result of this section has been numerically checked against known results for the chiral half of the stress-tensor supermultiplet. Translating NMHV form factors for more general operators to momentum space is more involved, as these contain derivative operators that do not commute with the Fourier transform to momentum space. Despite this, we still manage to compute an example of the NMHV form factor of an operator consisting of a fermion and $L-1$ scalar fields. As the previous chapter, the current chapter is once more based on \cite{Koster:2016fna} and contains significant overlap with this paper.

\section{NMHV Amplitudes in momentum twistor space}
\label{eq:NMHV amplitudes in momentum twistor space}
\newcommand{\rmom}{y}
\newcommand{\rsmom}{\Gamma}
\newcommand{\mtwistor}{\mathcal{W}}

In \cite{Adamo:2011cb} it was observed that expression \eqref{eq: NMHV amplitude in position twistor space final} is the known expression for NMHV amplitudes in momentum twistor space stripped of MHV factors:
\begin{equation}
\label{eq: NMHV amplitude in momentum twistors}
\mathscr{A}^{\text{NMHV}}(1,\dots,n)= \mathscr{A}^{\text{MHV}}(1,\dots,n)\sum_{1\leq j<k\leq n}[\mtwistor_{j},\mtwistor_{j+1},\star,\mtwistor_{k},\mtwistor_{k+1}]\eqncom
\end{equation}
where $\mathscr{A}^{\text{MHV}}$ was given in \eqref{eq:MHV amplitude} and the definition of the momentum twistors $\mtwistor$ is given below. In this section we prove this claim starting from formula \eqref{eq: NMHV amplitude in position twistor space final} for the NMHV amplitudes in position twistor space.\newline
Let us first briefly review the definition of momentum twistors \cite{Hodges:2009hk}.
Starting with $n$ supermomenta $\fP_i=(\fp_i,\eta_i)$, we define dual, or region, momenta and supermomenta by
\begin{equation}
\label{eq: regional momenta and supermomenta}
 \begin{aligned}
  \fp_i^{\alpha \dot\alpha}=(\rmom_{i}-\rmom_{i-1})^{\alpha\dot\alpha}\eqncom \qquad  p_i^{\alpha}\eta_{i,a}=(\rsmom_{i}-\rsmom_{i-1})^{\alpha}_{\phantom{\alpha}a}\eqncom
 \end{aligned}
\end{equation}
see  Figure~\ref{fig:MomentumTwistors} for an illustration in the case $n=6$. 
The origin in dual (super)momentum space is arbitrary.
\begin{figure}[tbp]
 \centering
  \includegraphics[height=3.7cm]{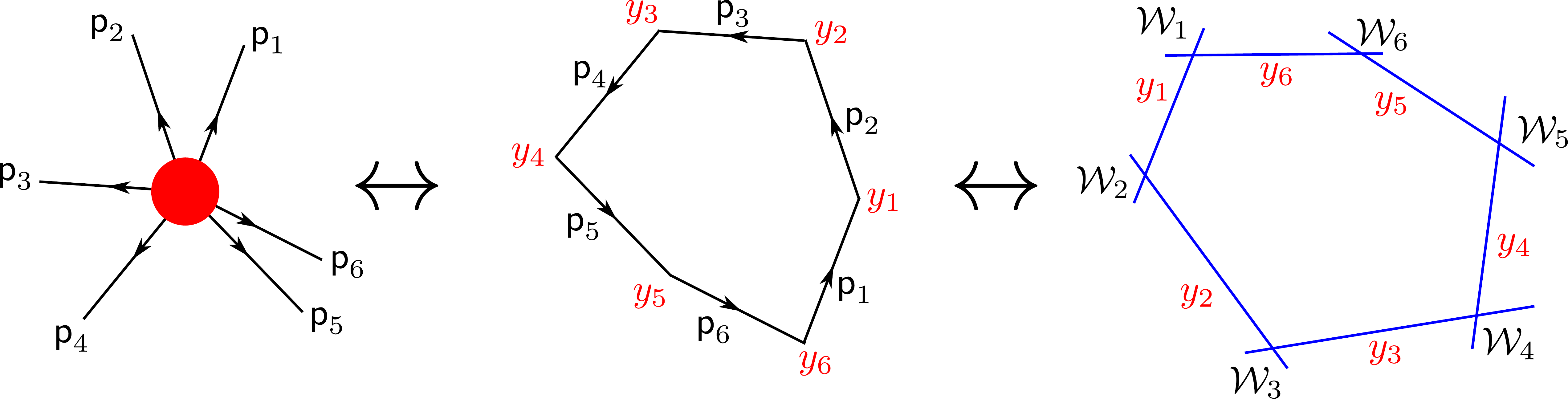}
  \caption{\it This figure illustrates the relationship between (super)momenta, region momenta and momentum twistors for $n=6$. We suppress the fermionic part.
  }
  \label{fig:MomentumTwistors}
\end{figure}
For example, we can choose
\begin{equation}
\label{eq:definitionofdualregionvariables}
 \rmom_{i}^{\alpha\dot\alpha}=\sum_{j=1}^i p_j^\alpha\bar{p}_j^{\dot\alpha}\eqncom \qquad (\rsmom_{i})^{\alpha}{}_a=\sum_{j=1}^i p_j^\alpha\eta_{j,a}
 \eqndot
\end{equation}
The momentum twistors $\mtwistor_j$ are then defined as the intersections of the lines $(\rmom_{j-1},\rsmom_{j-1})$ and $(\rmom_{j},\rsmom_{j})$ in twistor space:
\begin{equation}
\label{eq: definition momentum twistors}
 \mtwistor_j\equiv (\rmom_{j-1},\rsmom_{j-1})\cap(\rmom_{j},\rsmom_{j})=(p_{j\alpha},i \rmom_j^{\alpha\dot\alpha}p_{j \alpha},i (\rsmom_j)^{\alpha}_{\phantom{\alpha}a}p_{j\alpha})\equiv\calZ_{(\rmom_{j},\rsmom_{j})}(p_j)\eqndot
\end{equation}
As in the case of position twistors, we will frequently abbreviate $\calZ_{\rmom_{j}}(p_j)\equiv\calZ_{(\rmom_{j},\rsmom_{j})}(p_j)$.

Having introduced our notation for the momentum twistors, we can now compute the NMHV amplitudes and rephrase them in momentum twistor space. 
Recall that inserting the momentum eigenstates \eqref{eq:definitiononshellmomentumeigenstates} into \eqref{eq: NMHV amplitude in position twistor space} 
is effectively accomplished by the following integration
\begin{equation}
 \mathscr{A}^{\text{NMHV}}(1,\dots,n)=\int \Big[\prod_{s=1}^n\DD\calZ_s\frac{\calA_{\fP_s}(\calZ_s)}{2\pi i}\Big]\mathbb{A}^{\text{NMHV}}(1,\dots,n)\eqndot
\end{equation}
Now using the methods of e.g.\ \cite{Adamo:2011cb},
we find 
\begin{align}
\label{eq: amplitude}
\mathscr{A}^{\text{NMHV}}(1,\dots,n)
&=\nalpha\sum_{1\leq j<k\leq n}\int \frac{\dd^{4}z_1\,\dd^8\vartheta_1}{(2\pi)^4}\frac{ \dd^{4}z_2\,\dd^8\vartheta_2}{(2\pi)^4} \frac{1}{\prod_{i=1}^n\abra{p_i}{p_{i+1}}}\frac{\abra{p_j}{p_{j+1}}\abra{p_k}{p_{k+1}}}{\abra{p_{k+1}}{p_j}\abra{p_{j+1}}{p_k}}
\nonumber\\&\phaneq \times [\calZ_{(z_1,\vartheta_1)}(p_{k+1}),\calZ_{(z_1,\vartheta_1)}(p_j),\star,\calZ_{(z_2,\vartheta_2)}(p_{j+1}),\calZ_{(z_2,\vartheta_2)}(p_k)]\\&\phaneq
\times \e^{i\sum_{s={k+1}}^j(z_1\fp_s+\vartheta_1 p_s\eta_s)}\e^{i\sum_{s={j+1}}^k(z_2\fp_s+\vartheta_2 p_s\eta_s)}\eqndot \nonumber
\end{align}
Using \eqref{eq:Rinvariantforxandxprime},
we see that the five-bracket only depends on the differences $z_2-z_1$ and $\vartheta_2-\vartheta_1$.
Thus, we shift the integration variables as $z_2\to z_1+z_2$ and $\vartheta_2\to \vartheta_1+\vartheta_2$.
As a result, the five-bracket no longer depends on $z_1$ and $\vartheta_1$.
Moreover, considering the explicit form \eqref{eq:fivebracket}
of the five-bracket, the replacement allows us to use a Fourier-type integration identity derived in Appendix~\ref{app:importantFourier} for $(z_2,\vartheta_2)$.
In the five-bracket, the formula \eqref{eq:superFourierRinvariant} for the $(z_2,\vartheta_2)$ integration effectively replaces 
\begin{equation}
\label{eq:replacementzvarthetawithmomentumtwistors}
 z_2\to\sum_{s=j+1}^{k}p_s\bar{p}_s = \rmom_k-\rmom_j\,,\qquad  \vartheta_2\to\sum_{s=j+1}^{k}p_s\eta_s = \rsmom_k-\rsmom_j 
\end{equation}
and generates an overall factor of $\tfrac{1}{\nalpha}$.
Hence, \eqref{eq: amplitude} becomes
\begin{equation}
\label{eq: amplitude after integration}
\begin{aligned}
\mathscr{A}^{\text{NMHV}}(1,\dots,n)&= \sum_{1\leq j<k\leq n}\int \frac{\dd^{4}z_1\dd^8\vartheta_1}{(2\pi)^4} 
\frac{\e^{i(z_1\sum_{s=1}^n\fp_s+\vartheta_1\sum_{s=1}^n p_s\eta_s)}}{\prod_{i=1}^n\abra{p_i}{p_{i+1}}}\frac{\abra{p_j}{p_{j+1}}\abra{p_k}{p_{k+1}}}{\abra{p_{k+1}}{p_j}\abra{p_{j+1}}{p_k}}\\
&\phaneq \times \big[\calZ_{0}(p_{k+1}),\calZ_{0}(p_j),\star,\calZ_{\rmom_k-\rmom_j}(p_{j+1}),\calZ_{\rmom_k-\rmom_j}(p_k)\big]\,, 
\end{aligned}
\end{equation}
where we keep in mind that $\star=(0,\zeta,0)$.
The only remaining dependence on the integration variables $z_1$ and $\vartheta_1$ is in the exponential factor, which reduces to the momentum- and supermomentum-conserving delta functions.
Hence, we find 
\begin{align}
\label{eq: amplitude simplified}
&\mathscr{A}^{\text{NMHV}}(1,\dots,n)\nonumber\\&=\sum_{1\leq j<k\leq n}\frac{\delta^{4|8}(\sum_{i=1}^n\fP_i)}{\prod_{i=1}^n\abra{p_i}{p_{i+1}}}
\frac{\abra{p_j}{p_{j+1}}\abra{p_k}{p_{k+1}}}{\abra{p_{k+1}}{p_j}\abra{p_{j+1}}{p_k}}[\calZ_{0}(p_{k+1}),\calZ_{0}(p_j),\star,\calZ_{\rmom_k-\rmom_j}(p_{j+1}),\calZ_{\rmom_k-\rmom_j}(p_{k})]\nonumber\\
&=\mathscr{A}^{\text{MHV}}(1,\dots,n)
\sum_{1\leq j<k\leq n}[\calZ_{0}(p_j),\calZ_{0}(p_{j+1}),\star,\calZ_{\rmom_k-\rmom_j}(p_k),\calZ_{\rmom_k-\rmom_j}(p_{k+1})]
\eqncom
\end{align}
where in the last step we made use of \eqref{eq:Rinvariantforxandxprime} and the total antisymmetry of the five-bracket. 
The prefactor in the last step of \eqref{eq: amplitude simplified} is exactly the MHV amplitude \eqref{eq:MHV amplitude}, while the five-bracket equals 
\begin{equation}
 \label{eq: amplitude five-bracket simplified}
\begin{aligned}
&[\calZ_{0}(p_j),\calZ_{0}(p_{j+1}),\star,\calZ_{\rmom_k-\rmom_j}(p_k),\calZ_{\rmom_k-\rmom_j}(p_{k+1})]\\
&=[\calZ_{\rmom_j}(p_j),\calZ_{\rmom_j}(p_{j+1}),\star,\calZ_{\rmom_k}(p_k),\calZ_{\rmom_k}(p_{k+1})]\\
&=[\calZ_{\rmom_j}(p_j),\calZ_{\rmom_{j+1}}(p_{j+1}),\star,\calZ_{\rmom_k}(p_k),\calZ_{\rmom_{k+1}}(p_{k+1})]\\
&\equiv[\mtwistor_{j},\mtwistor_{j+1},\star,\mtwistor_{k},\mtwistor_{k+1}]\,.
\end{aligned}
\end{equation}
In the first step of \eqref{eq: amplitude five-bracket simplified}, we have used the invariance of the five-bracket under shifts \eqref{eq:Rinvariantforxandxprime} and in the second step the fact that, due to the definition \eqref{eq:definitionofdualregionvariables}, we have the relations
\begin{equation}
 p_{i+1}\rmom_{i+1}=p_{i+1}\rmom_{i}\eqncom \qquad p_{i+1}\rsmom_{i+1}=p_{i+1}\rsmom_{i}\eqndot
\end{equation}
Thus, we indeed obtain the desired amplitude in momentum twistor space \eqref{eq: NMHV amplitude in momentum twistors}.

\section{NMHV form factors for operators without \texorpdfstring{$\dot\alpha$}{dotted} indices}
\label{subsec:NMHV form factors in momentum twistor space}

Having understood how to move from position to momentum twistor space for amplitudes in the previous section, we now want to do the same for the NMHV form factors of section~\ref{NMHVformfactortwistorspace}. 
In the case of form factors, the forming operator \eqref{eq:definitionformingfactoronshellstates} occurs inside of the (super) Fourier-type integrals over the positions of the vertices and the operator. While the fermionic $\theta_i$ derivatives in the forming operator trivially commute with the fermionic integrals, this is in general not the case for the bosonic $x_i$ derivatives and the space-time integrals. The details will be deferred to Appendix~\ref{app:importantFourier}.
In the second half of this subsection, we restrict ourselves to operators without $\dot\alpha$ indices, i.e.\ to forming operators without $x_i$ derivatives.

One difference with respect to the amplitude case lies in the definition of the momentum twistors for form factors. In contrast to the amplitude case, for form factors the on-shell momenta $\fp_i$ of the $n$ external on-shell states do not add up to zero but to the off-shell momentum $\fq$ of the composite operator. Hence, the contour they form in the space of region momenta is periodic instead of closed \cite{Alday:2007he}:
\begin{equation}
 \rmom_{i+n}=\rmom_i +\fq \eqncom
\end{equation}
see  Figure~\ref{fig:NonperiodicMomentumtwistors}.
One way to define momentum twistors in this case is to use two periods of the contour\footnote{Alternatively, the momentum twistors can be defined by closing a single period of the contour via two auxiliary on-shell momenta as done in \cite{Frassek:2015rka}. This results in $n+2$ momentum twistors but requires to consider the different possible ways to close the contour.}, which results in $2n$ momentum twistors \cite{Brandhuber:2011tv,Bork:2014eqa} still defined via \eqref{eq: definition momentum twistors} but with $i=1,\dots,2n$.
\begin{figure}[tbp]
 \centering
  \includegraphics[height=3.1cm]{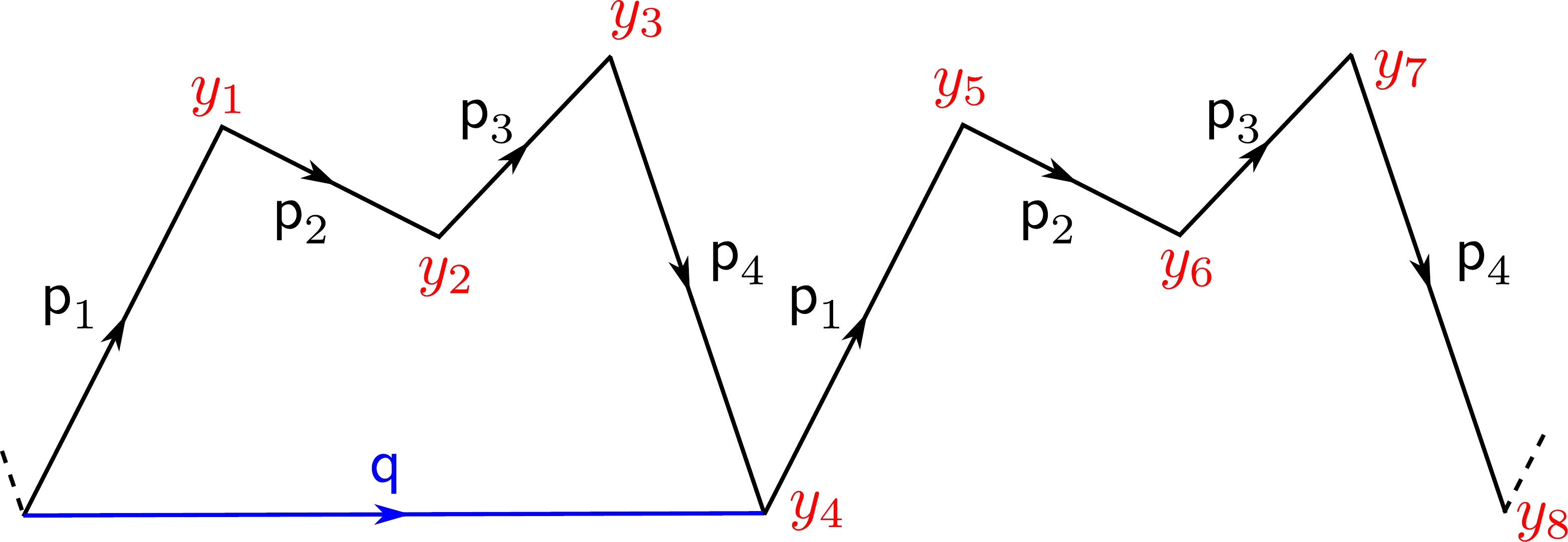}
  \caption{\it An illustration of the region momenta used for form factors. Here, we consider a case with $n=4$. }
  \label{fig:NonperiodicMomentumtwistors}
\end{figure}

The computation of the NMHV form factor parallels the calculations done for amplitudes in the preceding section. Inserting momentum eigenstates into $\mathbb{F}^{\text{NMHV}}$ of  \eqref{eq:NMHVWilsonloopformfactor}, including the forming operator \eqref{eq:definitionformingfactoronshellstates} and taking the operator limit, we find   
\begin{equation}
\label{eq:NMHVWilsonloopformfactor momentumspace}
\begin{aligned}
\mathscr{F}^{\text{NMHV}}_{\calO}(1,\dots,n;\fq) &=\nalpha
\sum_{X}
\sum_{a\leq j<k\leq b}
\int \frac{\dd^{4}x}{(2\pi)^4}\e^{-ix\fq}\int\frac{ \dd^{4}z\,\dd^8\vartheta}{(2\pi)^4} \lim_{\hexagon\rightarrow \xdot}\bigg\{\PPP_{\mathcal{O}}
\\
&\phaneq\times \frac{1}{\prod_{i=a+1}^{b}\abra{p_i}{p_{i+1}}}
\frac{\abra{(p_j\stackrel{j=a}{\longrightarrow }\lambda_{X1})}{p_{j+1}}\abra{p_k}{(p_{k+1}\stackrel{k=b}{\longrightarrow }\lambda_{X2})}}{\abra{(p_j\stackrel{j=a}{\longrightarrow }\lambda_{X1})}{(p_{k+1}\stackrel{k=b}{\longrightarrow }\lambda_{X2})}\abra{p_k}{p_{j+1}}}
\\&\phaneq \times 
[\calZ_{X}(p_{k+1}\stackrel{k=b}{\longrightarrow }\lambda_{X2}),\calZ_{X}(p_j\stackrel{j=a}{\longrightarrow }\lambda_{X1}),\star,\calZ_{(z,\vartheta)}(p_{j+1}),\calZ_{(z,\vartheta)}(p_k)]
\\&\phaneq \times \e^{i\sum_{s={a+1}}^j(X\fp_s+\theta p_s\eta_s)}\e^{i\sum_{s={j+1}}^k(z\fp_s+\vartheta p_s\eta_s)}\e^{i\sum_{s={k+1}}^b(X\fp_s+\theta p_s\eta_s)}\\
&\phaneq\times \prod_{X'\neq X}\text{contributions from edge }X'\bigg\}
\bigg{|}_{\theta=0}\,,
\end{aligned}
\end{equation}
where we have suppressed the contributions from the other edges.
In general, we have to act with the forming operator on the last four lines before doing the integration.
In the special case that the forming operator does not contain $x_i$ derivatives, it does however commute with the integration; we can then do the integral as in the amplitude case.
We will focus on this special case for the rest of this section and treat a simple example of the general case in the next section. The fact that the integral and the $x_i$ derivatives do in general not commute is exemplified in Appendix~\ref{app:importantFourier}.

Let us restrict ourselves to forming operators without $x_i$ derivatives, i.e.\ to composite operators built of irreducible fields whose forming operators \eqref{eq:definitionformingfactoronshellstates} contain only $\theta$ derivatives.
These are the scalars $\phi$, the fermions $\psi$ and the self-dual part of the field strength $F$; they require $N=n_\theta=2,3,4$ $\theta$ derivatives, respectively.
We can then commute the $(z,\vartheta)$ integral past the forming operator and evaluate this integral in complete analogy to the amplitude case treated in the previous section.
The full calculation involves the sum over all distributions of the external fields on the different edges, which replace $a$ and $b$ in \eqref{eq:NMHVWilsonloopformfactor momentumspace}.
We parametrize it via  $1\leq m_1\leq m_1'\leq m_1''< m_2\leq m_2'\leq m_2''<m_3\leq \cdots <m_L\leq m_L'\leq m_L''< m_1+n\leq 2n$ and abbreviate this set as $\{m_i,m_i',m_i''\}$.
The details of this calculation are quite technical, and we relegate them to Appendix~\ref{app:NMHVformfactors}.
Here, we just present the result:
\begin{equation}
\label{eq:NMHFformfactorMomentumPart2}
\begin{aligned}
\mathscr{F}^{\text{NMHV}}_{\calO}(1,\dots,n;\fq)\
= {}&{}\frac{\delta^{4}\big(\sum_{r=1}^n\fp_r-\fq\big)}{\prod_{r=1}^n\abra{p_r}{p_{r+1}}}\sum_{\{m_i,m_i',m_i''\}}
\widetilde{\mathscr{F}}_{\calO}(\{m_i,m_i',m_i''\})
\\
\phaneq\times \sum_{i=1}^L \Biggl\{
\sum_{m_{i-1}''\leq j<k\leq m_i-1}&\Big[\calW_j\stackrel{j=m_{i-1}''}{\longrightarrow }\calZ_{\rmom_j}(\textbf{n}),\calW_{j+1},\star, \calW_{k},\calW_{k+1}\stackrel{k=m_i-1}{\longrightarrow }\calZ_{\rmom_k}(\tau_i)\Big]\\
+ \sum_{m_i-1\leq j<k\leq m_i'} &\Big[\calW_j\stackrel{j=m_i-1}{\longrightarrow }\calZ_{\rmom_j}(\tau_i),\calW_{j+1},\star, \calW_{k},\calW_{k+1}\stackrel{k=m_i'}{\longrightarrow }\calZ_{\rmom_k}(\tau_i)\Big]\\
+ \sum_{m_{i}'\leq j<k\leq m_i''}&\Big[\calW_j\stackrel{j=m_{i}'}{\longrightarrow }\calZ_{\rmom_j}(\tau_i),\calW_{j+1},\star, \calW_{k},\calW_{k+1}\stackrel{k=m_i''}{\longrightarrow }\calZ_{\rmom_k}(\textbf{n})\Big]
\Biggr\}\,,
\end{aligned}
\end{equation}
where we have defined
\begin{align}
\widetilde{\mathscr{F}}_{\calO}(\{m_i,m_i',m_i''\})&= \prod_{r=1}^L 
\left\{\delta_{m_{r-1}''+1,m_r}+(1-\delta_{m_{r-1}''+1,m_r})\frac{\abra{ \n}{\tau_r}}{\abra{\n}{p_{m_{r-1}''+1}}\abra{p_{m_r-1}}{\tau_r}} \right\}
\nonumber\\&\quad \times \abra{p_{m_r-1}}{p_{m_r}}
\frac{\left(\sum_{s=m_r}^{m_r'}\abra{\tau_r}{p_s}\{\xi_r\eta_s\}\right)^{N_r}
}{\abra{p_{m_r}}{\tau_r}\abra{p_{m_r'}}{\tau_r}}\abra{p_{m_r'}}{p_{m_r'+1}}
\label{eq: F tilde}
\\&\quad \times\left\{\delta_{m_r',m_r''}+(1-\delta_{m_r',m_r''})
\frac{\abra{\tau_r}{\n}}{\abra{\tau_r}{p_{m_r'+1}}\abra{p_{m_r''}}{\n}}
\right\}\abra{p_{m_r''}}{p_{m_r''+1}}\nonumber\,.
\end{align}

Two remarks are in order. First, the result \eqref{eq:NMHFformfactorMomentumPart2} still contains the auxiliary spinor  $\textbf{n}$ from the operator limit \eqref{eq:oplimit2}.
For the MHV form factor 
\begin{equation}
\begin{aligned}
\label{eq:MHVformfactorMomentum}
\mathscr{F}^{\text{MHV}}_{\calO}(1,\ldots, n;\fq)
&=\frac{\delta^{4}\big(\sum_{r=1}^n\fp_r-\fq\big)}{\prod_{r=1}^n\abra{p_r}{p_{r+1}}}\sum_{\{m_i,m_i',m_i''\}}
\widetilde{\mathscr{F}}_{\calO}(\{m_i,m_i',m_i''\})\,,
\end{aligned}
\end{equation}
we have explicitly shown in Section~\ref{app: derivation} how $\textbf{n}$ drops out thanks to a repeated application of the Schouten identity. 
In the $n=(L+1)$-point case,  the spinor $\textbf{n}$ trivially drops out of the NMHV form factor due to the summation ranges and the Kronecker deltas in \eqref{eq:NMHFformfactorMomentumPart2} and \eqref{eq: F tilde}.
We leave a proof of the independence of $\mathscr{F}^{\text{NMHV}}_{\calO}$ on $\textbf{n}$ in the general case for future work.
Moreover, \eqref{eq:NMHFformfactorMomentumPart2} contains the reference twistor $\calZ_\star$. As in the case of general amplitudes, it is not immediate to see how $\calZ_\star$ drops out.

Second, we have numerically checked in several cases that \eqref{eq:NMHFformfactorMomentumPart2} reproduces the known result \cite{Brandhuber:2011tv,Bork:2014eqa} for the chiral half of the stress-tensor supermultiplet $T$:
\begin{equation}
\label{eq: T NMHV form factor}
 \mathscr{F}^{\text{NMHV}}_{T}(1,\ldots,n;\fq)= \mathscr{F}^{\text{MHV}}_{T}(1,\ldots,n;\fq) \sum_{j=1}^{n}\sum_{k=j+2}^{n+j-1} \big[\calW_j,\calW_{j+1},\star,\calW_k,\calW_{k+1}\big]\eqncom
\end{equation}
which factorizes similarly to the amplitude \eqref{eq: NMHV amplitude in momentum twistors}. 
It would be desirable to find a factorized form of \eqref{eq:NMHFformfactorMomentumPart2} also for general operators\footnote{For more general operators, the immediate generalization of \eqref{eq: T NMHV form factor} is \emph{not} true.
}.

\section{NMHV form factors for general operators}
\label{subsec: non-chiral operators}

We now turn to form factors of general operators, for which also some $x_i$ derivatives occur in the forming operator \eqref{eq:definitionformingfactoronshellstates}.
 We cannot interchange these derivatives with the integral, which complicates the calculation. This is shown in Appendix~\ref{app:importantFourier}.
To demonstrate that our formalism nevertheless continues to work in this case, we consider a simple example.

We study the next-to-minimal (i.e.\ $(L+1)$-point) NMHV form factor of the operator $\Tr(\bar\psi_2(\phi_{13})^{L-1})$ and consider the external state $1^{\phi_{12}}2^{\psi_{234}}3^{\phi_{13}}\dots(L+1)^{\phi_{13}}$.
The advantage of this example is that only one diagram contributes, in which $\bar\psi_2\rightarrow \phi_{12}\psi_{234}$, and in particular this single diagram must be gauge-invariant.
This diagram is the counterpart of \eqref{eq:NMHVWilsonloopformfactor momentumspace} with $a=j=n=L+1$, $j+1=1$ and $k=b=2$.
Picking up the calculation at this point, we find
\begin{align}
\label{eq: start of calculation}
&\calF_{\Tr(\bar\psi_2(\phi_{13})^{L-1})}(1^{\phi_{12}}2^{\psi_{234}}3^{\phi_{13}}\dots(L+1)^{\phi_{13}};x)\nonumber\\
&=\nalpha\lim_{\hexagon\rightarrow \xdot}\Bigg\{\frac{\partial^5}{\partial \eta_1^1\partial \eta_1^2\partial \eta_2^2\partial \eta_2^3\partial\eta_2^4}\int \frac{\dd^{4}z\dd^8\vartheta}{(2\pi)^4} \frac{\e^{iz(\fp_1+\fp_2)+i\vartheta(p_1\eta_1+p_2\eta_2)}}{\abra{p_1}{p_2}\abra{p_2}{p_1}}\nonumber\\
&\phaneq\times
\left(-\frac{\lambda_{X1}^\alpha \lambda_{X2}^{\beta}}{\abra{\lambda_{X1}}{\lambda_{X2}}}\right)
\left(-i\bar{\tau}^{\dot\alpha}\frac{\partial}{\partial x^{\alpha\dot{\alpha}}}\right)
\left(-i\xi^a\frac{\partial}{\partial \theta^{\beta a }}\right)\\
&\phaneq\times
 \big[\calZ_{(x,\theta)}(\la_{X2}),\calZ_{(x,\theta)}(\la_{X1}),\star,\calZ_{(z,\vartheta)}(p_1),\calZ_{(z,\vartheta)}(p_2)\big] \Bigg\}\Bigg|_{\xi^a=\delta^{a2},\theta=0} \e^{ix\sum_{i=3}^{L+1}\fp_i}\nonumber\\
&=- \nalpha \int\frac{\dd^{4}z}{(2\pi)^4} \e^{iz(\fp_1+\fp_2)}\left[\left(-i\tau^{\alpha}\bar{\tau}^{\dot\alpha}\frac{\partial}{\partial x^{\alpha\dot\alpha}}\right)\frac{1}{(x-z)^2}\frac{\bradot{p_2}{x-z}{\zeta}}{\bradot{\tau}{x-z}{\zeta}}\right]\e^{ix\sum_{i=3}^{L+1}\fp_i}\,,\nonumber
\end{align}
where the $\eta$ derivatives serve to select the $\phi_{12}\psi_{234}$ component of the super form factor and the only effective contribution of the other edges is the phase $\e^{ix\sum_{i=3}^{L+1}\fp_i}$. 
The polarization vectors $\tau$ and $\bar\tau$ correspond to the field $\bar\psi_2$ and we have dropped the index as the polarization vectors of all other fields have already dropped out. We remind that in the operator limit $\lambda_{X1}\rightarrow \tau$  and $\lambda_{X2}\rightarrow \tau$.
Employing the following identity,
\begin{equation}
\label{eq:cuteidentity}
\bradot{A}{x}{C}\bradot{B}{x}{D}-\bradot{A}{x}{D}\bradot{B}{x}{C}=-x^2\abra{A}{B}\sbra{C}{D}\,,
\end{equation}
we obtain after acting with the derivative inside the $z$ integral
\begin{align}
\label{eq:fpsi1}
&\calF_{\Tr(\bar\psi_2(\phi_{13})^{L-1})}(1^{\phi_{12}}2^{\psi_{234}}3^{\phi_{13}}\dots(L+1)^{\phi_{13}};x)\nonumber\\
&=-(4\pi)^2\int \frac{\dd^4z}{(2\pi)^4}\frac{\e^{iz(\fp_1+\fp_2)}}{(x-z)^4}\frac{(x-z)^2\abra{\tau}{p_2}\sbra{\zeta}{\bar\tau}-\bradot{\tau}{x-z}{\bar\tau}\bradot{p_2}{x-z}{\zeta}}{\bradot{\tau}{x-z}{\zeta}}\e^{ix\sum_{i=3}^{L+1}\fp_i}\nonumber\\
&=-(4\pi)^2\int \frac{\dd^4z}{(2\pi)^4} \e^{iz(\fp_1+\fp_2)}\frac{\bradot{p_2}{z-x}{\bar\tau}}{(z-x)^4}\e^{ix\sum_{i=3}^{L+1}\fp_i}\nonumber\\
&=-\frac{\e^{ix\sum_{i=1}^{L+1}\fp_i}}{(\fp_1+\fp_2)^2} \bradot{p_2}{\fp_1+\fp_2}{\bar\tau}\,,
\end{align}
where we have used \eqref{eq:IxMinkowskiFinalResult}. 
Fourier transforming in $x$ and using $\bradot{p_2}{\fp_1+\fp_2}{\bar\tau}=\abra{p_1}{p_2}\sbra{\bar \tau}{\bar p_1}$ as well as  $(\fp_1+\fp_2)^2=-\abra{p_1}{p_2}\sbra{\bar p_1}{\bar p_2}$, we find
\begin{equation}
\calF_{\Tr(\bar\psi_2(\phi_{13})^{L-1})}(1^{\phi_{12}}2^{\psi_{234}}3^{\phi_{13}}\dots(L+1)^{\phi_{13}};\fq)=\delta^4\left(\fq-\sum_{i=1}^{L+1}\fp_i\right)\frac{\sbra{\bar\tau}{\bar{p}_1}}{\sbra{\bar{p}_1}{\bar{p}_2}}\, .
\end{equation}
This is indeed the expected gauge-invariant expression; it can be obtained e.g.\ from the MHV form factor $\calF_{\Tr(\psi_{134}(\phi_{24})^{L-1})}(1^{\phi_{34}}2^{\bar\psi_{1}}3^{\phi_{24}}\dots(L+1)^{\phi_{24}};\fq)$ by conjugation.

In the above computation, it was mandatory to first act with the space-time derivative in the forming operator before integrating out the interaction, so that we may simplify the expression and in particular get rid of the reference twistor through \eqref{eq:cuteidentity}. In Appendix~\ref{subsec:non-chiralFourierAppendix}, we show that reversing the order of integration and derivative does not work naively. 
It would be desirable to be able to evaluate the integrals occurring in \eqref{eq:NMHVWilsonloopformfactor momentumspace} for general operators or to find a general form for the commutator of the forming operator and the integrals that allows for a more efficient evaluation.
We leave this for future work.

\chapter{The one-loop dilatation operator in the SO$(6)$ sector}
\label{corrfunc}
In the previous chapters we found the expressions for all local composite operators in twistor space. Furthermore, we successfully computed the partially off-shell form factors for all operators at MHV level and for certain operators at more general N$^k$MHV level. In this chapter we move on to the computation of the completely off-shell correlation functions. In particular, we compute the one-loop correlation functions for operators in the SO$(6)$ sector in twistor space, from which we extract the one-loop dilatation operator in this sector. In Section~\ref{sectreelevelcorr} we compute the tree-level correlation functions that are relevant for this process. Subsequently, in Section~\ref{seconeloop} the one-loop diagrams are calculated. We start by considering the diagram that contains a four-vertex, which as we shall see, gives the full contribution to the one-loop dilatation operator. This section follows the train of thought of Section~\ref{oneloopdilspacetime}, where we argued that one only needed to determine the coefficients of certain SU$(4)$ invariants. Interestingly, in the operator limit there appear spurious singularities that require careful treatment. In the remainder of this section we show that all other one-loop diagrams do not contribute to the dilatation operator. This chapter is based on and contains overlap with \cite{Koster:2014fva,Koster:2016fna}. The paper \cite{Wilhelm:2014qua} that appeared on the same day as \cite{Koster:2014fva}, contains the derivation of the complete one-loop dilatation operator from unitarity. This was the first field theoretic derivation of the complete one-loop dilatation operator. Following \cite{Koster:2014fva}, the equivalent computation using the closely related MHV diagrams method was done in \cite{Brandhuber:2014pta}. The same authors subsequently reproduced the results of \cite{Wilhelm:2014qua} only for the SO$(6)$- and SU$(2|3)$ sectors from generalized unitarity in \cite{Brandhuber:2015boa}.
\section{Tree-level correlation functions in twistor space}
\label{sectreelevelcorr}
Recall that the computation of the one-loop dilatation operator involves computing the UV-divergent contributions to the one-loop two-point correlation function in twistor space. These can then be expressed as the action of the dilatation operator on the tree-level correlation function. Therefore, a necessary preliminary is to compute the tree-level correlator of two operators consisting only of scalar fields. Let us first consider tree-level correlation functions for general operators. Classically, the correlation function of two local composite operators in twistor space before taking the operator limit is shown in  Figure~\ref{fig:TreeLevelCorrelation}. \begin{figure}[htbp]
 \centering
  \includegraphics[height=3cm]{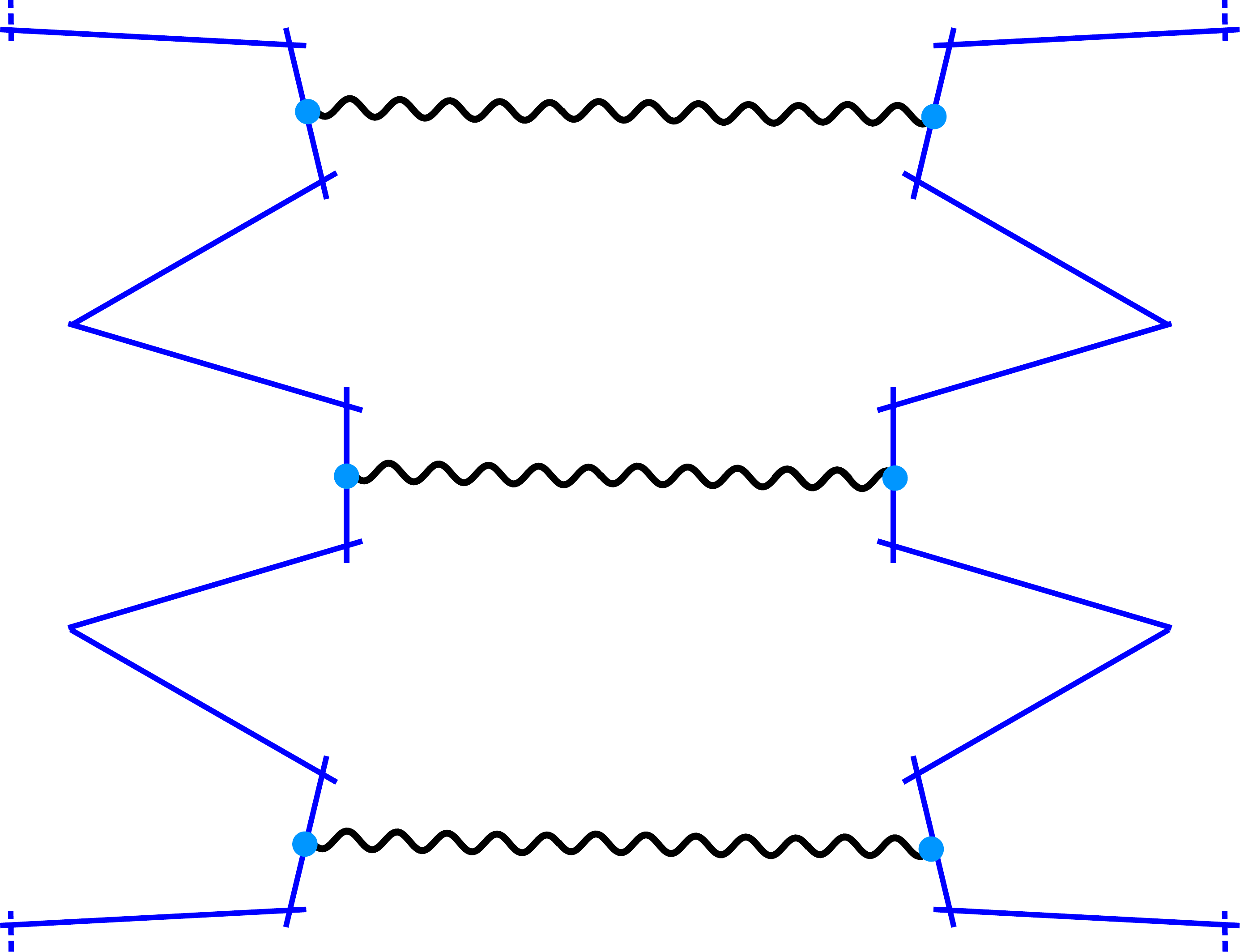}
  \caption{\it A diagram for the tree-level correlation functions.}
  \label{fig:TreeLevelCorrelation}
\end{figure}Tree-level correlators of operators of different length vanish trivially. Furthermore, by an elementary counting of Gra\ss{}mann numbers, one sees that at tree level only the minimal operator vertices contribute\footnote{These are given in \eqref{eq: vertex for minimal valency}.}.
For any two operators of the same length $L$, the tree-level correlation function can be calculated via the inverse soft limit. Using underlined symbols for the second operator, we find
\begin{equation}
 \langle \calO(x) \underline{\calO}(y)\rangle_{\text{tree}} = \lim_{\hexagon\rightarrow \xdot_x,\underline{\hexagon}\rightarrow \xdot_y}\PPP_{\mathcal{O}}\underline{\PPP}_{\underline{\mathcal{O}}}
 \sum_{\sigma}
 \prod_{i=1}^L\big[\calZ_i,\calZ_i',\star,\underline{\calZ}_{\sigma(i)},\underline{\calZ}_{\sigma(i)}'\big]
 {\big{|}}_{\theta,\underline{\theta}=0}\eqncom
\end{equation}
where the permutation $\sigma$ is cyclic in the planar limit. We immediately see that this is zero unless complementary fermionic derivatives act on all pairs $i$ and $\sigma(i)$. 
Restricting to the SO$(6)$ sector, the correlator $\vac{\mathcal{O}(x)\underline\calO(y)}_{\text{tree}}$ reduces to cyclic products over simple Wick contractions of the form (we suppress the color indices)
\begin{align}
\label{eq:treelevelpropagator}
\langle \phi_{ab}(x_i)\phi_{a'b'}(y_i)\rangle_{\text{tree}} &=\int\DD\la_i\DD\underline\la_i \vac{\frac{\partial^2 \AAA(\calZ_i)}{\partial\chi_i^a\partial \chi_i^b}\frac{\partial^2 \AAA(\underline\calZ_i)}{\partial \underline\chi_i^{a'}\partial \underline\chi_i^{b'}}}|_{\theta = \underline\theta =0}
\nonumber\\
&=\int\DD\la_i \int\DD\underline\la_i
\frac{\partial^2}{\partial\chi_{i}^a\partial \chi_{i}^b}\frac{\partial^2}{\partial\underline\chi_{i}^{a'}\partial \underline\chi_{i}^{b'}}\bar{\delta}^{2|4}(\calZ_i,\star,\underline\calZ_i)=\frac{1}{(2\pi)^2}\frac{2\epsilon_{aba'b'}}{|x_i-y_i|^2}\, \eqndot
\end{align} This result is up to a scaling factor of $2$ that can be absorbed into the definition of the fields, identical to the tree-level result that one obtains in space-time \eqref{eq:treelevelpropagatorposspace}.

\section{One-loop diagrams}
\label{seconeloop}
Recall that at one-loop order in space-time, computing the two-point correlation functions of the operators $\calO$ reduced to the calculation of the subcorrelator
\begin{equation}\label{subcor}
\langle\big( \phi_{ab}\phi_{cd}\big)^i_{\phantom{i}j}(x) \big(\phi_{a'b'}\phi_{c'd'}\big)^k_{\phantom{k}l}(y) \rangle_{\mathrm{one-loop}}\, ,
\end{equation} where $a,b,c,d,a',b',c',d'$ are SU$(4)$-flavor indices and $i,j,k,l$ are color indices. In twistor space, we can also suffice by computing a similar subcorrelator. However, we have to use our Wilson loop construction and forming operators for the scalar fields as defined in Chapter~\ref{chap:}. Only at the very end of our calculation we take the operator limit to shrink the Wilson loop to a point. The diagrams we consider are therefore two cogwheel Wilson loops connected to each other and to precisely one interaction vertex from the twistor action.
A priori, we could have an infinite number of different diagrams, as there are infinitely many vertices in the twistor action. 
However, all but a few vanish due to the integration over the Gra\ss{}mann variables at the interaction vertex.
The first non-vanishing diagram contains the four-point interaction vertex connected to minimal operator vertices, shown in  Figure~\ref{fig:oneloopCorrelation4}.
The second diagram consists of a three-point interaction vertex which is connected to a minimal operator vertex and a next-to-minimal operator vertex, which is exemplified on the right-hand side of  Figure~\ref{fig:oneloopCorrelation3}.
The third possibility consists of a two-point interaction vertex connected to two next-to-minimal operator vertices, an example of which is shown in on the left-hand side of  Figure~\ref{fig:oneloopCorrelation3}. The fourth and last possibility is a two-point vertex connected to a next-to-next-to-minimal operator vertex, depicted in the middle of  Figure~\ref{fig:oneloopCorrelation3}.
Note that the figures show the correlation functions before taking the operator limit. Therefore, the two- and three- vertices can be connected to several different edges of the Wilson loop, of which only one is shown in Figure~\ref{fig:oneloopCorrelation3}.
We will see that the dilatation operator is given only by the four-point vertex of  Figure~\ref{fig:oneloopCorrelation4}. We start therefore by computing this diagram and finally argue why all other diagrams, containing a two- or three- vertex do not contribute to the dilatation operator. 
\begin{figure}[htbp]
 \centering
  \includegraphics[height=3cm]{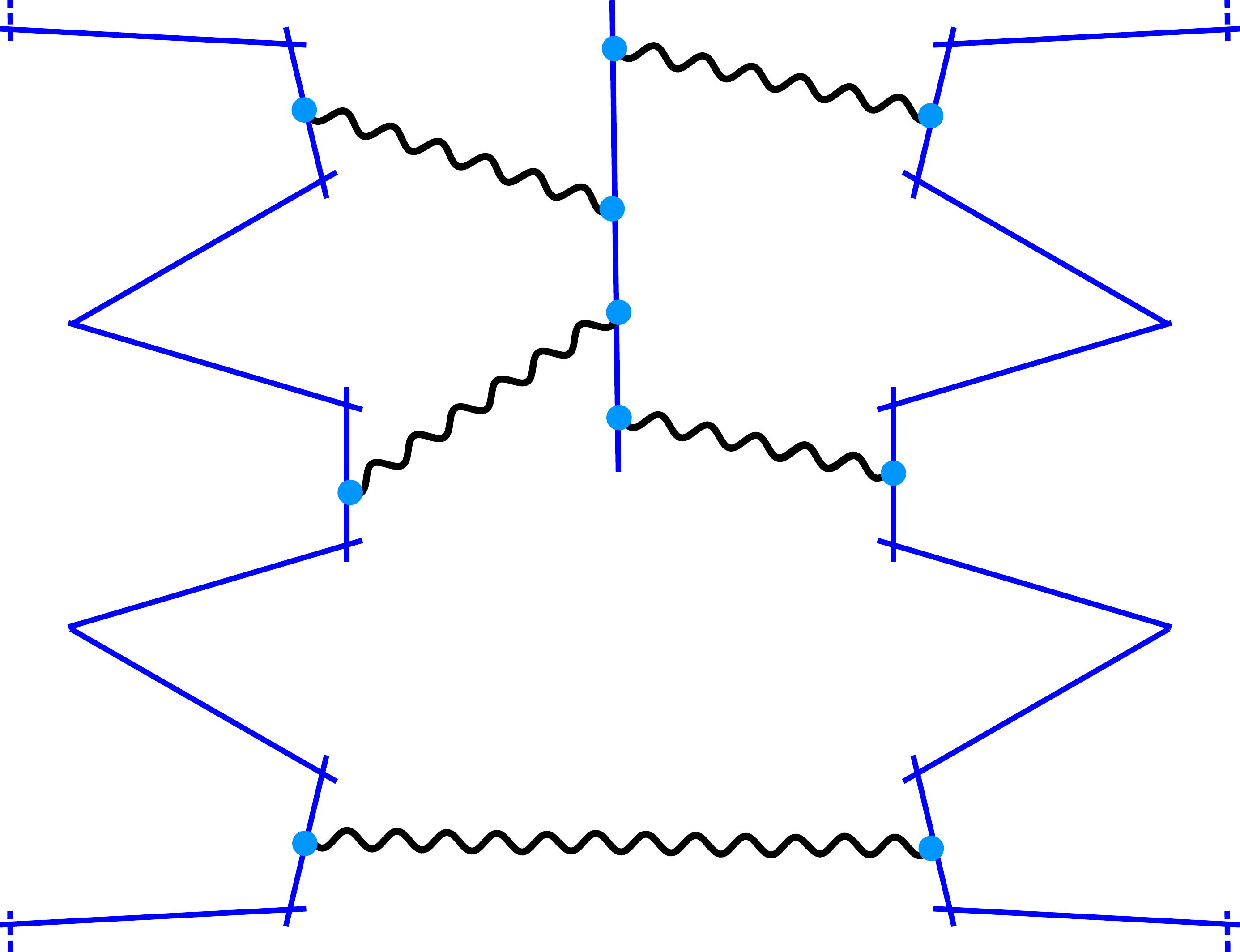}
  \caption{\it A diagram with a four-point vertex contributing to the one-loop two-point correlation function.}
  \label{fig:oneloopCorrelation4}
\end{figure}
\begin{figure}[htbp]
 \centering
  \includegraphics[height=2.4cm]{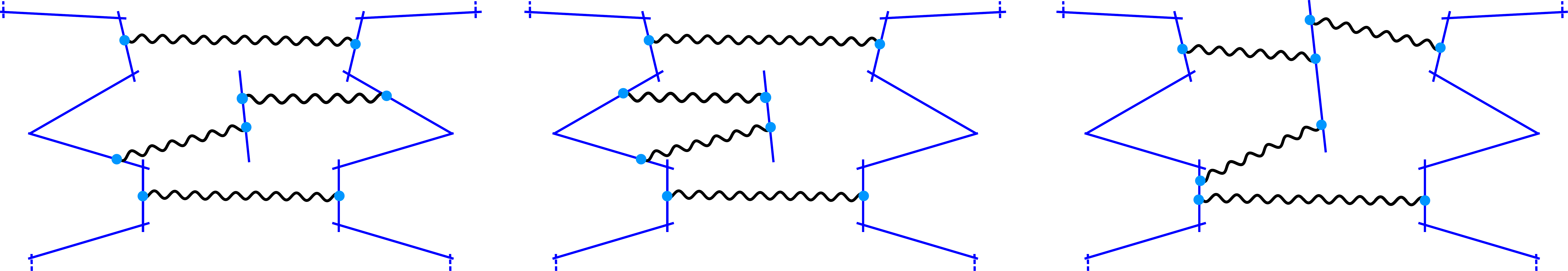}
  \caption{\it Two diagrams with a two-point vertex (left and middle) and one with a three-point vertex (right) contributing to the one-loop two-point correlation function.}
  \label{fig:oneloopCorrelation3}
\end{figure}
\paragraph{The four-vertex as the dilatation operator} In this paragraph, we compute the diagram shown in   Figure~\ref{fig:oneloopCorrelation4}, where four scalar fields are connected to a four-vertex. For each of the four scalar fields we use the definition of the vertex $\bf{W}_{\phi(x)}$ given by \eqref{eq:secondattempt}. A straightforward counting of Gra\ss mann variables shows that for each scalar only the first line in \eqref{eq:secondattempt}, corresponding to the minimal operator vertex, contributes. Thus we find
\begin{equation}
\label{eq:V4part1}
\begin{aligned}
V_4=&\int\frac{\dd^{4}z\dd^8\vartheta}{(2\pi)^4}\int \DD\la_1\DD\la_{2}\DD \underline{\la}_1\DD \underline{\la}_{2}\frac{\DD\rho_1\DD\rho_2\DD\rho_3\DD\rho_4}{\abra{\rho_1}{\rho_2}\abra{\rho_2}{\rho_3}\abra{\rho_3}{\rho_4}\abra{\rho_4}{\rho_1}}\\
&\frac{\partial^2}{\partial\chi_1^a\partial\chi_1^b}\bar{\delta}^{2|4}\big(\calZ_{x_i}(\la_1),\star,\calZ_{z}(\rho_2)\big)\frac{\partial^2}{\partial\chi_2^c\partial\chi_2^d}\bar{\delta}^{2|4}\big(\calZ_{x_{i+1}}(\la_2),\star,\calZ_{z}(\rho_1)\big) \notag\\
&\frac{\partial^2}{\partial\underline\chi_2^{c'}\partial\underline\chi_2^{d'}}\bar{\delta}^{2|4}\big(\calZ_{z}(\rho_3),\star,\calZ_{y_{i+1}}(\underline\la_2)\big)\frac{\partial^2}{\partial\underline\chi_1^{a'}\partial\underline\chi_1^{b'}}\bar{\delta}^{2|4}\big(\calZ_{z}(\rho_4),\star,\calZ_{y_i}(\underline\la_1)\big)|_{\theta=0}\eqndot
\end{aligned}
\end{equation}
In analogy to \eqref{SU4structure}, this correlation function can be written in terms of SU$(4)$ invariants, for which we only need to compute the coefficients,
\begin{equation}\label{su4structuretwistor}
(\textsf{A} \epsilon_{abcd}\epsilon_{a'b'c'd'}+\textsf{B}\epsilon_{aba'b'}\epsilon_{cdc'd'}+\textsf{C}\epsilon_{abc'd'}\epsilon_{a'b'cd})\times \int \dd^4z \frac{1}{|x_i-z|^2|x_{i+1}-z|^2|y_i-z|^2|y_{i+1}-z|^2}
\end{equation} for some coefficients $\textsf{A}$, $\textsf{B}$ and $\textsf{C}$ that are to be determined\footnote{Note that the space-time integral in \eqref{su4structuretwistor} is not UV-divergent before taking the operator limit. Only when $x_i\rightarrow x_{i+1}$, $y_i\rightarrow y_{i+1}$, the correct UV-divergent integral is recovered.}. These can be determined by making specific choices for the flavor indices, e.g.\ setting $(abcd)=(1234)$, $(a'b'c'd')=(1342)$ we pick up only $\textsf{A}$, $(abcd)=(1213)$,  $(a'b'c'd')=(3442)$ gives $\textsf{B}$ and finally $(abcd)=(1213)$ and $(a'b'c'd')=(4234)$ yields $\textsf{C}$. We integrate over $\dd^8\vartheta$ using Nair's lemma \cite{Nair:1988bq}, which results in a fraction of angular brackets containing the $\rho$ variables. The integrals and the integration over $\dd^8\vartheta$ gives for
\begin{equation}
\textsf{B}=\frac{\abra{1}{2}\abra{2}{3}\abra{3}{4}\abra{1}{4}}{\abra{1}{2}\abra{2}{3}\abra{3}{4}\abra{4}{1}}=-1\eqncom
\end{equation}
where $\abra{i}{j}\equiv \abra{\rho_i}{\rho_j}$.
Similarly, for $\textsf{C}$ we find
\begin{equation}
\textsf{C}=\frac{\abra{1}{2}\abra{2}{4}\abra{4}{3}\abra{1}{3}}{\abra{1}{2}\abra{2}{3}\abra{3}{4}\abra{4}{1}}=\frac{-\abra{2}{4}\abra{1}{3}}{\abra{2}{3}\abra{4}{1}}\eqndot
\end{equation}
Finally for $\textsf{A}$ we obtain
\begin{equation}\label{singualrA}
\textsf{A} = \frac{\abra{1}{3}\abra{3}{2}\abra{2}{4}\abra{1}{4}}{\abra{1}{2}\abra{2}{3}\abra{3}{4}\abra{4}{1}}=\frac{\abra{1}{3}\abra{2}{4}}{\abra{1}{2}\abra{3}{4}}\eqndot
\end{equation}
Subsequently, we perform the fiber integrals over the variables $\la_i$ and $\underline\la_i$ and $\rho_i$, using the delta functions which completely localize the integrals as
\begin{align}
\label{eq:localizationspinorfourvertex}
&\la_{1\alpha}=\rho_{2\alpha}=\frac{i(x_i-z)_{\alpha\dot\alpha}\zeta^{\dot\alpha}}{(x_i-z)^2}\,,&
&\la_{2\alpha}=\rho_{1\alpha}=\frac{i(x_{i+1}-z)_{\alpha\dot\alpha}\zeta^{\dot\alpha}}{(x_{i+1}-z)^2}\,,&\nonumber\\
&\underline{\la}_{1\alpha}=\rho_{4\alpha}=\frac{i(y_i-z)_{\alpha\dot\alpha}\zeta^{\dot\alpha}}{(y_i-z)^2}\,,& &\underline{\la}_{2\alpha}=\rho_{3\alpha}=\frac{i(y_{i+1}-z)_{\alpha\dot\alpha}\zeta^{\dot\alpha}}{(y_{i+1}-z)^2}\,\eqndot&
\end{align}
In the operator limit where $x_i\to x_{i+1}$ and $y_i\to y_{i+1}$ it follows straightforwardly that $\textsf{C}=1$. However, in this limit we encounter an unexpected spurious $1/0$-like divergence for $\textsf{A}$. Spurious singularities are a common feature, or rather bug, of axial gauges and deserve some more careful treatment. Here, we observe that in the operator limit, $\rho_1$ and $\rho_2$ (resp. $\rho_3$ and $\rho_4$) are both evaluated at the point $\tfrac{i(x-z)_{\alpha\dot\alpha}\zeta^{\dot\alpha}}{(x-z)^2}$ (resp. $\tfrac{i(z-y)_{\alpha\dot\alpha}\zeta^{\dot\alpha}}{(z-y)^2}$), and thus the evaluation is symmetric in the two variables. More precisely, in the operator limit the four variables become pairwise indistinguishable. Therefore, if the evaluation is well defined we should be allowed to first symmetrize the function \eqref{singualrA} with respect to $\rho_1$ and $\rho_2$ first and then evaluate it at $\tfrac{i(x-z)_{\alpha\dot\alpha}\zeta^{\dot\alpha}}{(x-z)^2}$. This procedure yields
\begin{equation}
\textsf{A}=\frac{1}{2} \left(\frac{\abra{1}{3}\abra{2}{4}}{\abra{1}{2}\abra{3}{4}}+ \frac{\abra{2}{3}\abra{1}{4}}{\abra{2}{1}\abra{3}{4}}\right)=\frac{1}{2}\frac{\abra{1}{2}\abra{3}{4}}{\abra{1}{2}\abra{3}{4}}=\frac{1}{2}\eqncom
\end{equation}
where we used the Schouten identity in the second equation.
In the next sections we will show that the other diagrams, which contain a two-vertex or a three-vertex do not contribute. Therefore, we state here the result that 
one finds the one-loop dilatation operator in the SO$(6)$ sector by only computing one correlation function, containing the four-vertex
\begin{align}\label{loop-integral}
&\langle\big( \phi_{ab}\phi_{cd}\big)^i_{\phantom{i}j}(x) \big(\phi_{a'b'}\phi_{c'd'}\big)^k_{\phantom{k}l}(y) \rangle_{1-\mathrm{loop}}=\\
&=\frac{16 \delta^{i}_{l}\delta^{k}_{j}g^2_{YM}N^2}{4 (2\pi)^8}   \int\frac{\dd^{4}z}{|x-z|^4|z-y|^4} \left(\frac{1}{2} \epsilon_{abcd}\epsilon_{a'b'c'd'}-\epsilon_{aba'b'}\epsilon_{c'd'cd}+\epsilon_{abc'd'}\epsilon_{a'b'cd}\right).\nonumber
\end{align}
Note the factor of $16$ in the numerator coming from the four factors of $2$ in the propagator \eqref{eq:treelevelpropagator}. In extracting the anomalous dimensions, two factors of $2/(2\pi)^2$ vanish due to the normalization of the tree-level propagator \eqref{eq:treelevelpropagator}.
Furthermore, we obtain an extra factor of $2\pi^2$ due to the regularization of the UV-divergent integral. Finally, we then extract from \eqref{loop-integral} the one-loop dilatation operator in the SO$(6)$ sector
\beq
\Gamma=\frac{g_{YM}^2N}{8\pi^2}\sum_{\ell=1}^L\left(\text{Id}-P_{\ell,\ell+1}+\frac{1}{2}K_{\ell,\ell+1}\right)\eqncom
\eeq
where $\text{Id}$, $P$ and $K$ are the Identity, Permutation and Trace operator that were defined above \eqref{dilintermsoftr}. In the remainder of this chapter we show that all contributions of the other diagrams vanish.
\paragraph{The two-point vertex connected to both sides of the diagram}
We compute diagrams of the type depicted on the left-hand side of  Figure~\ref{fig:oneloopCorrelation3}, which is reproduced in  Figure~\ref{fig:appCpar1} including labels. 
The two-point vertex is attached to two gluons $g^+$ emitted from the positions $\tilde{\la}_1$ and $\tilde{\underline{\la}}_1$. We have to sum over all the planar ways of attaching the vertex, so that the attachment line $\tilde{x}$ has to be summed over the possible lines $x_i$, $x_{i}''$, $x_{i+1}'$ and $x_{i+1}$ and similarly for $\tilde{y}$. Here, for the $i$-th cog, the left edge is denoted by $x_i'$ the middle (and operator bearing) edge by $x_i$ and the right one by $x_i''$ as in    Figure~\ref{fig:CogwheelZoom1}. Accordingly, the spinors $\tilde{\lambda}_0$, $\tilde{\lambda}_2$, $\tilde{\underline{\lambda}}_0$ and $\tilde{\underline{\lambda}}_2$ take the appropriate values of the vertices of the cogwheel Wilson loops, see  Figure~\ref{fig:CogwheelZoom1}. 
\begin{figure}[tbp]
 \centering
  \includegraphics[height=3.5cm]{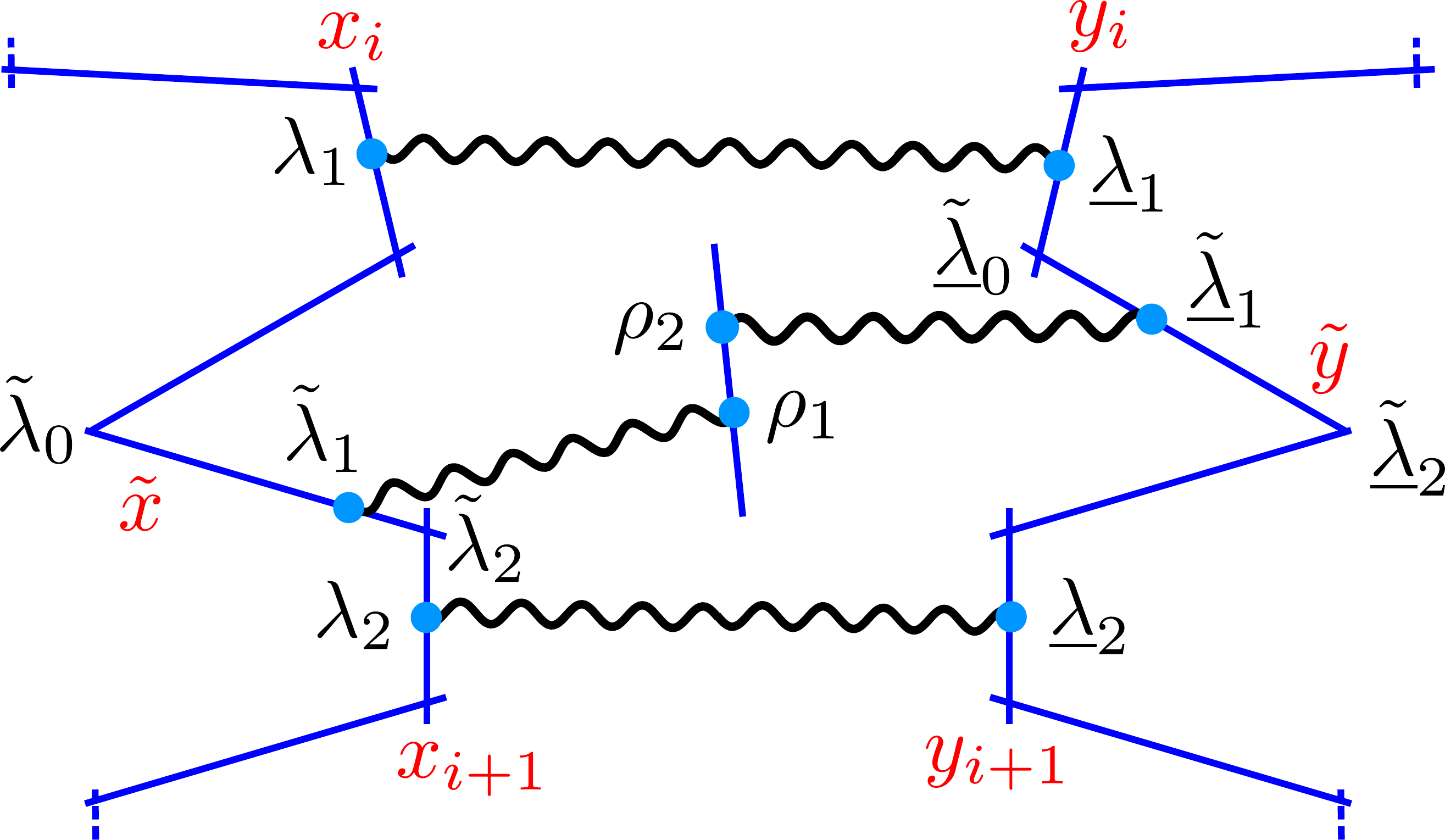}
  \caption{\it A one-loop correlation function diagram with the two-point vertex attached to both operators.}
  \label{fig:appCpar1}
\end{figure}
The result of the particular diagram shown in  Figure~\ref{fig:appCpar1} is 
\begin{equation}
\label{eq:V2part1}
\begin{aligned}
V_2=&\int\frac{\dd^{4}z\dd^8\vartheta}{(2\pi)^4}\int \DD\la_1\DD\la_{2}\DD \underline{\la}_1\DD \underline{\la}_{2} \DD\tilde{\la}_1\DD\tilde{\underline{\la}}_1\frac{\DD\rho_1\DD\rho_2}{\abra{\rho_1}{\rho_2}\abra{\rho_2}{\rho_1}}\\
&\times \bar{\delta}^{2}\big(\calZ_{x_i}(\la_1),\star,\calZ_{y_i}(\underline{\la}_1)\big)\bar{\delta}^{2}\big(\calZ_{x_{i+1}}(\la_2),\star,\calZ_{y_{i+1}}(\underline{\la}_2)\big) F(\la_1,\la_2,\underline{\la}_1,\underline{\la}_2)
\\
&\times 
\frac{\abra{\tilde{\la}_0}{\tilde{\la}_2}}{\abra{\tilde{\la}_0}{\tilde{\la}_1}\abra{\tilde{\la}_1}{\tilde{\la}_2}}
\frac{\abra{\tilde{\underline{\la}}_0}{\tilde{\underline{\la}}_2}}{\abra{\tilde{\underline{\la}}_0}{\tilde{\underline{\la}}_1}\abra{\tilde{\underline{\la}}_1}{\tilde{\underline{\la}}_2}}
\bar{\delta}^{2|4}\big(\calZ_{\tilde{x}}(\tilde{\la}_1),\star,\calZ_z(\rho_1)\big)\bar{\delta}^{2|4}\big(\calZ_{\tilde{y}}(\tilde{\underline{\la}}_1),\star,\calZ_z(\rho_2)\big)\,,
\end{aligned}
\end{equation}
where $F$ is a function of the appropriate homogeneity that takes into account which operators are inserted at the operator-bearing edges; its precise form does not matter in the present discussion.
The fraction of angular brackets comes from the expansion of the two frames $U$.
We recall that according to our prescription we are to take the integral over $z$ in the interaction vertex last. The fermionic integration over $\vartheta$ can however be taken immediately, which leads to an additional factor of $\abra{\rho_1}{\rho_2}^4$ and removes the fermionic pieces from the $\bar{\delta}^{2|4}$ functions. We can replace 
\begin{equation}
\bar{\delta}^{2}(Z_1,Z_2,Z_3)=\int \frac{\dd s\dd t}{s t}\bar{\delta}^4(Z_1+sZ_2+tZ_3)\,,
\end{equation}
where one has to be careful that the above is homogeneous of degree 0 in $Z_2$ and $Z_3$ but of degree $-4$ in $Z_1$.
The appropriate form of the delta function has to be taken depending on the homogeneity of the integrand in each of the $Z_i$ so that the integral is homogeneous of degree 0. Thus, we obtain
\begin{equation}
\begin{aligned}
\label{eq:V2part2}
V_2\propto \frac{F(\la_1,\la_2,\underline{\la}_1,\underline{\la}_2) \abra{\tilde{\la}_0}{\tilde{\la}_2}\abra{\tilde{\underline{\la}}_0}{\tilde{\underline{\la}}_2}}{|x_i-y_i|^2|x_{i+1}-y_{i+1}|^2}
\int  \frac{\dd^4 z}{|\tilde{x}-z|^2|\tilde{y}-z|^2}\frac{\abra{\rho_1}{\rho_2}^2}{\abra{\tilde{\la}_0}{\tilde{\la}_1}\abra{\tilde{\la}_1}{\tilde{\la}_2}\abra{\tilde{\underline{\la}}_0}{\tilde{\underline{\la}}_1}\abra{\tilde{\underline{\la}}_1}{\tilde{\underline{\la}}_2}}\,,
\end{aligned}
\end{equation}
where the spinors that we integrate over in \eqref{eq:V2part1} have been \emph{fixed} by the delta function to the values
\begin{align}
\label{eq:localizationspinorpart1}
&\la_{1\alpha}=-\underline{\la}_{1\alpha}=\frac{i(x_i-y_i)_{\alpha\dot\alpha}\zeta^{\dot\alpha}}{(x_i-y_i)^2}\,,&
&\la_{2\alpha}=-\underline{\la}_{2\alpha}=\frac{i(x_{i+1}-y_{i+1})_{\alpha\dot\alpha}\zeta^{\dot\alpha}}{(x_{i+1}-y_{i+1})^2}\,,&\nonumber\\
&\tilde{\la}_{1\alpha}=-\rho_{1\alpha}=\frac{i(\tilde{x}-z)_{\alpha\dot\alpha}\zeta^{\dot\alpha}}{(\tilde{x}-z)^2}\,,& &\tilde{\underline{\la}}_{1\alpha}=-\rho_{2\alpha}=\frac{i(\tilde{y}-z)_{\alpha\dot\alpha}\zeta^{\dot\alpha}}{(\tilde{y}-z)^2}\,,&
\end{align}
and the factors of $|\tilde{x}-z|^2|\tilde{y}-z|^2$ and $|x_i-y_i|^2|x_{i+1}-y_{i+1}|^2$ in the denominator arose due to the Jacobians of the integration. Here, as elsewhere,  $\zeta^{\dot\alpha}$ is the lower component of the (bosonic) reference twistor $Z_\star$. 
We now have to take the sum over the insertions points and the operator limit, \emph{before} we perform the integration over $z$. The contributions from the other insertion points are similar and the sum can be done using the Schouten identity in the same way as described in figure 8 of \cite{Koster:2016loo}. It effectively leads to replacing in \eqref{eq:V2part2} $\tilde{\la}_0$ by $\la_1$, $\tilde{\la}_2$ by $\la_2$, $\tilde{\underline{\la}}_0$ by $\underline{\la}_1$ and $\tilde{\underline{\la}}_2$ by $\underline{\la}_2$. Finally, in the operator limit we have $x_i,x_{i+1}\rightarrow x$ and $y_j,y_{j+1}\rightarrow  y$, so that the prefactor $\abra{\la_1}{\la_2}\abra{\underline{\la}_1}{\underline{\la}_2}$ in front of the integral vanishes and hence the contribution is zero. 
\paragraph{The two-point vertex connected to one side of the diagram}
For the other diagram containing a two-point vertex, shown in the middle of  Figure~\ref{fig:oneloopCorrelation3} and in more detail in  Figure~\ref{fig:2vertex1LoopDiagram}, we have to connect the two propagators coming from the vertex to any combination of the edges between $\la_1$ and $\la_2$. 
Below, we compute just the combination shown in  Figure~\ref{fig:2vertex1LoopDiagram}.
We find that, 
after the $\vartheta$ integration and the integration over $\lambda_1$, $\lambda_2$, $\underline{\la}_1$  and $\underline{\la}_2$ which localizes these spinors in a similar way as \eqref{eq:localizationspinorpart1},
 it contributes
\begin{align}
\label{twovertex}
V_2'&\propto\frac{F(\la_1,\la_2,\underline{\la}_1,\underline{\la}_2)}{|x_i-y_i|^2|x_{i+1}-y_{i+1}|^2}\int \frac{\dd^4z}{(2\pi)^4}\DD\rho_1\DD\rho_2\DD\tilde\la_1\DD\tilde\la_1'\Bigg[ \frac{\abra{\rho_1}{\rho_2}^4}{\abra{\rho_1}{\rho_2}^2} 
\frac{\abra{\tilde\la_0}{\tilde\la_{2}}}{\abra{\tilde\la_{0}}{\tilde\la_1}\abra{\tilde\la_{1}}{\tilde\la_2}}
\frac{\abra{\tilde\la_0'}{\tilde\la_2'}}{\abra{\tilde\la_0'}{\tilde\la_1'}\abra{\tilde\la_1'}{\tilde\la_2'}} \nonumber\\
&\phaneq\times \bar{\delta}^{2}\big(Z_{\tilde{x}}(\tilde{\la}_1),\star,Z_z(\rho_1)\big)\bar{\delta}^{2}\big(Z_{\tilde{x}'}(\tilde{\la}_1'),\star,Z_z(\rho_2)\big)\Bigg]\,.
\end{align} 
\begin{figure}[tbp]
 \centering
 \includegraphics[height=4cm]{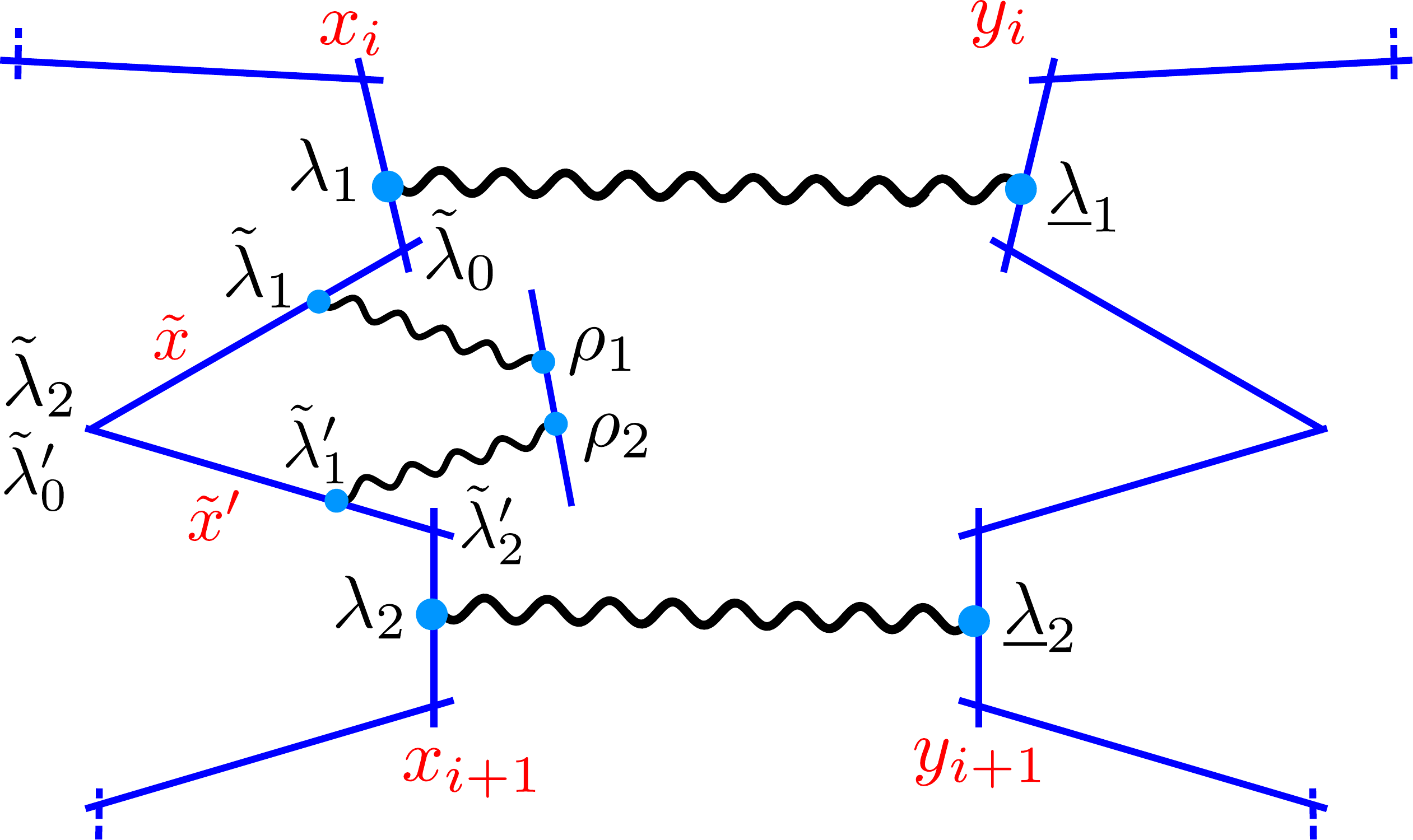}
  \caption{\it A one-loop diagram with the two-point vertex attached to one of the two cogwheel Wilson loops.
  }
  \label{fig:2vertex1LoopDiagram}
\end{figure}The term $\abra{\rho_1}{\rho_2}^2$ in the denominator comes directly from the interaction vertex in the action, while the factor $\abra{\rho_1}{\rho_2}^4$ in the numerator is the result of the $\dd^8\vartheta$ integration as for the previous two-point vertex.
Doing the remaining integrations and summing over the possible ways to connect the diagram leads in the operator limit (implying $\tilde{x}, \tilde{x}'\rightarrow x$) to
\begin{equation}
\sum_{\text{connections}} V_2' \propto \frac{F(\la_1,\la_2,\underline{\la}_1,\underline{\la}_2)}{|x-y|^4}\int \frac{\dd^4z}{|x-z|^4} \frac{\abra{\la_1}{\la_{2}}\abra{\rho_1}{\rho_2}^{\cancel{2}}}{\abra{\la_{1}}{\tilde\la_1}\cancel{\abra{\tilde\la_{1}}{\tilde\la_1'}}\abra{\tilde\la_1'}{\tilde\la_2}}\eqncom
\end{equation}
where the cancellation happens because the spinors are \textit{localized}
\begin{equation}
\begin{aligned} 
&\la_{1\alpha}=\underline\la_{1\alpha}\propto(x-y)_{\alpha\dot\alpha}\zeta^{\dot\alpha}\,,&
&\la_{2\alpha}=\underline\la_{2\alpha}\propto(x-y)_{\alpha\dot\alpha}\zeta^{\dot\alpha}\,,&\\
&\tilde\la_{1\alpha}=- \rho_{1\alpha}\propto(x-z)_{\alpha\dot\alpha}\zeta^{\dot\alpha}\,,&
&\tilde\la_{1\alpha}'=-\rho_{2\alpha}\propto(x-z)_{\alpha\dot\alpha}\zeta^{\dot\alpha}\,,&
\end{aligned}
\end{equation}
by the delta functions of the propagators. Hence, we see that the contribution becomes zero due to the vanishing of the product of brackets $\abra{\la_1}{\la_2}\abra{\rho_1}{\rho_2}$ in the operator limit, which we take before performing the integration over $z$.
 
\paragraph{Diagrams with a three-point vertex} 
Finally, we consider the diagrams with a three-point interaction, where we take the operator vertex up to the second term in the expansion of the parallel propagator $U_{(x_i,\theta)}(\calZ_i,\calZ_{i+1})$, see  Figure~\ref{fig:1loopdiagram}. The particles emitted from the edge $x_i$ of the loop can be either a scalar $\phi$ and a positive-helicity gluon $g^+$ or two anti-fermions $\bar\psi$. 
Therefore, we cannot factor out the forming operator as in the cases with the two-point vertex. 
In the case of two anti-fermions $\bar\psi$, both are attached to the operating-bearing edge $x_i$ of the Wilson loop. However, when there is a scalar $\phi$ and a positive-helicity gluon $g^+$, the gluon can be attached to many different edges and we have to sum over all of them. Here, we will just treat the attachment shown in  Figure~\ref{fig:1loopdiagram} in detail and then argue that the other terms are similar and can be combined using the Schouten identity after the operator limit is taken. 
We obtain for the vertex on the cog $x_i$ (from \eqref{eq:secondattempt})
\begin{figure}[tbp]
 \centering
 \includegraphics[height=4cm]{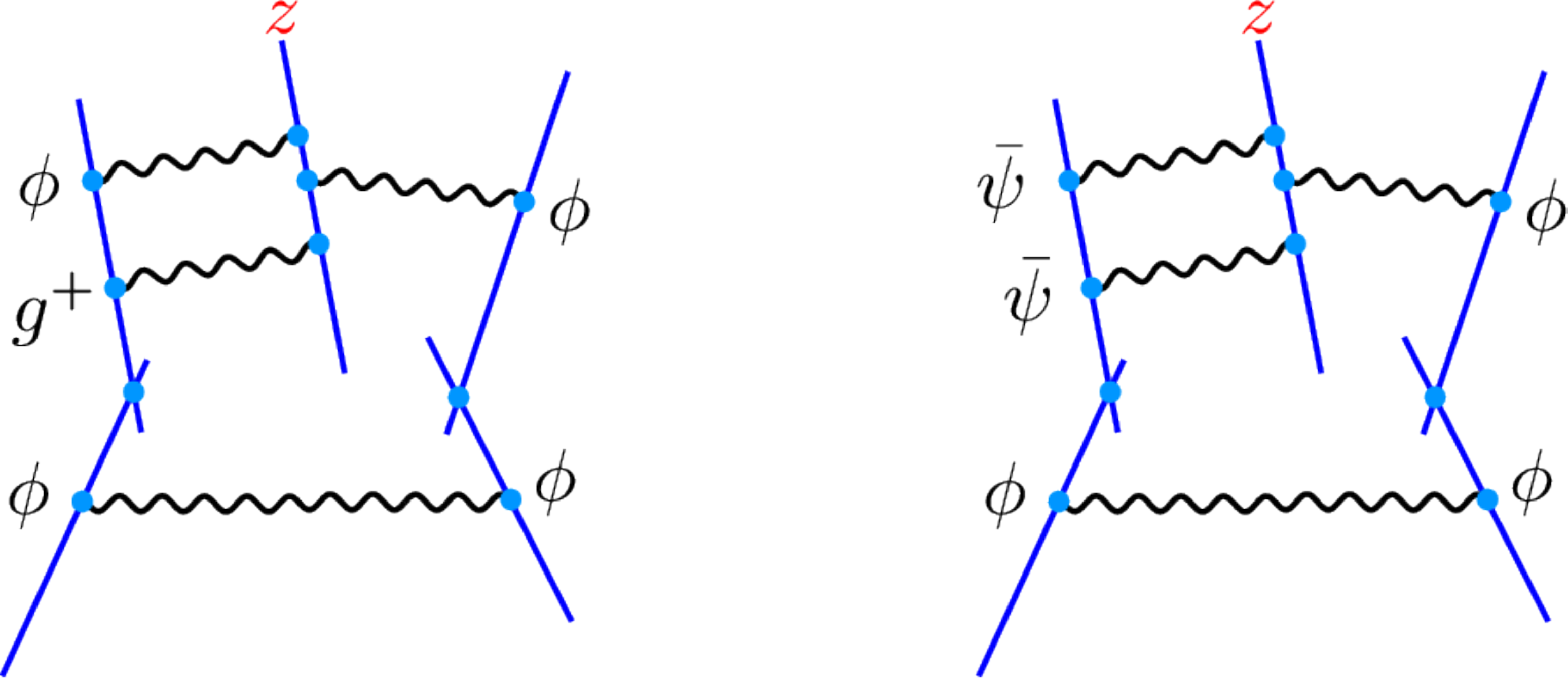}
  \caption{\it Three-point vertex contributions to the one-loop two-point correlation function.
  }
  \label{fig:1loopdiagram}
\end{figure}
\begin{equation}
\label{scalar field vertex}
\begin{aligned}
\textbf{W}_{\phi}&=\int \DD{\la}_1\DD{\la}_1'
\Bigg[ \frac{\partial^2\AAA(\calZ_{(x,\theta)}({\la}_1))}{\partial{\chi}_1^a\partial{\chi}_1^b} \frac{\abra{{\la}_1}{ \la_{i}'}}{\abra{{\la}_1}{{\la}_1'}\abra{{\la}_1'}{\la_{i}'}} \AAA(\calZ_{(x,\theta)}({\la}_1'))\notag\\
&\qquad\qquad\qquad \qquad+\AAA(\calZ_{(x,\theta)}({\la}_1))\frac{\abra{{\la}_1}{ \la_{i}'}}{\abra{{\la}_1}{{\la}_1'}\abra{{\la}_1'}{\la_{i}'}} \frac{\partial^2\AAA(\calZ_{(x,\theta)}({\la}_1'))}{\partial{\chi'}_1^a\partial{\chi'}_1^b}
\\&\qquad\qquad\qquad \qquad + \frac{\partial\AAA(\calZ_{(x,\theta)}({\la}_1))}{\partial{\chi}_1^a}\frac{1}{\abra{{\la}_1}{\la_1'}}\frac{\partial\AAA(\calZ_{(x,\theta)}({\la}_1'))}{\partial{\chi}_1'^b}\notag\\
&\qquad\qquad\qquad \qquad - \frac{\partial\AAA(\calZ_{(x,\theta)}({\la}_1))}{\partial{\chi}_1^b}\frac{1}{\abra{{\la}_1}{\la_1'}}\frac{\partial\AAA(\calZ_{(x,\theta)}({\la}_1'))}{\partial{\chi}_1'^a}  \left.\Bigg]\right\vert_{\theta=0}\,\eqndot
\end{aligned}
\end{equation}
The first term corresponds to a scalar $\phi$ and a positive-helicity gluon $g^+$ being emitted from $x_i$, in the second term this is reversed, the third term to two anti-fermions $\bar\psi$ being emitted and the fourth one the two two anti-fermions are reversed. The first term equals the second and the third term equals the last, so we just show that the first term cancels the third.
On the opposite side of the correlation function, we have the operator vertex
\begin{equation}
\textbf{W}_{\underline{\phi}}=\left.\int \DD\underline\la_1\frac{\partial^2\AAA(\calZ_{(y,\underline\theta)}(\underline\la_1))}{\partial\underline\chi_1^a\partial\underline\chi_1^b}\right\vert_{\underline\theta=0}\eqndot
\end{equation}
The three $\AAA$ fields, two at line $x$ and one at line $y$, are connected to the three-point vertex
\begin{equation}
\textbf{V}_3=\int \frac{\dd^{4}z\dd^8\vartheta}{(2\pi)^4} \int \frac{\DD\rho_1\DD\rho_2\DD\rho_3}{\abra{\rho_1}{\rho_2}\abra{\rho_2}{\rho_3}\abra{\rho_3}{\rho_1}} \Tr \left(\AAA(\calZ_{(z,\vartheta)}(\rho_1))\AAA(\calZ_{(z,\vartheta)}(\rho_2))\AAA(\calZ_{(z,\vartheta)}(\rho_3))\right) \eqndot
\end{equation}
The expression for the diagram corresponding to the first term in \eqref{scalar field vertex} therefore yields
\begin{align}
&\frac{F(\la_2,\underline\la_2)}{|x_{i+1}-y_{i+1}|}\int \DD{\la}_1\DD{\la}_1' \frac{\dd^{4}z\dd^8\vartheta}{(2\pi)^4}  \frac{\DD\rho_1\DD\rho_2\DD\rho_3}{\abra{\rho_1}{\rho_2}\abra{\rho_2}{\rho_3}\abra{\rho_3}{\rho_1}} 
 \frac{\partial^2}{\partial{\chi}_1^a\partial{\chi}_1^b}\bar\delta^{2|4}(\calZ_{(x_i,\theta)}({\la}_1),\star, Z_{(z,\vartheta)}(\rho_1)) \notag\\
 &\frac{\abra{{\la}_1}{ \la_{i}'}}{\abra{{\la}_1}{{\la}_1'}\abra{{\la}_1'}{\la_{i}'}} 
 \bar\delta^{2|4}(\calZ_{(x_i,\theta)}({\la}_1'),\star, \calZ_{(z,\vartheta)}(\rho_3))\frac{\partial^2}{\partial\underline\chi_1^a\partial\underline\chi_1^b}
  \bar\delta^{2|4}(\calZ_{(z,\vartheta)}({\rho}_2),\star,\calZ_{(y_i,\underline\theta)}(\underline\la_1))\eqncom
\end{align}
where we have localized two spinors as $\la_{2\alpha}\propto \underline{\la}_{2\alpha}\propto (x_{i+1}-y_{i+1})_{\alpha\dot{\alpha}}\zeta^{\dot{\alpha}}$ and where $F(\la_2,\underline\la_2)$ is a function that takes into account what operator are inserted at $\la_2$ and $\underline\la_2$.
Integrating over $\dd^8\vartheta$ gives a factor of $\abra{\rho_3}{\rho_1}^2\abra{\rho_3}{\rho_2}^2$. Evaluating the remaining integrations and summing over the possible ways to connect the diagram leads in the operator limit to
\begin{equation}
\label{eq:3vertexcomputationterm1}
\begin{aligned}
\frac{F(\la_2,\underline\la_2)}{|x-y|^2}\int \frac{\dd^{4}z}{(2\pi)^4} \frac{1}{|x-z|^4|y-z|^2} \Bigg[
 \underbrace{\frac{\abra{\rho_2}{ \la_{i}'}}{\abra{\rho_2}{\rho_1}\abra{\rho_1}{\la_{i}'}} \frac{\abra{\rho_2}{\rho_1}^2\abra{\rho_3}{\rho_1}^2}{\abra{\rho_1}{\rho_2}\abra{\rho_2}{\rho_3}\abra{\rho_3}{\rho_1}}}_{= \frac{\abra{\rho_2}{ \la_{i}'}\abra{\rho_1}{\rho_3}}{\abra{\rho_1}{\la_{i}'}\abra{\rho_2}{\rho_3}}\stackrel{\hexagon\rightarrow \xdot}=1}\Bigg]\eqncom
\end{aligned}
\end{equation}
due to the fact that the spinorial integration leads to
\begin{equation}\label{localizedspinors} \la_1=\rho_2,\quad \la_1'=\rho_1,\quad \underline\la_1=\rho_3\eqndot
\end{equation} 
Similarly, the third term in \eqref{scalar field vertex} yields
\begin{align}
&F(\la_2,\underline\la_2)\int \DD{\la}_1\DD{\la}_1' \frac{\dd^{4}z\dd^8\vartheta}{(2\pi)^4}  \frac{\DD\rho_1\DD\rho_2\DD\rho_3}{\abra{\rho_1}{\rho_2}\abra{\rho_2}{\rho_3}\abra{\rho_3}{\rho_1}} 
 \frac{\partial}{\partial{\chi}_1^a}\bar\delta^{2|4}(\calZ_{(x_i,\theta)}({\la}_1),\star, Z_{(z,\vartheta)}(\rho_1)) \notag\\
 &\frac{1}{\abra{{\la}_1}{{\la}_1'}} 
 \frac{\partial}{\partial{\chi}_1'^b}\ \bar\delta^{2|4}(\calZ_{(x_i,\theta)}({\la}_1'),\star, \calZ_{(z,\vartheta)}(\rho_3))\frac{\partial^2}{\partial\underline\chi_1^a\partial\underline\chi_1^b}
  \bar\delta^{2|4}(\calZ_{(z,\vartheta)}({\rho}_2),\star,\calZ_{(y_i,\underline\theta)}(\underline\la_1))\eqndot
\end{align}   
Integrating over $\dd^8\vartheta$ yields a factor of $-\abra{\rho_3}{\rho_2}\abra{\rho_3}{\rho_1}\abra{\rho_2}{\rho_1}^2$ so that in the operator limit we find
\begin{equation}
\label{eq:3vertexcomputation}
\begin{aligned}
&\frac{F(\la_2,\underline\la_2)}{|x-y|^2}\int \frac{\dd^{4}z}{(2\pi)^4} \frac{1}{|x-z|^4|y-z|^2} \Bigg[
- \underbrace{\frac{1}{\abra{\rho_2}{\rho_1}}\frac{\abra{\rho_3}{\rho_2}\abra{\rho_3}{\rho_1}\abra{\rho_1}{\rho_2}^2}{\abra{\rho_1}{\rho_2}\abra{\rho_2}{\rho_3}\abra{\rho_3}{\rho_1}}}_{=1}\Bigg]\eqncom
\end{aligned}
\end{equation}
where also here the spinors are localized according to \eqref{localizedspinors}. We observe that the contributions \eqref{eq:3vertexcomputationterm1} and \eqref{eq:3vertexcomputation} precisely cancel. Hence, the three-point vertex does not contribute to the one-loop UV divergent part of the correlation function. This completes our treatment of the one-loop dilatation operator in the SO$(6)$ sector from twistor space.
\chapter{Conclusions and Outlook}
Over the past decades, there has been enormous interest in computing observables in $\calN=4$ SYM. Two types of quantities in particular have been studied very extensively and with tremendous success. These are the on-shell scattering amplitudes and the off-shell correlation functions of local composite operators. Two independent directions of research have been largely responsible for this: the development of various on-shell techniques for scattering amplitudes and the study of integrability for correlation functions. Both fields of research exploit the many symmetries of $\calN=4$ SYM, which are often obscured in conventional Feynman diagram computations. Particularly in the twistor formulation of the theory, which can be identified with the on-shell method CSW, it is straightforward to compute amplitudes, at least at tree-level and at lowest MHV degree. In this thesis we have investigated whether twistor methods provide a similar improvement over space-time Feynman diagram calculations for certain (partially) off-shell quantities, namely form factors and correlation functions. A form factor is a hybrid of a scattering amplitude and a correlation function. This, together with the fact that it is very similar  in structure to an amplitude, makes it ideal for testing whether on-shell methods can be applied to off-shell quantities as well. Indeed, several on-shell methods have already successfully been used to calculate form factors. However, a twistor description of form factors was up to this point still missing. In this thesis we have made large steps towards filling this void. Furthermore, we used the twistor formalism to compute correlation functions, which allowed us to rederive the one-loop dilatation operator in the scalar sector of the theory.\newline\\
First, we found expressions for all gauge-invariant local composite operators and their operator vertices in twistor variables. These operator vertices were obtained by acting with derivative operators, called forming operators, on light-like Wilson loops and subsequently shrinking the Wilson loop to a line in twistor space. This limit was called the operator limit. In the current work we used a cogwheel-shaped Wilson loop that can be used for any operator, but this choice is not unique and may in fact not be the most efficient one. It would be interesting to see if there exists a simpler Wilson loop that can generate all operator vertices. The operator vertices played the same role in the computation of form factors as interaction vertices did for amplitudes. Namely, inserting on-shell external states directly into the interaction vertices yielded all MHV amplitudes straightforwardly. Likewise, inserting these states into the operator vertices gave all MHV form factors immediately. This expression for all tree-level MHV form factors had previously been unknown and is one of the main results of this thesis. It shows a major advantage of twistor methods over ordinary space-time or momentum space techniques for form factors at MHV level. \newline\\
Although the method that is presented in this work proved much more efficient than conventional techniques for computing MHV form factors, at N$^k$MHV level several subtleties occurred that made it less effective. We presented a general construction of all N$^k$MHV form factors in twistor space in Chapter~\ref{NMHV}. This derivation used an inverse soft limit for form factors. Thus, in \textit{position twistor space}, form factors of general local composite operators can be calculated using the twistor action in complete analogy to the amplitude case of \cite{Adamo:2011cb}. In that paper, some problems concerning the boundary-boundary case were left open for the expression of the amplitude beyond NMHV level in position twistor space. The same issues arise for form factors at next-to-next-to-MHV level and higher. Solving these problems both for form factors and for amplitudes is left for future work. At NMHV level we found a general expression for all form factors in twistor space before acting with the forming operator and taking the operator limit. This expression was however still in position twistor space. To make contact with the results in the literature we made steps in translating it to momentum space in Chapter~\ref{momentumspace}. To this end, we first Fourier transformed the general NMHV amplitude to momentum twistor space, proving an observation made in \cite{Adamo:2011cb} and thus filling a gap in the literature. Fourier transforming is easier for the NMHV amplitude than for the NMHV form factor as the latter contains derivatives in the forming operator that in general do not commute with the Fourier transform. However, for operators without $\dot\alpha$ indices, the forming operator does not contain any space-time derivatives and can be interchanged with the Fourier integral. For these operators we successfully transformed the NMHV form factor to momentum twistor space. Our final result still had explicit dependence on the reference twistor and an auxiliary spinor that was introduced in the operator limit. Clearly, the physical answer must be independent of these two quantities. For a form factor with $L+1$ external fields, where $L$ is the length of the operator, the dependence on the auxiliary spinor trivially drops out. For higher number of external particles this remains to be shown, which is left for future work. Furthermore, it would be desirable to factor the MHV form factor out of the expression for the NMHV form factor. For the chiral half of the stress-tensor multiplet such a factorization is known. We have numerically checked that our result for this operator agrees with this known expression, but an analytical derivation would be preferable. \newline\\ For NMHV form factors of more general operators with $\dot\alpha$ indices however, the commutator of the forming operator and the space-time integral does not vanish. This is a major drawback since this forbids us to first Fourier transform and act with the forming operator only in the final steps of the calculation. This means that we have to treat these NMHV form factors in a case-by-case manner, which greatly complicates finding a general expression. However, applying the forming operator inside the space-time integral the computation of these form factors, albeit more tedious, can still be done for each operator separately. We exemplify this by computing an NMHV form factor of an operator consisting of an arbitrary number of scalars and one fermion. It would however be better to find a closed expression for the commutator of the space-time derivative and the space-time integral. This would allow us to first Fourier transform the expression and then apply the forming operator by adding a correction term. We leave this for future work. Another possible research direction is to use the construction presented here for computing form factors at loop level. This might prove to be challenging since the complications concerning the Fourier transform that arise for NMHV form factors at tree-level are also expected to occur in loop diagrams. Furthermore, at loop level one also encounters boundary and boundary-boundary diagrams. In addition, loop diagrams exhibit divergencies that require regularization. One should investigate how this can be done in twistor space. 
\newline\\
In the last chapter of this thesis, we used our formalism to compute a completely off-shell quantity, namely the planar one-loop correlation function in the scalar sector. From this, we extracted the one-loop dilatation operator from an interaction that is mediated by a four-vertex. However, in the operator limit the diagram containing the four-vertex suffers from spurious singularities. The origin of these singularities is still unclear. They might be an artefact of the axial gauge that was chosen. It is important to fully understand why they appear and how they can be removed in general, as they might occur also in more complicated computations. Here, we explicitly used the symmetry of the problem to cancel them. Thus we obtained the well-known UV-divergent space-time integral, whose coefficient is the one-loop dilatation operator in the SO$(6)$ sector. This points to another direction for further research. Namely, it would be preferable to perform the computation entirely in twistor space instead of integrating out the additional twistor space degrees of freedom to retrieve the UV-divergent space-time integral. This means that we should do the regularization completely in twistor space, which requires finding a suitable scheme. Dimensional regularization is not the obvious candidate as twistor space is restricted to precisely four dimensions. Moreover, the computation in this work was done only for the SO$(6)$ sector. We should not only extend this work to all other sectors, but compute it in complete generality without specifying which operators are involved. Our formalism with a generic forming operator could allow for such a derivation. One might be able to make a connection with the form of the dilatation operator of \cite{Zwiebel:2011bx} in spinor-helicity variables, which is essentially the four-vertex supplemented by a regulating piece. It would be interesting to find out what this additional regulating term translates to in twistor space and whether it is related to the spurious singularities that appeared in the computation of the dilatation operator in the SO$(6)$ sector presented here.\newline\\
Finally, one could try to link the current formalism to other work done on form factors, correlation functions and/or integrability. For example, it would be interesting to see whether we can do BCFW recursion for form factors in twistor space. Another direction to explore is the connected prescription of \cite{He:2016dol,Brandhuber:2016xue,He:2016jdg}. Another intriguing question is how results on Gra\ss{}mannian integrals for form factors \cite{Bork:2015fla,Frassek:2015rka,Bork:2016hst, Bork:2016xfn, Bork:2016egt} would relate to the current work. Moreover, we could learn more by linking the methods presented here to the LHC formalism of \cite{Chicherin:2016fac, Chicherin:2016fbj}. The complete expression for all MHV form factors was confirmed in \cite{Chicherin:2016qsf} using LHC space and it might be that those techniques offer some improvement over the twistor-space ones at NMHV level. Curiously, in \cite{Chicherin:2016ybl} similar Fourier integrals were encountered suggesting that the methods presented there might provide some insight in the non-commutativity of the integral and derivative in  Section~\ref{subsec: non-chiral operators}. Also, the duality described in \cite{Chicherin:2016ybl} between $n$-legged form factors of $m$-sided Wilson loops and form factors of the same kind but with the number of external legs and edges of the Wilson loop interchanged should have a twistor analogue that might be worth exploring. 
\newline\\
In conclusion, we have investigated whether twistor methods are suitable for computing correlation functions and form factors. We have incorporated all gauge-invariant local composite operators into the twistor formalism. Using this we found a closed form expression for all tree-level MHV form factors, which had previously been unknown and constitutes the main success of the formalism presented in this thesis. The construction was also used to compute form factors at NMHV level and beyond, however, some issues arose there that made the computations less straightforward, which should be investigated further. The most immediate direction for future research that builds on this work is to find the general NMHV form factor for operators without space-time derivatives in momentum space in a factorized manner. Furthermore, one should solve the issues related to the non-commutativity of the derivative and integral for form factors of general operators, to derive a closed form expression for the NMHV form factor at tree-level in momentum space.
We thus conclude that for form factors, our formalism provides a major improvement over Feynman diagram methods, albeit so far only at tree-level and most eminently at lowest MHV degree. Finally, we showed that the twistor formalism can be used to compute also correlation functions by rederiving the one-loop dilatation operator in the scalar sector. This was the first quantum integrability related computation in twistor space and one of the most obvious future directions is extending it to all sectors. Furthermore, it points to many interesting open questions that are worth answering, the most prominent one being how to treat divergences, both spurious singularities and UV divergences, in twistor space. 
\cleardoublepage
\phantomsection
\addcontentsline{toc}{chapter}{Acknowledgements}
\chapter*{Acknowledgements}
\markboth{Acknowledgements}{Acknowledgements}
First and foremost I thank my adviser Matthias Staudacher for giving me the opportunity to do my PhD under his supervision. I am grateful for all the opportunities you created, from traveling around the world, introducing me to the worlds experts in the field to creating countless possibilities for me to present my work. With great pleasure I look back at the many physics chats we had. Besides physics I have greatly valued your advice in difficult times. After Matthias it is only right I thank Vladimir Mitev, for collaborating with me for the entire duration of the PhD program. I am grateful for all the time, energy and patience you had and for all the physics you taught me along the way. Thirdly, Matthias Wilhelm I thank you for a great collaboration that I have thoroughly enjoyed. I am indebted to Lionel Mason for countless twistor discussions. I am grateful to Brenda Penante, James McGrane, papa and Vladimir Mitev for proofreading parts of this thesis. I thank Leo Zippelius and Felix Paul for translating the abstract to German. I thank my officemates: Vladimir Mitev, Yumi Ko and Brenda Penante. I thank all my colleagues, among those in particular the friends I made: Brenda Penante, Christian Marboe, Leo Zippelius, Lorenzo Bianchi, Pedro Liendo, Rouven Frassek, Stijn van Tongeren, Vladimir Mitev and Wadim Wormsbecher for support. Special thanks go to Marco Bianchi and Mat\'{i}as Leoni for the unforgettable introduction to the concept of a PhD school. I will think back with pleasure to interesting discussions about physics and beyond with Anne Spiering, Ben Hoare, Burkhard Eden, Christoph Sieg, David Meidinger, Dennis M\"{u}ller, Dhritiman Nandan, Edoardo Vescovi, Florian Loebbert, Gang Yang, Hagen M\"{u}nkler, Jan Fokken, Jan Plefka, Johannes Broedel, Matteo Rosso, Ruth Shir, Sergey Frolov, Sourav Sarkar, Thomas Klose, Tim Adamo, Valentina Forini and Yumi Ko. I am indebted also to Jenny Collard and Sylvia Richter for countless help with bureaucratic issues. I thank the friends that I made in Berlin outside of the institute, especially Laura, Julia and Daiana. I thank the people at the CRST at Queen Mary, especially Gabriele Travaglini for inviting me to London. \newline
I am grateful to Stef and Josh for taking me into their home and being dear friends. Bedankt aan al mijn lieve vrienden thuis: Floor, Sanne, de jaarclub en de natuurkunde meisjes. Ik mag me gelukkig prijzen met de vriendschap van de nerds Watse, Mathijs, Bram, George en Peter, zonder wie mijn natuurkunde studie er een stuk saaier had uitgezien. Bedankt aan al mijn familie voor steun en nooit ophoudende interesse. Speciaal dankjewel aan Wilma, die ik ook tot familie reken.\newline
Warmest gratitude goes to the loft, my surrogate family for the past three years. I greatly thank Jakob, Omid, Katharina, Diane, Sasha, Sonja and Amy. Without all of you I simply would not have sat it out until the end. I thank you also for showing me the world of Berlin and broadening my horizon in many ways. I come out not only a better physicist but also a more experienced person.\newline
I deeply thank you James for all your support, patience and love. \newline
Tenslotte gaat mijn grootste dank uit naar Maria, mama en papa. Ik dank jullie voor alle steun zowel tijdens deze drie jaar als ervoor. Ik heb ervaren dat in het buitenland wonen alleen gaat in de wetenschap dat er een thuis is waar je altijd naartoe terug kunt keren. Ik bedank jullie alle drie voor het grote voorbeeld dat jullie voor mij zijn. In het bijzonder bedank ik dan nog papa voor ontelbare wis- en natuurkunde lessen; van met negatieve getallen rekenen, naar meetkunde lessen, via talloze gesprekken over de uitgestrektheid van het heelal, uren staren naar de sterren en een sterrenkunde cursus in Artis, tot uiteindelijk samen college volgen in Utrecht. Zonder dat alles was de natuurkunde vrijwel zeker aan mij voorbij gegaan.\\
\newline
This research was supported in part by the SFB 647 \emph{``Raum-Zeit-Materie. Analytische und Geometrische Strukturen''} and in part by the DFG-funded Graduate School GK 1504. I also acknowledge the support of the Marie Curie network GATIS (gatis.desy.eu) of the European
Union's Seventh Framework Programme FP7/2007-2013/ under REA Grant Agreement No 317089 and the Marie Curie International Research Staff Exchange Network UNIFY (FP7-People-2010-IRSES under Grant Agreement No 269217).
\appendix

\chapter{Harmonic forms and Hodges theorem}
\label{appA}
In this appendix we prove the specific form of components of the twistor field in the harmonic gauge used in Section~\ref{twistoraction}. 
Namely, in that section we used that the components $g_0^+$, $\bar\psi_0$, $\phi_0$, $\psi_0$ and $g^-_0$ are in the first cohomology class $H^1(\mathbb{CP}^1, \mathcal{O}(k))$ for some $k$ as a consequence of Hodge theorem. In this appendix we argue why this is so and furthermore show that the elements of the first cohomology class $H^1(\mathbb{CP}^1, \mathcal{O}(k))$ vanish for $k\ge -1$ and that for $k<-1$, the corresponding cohomology class is just $\mathbb{C}^{-k-1}$. This appendix assumes some familiarity with differential geometry. \newline\\The gauge condition \eqref{gauge} states that the components $g_0^+$, $\bar\psi_0$, $\phi_0$, $\psi_0$ and $g^-_0$ are co-closed along the fibers $\mathbb{CP}^1$. Since $\mathbb{CP}^1$ is $1$-complex dimensional, $g_0^+\bar{e}^0$, $\bar\psi_0\bar{e}^0$, $\phi_0\bar{e}^0$, $\psi_0\bar{e}^0$ and $g^-_0\bar{e}^0$ are also closed and therefore harmonic. Note that their respective homogeneities imply that these $(0,1)$-forms take value in $\mathcal{O}(k)\otimes \text{End}(\mathcal{E})$, for $k=0,-1,-2,-3,-4$ respectively. Now we can apply Hodge theorem, which states that for $X$ a compact complex manifold and $E \rightarrow X$ a holomorphic vector bundle,
\begin{equation}
\text{Harm}^{p,q}(X,E)\cong H_{\bar\partial}^{p,q}(X,E)\eqndot
\end{equation}
In words, the space of harmonic $(p,q)$-forms on $X$ taking values in $E$ is isomorphic to the first $\bar\partial$-cohomology group with values in $E$. By Dolbeault's theorem, the latter space is also isomorphic to $q^{\text{th}}$ \v{C}ech cohomology, $H^q(X,\Lambda^p T^* X\otimes E)$. In our case, these theorems imply that the $(0,1)$-forms $g_0^+\bar{e}^0$, $\bar\psi_0\bar{e}^0$, $\phi_0\bar{e}^0$, $\psi_0\bar{e}^0$ and $g^-_0\bar{e}^0$ take values in $H^1(\mathbb{CP}^1,\mathcal{O}(k)\otimes\text{End}(\mathcal{E}))$, for $k=0,\dots,-4$.\newline\\
In the remainder of this section, we show that $H^1(\mathbb{CP}^1,\mathcal{O}(k)\otimes\text{End}(\mathcal{E}))=0$, for $k=0,-1$ and $H^1(\mathbb{CP}^1,\mathcal{O}(k)\otimes\text{End}(\mathcal{E}))=\mathbb{C}^{-k-1}$ for $k<-1$. \newline\\
Let $\la_{\alpha}=(\la_1,\la_2)$ be homogeneous coordinates on $\mathbb{CP}^1$ and define two charts 
\begin{equation}
V_1=\{ (z^{-1},1)\equiv (\la_2/\la_1)\, |\, \la_1\neq 0\}  \quad \text{and }\quad V_2=\{ (1,z)\equiv (1,\la_1/\la_2)\, |\, \la_2\neq 0\} \eqndot
\end{equation}
An element of this first cohomology group assigns a holomorphic section on the intersection $V_1\cap V_2$. This element is defined up to addition of a holomorphic section on $V_1$ or addition of a holomorphic section on $V_2$.
Now, let $g$ be in $H^1(V_1\cap V_2,\mathcal{O}(k))$, then in the chart $V_2$, $g$ can be represented by a holomorphic function $f_2$ on $\mathbb{C}-0$ and in the chart $ V_1$ by $ f_1 = z^{-k} f_2$. Now we can write for $k\ge -1$
\begin{equation}
f_2 =\sum_{-\infty}^{\infty} a_i z^i = \sum_{0}^{\infty} a_i z^i +\sum_{1}^{\infty} a_{-i} z^{-i}= \sum_{0}^{\infty} a_i z^i +z^k\sum_{1}^{\infty} a_{-i} z^{-i-k}\eqndot
\end{equation}
The term  $\sum_{0}^{\infty} a_i z^i $  is holomorphic on $V_2$ and $z^k\sum_{1}^{\infty} a_{-i} z^{-i-k}$ is holomorphic on $V_1$. Therefore, $g\sim0$ and in particular $H^1(\mathbb{CP}^1,\mathcal{O}(-1))=H^1(\mathbb{CP}^1,\mathcal{O}(0)=0$. For $k<-1$ we can write
\begin{equation}
f_2 =\sum_{-\infty}^{\infty} a_i z^i = \sum_{0}^{\infty} a_i z^i +\sum_{1}^{\infty} a_{-i} z^{-i}= \sum_{0}^{\infty} a_i z^i +z^k\sum_{-k}^{\infty} a_{-i} z^{-i-k}+\sum_{1}^{-k-1} a_{-i} z^{-i}\eqncom
\end{equation}
where the last term is holomorphic on $V_1\cap V_2$. This means that $g$ is in the same class as the section given by $\sum_{1}^{-k-1} a_{-i} z^{-i}$ and therefore, $H^1(\mathbb{CP}^1,\mathcal{O}(K))=\mathbb{C}^{-k-1}$ for $k<-1$.

\chapter{Integral identity}
\label{kahlerintegral}
In this appendix we compute certain integrals over the fibers $\mathbb{CP}^1$ in Euclidean twistor space. These identities together lead to formula \eqref{joehoe}, 
\begin{equation}
\label{326}
\int K_1 \frac{\lambda_{1{\alpha}}\lambda_{1\beta}}{\abra{1}{2}\abra{3}{1}}= - \frac{\pi i}{\abra{2}{3}}\left( \frac{\lambda_{2{\alpha}}\hat{\lambda}_{2\beta}+\hat{\lambda}_{2{\alpha}}\lambda_{2\beta}}{\abra{2}{\hat 2}}-\frac{\lambda_{3{\alpha}}\hat{\lambda}_{3\beta}+\hat{\lambda}_{3{\alpha}}\lambda_{3\beta}}{\abra{3}{\hat 3}}\right)\eqncom
\end{equation}
which is a correction of Eq. (3.26) of \cite{Boels:2006ir}.
Here, we remind the reader that $K_i$ denotes the K\"{a}hler metric on $\mathbb{CP}^1$,
\begin{equation}
K_i=\frac{\abra{\la_i}{\dd\la_i}\abra{\hat\la_i}{\dd\hat\la_i}}{\abra{\la_i}{\hat\la_i}^2}\eqndot
\end{equation}
The identity \eqref{326} is used in Section~\ref{twistoraction} and is a correction of Eq. (3.26) of \cite{Boels:2006ir}. 
To prove it, we use affine, or inhomogeneous coordinates on $\mathbb{CP}^1$. We take the charts where $\lambda_{i1}\neq 0$ for $i=1,2,4$. In these coordinates
\begin{align}
(1, \lambda_{12}/\lambda_{11})&=:(1,z),\notag\\
(1, \lambda_{22}/\lambda_{21})&=:(1,u),\notag\\
(1, \lambda_{32}/\lambda_{31})&=:(1,v).
\end{align}
Furthermore, the conjugate for example of $\la_1$ is defined as $\bar{\la}_1=(-\bar{z},1)$.
Then the lefthand side of \eqref{326} as a $2\times 2$-matrix has entries of the form 
\begin{equation}\label{327}
\int\frac{\dd z\dd\bar{z}}{(1+|z|^2)^2}\frac{z^p}{(z-u)(v-z)},
\end{equation}
where $p=0,1,2$ depending on the values of $\alpha$ and $\beta$, e.g.\ for $\alpha=\beta=1$, $p=0$.
We work out \eqref{327} for the three cases separately. First, we write $u=x e^{i{\alpha}}$ and $v=y e^{i\beta}$.
Without loss of generality we assume that $x<y$ so that we can rewrite \eqref{327} as
\begin{align}
&2i \int_0^{\infty}\frac{r\dd r}{(1+r^2)^2}\int_0^{2\pi}\dd\phi \frac{(r e^{i\phi})^p}{(r e^{i\phi}-x e^{i{\alpha}})(r e^{i\phi}-y e^{i\beta})}\notag\\
&=2i \left(\int_0^x\frac{r\dd r}{(1+r^2)^2}+\int_x^y\frac{r\dd r}{(1+r^2)^2}+\int_y^{\infty}\frac{r\dd r}{(1+r^2)^2}\right) \int_0^{2\pi}\dd\phi \frac{(r e^{i\phi})^p}{(r e^{i\phi}-x e^{i{\alpha}})(r e^{i\phi}-y e^{i\beta})}\eqndot
\end{align}
Furthermore, for later convenience we compute
\begin{align}
-\int_0^{x}\frac{r\dd r}{(1+r^2)^2}&=\frac{1}{2}\left(\frac{1}{(1+x^2)}-1\right),\notag\\
-\int_x^{y}\frac{r\dd r}{(1+r^2)^2}&=\frac{1}{2}\left(\frac{1}{(1+y^)}-\frac{1}{(1+x^2)}\right),\notag\\
-\int_y^{\infty}\frac{r\dd r}{(1+r^2)^2}&=\frac{1}{2}\frac{1}{(1+y^2)}\eqndot
\end{align}
Now, we study the three separate cases.
\paragraph{$p=0$}
We compute first the integral over $r$ from $0$ to $x$,
\begin{align}
&2i \int_0^x\frac{r\dd r}{(1+r^2)^2 }\int_0^{2\pi}\dd\phi \frac{1}{(r e^{i\phi}-x e^{i{\alpha}})(r e^{i\phi}-y e^{i\beta})}\notag\\
&=2i \int_0^x\frac{r\dd r}{(1+r^2)^2 }\int_0^{2\pi}\dd\phi \frac{1}{(r/x e^{i(\phi-{\alpha})}-1)(r/y e^{i(\phi-\beta)}-1)}\frac{1}{uv}\notag\\
&=2i \int_0^x\frac{r\dd r}{(1+r^2)^2 }\int_0^{2\pi}\dd\phi \frac{1}{uv}\sum_{m=0}^{\infty}\sum_{n=0}^{\infty} \left(\frac{r}{x}\right)^m \left(\frac{r}{y}\right)^n e^{im\phi}e^{i n \phi}e^{-i m{\alpha}}e^{-in\beta}\notag\\
&=2i \int_0^x\frac{r\dd r}{(1+r^2)^2 } \frac{2\pi}{uv}\notag\\
&=-i\left(\frac{1}{(1+|u|^2)}-1\right)\frac{2\pi}{uv}\eqndot
\end{align}
Then we compute the integral from $x$ to $y$,
\begin{align}
&2i \int_x^y\frac{r\dd r}{(1+r^2)^2 }\int_0^{2\pi}\dd\phi \frac{1}{(r e^{i\phi}-x e^{i{\alpha}})(r e^{i\phi}-y e^{i\beta})}\notag\\
&=-2i \int_x^y\frac{r\dd r}{(1+r^2)^2 }\int_0^{2\pi}\dd\phi  \frac{1}{(1-x/r e^{i({\alpha}-\phi)})(1-r/y e^{i(\phi-\beta)})}\frac{1}{v}\frac{1}{r e^{i\phi}}\notag\\
&= -2i\int_x^y\frac{r\dd r}{(1+r^2)^2 }\int_0^{2\pi}\dd\phi \frac{1}{v} \sum_{m=0}^{\infty} \left(\frac{x}{r}\right)^m e^{i{\alpha} m}\sum_{n=0}^{\infty}\left(\frac{r}{y}\right)^n e^{-i\beta n}e^{i\phi(n-m)}\frac{1}{re^{i\phi}}\notag\\
&= -2i \int_x^y\frac{r\dd r}{(1+r^2)^2 }\int_0^{2\pi}\dd\phi \frac{1}{v} \sum_{m=0}^{\infty} \left(\frac{x}{r}\right)^m e^{i{\alpha} m}\left(\frac{r}{y}\right)^{m+1} e^{-i\beta (m+1)}\frac{1}{r}\notag\\
&=-2i \int_x^y \frac{r\dd r}{(1+r^2)^2}\int\dd \phi \left(\frac{1}{1-\frac{u}{v}}\right)\frac{1}{v^2}\notag\\
&=-2i \frac{1}{2}\left(\frac{1}{1+|u|^2}-\frac{1}{1+|v|^2}\right)2\pi \frac{1}{v^2-uv}\eqndot
\end{align}
Finally, we integrate from $y$ to $\infty$,
\begin{align}
&2i\int_y^{\infty}\frac{r\dd r}{(1+r^2)^2 }\int_0^{2\pi}\dd\phi \frac{1}{(r e^{i\phi}-x e^{i{\alpha}})(r e^{i\phi}-y e^{i\beta})}\notag\\
&=2i \int_y^{\infty}\frac{r\dd r}{(1+r^2)^2 }\int_0^{2\pi}\dd\phi  \frac{1}{(1-x/r e^{i({\alpha}-\phi)})(1-y/r e^{i(\beta-\phi)})}\frac{1}{r^2 e^{i2\phi}}\notag\\
&=2i \int_y^{\infty}\frac{r\dd r}{(1+r^2)^2 }\int_0^{2\pi}\sum_{m=0}^{\infty}\left(\frac{x}{r}\right)^me^{i{\alpha} m}\sum_{n=0}^{\infty}\left(\frac{y}{r}\right)^ne^{i n \beta}e^{(-m-n-2)i\phi}\notag\\
&=0\eqndot
\end{align}
Collecting the terms we see that for $p=0$, \ref{327} equals
\begin{align}
&(-2\pi i)\left(-\frac{1}{uv}+\frac{1}{u}\frac{1}{v-u}\frac{1}{1+|u|^2}-\frac{1}{v}\frac{1}{v-u}\frac{1}{1+|v|^2}\right)\notag\\
&=(-2\pi i)\frac{1}{v-u}\left(\frac{-\bar{u}}{1+|u|^2}-\frac{-\bar{v}}{1+|v|^2}\right)\eqndot
\end{align}
\paragraph{$p=1$}
We use the same techniques as in the $p=0$ case, so we first integrate from $0$ to $x$,
\begin{align}
&2 i \int_0^x\frac{r\dd r}{(1+r^2)^2}\int_0^{2\pi}\dd\phi \frac{re^{i\phi}}{(\frac{r}{x}e^{i(\phi-{\alpha})}-1)}\frac{1}{(\frac{r}{y}e^{i(\phi-\beta)}-1)}\frac{1}{uv}\notag\\
&=2i \int_0^x\frac{r\dd r}{(1+r^2)^2}\int\dd\phi \frac{1}{uv}\sum_{m=0}^{\infty}\sum_{n=0}^{\infty}\left(\frac{r}{x}\right)^m\left(\frac{r}{y}\right)^n e^{i(m+n+1)\phi}r\notag\\
&=0\eqndot
\end{align}
Next we integrate from $x$ to $y$,
\begin{align}
&2i \int_x^y\frac{\dd r}{(1+r^2)^2}\int_0^{2\pi}\dd \phi \frac{re^{i\phi}}{(\frac{x}{r}e^{-i(\phi-{\alpha})}-1)}\frac{1}{(\frac{r}{y}e^{i(\phi-\beta)}-1)}\frac{1}{vr e^{i\phi}}\notag\\
&=-2i \int_x^y\frac{\dd r}{(1+r^2)^2}\int_0^{2\pi}\dd \phi\frac{1}{v}\sum_{m=0}^{\infty}\left(\frac{x}{r}\right)^m e^{im{\alpha}}\sum_{n=0}^{\infty}\left(\frac{r}{y}\right)^n e^{i n\beta}e^{i(n-m)\phi}\notag\\
&=-2i \int_x^y\frac{\dd r}{(1+r^2)^2}\int_0^{2\pi}\dd \phi\frac{1}{v}\frac{1}{1-\frac{u}{v}}\notag\\
&=-2i \pi \left(\frac{1}{1+|u|^2}-\frac{1}{1+|v|^2}\right)\frac{1}{v}\frac{1}{1-\frac{u}{v}}\notag\\
&=-2i \pi \left(\frac{1}{1+|u|^2}-\frac{1}{1+|v|^2}\right)\frac{1}{v-u}\eqndot
\end{align}
Finally the integral from $y$ to $\infty$,
\begin{align}
&2i \int_y^{\infty}\frac{r\dd r}{(1+r^2)^2}\int_0^{2\pi}\dd \phi \frac{r e^{i\phi}}{(1-\frac{x}{r} e^{i({\alpha}-\phi)})}\frac{1}{(1-\frac{y}{r} e^{i(\beta-\phi)})}\frac{1}{r^2e^{2i\phi}}\notag\\
&=2i \int_y^{\infty}\frac{r\dd r}{(1+r^2)^2}\int_0^{2\pi}\dd \phi \sum_{m=0}^{\infty}\left(\frac{x}{r}\right)^m e^{i m{\alpha}}\sum_{n=0}^{\infty}\left(\frac{y}{r}\right)^n e^{i n\beta} e^{i (-n-m-1)\phi}\notag\\
&=0\eqndot
\end{align}
We conclude that for $p=1$ the integral equals
\begin{equation}
\frac{-2\pi i}{v-u}\left(\frac{1}{1+|u|^2}-\frac{1}{1+|v|^2}\right).
\end{equation}
\paragraph{$p=2$}
By the same arguments as in the previous section, the integral from $0$ to $x$ is zero.
We proceed by computing the integral from $x$ to $y$,
\begin{align}
&-2i \int_x^y \frac{r\dd r}{(1+r^2)^2}\int_0^{2\pi} \dd \phi \frac{r^2e^{2i\phi}}{(1-\frac{x}{r}e^{i({\alpha}-\phi)})}\frac{1}{(1-\frac{r}{y}e^{-i(\beta-\phi)})}\frac{1}{re^{i\phi}ye^{i\beta}}\notag\\
&=-2i \int_x^y \frac{\dd r }{(1+r^2)^2}\int_0^{2\pi}\dd \phi\sum_{m=0}^{\infty}\left(\frac{x}{r}\right)^m e^{i m{\alpha}}\sum_{n=0}^{\infty} \left(\frac{r}{y}\right)^n e^{-in\beta}r e^{i\phi(1-m+n)}\frac{1}{v}\notag\\
&=-2i\int_x^y \frac{r\dd r}{(1+r^2)^2}\int_0^{2\pi} \dd \phi \sum_{n=0}^{\infty}\left(\frac{x}{r}\right)^{n+1}e^{i(n+1){\alpha}}\left(\frac{r}{y}\right)^n e^{-in\beta} r\frac{1}{v}\notag\\
&=-2i\int_x^y \frac{r\dd r}{(1+r^2)^2}\int\dd \phi \left(\frac{u}{v}\right)^n \frac{u}{v}\notag\\
&=-2i\int_x^y \frac{r\dd r}{(1+r^2)^2}\int_0^{2\pi}\dd \phi\frac{1}{1-\frac{u}{v}}\frac{u}{v}\notag\\
&=-4\pi i \int_x^y \frac{r\dd r}{(1+r^2)^2}\frac{u}{v-u}\notag\\
&= -2 \pi i \left(\frac{1}{1+|u|^2}-\frac{1}{1+|v|^2}\right) \frac{u}{v-u}\eqndot
\end{align}
Finally, we do the remaining integral from $y$ to $\infty$,
\begin{align}
&2i\int_y^{\infty} \frac{r\dd r}{(1+r^2)^2} \int_0^{2\pi} \dd \phi \frac{r^2 e^{2i\phi}}{(1-\frac{x}{r}e^{i({\alpha}-\phi)})}\frac{1}{(1-\frac{y}{r}e^{i(\beta-\phi)})}\frac{1}{r^2e^{2i\phi}}\notag\\
&=2i\int_y^{\infty} \frac{r\dd r}{(1+r^2)^2} \int_0^{2\pi} \dd \phi \sum_{m=0}^{\infty}\left(\frac{x}{r}\right)^m e^{i m {\alpha}} \sum_{n=0}^{\infty} \left(\frac{y}{r}\right)^n e^{i n\beta}e^{i(-n-m)\phi}\notag\\
&= 2i\int_y^{\infty} \frac{r\dd r}{(1+r^2)^2} \int_0^{2\pi} \dd \phi\notag\\
&=2i\pi \frac{1}{1+|v|^2}\eqndot
\end{align}
Taking all these terms together we see that the integral for $p=2$ equals,
\begin{equation}
\frac{-2\pi i}{v-u}\left(\frac{u}{1+|u|^2}-\frac{v}{1+|v|^2}\right).
\end{equation}
Now we recognize that 
\begin{equation}
\int K_1 \frac{\lambda_{1{\alpha}}\lambda_{1\beta}}{\abra{1}{2}\abra{3}{1}}= - \frac{\pi i}{\abra{2}{3}}\left( \frac{\lambda_{2{\alpha}}\hat{\lambda}_{2\beta}+\hat{\lambda}_{2{\alpha}}\lambda_{2\beta}}{\abra{2}{\hat 2}}-\frac{\lambda_{3{\alpha}}\hat{\lambda}_{3\beta}+\hat{\lambda}_{3{\alpha}}\lambda_{3\beta}}{\abra{3}{\hat 3}}\right)\eqncom
\end{equation}
and we have proven identity \eqref{326}.

\chapter{Wilson loop and operator limit}
\label{app:geometry}

In this appendix, we present explicitly the  geometry of the Wilson loop that is used in the construction of the composite operators of Section~\ref{sec:construction}. Specifically, we start with a list of requirements that the Wilson loop needs to satisfy, then show our solution and finish with the operator limit -- the procedure that sends the Wilson loop to a point. This appendix contains overlap with Appendix A of \cite{Koster:2016loo}.
\section{The geometry of the Wilson loop}
As explained in Section~\ref{secwilsonscalar} for the scalar field operators, we need to make sure that the local operators do not depend on the edges of the Wilson loop $\la_i$ or $\la_i'$ after the loop has been shrunk. 
Moreover, it must not matter which derivatives in \eqref{eq:definitionformingfactoronshellstates} are contracted with $\la_i$, which with $\la'_i$ and which with the polarization vectors $\tau_i$ from \eqref{eq: alphabet of fields with polarizations}.
The solution is to shrink the Wilson loop in such a way that $\la_i\rightarrow \tau_i$ and $\la_i'\rightarrow \tau_i$. This needs to be done for each edge of the Wilson loop that is acted upon by derivatives. Therefore, we need to add extra edges to the Wilson loop that will not be acted on by derivatives, i.e.\ that will not carry any irreducible fields \eqref{eq: alphabet of fields with polarizations}. If we acted with derivatives on each edge of the Wilson loop, then in the operator limit, all the corners of the loop would have to be identical. This would then imply that we have just one independent supertwistor on the loop in that limit, see Figure~\ref{fig:operatorlimit}, but we need to have at least two in order for the line $x$ to be well defined.
Finally, the geometry of the Wilson loop must allow us to take derivatives of $x_i$ in the direction $\tau_i$, i.e.\ to infinitesimally vary $x_i$ in this direction without destroying the light-like nature of the Wilson loop. This constraints the relative positions of the points neighboring $x_i$.\newline\\
The simplest Wilson loop geometry that we found involves $3L$ edges for a general composite operator $\calO$ of length $L$ \eqref{eq:composite operator}. The shape of our Wilson loop is reminiscent of a cogwheel. Specifically, we consider a light-like Wilson loop as shown in  Figure~\ref{fig:CogwheelBig} with $3L$ points, or corners, in space-time. These points are labeled as $x_i$, $x'_i$ and $x''_i$ with $i=1,\ldots,L$, see   Figure~\ref{fig:CogwheelZoom}, and they are ordered as $x'_1,x_1,x''_1,\ldots, x'_L,x_L,x''_L$. 
The (super) light-like condition implies that any two neighboring points $(x,\theta)$ and $(y, \vartheta)$ must satisfy $(x-y)^2=0$ and $(x-y)^{\alpha\dot\alpha}(\theta-\vartheta)_{\alpha}^{\phantom{\alpha}a}=0$. 
For the cogwheel Wilson loop, we solve these constraints as follows. We let $(x,\theta)$ be the center of the loop and parametrize
\begin{equation}
\label{eq:Cogwheelparametrization}
\begin{aligned}
&x_i^{\alpha\dot{\alpha}}=x^{\alpha\dot{\alpha}}+\m_i^{\alpha}\bm_i^{\dot{\alpha}}\eqncom&\quad &{x_i'}^{\alpha\dot{\alpha}}=x^{\alpha\dot{\alpha}}+\n_i^{\alpha}\bm_i^{\dot{\alpha}}\eqncom&\quad &{x_i''}^{\alpha\dot{\alpha}}=x^{\alpha\dot{\alpha}}+\n_{i+1}^{\alpha}\bm_i^{\dot{\alpha}}\eqncom&
\\
&\theta_i^{\alpha a }= \theta + \m_i^{\alpha} \xi_i^a\eqncom&  &{\theta_i'}^{\alpha a }=\theta + \n_i^{\alpha} \xi_i^a\eqncom&  &{\theta_i''}^{\alpha a }= \theta + \n_{i+1}^{\alpha} \xi_i^a\eqncom&
\end{aligned}
\end{equation}
where $\m_i$, $\bm_i$ and $\n_i$ are complex spinors and $\xi_i$ are Gra\ss mann parameters.
\begin{figure}[htbp]
 \centering
  \includegraphics[height=3.2cm]{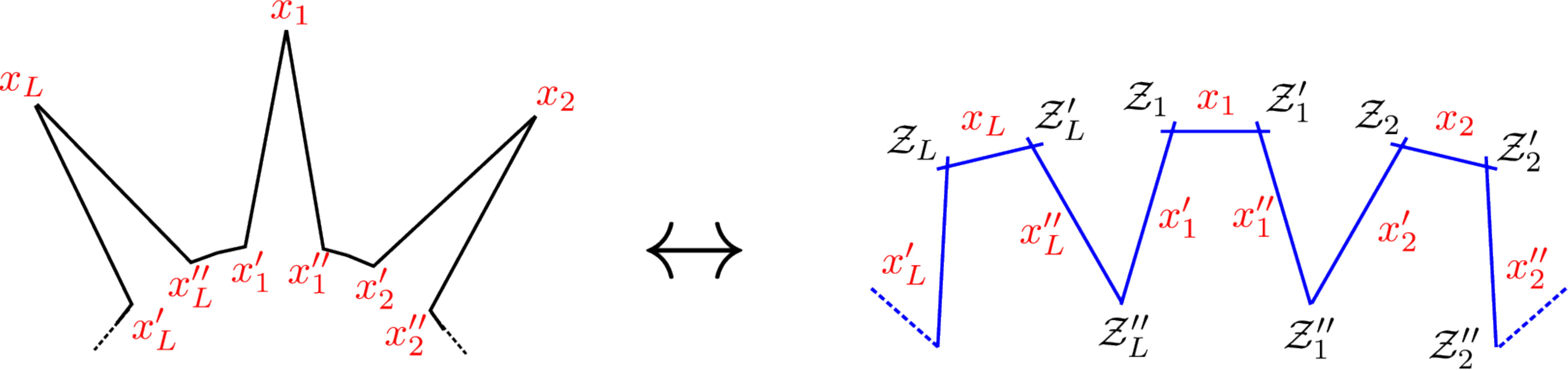}
  \caption{\it The geometry of the light-like Wilson loop. }
  \label{fig:CogwheelZoom}
\end{figure}
Thus, \eqref{eq:Cogwheelparametrization} ensures that the loop is light-like. We express the differences between the points as
\begin{equation}
\begin{split}
&\lambda_i\bar{\lambda}_i=x'_{i}-x_i=(\n_i-\m_i)\bm_i\eqncom\\
&\lambda_i'\bar{\lambda}_i'=x_{i}-x''_i=(\m_i-\n_{i+1})\bm_i\eqncom\\
&\lambda_i''\bar{\lambda}_i''=x''_{i}-x'_{i+1}=\n_{i+1}(\bm_i-\bm_{i+1})\eqndot
\end{split}
\end{equation}
We can choose, up to rescaling, to satisfy the above equations via
\begin{align}
\label{eq:definitionoflamdalambdaprimeandsecond}
&\lambda_i=\n_i-\m_i\eqncom& &\lambda_i'=\m_i-\n_{i+1}\eqncom& &\lambda_i''=\n_{i+1}\eqncom&\nonumber\\ &\bar{\lambda}_i=\bm_i\eqncom&
 &\bar{\lambda}_i'=\bm_i\eqncom&
 &\bar{\lambda}_i''=\bm_i-\bm_{i+1}\eqndot&
\end{align}
The twistors that correspond to the intersection of the lines are then
\begin{equation}
\label{eq:intersectiontwistorsdefinition}
\begin{split}
x'_i\cap x_i&=\calZ_i=(\lambda_i, i(x+\m_i\bm_i) \lambda_i, i(\theta+\m_i\xi_i)\lambda_i)\eqncom\\
x_i\cap x''_i&=\calZ_i'=(\lambda_i', i(x+\m_i\bm_i) \lambda_i',i(\theta+\m_i\xi_i)\lambda_i')\eqncom\\
x''_i\cap x'_{i+1}&=\calZ_i''=(\lambda_i'', i(x+\n_{i+1}\bm_i) \lambda_i'',i(\theta+\n_{i+1}\xi_i)\lambda_k'')\,,
\end{split}
\end{equation}
where the index contractions are left implicit. Putting everything together, we write down our cogwheel Wilson loops as
\begin{equation}
\label{eq:finaldefinitionWilsonloop}
\begin{split}
 \WWW(x_1',x_1,x_1'',\ldots, x_L',x_L,x_L'')= \Tr\Big[&U_{x_1'}(\calZ_L'',\calZ_1) U_{x_1}(\calZ_1,\calZ_1')U_{x_1''}(\calZ_1',\calZ_1'')\\&  U_{x_2'}(\calZ_1',\calZ_2)U_{x_2}(\calZ_1,\calZ_1')U_{x_2''}(\calZ_1',\calZ_2'') \cdots \\&\cdots U_{x_L'}(\calZ_{L-1}'',\calZ_L) U_{x_L}(\calZ_{L},\calZ_L') U_{x_L''}(\calZ_{L}',\calZ_L'') \Big]\eqncom
\end{split}
\end{equation}
cf.\   Figure~\ref{fig:CogwheelZoom}. We act on this Wilson loop with derivatives, as in \eqref{derivativeonUU}, but only on the unprimed edges.\newline\\\begin{figure}[htbp]
 \centering
  \includegraphics[height=3cm]{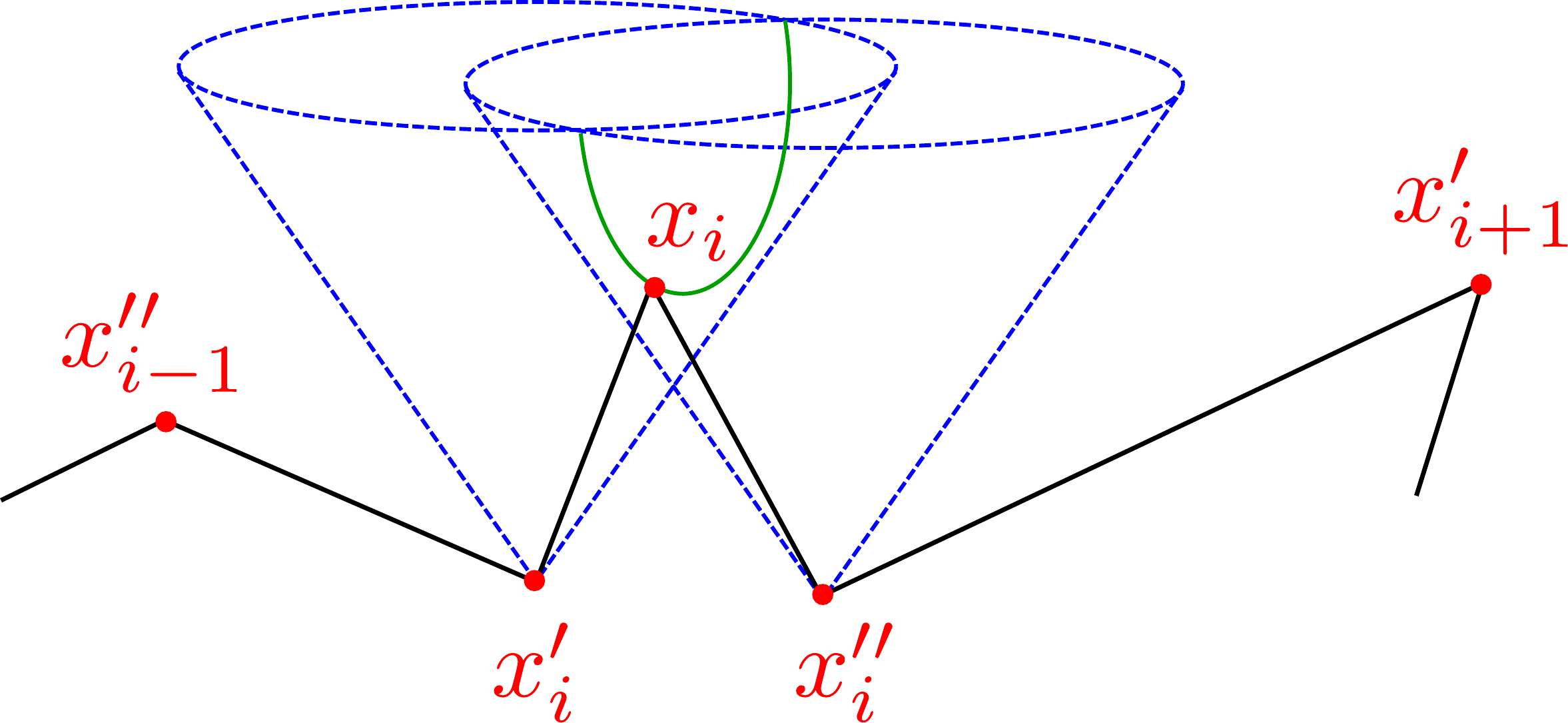}
  \caption{\it The variation of the Wilson loop has to preserve its light-like structure, thus constraining the point $x_i$ to lie on the intersection of the two light cones emanating from $x_{i-1}$ and $x_{i+1}$, depicted here in green.}
  \label{fig:LightLikeDeformation}
\end{figure}In order to include operators with covariant derivatives, we need to consider infinitesimal variations of the loop, specifically of the points $x_i$. To preserve the light-like nature of the loop at first order, see  Figure~\ref{fig:LightLikeDeformation}, the variations need to be of the type
\begin{equation}
\label{eq:definitionofthepolarizationvectors}
\delta x_i=a_i \lambda_i\bar{\lambda}_i'+b_i\lambda_i'\bar{\lambda}_i=\big(a_i(\n_i-\m_i)+b_i(\m_i-\n_{i+1})\big)\bm_i\equiv c_i \tau_i\bar{\tau}_i\eqncom
\end{equation}
where $a_i$, $b_i$ and $c_i$ are arbitrary infinitesimal parameters. The spinors $\tau_i$ and $\bar{\tau}_i$ are identified as the polarization vectors \eqref{eq: alphabet of fields with polarizations}. Equation \eqref{eq:definitionofthepolarizationvectors} can be used to solve for $\m_i$, $\n_i$ and $\bm_i$ as functions of $\tau_i$ and $\bar{\tau}_i$. There are clearly many solutions, but since we are only interested in them in the limit in which the Wilson loop shrinks to a point, we shall abstain from presenting them here.
\section{The operator limit}
\label{The operator limit}
We now want to discuss how the loop is to be shrunk to a point, i.e.\ \textit{the operator limit}. We first rescale the spinors $\m_i$, $\bm_i$ and $\n_i$ by $u$. For the intersection twistors $\calZ_i=(\lambda_i,\mu_i,\chi_i)$ of \eqref{eq:intersectiontwistorsdefinition}, this has the effect
\begin{equation}
\lambda_i\rightarrow u \lambda_i\eqncom \qquad \mu_i\rightarrow i(x+u^2\m_i\bm_i)u\lambda_i\eqncom \qquad \chi_i\rightarrow i(\theta+u^2\m_i\xi_i)u\lambda_i\eqncom
\end{equation}
and similarly for the $\calZ_i'$ and $\calZ_i''$.
Since the twistors  are projective quantities, the overall $u$ is irrelevant. Hence, rescaling the $\calZ_i$ leaves the $\lambda_i$ invariant, and in the limit $u\rightarrow 0$ we get
\begin{equation}
\label{eq:oplimit1}
\calZ_i\rightarrow (\lambda_i, ix\lambda_i, i\theta \lambda_i)\eqncom \qquad \calZ_i'\rightarrow (\lambda_i', ix\lambda_i',i \theta\lambda_i')\eqncom\qquad \calZ_i''\rightarrow (\lambda_i'', ix\lambda_i'', i\theta \lambda_i'')\eqncom
\end{equation}
i.e.\ all the intersection twistors lie on the same line.
\begin{figure}[htbp]
 \centering
  \includegraphics[height=4.5cm]{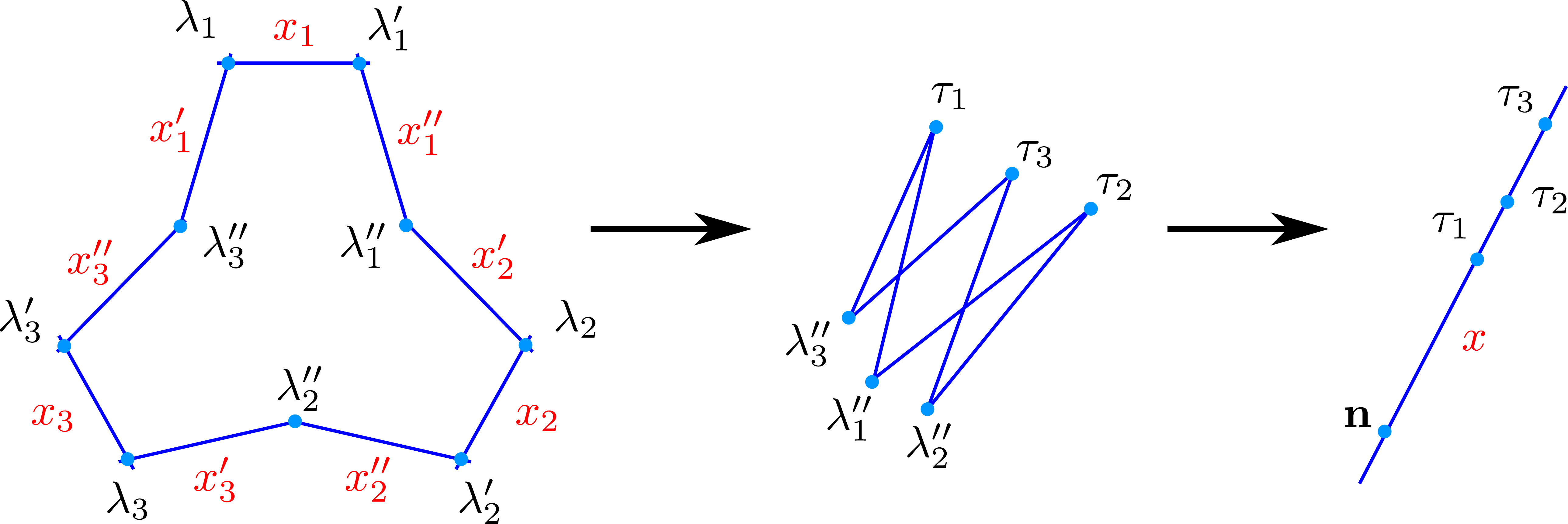}
  \caption{\it This figure illustrates the operator limit for $L=3$. To visualize the process a bit better, as a first step, we set $\lambda_i=\la_i'=\tau_i$ while bringing the $x_i$ closer to each other. The second step then just sends all $x_i$ to $x$ and $\lambda_{i}''\rightarrow \n$. 
  }
  \label{fig:operatorlimitapp}
\end{figure}
The above limit does not yet realize $\lambda_i\parallel\lambda_i'\parallel\tau_i$.
This can be achieved by setting all spinors $\n_i=\n$ to be equal. 
Due to \eqref{eq:definitionoflamdalambdaprimeandsecond} and \eqref{eq:definitionofthepolarizationvectors}, this sets $\lambda_i$, $\lambda_i'$ and  $\tau_i$ equal up to rescaling and we choose them to be equal. Summarizing, we have in the operator limit:
\begin{equation}
\label{eq:oplimit2}
\lambda_i\rightarrow \tau_i\,,\qquad  \lambda_i' \rightarrow \tau_i\,,\qquad \lambda_i''\rightarrow \n \eqndot
\end{equation}
In addition, due to \eqref{eq:definitionofthepolarizationvectors}, in the operator limit we have
\begin{equation}
\tau_i=\m_i-\n\eqncom\qquad \bar{\tau}_i=\bm_i\eqndot
\end{equation}
We illustrate the operator limit geometrically in  Figure~\ref{fig:operatorlimitapp}. 

\chapter{Operator vertices}
\label{app:formfactordatamine}

In this appendix, for the convenience of the reader, we have worked out the results of applying the derivative operators \eqref{eq:definitionformingfactoronshellstates} and the operator limit \eqref{eq:oplimit1}, \eqref{eq:oplimit2} to obtain the field vertices $\mathbf{W}_{D^{k_i}\Phi_i(x)}$ in several explicit examples.
Multiplying them together according to \eqref{eq: operator vertex} leads to the corresponding operator vertices. This appendix contains overlap with Appendix B of \cite{Koster:2016loo}.
For a fermion, written as $\bar{\psi}=\bar{\tau}^{\dot\alpha}\xi^a\bar{\psi}_{a\dot\alpha}(x)$, the field vertex reads 
\begin{equation}\label{eq: anitfermion}
\begin{aligned}
 \mathbf{W}_{\bar{\psi}}=&\int \DD\la h_x^{-1}(\la) \bar{\tau}^{\dot\alpha}\xi^a\frac{\partial^2 \AAA(\la)}{\partial \chi^a \partial\mu^{\dot\alpha}}h_x(\la)\notag\\
& +\int \DD\la\DD\la' h_x^{-1}(\la) \xi^a\frac{\partial \AAA(\la)}{\partial\chi^a}\frac{U_x(\la,\la')}{\abra{\la}{\la'}}\bar{\tau}^{\dot\alpha}\frac{\partial \AAA(\la')}{\partial\mu'^{\dot\alpha}}h_x(\la')\\
 &+\int \DD\la\DD\la' h_x^{-1}(\la) \bar{\tau}^{\dot\alpha}\frac{\partial \AAA(\la)}{\partial\mu^{\dot\alpha}}\frac{U_x(\la,\la')}{\abra{\la}{\la'}}\xi^a\frac{\partial \AAA(\la')}{\partial\chi'^{a}}h_x(\la')
 \eqndot
\end{aligned}
\end{equation}
The self-dual part of the field strength $\bar{F}=\bar{\tau}^{\dot\alpha}\bar{\tau}^{\dot\beta}\bar{F}_{\dot\alpha\dot\beta}(x)$ has the vertex
\begin{equation}\label{eq: asdfieldstrength}
\begin{aligned}
 \mathbf{W}_{\bar{F}}= &\int \DD\la h^{-1}_x(\la) \bar{\tau}^{\dot\alpha}\bar{\tau}^{\dot\beta}\frac{\partial^2 \AAA(\la)}{\partial\mu^{\dot\alpha} \partial\mu^{\dot\beta}}h_x(\la)
 \\
  &
  +2\int \DD\la\DD\la' h^{-1}_x(\la) \bar{\tau}^{\dot\alpha}\frac{\partial \AAA(\la)}{\partial\mu^{\dot\alpha}}\frac{U_x(\la,\la')}{\abra{\la}{\la'}}\bar{\tau}^{\dot\beta}\frac{\partial \AAA(\la')}{\partial\mu'^{\dot\beta}}h_x(\la')\eqndot
\end{aligned}
\end{equation}
The vertex of an antifermion $\psi=-\tau^\alpha\xi^a\xi^b\xi^c\psi_{abc\alpha}(x)$ is
\begin{equation}\label{eq: fermion}
\begin{aligned}
& \mathbf{W}_{\psi}=-\int \DD\la h^{-1}_x(\la) \abra{\tau}{\la}\xi^a\xi^b\xi^c\frac{\partial^3 \AAA(\la)}{\partial \chi^a \partial \chi^b\partial \chi^c}h_x(\la)\\ 
 &-3\int \DD\la\DD\la' h^{-1}_x(\la) \abra{\tau}{\la}\xi^a\xi^b\frac{\partial^2 \AAA(\la)}{\partial\chi^a\partial\chi^b}\frac{U_x(\la,\la')}{\abra{\la}{\la'}}\xi^c\frac{\partial \AAA(\la')}{\partial\chi'^{c}}h_x(\la')\\
 &-3\int \DD\la\DD\la' h^{-1}_x(\la) \xi^a\frac{\partial \AAA(\la)}{\partial\chi^a}\frac{U_x(\la,\la')}{\abra{\la}{\la'}}\abra{\tau}{\la'}\xi^b\xi^c\frac{\partial^2 \AAA(\la')}{\partial\chi'^{b}\partial\chi'^{c}}h_x(\la')\\
 &+6\int \DD\la\DD\la'\DD\la'' h^{-1}_x(\la) \xi^a\frac{\partial \AAA(\la)}{\partial\chi^{a}}\frac{U_x(\la,\la')}{\abra{\la}{\la'}}\abra{\tau}{\la'}\xi^b\frac{\partial \AAA(\la')}{\partial\chi'^{b}}\frac{U_x(\la',\la'')}{\abra{\la'}{\la''}}\xi^c\frac{\partial \AAA(\la'')}{\partial\chi''^{c}}h_x(\la'')
 \eqndot
\end{aligned}
\end{equation}
\allowdisplaybreaks
The vertex of the anti-self-dual part of the field strength $F=\tau^\alpha\tau^\beta\xi^a\xi^b\xi^c\xi^dF_{\alpha\beta abcd}(x)$ equals
\begin{align}
\nonumber
 &\mathbf{W}_{F}= \int \DD\la h^{-1} \abra{\tau}{\la}^2\xi^a\xi^b\xi^c\xi^d\frac{\partial^4 \AAA(\la)}{\partial \chi^a \partial \chi^b\partial \chi^c \partial \chi^d}h\\ \nonumber
 &+6\int \DD\la\DD\la'h^{-1} \abra{\tau}{\la}\xi^a\xi^b\frac{\partial^2 \AAA(\la)}{\partial\chi^a\partial\chi^b}\frac{U_x(\la,\la')}{\abra{\la}{\la'}}\abra{\tau}{\la'}\xi^c\xi^d\frac{\partial^2 \AAA(\la')}{\partial\chi'^c\partial\chi'^{d}}h'\\ \nonumber
 &-4\int \DD\la\DD\la'h^{-1} \abra{\tau}{\la}^2\xi^a\xi^b\xi^c\frac{\partial^3 \AAA(\la)}{\partial\chi^a\partial\chi^b\partial\chi^c}\frac{U_x(\la,\la')}{\abra{\la}{\la'}}\xi^d\frac{\partial \AAA(\la')}{\partial\chi'^{d}}h'\\ \nonumber
 &-4\int \DD\la\DD\la'h^{-1} \xi^a \frac{\partial \AAA(\la)}{\partial\chi^a}\frac{U_x(\la,\la')}{\abra{\la}{\la'}}\abra{\tau}{\la'}^2\xi^b\xi^c\xi^d\frac{\partial^3 \AAA(\la')}{\partial\chi'^{b}\partial\chi'^{c}\partial\chi'^{d}}h'\\ \nonumber
 &-12\int \DD\la\DD\la'\DD\la''h^{-1} \abra{\tau}{\la}\xi^a\xi^b\la^{\alpha}\frac{\partial^2 \AAA(\la)}{\partial\chi^{a}\partial\chi^{b}}\frac{U_x(\la,\la')}{\abra{\la}{\la'}}\abra{\tau}{\la'}\xi^c\frac{\partial \AAA(\la')}{\partial\chi'^{c}}\frac{U_x(\la',\la'')}{\abra{\la'}{\la''}}\xi^d\frac{\partial \AAA(\la'')}{\partial\chi''^{d}}h''\\ \nonumber
 &-12\int \DD\la\DD\la'\DD\la''h^{-1} \xi^a\frac{\partial \AAA(\la)}{\partial\chi^{a}}\frac{U_x(\la,\la')}{\abra{\la}{\la'}}\abra{\tau}{\la'}^2\xi^b\xi^c\frac{\partial^2 \AAA(\la')}{\partial\chi'^{b}\partial\chi'^{c}}\frac{U_x(\la',\la'')}{\abra{\la'}{\la''}}\xi^d\frac{\partial \AAA(\la'')}{\partial\chi''^{d}}h''\\ \nonumber
 &-12\int \DD\la\DD\la'\DD\la''h^{-1} \xi^a\frac{\partial \AAA(\la)}{\partial\chi^{a}}\frac{U_x(\la,\la')}{\abra{\la}{\la'}}\abra{\tau}{\la'}\xi^b\frac{\partial \AAA(\la')}{\partial\chi'^{b}}\frac{U_x(\la',\la'')}{\abra{\la'}{\la''}}\abra{\tau}{\la''}\xi^c\xi^d\frac{\partial^2 \AAA(\la'')}{\partial\chi''^{c}\partial\chi''^{d}}h''\\ \nonumber
 &+24\int \DD\la\DD\la'\DD\la''\DD\la''' h^{-1} \xi^a\frac{\partial \AAA(\la)}{\partial\chi^{a}}\frac{U_x(\la,\la')}{\abra{\la}{\la'}}\abra{\tau}{\la'}\xi^b\frac{\partial \AAA(\la')}{\partial\chi'^{b}}\frac{U_x(\la',\la'')}{\abra{\la'}{\la''}}\abra{\tau}{\la''}\\&\qquad \qquad \times \xi^c\frac{\partial \AAA(\la'')}{\partial\chi''^{c}}\frac{U_x(\la'',\la''')}{\abra{\la''}{\la'''}}\xi^d\frac{\partial \AAA(\la''')}{\partial\chi'''^{d}}h'''
 \eqncom
 \label{eq: sdfieldstrength}
\end{align}
where we abbreviated $h\equiv h_x(\lambda)$, $h'\equiv h_x(\lambda')$ and so on.
For a scalar with covariant derivative $D\phi=-\tau^\alpha\bar{\tau}^{\dot\alpha}\xi^a\xi^bD_{\alpha\dot\alpha}\phi_{ab}$, we find
\begin{equation}
\label{eq:covariantderivativephi}
\begin{aligned}
 \mathbf{W}_{D\phi}= &-\int \DD\la h^{-1} \abra{\tau}{\la}\bar{\tau}^{\dot\alpha}\xi^a\xi^b\frac{\partial^3 \AAA(\la)}{\partial\mu^{\dot\alpha}\partial\chi^a\partial\chi^b}h\\
 &+2\int \DD\la\DD\la' h^{-1} \abra{\tau}{\la}\bar{\tau}^{\dot\alpha}\xi^a\frac{\partial^2 \AAA(\la)}{\partial\mu^{\dot\alpha}\partial\chi^a}\frac{U_x(\la,\la')}{\abra{\la}{\la'}}\xi^b\frac{\partial \AAA(\la')}{\partial\chi'^{b}}h'\\
 &+2\int \DD\la\DD\la' h^{-1} \xi^a\frac{\partial \AAA(\la)}{\partial\chi^a}\frac{U_x(\la,\la')}{\abra{\la}{\la'}}\abra{\tau}{\la'}\bar{\tau}^{\dot\alpha}\xi^b\frac{\partial^2 \AAA(\la')}{\partial\mu'^{\dot\alpha}\partial\chi'^{b}}h'\\
 &-\int \DD\la\DD\la' h^{-1} \abra{\tau}{\la}\xi^a\xi^b\frac{\partial^2 \AAA(\la)}{\partial\chi^a\partial\chi^{b}}\frac{U_x(\la,\la')}{\abra{\la}{\la'}}\bar{\tau}^{\dot\alpha}\frac{\partial \AAA(\la')}{\partial\mu'^{\dot\alpha}}h'\\
 &-\int \DD\la\DD\la' h^{-1} \bar{\tau}^{\dot\alpha}\frac{\partial \AAA(\la)}{\partial\mu^{\dot\alpha}}\frac{U_x(\la,\la')}{\abra{\la}{\la'}}\abra{\tau}{\la'}\xi^a\xi^b\frac{\partial^2 \AAA(\la')}{\partial\chi'^a\partial\chi'^{b}}h'\\
 &+2\int \DD\la\DD\la'\DD\la'' h^{-1} 
 \bar{\tau}^{\dot\alpha}\frac{\partial \AAA(\la)}{\partial\mu^{\dot\alpha}}\frac{U_x(\la,\la')}{\abra{\la}{\la'}}\abra{\tau}{\la'}\xi^a\frac{\partial \AAA(\la')}{\partial\chi'^a}\frac{U_x(\la',\la'')}{\abra{\la'}{\la''}}\xi^b\frac{\partial \AAA(\la'')}{\partial\chi''^b}h''\\
 &+2\int \DD\la\DD\la'\DD\la'' h^{-1} \xi^a\frac{\partial \AAA(\la)}{\partial\chi^a}\frac{U_x(\la,\la')}{\abra{\la}{\la'}}\abra{\tau}{\la'}\bar{\tau}^{\dot\alpha}\frac{\partial \AAA(\la')}{\partial\mu'^{\dot\alpha}}\frac{U_x(\la',\la'')}{\abra{\la'}{\la''}}\xi^b\frac{\partial \AAA(\la'')}{\partial\chi''^b}h''\\
 &+2\int \DD\la\DD\la'\DD\la'' h^{-1} \xi^a\frac{\partial \AAA(\la)}{\partial\chi^{a}}\frac{U_x(\la,\la')}{\abra{\la}{\la'}}\abra{\tau}{\la'}\xi^b\frac{\partial \AAA(\la')}{\partial\chi'^b}\frac{U_x(\la',\la'')}{\abra{\la'}{\la''}}\bar{\tau}^{\dot\alpha}\frac{\partial \AAA(\la'')}{\partial\mu''^{\dot\alpha}}h''
 \eqndot
\end{aligned}
\end{equation}
In all the above expressions, it is understood that $\calA(\la)\equiv\calA(\calZ_x(\la))$ and so on.

\chapter{Computation of the NMHV form factors}
\label{app:NMHVformfactors}

In this appendix, we expand on the more tedious aspects in the calculation of the NMHV form factors. We start in position twistor space in Section~\ref{appnmhvpos}, which is based on and contains overlap with Appendix of A.1 of \cite{Koster:2016fna}.  We transform to momentum twistor space in Section~\ref{appnmhvmom}, which has significant overlap with Appendix A.2 of \cite{Koster:2016fna}.

\section{Position twistor space}
\label{appnmhvpos}
In order to compute the form factor, we use the cogwheel-shaped Wilson loop that we described in detail in Appendix~\ref{app:geometry}.
It has three families of edges $x_i$, $x_i'$ and $x_i''$, cf.\  Figure~\ref{fig:CogwheelZoom}.
Each of these edges gives rise to a vertex $\mathbf{W}$.
In order to obtain an $n$-point form factor, we have to sum over all ways to distribute the $n$ external fields on the edges.
Let $m_i$ be the first external field emitted on $x_i$, $m_i'$ the last external field emitted on $x_i$ and $m_i''$ the last external field emitted on\footnote{External fields that are emitted on an interaction line $z$ connected to $x_i$ by a propagator are also considered to be emitted on the edge $x_i$ for this purpose, and analogously for the other edges and the case of several interaction lines.} $x_i''$.%
We then have to sum over 
\begin{equation}
\label{eq:constraintsforthems}
 1\leq m_1\leq m_1'\leq m_1''< m_2\leq m_2'\leq m_2''<m_3\leq \cdots <m_L\leq m_L'\leq m_L''< m_1+n\leq 2n\eqncom
\end{equation}
where $m_i\leq m_i'$ ensures that there is at least one external field emitted from $x_i$; contributions with no  field on $x_i$ are annihilated by the forming operator \eqref{eq:definitionformingfactoronshellstates}.

For each cog, we divide the computation of the NMHV form factors into three pieces, illustrated in  Figure~\ref{fig:AllTermsNMHV1} and  Figure~\ref{fig:AllTermsNMHV2}. In the first diagram shown in  Figure~\ref{fig:AllTermsNMHV1}, we need to restrict the range of the parameters as $m_i-1\leq j<k\leq m_i'$, where $j=m_i-1$ implies that the propagator is attached to the leftmost twistor on the line $x_i$, while $k=m_i'$ means that the propagator is attached to the rightmost one. In addition, $k=j+1$ means that we are dealing with a two-point interaction vertex; as in the case of amplitudes, this can be dropped.
Similarly, we need to have $m_{i-1}''\leq j<k\leq m_i-1$ for the left diagrams in  Figure~\ref{fig:AllTermsNMHV2}, while for the right one we need $m_i'\leq j<k\leq m_i''$. 

\begin{figure}[tbp]
 \centering
  \includegraphics[height=3.7cm]{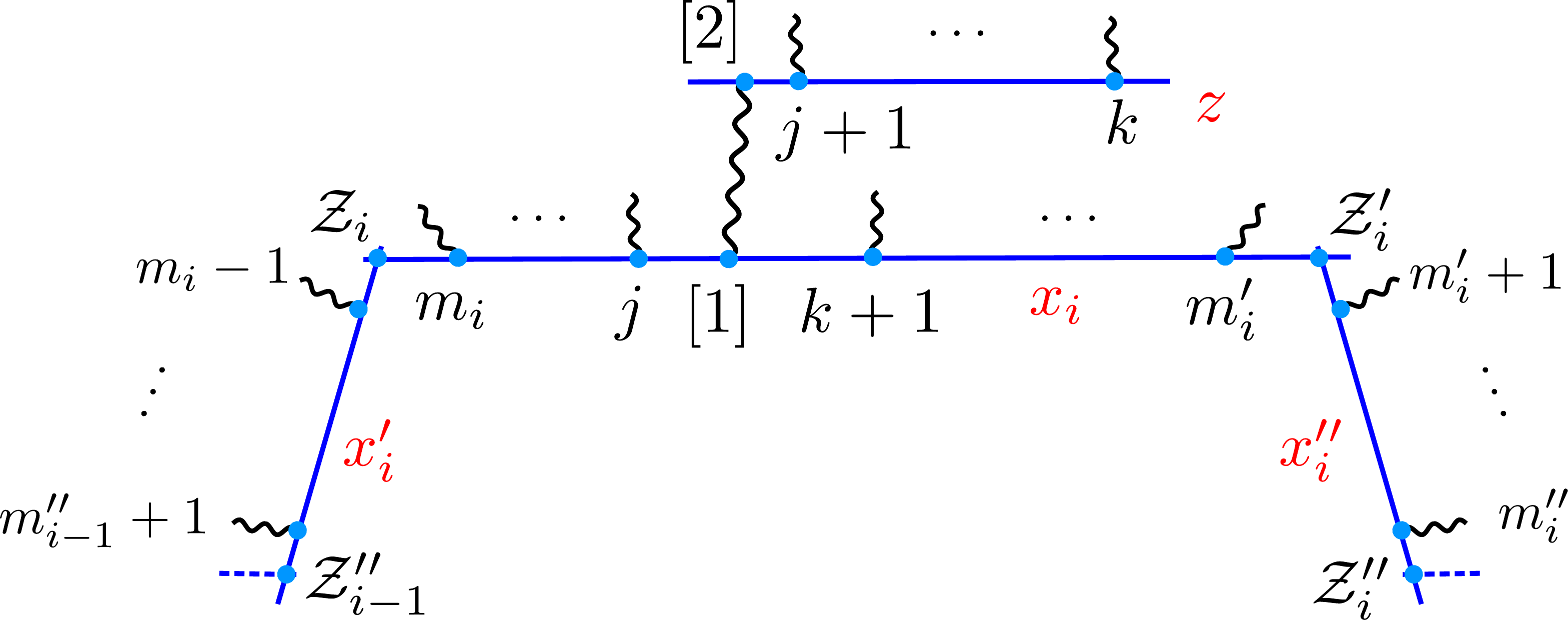}
  \caption{\it An NMHV diagram of type $\bbA$ -- the interaction vertex is connected to the operator-bearing edge $x_i$ of the Wilson loop.}
  \label{fig:AllTermsNMHV1}
\end{figure}
\begin{figure}[tbp]
 \centering
  \includegraphics[height=4.2cm]{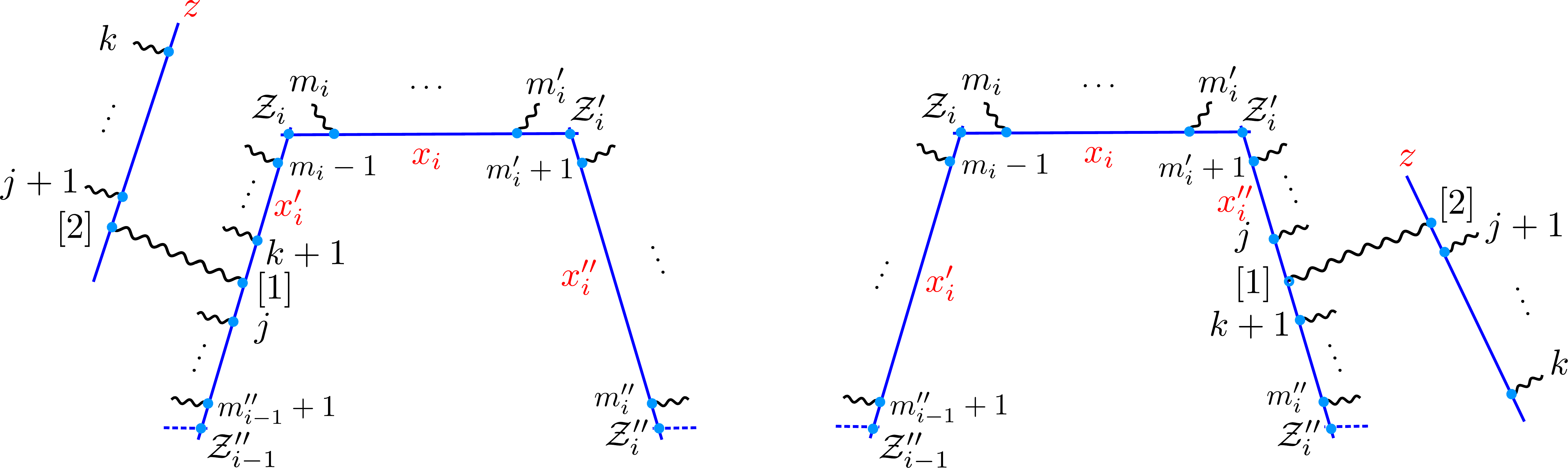}
  \caption{\it The NMHV diagrams of type $\bbB$ and $\bbC$ -- the interaction vertex is connected to the edges $x_i'$ and $x_i''$ of the Wilson loop, which do not carry the operator.}
  \label{fig:AllTermsNMHV2}
\end{figure}

In \eqref{eq:NMHVWilsonloopformfactor}, we gave an expression for the interaction vertex connected to a general edge. Specifying \eqref{eq:NMHVWilsonloopformfactor} to the three cases above and suppressing the vertices of the other cogs,
the diagrams of  Figure~\ref{fig:AllTermsNMHV1} contribute as
\begin{equation}
\label{eq:AcontributiontoNMHVWilsonloopformfactor}
\begin{aligned}
\bbA_i&=\nalpha\sum_{m_i-1\leq j<k\leq m_i'}\verop{x_{i}'}{\calZ_{i-1}'',\calZ_{i}}{\calZt_{m_{i-1}''+1},\ldots, \calZt_{m_i-1}}\\
&\phaneq\times \verop{x_i}{\calZ_i,\calZ_i'}{\calZt_{m_i},\ldots, \calZt_{j},\calZt_{k+1},\ldots,\calZt_{m_i'}}\verop{x_{i}''}{\calZ_{i}',\calZ_{i}''}{\calZt_{m_{i}'+1},\ldots, \calZt_{m_i''}}\\
&\phaneq\times\ver{\calZt_{j+1},\ldots, \calZt_k}[(\calZt_{k+1}\stackrel{k=m_i'}{\longrightarrow}\calZ_i'),(\calZt_j\stackrel{j=m_i-1}{\longrightarrow}\calZ_i),\star,\calZt_{j+1},\calZt_k]\,,
\end{aligned}
\end{equation}
while those on the left- and right-hand side of  Figure~\ref{fig:AllTermsNMHV2}, called $\bbB_i$ and $\bbC_i$ respectively, contribute 
\begin{equation}
\label{eq:BCcontributiontoNMHVWilsonloopformfactor}
\begin{aligned}
\bbB_i&= \nalpha\sum_{m_{i-1}''\leq j<k\leq m_i-1}\verop{x_{i}'}{\calZ_{i-1}'',\calZ_{i}}{\calZt_{m_{i-1}''+1},\ldots,\calZt_{j},\calZt_{k+1},\ldots, \calZt_{m_i-1}}\\
&\phaneq\times \verop{x_i}{\calZ_i,\calZ_i'}{\calZt_{m_i},\ldots, \calZt_{m_i'}}\verop{x_{i}''}{\calZ_{i}',\calZ_{i}''}{\calZt_{m_{i}'+1},\ldots, \calZt_{m_i''}}\\
&\phaneq\times\ver{\calZt_{j+1},\ldots, \calZt_k}[(\calZt_{k+1}\stackrel{k=m_i-1}{\longrightarrow}\calZ_i),(\calZt_j\stackrel{j=m_{i-1}''}{\longrightarrow}\calZ_{i-1}''),\star,\calZt_{j+1},\calZ_k]\,,\\
\bbC_i &=\nalpha\sum_{m_{i}'\leq j<k\leq m_i''}\verop{x_{i}'}{\calZ_{i-1}'',\calZ_{i}}{\calZt_{m_{i-1}''+1},\ldots, \calZt_{m_i-1}}\\
&\phaneq\times \verop{x_i}{\calZ_i,\calZ_i'}{\calZt_{m_i},\ldots, \calZt_{m_i'}}\verop{x_{i}''}{\calZ_{i}',\calZ_{i}''}{\calZt_{m_{i}'+1},\ldots,\calZt_{j},\calZt_{k+1},\ldots, \calZt_{m_i''}}\\
&\phaneq\times\ver{\calZt_{j+1},\ldots, \calZt_k}[(\calZt_{k+1}\stackrel{k=m_{i}''}{\longrightarrow}\calZ_i),(\calZ_j\stackrel{j=m_{i}'}{\longrightarrow}\calZ_{i-1}''),\star,\calZt_{j+1},\calZt_k]\,.
\end{aligned}
\end{equation}
Finally, the complete tree-level NMHV form factor of the Wilson loop in position twistor space is obtained by adding \eqref{eq:AcontributiontoNMHVWilsonloopformfactor} to the two terms in  \eqref{eq:BCcontributiontoNMHVWilsonloopformfactor}, multiplying with the contributions of the cogs of the Wilson loop that are not connected to the vertex $\mathbf{V}$ and then summing over the cogs $i$ that are as well as over the possible distributions of the integers $m_i,m_i'$ and $m_i''$ subject to \eqref{eq:constraintsforthems}:
\begin{equation}
\label{eq:wilsonloopformfactorinpositiontwistorspace}
\mathbb{F}^{\text{NMHV}}_{\calW}=\sum_{\{m_s,m_s',m_s''\}}\sum_{i=1}^L\Big(\prod_{r=1}^{i-1}\bbD_r\Big)(\bbA_i+\bbB_i+\bbC_i)\Big(\prod_{r=i+1}^{L}\bbD_r\Big)\,,
\end{equation}
where the contribution of a cog that is not connected to the interaction vertex is
\begin{equation}
\begin{aligned}
\bbD_i&= \verop{x_{i}'}{\calZ_{i-1}'',\calZ_{i}}{\calZt_{m_{i-1}''+1},\ldots, \calZt_{m_i-1}} \verop{x_i}{\calZ_i,\calZ_i'}{\calZt_{m_i},\ldots, \calZt_{m_i'}}
\\&\phaneq\times 
\verop{x_{i}''}{\calZ_{i}',\calZ_{i}''}{\calZt_{m_{i}'+1},\ldots, \calZt_{m_i''}}\,.
\end{aligned}
\end{equation}
To obtain the tree-level NMHV form factor of a local operator $\calO$ in position twistor space, we have to act with the forming operator $\mathbf{F}_\calO$ defined in \eqref{formingop} on the \emph{integrand} of \eqref{eq:wilsonloopformfactorinpositiontwistorspace}, i.e.\ \emph{before} doing the integrations over the vertex positions; see the discussion in the main text. 

\section{Momentum twistor space}
\label{appnmhvmom}

Let us now describe the transition to momentum (twistor) space. First, we recall some of the notation introduced in Chapter \ref{MHV}. For the operator $\calO$ of \eqref{eq:composite operator}, we denoted by $N_i$ the total number of derivatives (space-time and fermionic) required to generate the field $D^{k_i}\Phi_i$ in \eqref{eq:definitionformingfactoronshellstates}. Specifically, for $_i=\bar F, \bar \psi $ or $ \phi$, we had $N_i=k_i+2$. Otherwise, $N_i=k_i+3$ for $\Phi_i=\psi$ and $N_i=k_i+4$ for $\Phi_i=F$. 
Moreover, the number of $\frac{\partial}{\partial\theta_\ri}$ derivatives required to generate the field $D^{k_i}\Phi_i$ in \eqref{eq:definitionformingfactoronshellstates} was denoted by $n_{\theta_\ri}$. Concretely, $n_{\theta_\ri}=0,1,2,3,4$ for $\Phi_i=\bar F, \bar \psi, \phi, \psi, F $, respectively. 
For the operators without $\dot\alpha$ indices, we had $k_i=0$ and also $N_i=n_{\theta_i}=2,3,4$.
\newline\\
The computation of the NMHV form factor parallels the calculations done for amplitudes in Section~\ref{eq:NMHV amplitudes in momentum twistor space}. 
Starting from the position twistor expression \eqref{eq:wilsonloopformfactorinpositiontwistorspace}, we insert on-shell momentum eigenstates.
Inserting on-shell momentum eigenstates in the $\textbf{W}$ vertex gives%
\begin{equation}
\label{eq:definitionmathscrU}
\begin{aligned}
\mathscr{U}_x(\calZ_1,\calZ_2;\fP_1,\ldots, \fP_m)
&\equiv\verop{x}{\calZ_1,\calZ_2}{\calA_{\fP_1},\ldots, \calA_{\fP_m}}
\\&=\frac{\abra{\lambda_1}{\lambda_2}}{\abra{\lambda_1}{p_1}\left(\prod_{k=1}^{m-1}\abra{p_k}{p_{k+1}}\right)\abra{p_m}{\lambda_2}}\e^{i\sum_{k=1}^m(x\fp_k+\theta p_k\eta_k)}\,,
\end{aligned}
\end{equation}
for $m\geq 1$ and $\mathscr{U}_x(\calZ_1,\calZ_2;)=1$ for $m=0$; cf.\ \eqref{formula14.4}.
This leads to the general formula \eqref{eq:NMHVWilsonloopformfactor momentumspace}.
Analogously, we obtain for the contribution $\bbA_i$ \eqref{eq:AcontributiontoNMHVWilsonloopformfactor} in which the vertex is connected to the edge $x_i$ of the cog:
\begin{align}
\label{eq:hatAversion1}
\hat{\bbA}_i&=\nalpha \sum_{m_i-1\leq j<k\leq m_i'}\mathscr{U}_{x_{i}'}(\calZ_{i-1}'',\calZ_{i};\fP_{m_{i-1}''+1},\ldots, \fP_{m_i-1})
\mathscr{U}_{x_{i}''}(\calZ_{i}',\calZ_{i}'';\fP_{m_{i}'+1},\ldots, \fP_{m_i''})\nonumber\\
&\phaneq\times \int \frac{\dd^4z\,\dd^8\vartheta}{(2\pi)^4} \frac{\e^{i\sum_{s=j+1}^k(z \fp_s+\vartheta p_s\eta_s)}}{\big(\prod_{s=j+1}^{k-1}\abra{p_s}{p_{s+1}}\big)\abra{p_k}{p_{j+1}}}
 \\&\phaneq\times 
\PPP_{D^{k_i }\Phi_i}\mathscr{U}_{x_i}(\calZ_i,\calZ_i';\fP_{m_i},\ldots, \fP_{j},\fP_{k+1},\ldots,\fP_{m_i'})\nonumber\\&\phaneq\times[\calZ_{(x_i,\theta_i)}(p_{k+1}\stackrel{k=m_i'}{\longrightarrow }\lambda_i'),\calZ_{(x_i,\theta_i)}(p_j\stackrel{j=m_i-1}{\longrightarrow }\lambda_i),\star,\calZ_{(z,\vartheta)}(p_{j+1}),\calZ_{(z,\vartheta)}(p_k)]\,,\nonumber
\end{align}
where we have 
placed a $\hat{\phantom{\cdot}}$ on the contribution $\bbA_i$ from \eqref{eq:AcontributiontoNMHVWilsonloopformfactor} to indicate the fact that it uses the momentum on-shell external states \eqref{eq:definitiononshellmomentumeigenstates} and that we have included the forming operator \eqref{eq:definitionformingfactoronshellstates}.
Expressions similar to \eqref{eq:hatAversion1} can also be obtained for the contributions $\hat{\bbB}_i$ and $\hat{\bbC}_i$ coming from the other edges of the cog, see \eqref{eq:BCcontributiontoNMHVWilsonloopformfactor}. \newline\\
In order to proceed in the same way as we did for amplitudes, we need to change the order of the $z$ integration from the vertex $\mathbf{V}$ and the derivatives in the forming operator $\PPP_{D^{k_i }\Phi_i}$. 
For operators without space-time derivatives in the forming operator, i.e.\ $N_i=n_{\theta_i}$, this is possible.
As with amplitudes, we then shift $z\rightarrow z+x_i$ and $\vartheta\rightarrow \vartheta+\theta_i$. After doing the integration, we obtain
\begin{equation}
\label{eq:hatAi}
\begin{aligned}
\hat{\bbA}_i&= \sum_{m_i-1\leq j<k\leq m_i'} \mathscr{U}_{x_{i}'}(\calZ_{i-1}'',\calZ_{i};\fP_{m_{i-1}''+1},\ldots, \fP_{m_i-1})\mathscr{U}_{x_{i}''}(\calZ_{i}',\calZ_{i}'';\fP_{m_{i}'+1},\ldots, \fP_{m_i''})
\\&\phaneq\times \left(\PPP_{D^{k_i }\Phi_i}^{N_i=n_{\theta_i}}\mathscr{U}_{x_i}(\calZ_i,\calZ_i';\fP_{m_i},\ldots,\fP_{m_i'})\right)  \frac{\abra{(p_j\stackrel{j=m_i-1}{\longrightarrow }\lambda_i)}{p_{j+1}}\abra{p_k}{(p_{k+1}\stackrel{k=m_i'}{\longrightarrow }\lambda_i')}}{\abra{(p_j\stackrel{j=m_i-1}{\longrightarrow }\lambda_i)}{(p_{k+1}\stackrel{k=m_i'}{\longrightarrow }\lambda_i')}\abra{p_k}{p_{j+1}}}
\\&\phaneq\times [\calZ_{0}(p_{k+1}\stackrel{k=m_i'}{\longrightarrow }\lambda_i'),\calZ_{0}(p_j\stackrel{j=m_i-1}{\longrightarrow }\lambda_i),\star, \calZ_{\sum_{s=j+1}^{k}\fP_s}(p_{j+1}),\calZ_{\sum_{s=j+1}^{k}\fP_s }(p_{k})]\,,
\end{aligned}
\end{equation}
where we have used \eqref{eq:replacementzvarthetawithmomentumtwistors} and rearranged the second $\mathscr{U}$. 
As in the case of amplitudes, the five-bracket in \eqref{eq:hatAi} is independent of $x_i$ and $\theta_i$. This has the important consequence that the derivatives specifying the operator only act on the exponential functions contained in $\mathscr{U}_{x_i}$. We can now act with the forming operator
\eqref{formingop}, which affects only $\mathscr{U}_{x_i}$ and was computed in Section~\ref{app: derivation}. We then take the  operator limit, which sends all $x_i$ to $x$, $\lambda_i,\lambda_i'$ to $\tau_i$ and $\lambda_i''$ to the reference spinor  $\textbf{n}$; see  Figure~\ref{fig:operatorlimitapp} and also appendix A of \cite{Koster:2016loo}.
The computation is very similar to the one presented in Section~\ref{app: derivation} for the MHV form factor, up to the five-bracket that we need to treat carefully.
We find 
\begin{equation}
\begin{aligned}
\hat{\bbA}_i=& \e^{i\sum_{s=m_{i-1}''+1}^{m_i''}x\fp_s}\sum_{m_i-1\leq j<k\leq m_i'}\tilde{\calI}_i'(m_{i-1}'',m_i)\tilde{\calI}_i(m_i,m_i')\tilde{\calI}_i''(m_i',m_i'')\\
&\times  \Big[\calZ_{0}(p_j\stackrel{j=m_i-1}{\longrightarrow }\tau_i),\calZ_{0}(p_{j+1}),\star, \calZ_{\rmom_k-\rmom_j}(p_{k}),\calZ_{\rmom_k-\rmom_j}(p_{k+1}\stackrel{k=m_i'}{\longrightarrow }\tau_i)\Big]\,,
\end{aligned}
\end{equation}
where we have simplified the five-bracket by absorbing some angular brackets as in \eqref{eq: amplitude simplified}. The factors $\tilde{\calI}_i$, $\tilde{\calI}'_i$ and $\tilde{\calI}''_i$ were defined Section~\ref{app: derivation}. We reproduce them here for convenience: 
\begin{align}
\label{eq:definitiontildecalI}
\tilde{\calI}_j(m_j,m_j')&=-\frac{\left(\sum_{k=m_j}^{m_j'}\abra{\tau_j}{p_k}\sbra{\bar{p}_k}{\bar{\tau}_j}\right)^{N_j-n_{\theta_j}}\left(\sum_{k=m_j}^{m_j'}\abra{\tau_j}{p_k}\{\xi_j\eta_k\}\right)^{n_{\theta_j}}}{\abra{\tau_j}{p_{m_j}}\prod_{k=m_j}^{m_j'-1}\abra{p_k}{p_{k+1}}\abra{p_{m_j'}}{\tau_j}}\,,\nonumber\\
\tilde{\calI}_j'(m_{j-1}'',m_j)& =\frac{\abra{ \n}{\tau_{j}}}{\abra{\n}{p_{m_{j-1}''+1}}\prod_{k=m_{j-1}''+1}^{m_j-2}\abra{p_k}{p_{k+1}}\abra{p_{m_j-1}}{\tau_j}}\,,\\
\tilde{\calI}_j''(m_j',m_j'')&= \frac{\abra{ \tau_j}{\n}}{\abra{\tau_j}{p_{m_j'+1}}\prod_{k=m_j'+1}^{m_j''-1}\abra{p_k}{p_{k+1}}\abra{p_{m_j''}}{\n}}\,,\nonumber
\end{align}
where we recall that we are currently treating only the case $N_i=n_{\theta_i}$.
These objects are just the operator limit of \eqref{eq:definitionmathscrU} for each of the three edges of the cog with label $i$, stripped of the exponential factors. 
We need to keep in mind the condition \eqref{eq:constraintsforthems} on the integers $\{m_i,m_i',m_i''\}$. Furthermore, when no particle is emitted from a specific edge, the corresponding contribution equals one, i.e.\
\begin{equation}
\label{eq:definitiontildecalI2}
\tilde{\calI}_j'(m_{j-1}'',m_j=m_{j-1}'')=1 
\,,\qquad \tilde{\calI}_j''(m_j',m_j''=m_{j}')=1 
\,.
\end{equation}
In particular, the dependence on the operator $\calO$ is completely contained in the terms $\tilde{\calI}_i(m_i,m_i')$. Using the momentum supertwistors for the form factors and the same technique as in \eqref{eq: amplitude five-bracket simplified} for the five-brackets, we write
\begin{equation}
\label{eq:hatA}
\begin{aligned}
\hat{\bbA}_i&= \e^{i\sum_{s=m_{i-1}''+1}^{m_i''}x\fp_s}\tilde{\calI}_i'(m_{i-1}'',m_i)\tilde{\calI}_i(m_i,m_i')\tilde{\calI}_i''(m_i',m_i'')\\
&\phaneq\times  \sum_{m_i-1\leq j<k\leq m_i'} \Big[\calW_j\stackrel{j=m_i-1}{\longrightarrow }\calZ_{y_j}(\tau_i),\calW_{j+1},\star, \calW_{k},\calW_{k+1}\stackrel{k=m_i'}{\longrightarrow }\calZ_{y_k}(\tau_i)\Big]\,.
\end{aligned}
\end{equation}
Similarly, we obtain for $\hat{\bbB}_i$ and $\hat{\bbC}_i$ the formulae
\begin{equation}
\label{eq:hatBhatC}
\begin{aligned}
\hat{\bbB}_i&= \e^{i\sum_{s=m_{i-1}''+1}^{m_i''}x\fp_s}\tilde{\calI}_i'(m_{i-1}'',m_i)\tilde{\calI}_i(m_i,m_i')\tilde{\calI}_i''(m_i',m_i'')\\
&\phaneq\times  \sum_{m_{i-1}''\leq j<k\leq m_i-1}
\Big[\calW_j\stackrel{j=m_{i-1}''}{\longrightarrow }\calZ_{y_j}(\textbf{n}),\calW_{j+1},\star, \calW_{k},\calW_{k+1}\stackrel{k=m_i-1}{\longrightarrow }\calZ_{y_k}(\tau_i)\Big]
\,,\\
\hat{\bbC}_i&= \e^{i\sum_{s=m_{i-1}''+1}^{m_i''}x\fp_s}\tilde{\calI}_i'(m_{i-1}'',m_i)\tilde{\calI}_i(m_i,m_i')\tilde{\calI}_i''(m_i',m_i'')\\
&\phaneq\times  \sum_{m_{i}'\leq j<k\leq m_i''}
\Big[\calW_j\stackrel{j=m_{i}'}{\longrightarrow }\calZ_{y_j}(\tau_i),\calW_{j+1},\star, \calW_{k},\calW_{k+1}\stackrel{k=m_i''}{\longrightarrow }\calZ_{y_k}(\textbf{n})\Big]
\,.
\end{aligned}
\end{equation}
Finally, we also need the contributions from the cogs that are not attached to a vertex
\begin{equation}
\label{eq:hatD}
\hat{\bbD}_i=\e^{i\sum_{s=m_{i-1}''+1}^{m_i''}x\fp_s}\tilde{\calI}_i'(m_{i-1}'',m_i)\tilde{\calI}_i(m_i,m_i')\tilde{\calI}_i''(m_i',m_i'')\,.
\end{equation}
Taking everything together, the NMHV form factor is obtained after summing over all the cogs of the Wilson loop as well as over all the distributions of the integers $\{m_s, m_s',m_s''\}$ ordered as in \eqref{eq:constraintsforthems} and applying the Fourier transformation $\int \frac{\dd^4x}{(2\pi)^4}\e^{-i\fq x}$: 
\begin{equation}
\label{eq:NMHFformfactorMomentumPart1}
\mathscr{F}^{\text{NMHV}}_{\calO}(1,\ldots, n;\fq)=\int \frac{\dd^4x}{(2\pi)^4}\e^{-i\fq x}\sum_{\{m_s,m_s',m_s''\}}\sum_{i=1}^L\Big(\prod_{r=1}^{i-1}\hat{\bbD}_r\Big)(\hat{\bbA}_i+\hat{\bbB}_i+\hat{\bbC}_i)\Big(\prod_{r=i+1}^{L}\hat{\bbD}_r\Big)\,,
\end{equation}
where the quantities $\hat{\bbA}_r,\ldots, \hat{\bbD}_r$ are defined in \eqref{eq:hatA}, \eqref{eq:hatBhatC} and \eqref{eq:hatD}. 
Pulling out the Parke-Taylor denominator and the momentum-conserving delta function, we can bring \eqref{eq:NMHFformfactorMomentumPart1} to the form \eqref{eq:NMHFformfactorMomentumPart2}.

\chapter{Fourier-type integrals}
\label{app:importantFourier}

In this appendix, we derive several Fourier-type integral identities that are used in Chapter~\ref{momentumspace}.
Moreover, we demonstrate the non-commutativity of the integral with space-time derivatives coming from the forming operator in a given example. This appendix contains overlap with Appendix B of \cite{Koster:2016fna}.

\section{Fourier-type integrals with ratios of scalar products}
\label{app:mainfourier}

The main bosonic Fourier-type integral identity that we want to show reads
\begin{equation}
\label{eq:importantFourieridentity}
\int \frac{\dd^4x }{(2\pi)^4}\frac{1}{x^2}\frac{\langle s_1|x|\zeta]\cdots \langle s_k|x|\zeta]}{\langle t_{1}|x|\zeta]\cdots \langle t_{k}|x|\zeta]}\e^{i\fq x}=\frac{1}{i(4\pi)^2}\frac{1}{\fq^2}\frac{\langle s_1|\fq|\zeta]\cdots \langle s_k|\fq|\zeta]}{\langle t_{1}|\fq|\zeta]\cdots \langle t_{k}|\fq|\zeta]}\eqndot
\end{equation}
In contrast to common Fourier integrals, this integral, which stems from the bosonic part of the twistor-space propagator, is not only badly behaved for $x^2=0$ but also for $\langle t_{i}|x|\zeta]=0$.
Before calculating the integral, we should first find a prescription that makes it well-defined.
We address this issue with an explicit calculation of the case $k=1$.
Assuming that a similar prescription also exists for $k\geq2$, we then argue that the integral must be given by \eqref{eq:importantFourieridentity}.

\paragraph{Case $k=1$:}  Regularizing the integral \eqref{eq:importantFourieridentity} is not trivial, due to the fact that the term $\langle t_{1}|x|\zeta]$ introduces an extra pole in the integration over $x_0$. The guiding principle behind the regularization prescription that we choose is based on the following a-posteriori thinking: the expression \eqref{eq:importantFourieridentity} correctly translates the amplitudes from position twistor space to the known amplitudes in momentum space, i.e.\ the additional pole cannot contribute. A way to get rid of the extra pole starts by taking the vector $\ft_1\equiv t_1^{\alpha}\zeta^{\dot{\alpha}}$ and giving it a small mass. Then, we can boost it to be parallel to the $x_0$ direction and replace the $\ft_1 x$ in the denominator by $\ft_1 x+i \varepsilon\fq\ft_1=2\ft_{1,0}(x_0+i\varepsilon \fq_0)$. Since the sign of $\fq_0$ determines how to close the contour due to the exponent $\e^{i\fq x}$, i.e.\ whether we close the contour in the upper or lower $x_0$ plane, the extra pole is always avoided. One issue with this approach is that the intermediate steps of the computation break Lorentz invariance and it is quite non-trivial to see how it is restored in the end. Thus, we prefer to do the computation in Euclidean space and to Wick rotate in the end. 

Let us furthermore replace $s_1^{\alpha}\zeta^{\dot{\alpha}}$ and $t_1^{\alpha}\zeta^{\dot{\alpha}}$ by arbitrary vectors $\fs^{\alpha\dot{\alpha}}$ and $\ft^{\alpha\dot{\alpha}}$. Thus, we consider the integral
\begin{equation}
\label{eq:defIPanzer1}
\tilde{I}
=\int \frac{\dd^4x}{(2\pi)^4}\frac{1}{x^2}\frac{x \fs}{x \ft}\e^{i \fq x}
=\int \frac{\dd^4x}{(2\pi)^4}\frac{1}{x^2}\frac{2x \cdot\fs}{2x \cdot\ft}\e^{2i \fq \cdot x}
=\frac{1}{4}\int \frac{\dd^4x}{(2\pi)^4}\frac{1}{x^2}\frac{x\cdot \fs}{x\cdot \ft}\e^{i \fq\cdot x}\,,
\end{equation}
where we have rescaled $x$ to absorb the factor of 2 in the scalar product; recall that $xy\equiv x^{\alpha\dot\alpha}y_{\alpha\dot\alpha}=2x_\mu y^\mu\equiv2x\cdot y$.
We will be interested in the limit in which the vectors $\fs$ and $\ft$ become complex and obey $\fs^2=\ft^2=\fs \cdot\ft=0$, but for now we take them to be real.
The above integral \eqref{eq:defIPanzer1} is ill-defined due to the $x \cdot\ft$ in the denominator and we propose to define it as%
\footnote{We thank Erik Panzer for a very helpful discussion on this point.}
\begin{equation}
\label{eq:defIPanzer2}
\tilde{I}=\frac{1}{4}\lim_{\varepsilon\rightarrow 0}\int \frac{\dd^4x}{(2\pi)^4}\frac{1}{x^2}\frac{x \cdot \fs}{x \cdot \ft+ i\varepsilon (\fq  \cdot\ft)}\e^{i \fq \cdot  x}\,.
\end{equation}
We decompose $x$ as $x=x_t\hat{\ft}+x_{\perp}$ with $\hat{\ft}=\ft/|\ft|$ and $\ft \cdot x_\perp=0$. We see that we can compute the integration over $x_t$ in \eqref{eq:defIPanzer2} by doing a contour integral that we close in the upper (lower) half plane for $\fq  \cdot \hat{\ft}>0$ ($\fq \cdot  \hat{\ft}<0$). The poles are at $x_t=\pm i |x_\perp|$ and $x_t=-i \varepsilon \fq  \cdot \hat{\ft}$. Due to the way that we close the contour, the last pole never contributes, regardless of the value of $\fq \cdot \hat{\ft}$. Hence, we obtain after computing the residues and taking the limit $\varepsilon\rightarrow 0$
\begin{equation}
\begin{aligned}
\tilde{I}&=\frac{1}{4}\frac{2\pi i}{(2\pi)^4} \int \dd^3x_{\perp}\bigg[\Theta(\fq  \cdot \hat{\ft})\frac{(x_\perp+i |x_\perp|\hat{\ft})\cdot  \fs}{(i  |x_\perp| |\ft|)}\e^{-|x_\perp|\fq\cdot \hat{\ft}}
 \\
 &\hphantom{{}={}\frac{1}{4}\frac{2\pi i}{(2\pi)^4} \int \dd^3x_{\perp}\bigg[}
-\Theta(-\fq  \cdot \hat{\ft})\frac{(x_\perp-i |x_\perp|\hat{\ft}) \cdot \fs}{(i  |x_\perp| |\ft|)}\e^{|x_\perp|\fq \cdot\hat{\ft}}\bigg]\frac{\e^{i \fq \cdot  x_{\perp}}}{2i |x_\perp|}\\
&=\frac{1}{4}\frac{2\pi i}{(2\pi)^4} \int \dd^3x_{\perp}\left[\frac{\fs \cdot  \ft}{\ft^2}+\text{sgn}(\fq  \cdot \hat{\ft})\frac{\fs \cdot x_\perp}{i |\ft||x_\perp|}\right]\frac{\e^{i \fq   \cdot x_{\perp}-|x_\perp||\fq\cdot \hat{\ft}|}}{2i |x_\perp|}\,,
\end{aligned}
\end{equation}
where $\Theta$ denotes the Heaviside step function and $\text{sgn}(\fq  \cdot \hat{\ft})$ the sign of $\fq  \cdot \hat{\ft}$.
Going to spherical coordinates in the space perpendicular to $\ft$ 
via $x_\perp=r \hat{x}_\perp$, we find
\begin{equation}
\begin{aligned}
\tilde{I}&=\frac{1}{4}\frac{2\pi i}{(2\pi)^4}\int_{0}^\infty r^2\dd r\int_{S^2}\dd\hat{x}^2_\perp \frac{1}{2i r}\e^{-r(|\fq  \cdot \hat{\ft}|-i\fq \cdot \hat{x}_\perp )}\left(\frac{\fs \cdot \ft}{\ft^2}+\text{sgn}(\fq \cdot  \hat{\ft})\frac{\fs\cdot  \hat{x}_\perp}{i |\ft|}\right)\\
&=\frac{1}{4}\frac{\pi }{(2\pi)^4}\int_{S^2}\dd\hat{x}^2_\perp \frac{1}{\left(|\fq \cdot \hat{\ft}|-i\fq \cdot \hat{x}_\perp\right)^2}\left(\frac{\fs \cdot \ft}{\ft^2}+\text{sgn}(\fq  \cdot\hat{\ft})\frac{\fs \cdot \hat{x}_\perp}{i |\ft|}\right)\,.
\end{aligned}
\end{equation}
We now define $\fq_\perp=\fq-\ft\frac{\fq \cdot \ft}{\ft^2}$ and similarly for $\fs_\perp$. Setting $\hat{\fq}_\perp=\fq_\perp/|\fq_\perp|$, we decompose $\hat{x}_\perp$ as
\begin{equation}
\hat{x}_\perp=\cos(\theta) \hat{\fq}_\perp+\sin(\theta)\big(\cos(\varphi) \hat{z}+\sin(\varphi) \hat{w}\big)\,,
\end{equation}
where $\hat{z}$ and $\hat{w}$ are two orthonormal vectors spanning the plane orthogonal to $\ft$ and $ \hat{\fq}_\perp$. 
Then, $\dd\hat{x}^2_\perp=\sin(\theta) \dd\theta \dd\varphi$ and the $\varphi$ integral is easily done. It yields
\begin{equation}
\begin{aligned}
\tilde{I}=\frac{1}{4}\frac{2\pi^2}{(2\pi)^4}\int_{0}^\pi \frac{\sin(\theta) \dd\theta}{\left(|\fq  \cdot \hat{\ft}|+i|\fq_\perp|\cos(\theta) \right)^2}\left(\frac{\fs \cdot\ft}{\ft^2}+\text{sgn}(\fq  \cdot\hat{\ft})\cos(\theta)\frac{\fs_{\perp} \cdot\hat{\fq}_\perp}{i |\ft|}\right)\,.
\end{aligned}
\end{equation}
Integrating by parts and using $\log\frac{a+bi}{a-bi}=2i\arctan\frac{b}{a}$, we find
\begin{equation}
\begin{aligned}
\tilde{I}
=\frac{1}{(4\pi)^2\fq^2}\left[\frac{\fs \cdot \ft}{\ft^2}-\frac{(\fq \cdot \hat{\ft})(\fq_\perp \cdot \fs_\perp)}{|\ft|\fq_\perp^2}\right]+\frac{1}{(4\pi)^2}\text{sgn}(\fq\cdot\hat{\ft})\frac{\fq_{\perp}\cdot  \fs_\perp}{|\ft||\fq_\perp|^3}\arctan\frac{|\fq_\perp|}{|\fq \cdot \hat{\ft}|}\,.
\end{aligned}
\end{equation}
We now insert $|\fq_\perp|^2=\fq^2-(\fq\cdot   \ft)^2/\ft^2$ and $\fs_\perp\cdot  \fq_\perp=\fs  \cdot \fq -(\fq \cdot  \ft)(\fs\cdot \ft)/\ft^2$ so that after a couple of trivial manipulations
\begin{equation}
\label{eq:lastversionoftildeI}
\tilde{I}=\frac{1}{(4\pi)^2}\Bigg[\frac{1}{\fq^2}\frac{(\fs \cdot\fq)(\ft \cdot\fq)-(\fs\cdot\ft)\fq^2}{(\fq \cdot \ft)^2-\ft^2\fq^2}
+\frac{(\fs \cdot\ft)(\fq\cdot \ft)-\ft^2(\fs\cdot \fq)}{|(\fq\cdot  \ft)^2-\ft^2\fq^2|^{\frac{3}{2}}}\arctan\frac{\sqrt{|\ft^2\fq^2-(\fq \cdot \ft)^2|}}{|\fq \cdot \hat{\ft}|}\Bigg]\,.
\end{equation}
It is now obvious how to take the limit $\fs^2=\ft^2=\fs\cdot\ft=0$. 
In addition, if we now Wick rotate the integral, we pick up a factor of $-i$ from the measure, so that
\begin{equation}
\label{eq:lastexpressionforI1}
\int \frac{\dd^4x}{(2\pi)^4}\frac{1}{x^2}\frac{\bradot{s_1}{x}{\zeta}}{\bradot{t_1}{x}{\zeta}}\e^{i\fq x}=\frac{1}{\nalpha}\frac{1 }{\fq^2}\frac{\bradot{s_1}{\fq}{\zeta}}{\bradot{t_1}{\fq}{\zeta}}\,.
\end{equation}

\paragraph{Case $k\geq2$:} 
Let us now define $(\fs_i)_{\alpha\dot\alpha}=s_{i\alpha}\zeta_{\dot\alpha}$ and $(\ft_i)_{\alpha\dot\alpha}=t_{i\alpha}\zeta_{\dot\alpha}$ so that \eqref{eq:importantFourieridentity} is
\begin{equation}
\label{eq:importantFourieridentity2}
\int \frac{\dd^4x }{(2\pi)^4}\frac{1}{x^2}\frac{(\fs_1 x)\cdots (\fs_k x)}{(\ft_1 x)\cdots (\ft_k x)}\e^{i\fq x}=\frac{1}{i(4\pi)^2}\frac{1}{ \fq^2}\frac{(\fs_1 \fq)\cdots (\fs_k \fq)}{(\ft_1 \fq)\cdots (\ft_k \fq)}\eqncom
\end{equation}
for $\fs_i \fs_j=\fs_i \ft_j=\ft_i \ft_j=0$.
We shall assume that a prescription similar to the case $k=1$ exists that makes the integral well-defined.
We observe first that the left-hand side of \eqref{eq:importantFourieridentity2} is homogeneous of degree $1$ independently in each $\fs_i$ and of degree $-1$ in each $\ft_i$. Furthermore, a simple change of variables shows that it is homogeneous of degree $-2$ in $\fq$. Due to Lorentz invariance and the conditions that we wish to impose on the vectors $\fs_i$ and $\ft_i$, we can only use the scalar products $\fq^2$, $\fq \fs_i$ and $\fq \ft_i$ to build the answer. Using these ingredients, we cannot build a cross-ration that is invariant under all independent rescalings of the variables. Hence, up to a constant, the answer of the integral can only be equal to the right-hand side of \eqref{eq:importantFourieridentity2}. The constant, however, is fixed by considering the limit where $\fs_i\rightarrow \ft_i$ for $i=2, \dots k$, thus seeing that it is independent of $k$ and consequently given by the one found in the case $k=1$. This concludes the derivation of \eqref{eq:importantFourieridentity}.

\paragraph{Fourier transform of the R-invariants:} Armed with the identity \eqref{eq:importantFourieridentity}, we can prove an important result for the R-invariants, namely the following super Fourier-type identity:
\begin{multline}
\label{eq:superFourierRinvariant}
I_s\equiv \int \frac{\dd^4z\dd^8\vartheta}{(2\pi)^4} \e^{iz\fq+i\vartheta \Gamma}[\calZ_{(x,\theta)}(\lambda_1),\calZ_{(x,\theta)}(\lambda_2),\star,\calZ_{(z,\vartheta)}(\lambda_3),\calZ_{(z,\vartheta)}(\lambda_4)]\\=\frac{1}{\nalpha} \e^{ix\fq+i\theta\Gamma}[\calZ_{(0,0)}(\lambda_1),\calZ_{(0,0)}(\lambda_2),\star,\calZ_{(\fq,\Gamma)}(\lambda_3),\calZ_{(\fq,\Gamma)}(\lambda_4)]\eqncom
\end{multline}
where the eight variables $\Gamma\equiv \Gamma_{\alpha a}$ are fermionic.
\proof 
We start by using the identity 
\begin{equation}
\label{eq:Rinvariantforxandxprime}
[\calZ_{(x,\theta)}(\lambda_1),\calZ_{(x,\theta)}(\lambda_2),\star,\calZ_{(z,\vartheta)}(\lambda_3),\calZ_{(z,\vartheta)}(\lambda_4)]=-\frac{\abra{\lambda_1}{\lambda_2}\abra{\lambda_3}{\lambda_4}}{(x-z)^2}\frac{\prod_{a=1}^4\bradot{(\theta-\vartheta)^a}{x-z}{\zeta}}{\prod_{j=1}^4\bradot{\lambda_j}{x-z}{\zeta}}
\end{equation}
for the R-invariant and by shifting $z\rightarrow z+x$ and $\vartheta\rightarrow \vartheta+\theta$. After expressing $\vartheta$ through $\chi_3$ and $\chi_4$ as $\vartheta^{\alpha a}=\frac{-i}{\abra{\lambda_3}{\lambda_4}}(\lambda_3^{\alpha}\chi_4^a-\lambda_4^{\alpha}\chi_3^a)$, we obtain
\begin{equation}
I_s
=\e^{ix\fq+i\theta \Gamma}\int \frac{\dd^4z\dd^8\vartheta}{(2\pi)^4} \frac{(-1)\abra{\la_1}{\la_2}}{\abra{\lambda_3}{\lambda_4}^3}\frac{1}{z^2}\frac{\prod_{a=1}^4\left(\bradot{\la_3}{z}{\zeta}\chi_4^a-\bradot{\la_4}{z}{\zeta}\chi_3^a\right)}{\prod_{j=1}^4\bradot{\la_j}{z}{\zeta}}\eqndot 
\end{equation}
Using \eqref{eq:importantFourieridentity} to perform the integral over $z$ and expressing the result via $\vartheta$, we find
\begin{equation}
\label{eq:fourierofRinvariantproof}
\begin{aligned}
I_s&=\frac{\e^{ix\fq+i\theta\Gamma}}{\nalpha} \int \dd^8\vartheta \e^{i\vartheta \Gamma} \frac{(-1)\abra{\la_1}{\la_2}\abra{\la_3}{\la_4}}{\fq^2}\frac{\prod_{a=1}^4\bradot{\vartheta^a}{\fq}{\zeta}}{\prod_{j=1}^4\bradot{\la_j}{\fq}{\zeta}}\\
&=\frac{\e^{ix\fq+i\theta\Gamma}}{\nalpha} \int \dd^8\vartheta\frac{(\vartheta \Gamma)^4}{4!} \frac{(-1)\abra{\la_1}{\la_2}\abra{\la_3}{\la_4}}{\fq^2}\frac{\prod_{a=1}^4\bradot{\vartheta^a}{\fq}{\zeta}}{\prod_{j=1}^4\bradot{\la_j}{\fq}{\zeta}}\\
&=\frac{ \e^{ix\fq+i\theta\Gamma}}{\nalpha} \frac{(-1)\abra{\la_1}{\la_2}\abra{\la_3}{\la_4}}{\fq^2}\frac{\prod_{a=1}^4\bradot{\Gamma_a}{\fq}{\zeta}}{\prod_{j=1}^4\bradot{\la_j}{\fq}{\zeta}}\eqncom 
\end{aligned}
\end{equation}
where we used that only the fourth order in the expansion of $\e^{i\vartheta \Gamma}$ contributes as there are exactly four powers of $\vartheta$ in $\prod_{a=1}^4\bradot{\vartheta^a}{\fq}{\zeta}$.
Using an identity analogous to \eqref{eq:Rinvariantforxandxprime}, we see that \eqref{eq:fourierofRinvariantproof} is exactly the desired result \eqref{eq:superFourierRinvariant}. 

\section{An integral for a form factor with \texorpdfstring{$\dot\alpha$}{dotted} index}
\label{subapp: integral with derivative}

In Section~\ref{subsec: non-chiral operators}, we need to compute the following integral 
\begin{equation}
\int\frac{\dd^4z}{(2\pi)^4}\e^{i z \fq}\frac{\fs (z-x)}{((z-x)^2)^2}
=\e^{ix\fq }\int\frac{\dd^4z}{(2\pi)^4}\e^{i z \fq}\frac{\fs z}{(z^2)^2}
\,.
\end{equation}
We do this in Euclidean space, i.e.\ we calculate
\begin{equation}
I_x=\int\frac{\dd^4z}{(2\pi)^4}\e^{2i z \cdot\fq}\frac{2\fs\cdot z}{(z^2)^2}
\,,
\end{equation}
where we have also dropped the phase. 
We can rotate $z$ so that $\fs$ is parallel to the $z_0$ direction with component $\fs_0$. Then, we have $\fs\cdot z\equiv\fs_\mu z^\mu =\fs_0 z_0$. We write $\vec\fq$ for the component of $\fq$ that is perpendicular to the $0$-direction. Doing a contour integral for $z_0$, we find
\begin{equation}
I_x=\frac{2\fs_0 }{(2\pi)^4}\int \dd^3\vec{z} \e^{2i \vec{z}\cdot \vec\fq}\int_{-\infty}^{\infty} \dd z_0\frac{z_0\e^{2i z_0\fq_0}}{(z_0^2+|\vec{z}|^2)^2}=\frac{2\fs_0 }{(2\pi)^4}\int \dd^3\vec{z} \e^{2i \vec{z}\cdot\vec\fq}\frac{ \pi i \fq_0\e^{-2|\vec{z}||\fq_0|}}{|\vec{z}|}\,.
\end{equation}
Recognizing $\fs_0\fq_0$ as $\fs\cdot\fq$ and introducing polar coordinates for the remaining $\vec{z}$ integration, we find
\begin{equation}
\label{eq:IxEuclideanFinalResult}
I_x=\frac{\pi i 2\fs\cdot\fq}{(2\pi)^4}2\pi \int_{0}^\infty r^2\dd r\int_{-1}^1\dd u\e^{2i r  u |\vec\fq|}\frac{\e^{-2r|\fq_0|}}{r}
=\frac{\pi i }{(2\pi)^3}\frac{2\fs\cdot \fq}{2\fq^2}=\frac{i}{(4 \pi)^2} \frac{2\fs\cdot\fq}{\fq^2}\,.
\end{equation}
By Wick rotating to Minkowski space as before, which gives  an extra $-i$ factor, we obtain the final result:
\begin{equation}
\label{eq:IxMinkowskiFinalResult}
\int_{}\frac{\dd^4z}{(2\pi)^4}\e^{i z \fq}\frac{\fs (z-x)}{((z-x)^2)^2}=\frac{\fs \fq}{(4\pi)^2} \frac{\e^{i x \fq}}{\fq^2}\,.
\end{equation}

We remark that, unlike in the case that we shall present in Appendix~\ref{subsec:non-chiralFourierAppendix}, we can compute \eqref{eq:IxMinkowskiFinalResult} by exchanging integral and derivative\footnote{The reason why we can exchange the derivative and integral here but not in the next section is related to the additional factor $\ft z$ in the denominator.}. Namely, we can write after taking the derivative out of the integral and shifting $z$ by $x$:
\begin{equation}
\begin{aligned}
\int\frac{\dd^4z}{(2\pi)^4}\e^{i z \fq}\frac{\fs (z-x)}{((z-x)^2)^2}
&=\fs_{\alpha\dot\alpha}\frac{\partial}{\partial x_{\alpha\dot\alpha}} \int\frac{\dd^4z}{(2\pi)^4}\e^{i z \fq}\frac{1}{(z-x)^2}\\
&= \fs_{\alpha\dot\alpha}\frac{\partial}{\partial x_{\alpha\dot\alpha}}\e^{ix\fq}\int \frac{\dd^4z }{(2\pi)^4}\frac{\e^{iz \fq}}{z^2}=\frac{\fs \fq}{(4\pi)^2} \frac{\e^{i x \fq}}{\fq^2}\,,
\end{aligned}
\end{equation}
where we have used \eqref{eq:importantFourieridentity} for $k=0$ and the we used $x^2\equiv\frac{x^{\alpha\dot\alpha}x_{\alpha\dot\alpha}}{2}$ which follows directly from \eqref{explicitformofx}.

\section{Commutators of integrals and derivatives}
\label{subsec:non-chiralFourierAppendix}

In this section, we demonstrate that the derivative and integration in \eqref{eq: start of calculation} do not commute. Therefore, it is not possible to
first perform the Fourier transformations and take the space-time derivative only a the very end. As an example, we consider the integral in the last line of \eqref{eq: start of calculation}, defining
\begin{equation}
I_{nc}\equiv \int\frac{\dd^{4}z}{(2\pi)^4} \e^{iz(\fp_1+\fp_2)}\left[\tau^{\alpha}\bar{\tau}^{\dot\alpha}\frac{\partial}{\partial x^{\alpha\dot\alpha}}\frac{1}{(x-z)^2}\frac{\bradot{p_2}{x-z}{\zeta}}{\bradot{\tau}{x-z}{\zeta}}\right]\,.
\end{equation}
If we first evaluate the $x$ derivative and then evaluate the integral with the help of Appendix~\ref{subapp: integral with derivative}, we obtain
\begin{equation}
\label{eq: start of calculation appendix}
 I_{nc}=\frac{1}{\nalpha}\frac{\e^{ix(\fp_1+\fp_2)}}{(\fp_1+\fp_2)^2} \bradot{p_2}{\fp_1+\fp_2}{\bar\tau}\eqncom
\end{equation}
which led to the last line of \eqref{eq:fpsi1}.
Alternatively, if we pull out the derivative in front of the integral and then shift $z$ by $x$, we find
\begin{equation}
\begin{aligned}
\tilde{I}_{nc}&\equiv\tau^{\alpha}\bar{\tau}^{\dot\alpha}\frac{\partial}{\partial x^{\alpha\dot\alpha}}\int\frac{\dd^{4}z}{(2\pi)^4} \e^{iz(\fp_1+\fp_2)}\frac{1}{(x-z)^2}\frac{\bradot{p_2}{x-z}{\zeta}}{\bradot{\tau}{x-z}{\zeta}}\\
&=\tau^{\alpha}\bar{\tau}^{\dot\alpha}\frac{\partial}{\partial x^{\alpha\dot\alpha}}\int\frac{\dd^{4}z}{(2\pi)^4} \e^{i(z+x)(\fp_1+\fp_2)}\frac{1}{z^2}\frac{\bradot{p_2}{z}{\zeta}}{\bradot{\tau}{z}{\zeta}}\\&=\tau^{\alpha}\bar{\tau}^{\dot\alpha}\frac{\partial}{\partial x^{\alpha\dot\alpha}}\left\{\e^{i(\fp_1+\fp_2)x}\frac{1}{\nalpha}\frac{1}{(\fp_1+\fp_2)^2}\frac{\bradot{p_2}{\fp_1+\fp_2}{\zeta}}{\bradot{\tau}{\fp_1+\fp_2}{\zeta}}\right\}\,,
\end{aligned}
\end{equation} 
where we have used the integral formula \eqref{eq:lastexpressionforI1}. Evaluating the derivative and comparing the result with \eqref{eq: start of calculation appendix} yields
\begin{equation}
\begin{aligned}
\tilde{I}_{nc}&=\frac{1}{\nalpha}\frac{\e^{ix(\fp_1+\fp_2)} }{(\fp_1+\fp_2)^2}\frac{\bradot{\tau}{\fp_1+\fp_2}{\bar\tau}\bradot{p_2}{\fp_1+\fp_2}{\zeta}}{\bradot{\tau}{\fp_1+\fp_2}{\zeta}}
 \\
 &
=I_{nc}-\frac{1}{\nalpha}\e^{ix(\fp_1+\fp_2)} \frac{\abra{\tau}{p_2}\sbra{\zeta}{\bar\tau}}{\bradot{\tau}{\fp_1+\fp_2}{\zeta}}\,.
\end{aligned}
\end{equation}
where we have used \eqref{eq:cuteidentity} in the last line. Comparing this with \eqref{eq: start of calculation appendix} we conclude that we are not allowed to pull the $x$ derivative in front of the integral.

\cleardoublepage
\phantomsection
\addcontentsline{toc}{chapter}{Bibliography}
\bibliographystyle{JHEP}
\providecommand{\href}[2]{#2}\begingroup\raggedright\endgroup

\end{fmffile}
\end{document}